\documentclass[%
    paper=A4,               
    twoside=true,           
    openright,              
    parskip=half,           
    chapterprefix=true,     
    11pt,                   
    headings=normal,        
    bibliography=totoc,     
    listof=totoc,           
    titlepage=on,           
    captions=tableabove,    
    chapterprefix=false,    
    appendixprefix=false,    
    draft=false,            
]{scrreprt}%

\PassOptionsToPackage{utf8}{inputenc}
\usepackage{inputenc}

\newcommand{\thesisTitle}{It's a Wrap! Visualisations that Wrap Around Cylindrical, Toroidal, or Spherical Topologies}
\newcommand{\thesisName}{Kun-Ting Chen}
\newcommand{\thesisSubject}{Doctor of Philosophy}
\newcommand{\thesisDate}{August, 2022}

\newcommand{\thesisFirstReviewer}{Prof. Stephen Kobourov}
\newcommand{\thesisFirstReviewerUniversity}{\protect{University of Arizona}}
\newcommand{\thesisFirstReviewerDepartment}{Department of Computer Science}

\newcommand{\thesisSecondReviewer}{Prof. Melanie Tory}
\newcommand{\thesisSecondReviewerUniversity}{\protect{Northeastern University}}
\newcommand{\thesisSecondReviewerDepartment}{Data Visualization Research at the Roux Institute}

\newcommand{\thesisFirstSupervisor}{Prof. Tim Dwyer}
\newcommand{\thesisSecondSupervisor}{Prof. Kim Marriott}
\newcommand{\thesisThirdSupervisor}{Prof. Benjamin Bach}

\newcommand{\thesisUniversity}{\protect{Monash University}}
\newcommand{\thesisUniversityDepartment}{Department of Human-Centred Computing}
\newcommand{\thesisUniversityInstitute}{Data Visualisation \& Immersive Analytics Lab}

\newcommand{\thesisUniversityStreetAddress}{Wellington Road, Clayton, Victoria 3800, Australia}

\newcommand{\tstaticbar}{\textsc{StaticBar}}
\newcommand{\tinteractivebar}{\textsc{InteractiveBar}}
\newcommand{\tstaticpolar}{\textsc{StaticPolar}}
\newcommand{\tinteractivepolar}{\textsc{InteractivePolar}}

\newcommand{\tintervaltask}{\textsc{Trend Identification}}
\newcommand{\tmultiplebars}{\textsc{Pairwise Group Comparison}}
\newcommand{\tsinglebar}{\textsc{Pairwise Single Value Comparison}}

\newcommand{\tshortestpath}{\textsc{ShortestPath}}
\newcommand{\tneighbours}{\textsc{Neighbors}}
\newcommand{\tnodecount}{\textsc{NodeCount}}
\newcommand{\tlinkcount}{\textsc{LinkCount}}

\newcommand{\tnocontext}{\textsc{NoContext}}
\newcommand{\tfullcontext}{\textsc{FullContext}}
\newcommand{\tpartialcontext}{\textsc{PartialContext}}

\newcommand{\tfullcontextpan}{\textsc{FullContext-Pan}}
\newcommand{\tpartialcontextpan}{\textsc{PartialContext-Pan}}
\newcommand{\tnocontextpan}{\textsc{NoContext-Pan}}

\newcommand{\dsmall}{\textsc{Small}}
\newcommand{\dmedium}{\textsc{Medium}}
\newcommand{\dlarge}{\textsc{Large}}

\newcommand{\toruslayout}{\textsc{Torus}}

\newcommand{\deasy}{\textsc{Easy}}
\newcommand{\dhard}{\textsc{Hard}}
\newcommand{\dsmalleasy}{\textsc{Small+Easy}}
\newcommand{\dlargeeasy}{\textsc{Large+Easy}}
\newcommand{\dsmallhard}{\textsc{Small+Hard}}
\newcommand{\dlargehard}{\textsc{Large+Hard}}

\newcommand{\tpairwise}{\textsc{Pairwise}}
\newcommand{\twebcola}{\textsc{All-Pairs}}

\newcommand{\tclusteridentification}{\textsc{Cluster Number}}
\newcommand{\tbelongtocluster}{\textsc{Node Cluster}}

\newcommand{\fmap}{\textsc{Map Projection}}
\newcommand{\finteraction}{\textsc{Interaction}}
\newcommand{\fdifficulty}{\textsc{Difficulty}}

\newcommand{\mtime}{\textsc{Time}}
\newcommand{\merror}{\textsc{Error}}
\newcommand{\mconfidence}{\textsc{Confidence}}
\newcommand{\mlearnability}{\textsc{Learnability}}
\newcommand{\moverall}{\textsc{Overall}}
\newcommand{\mpref}{\textsc{Pref}}

\newcommand{\linefig}[1]{
  \includegraphics[height=\fontcharht\font`\B]{#1.png}%
}

\newcommand{\dstatic}{\textsc{Static}}
\newcommand{\dinteractive}{\textsc{Interactive}}

\newcommand{\tmollweide}{\textsc{Moll\-weide Hemisphere}  \linefig{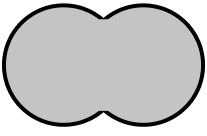}}
\newcommand{\thammer}{\textsc{Hammer} \linefig{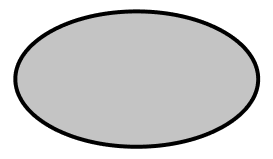}}
\newcommand{\torthographic}{\textsc{Orthographic Hemisphere} \protect\linefig{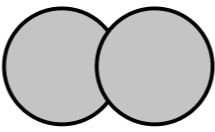}}
\protected\def\tequirectangular{\textsc{Equi\-rect\-angular} \linefig{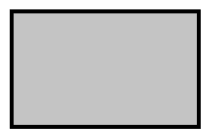}}

\newcommand{\tequalearth}{\textsc{Equal Earth} \protect\linefig{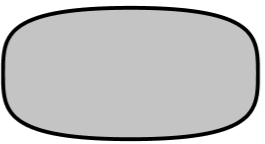}}
\newcommand{\tdistancecomparison}{\textsc{Distance Comparison} \linefig{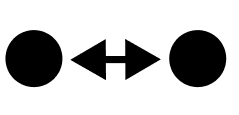}}
\newcommand{\tareacomparison}{\textsc{Area Comparison} \linefig{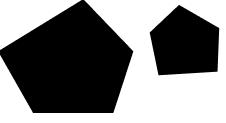}}
\newcommand{\tdirectionestimation}{\textsc{Direction estimation} \linefig{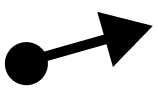}}

\newcommand{\ttorus}{\textsc{Torus} \linefig{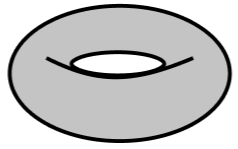}}
\newcommand{\tnodelink}{\textsc{Flat} \linefig{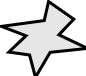}}

\newcommand{\clusteridentification}{\textsc{Cluster Number} \protect\linefig{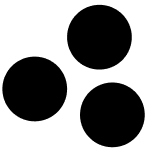}}
\newcommand{\shortestpathnumber}{\textsc{Shortest Path Number} \protect\linefig{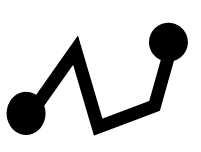}}

\newcommand{\rev}[1]{\textcolor{blue}{#1}}

\newcommand{\nodelinklayout}{\textsc{NoTorus}}

\newcommand{\tshortestpathnumber}{\textsc{Shortest Path Number}}

\usepackage[]{algorithm2e}
\usepackage{enumitem}
\usepackage{subfigure}
\RequirePackage[subfigure]{tocloft}
\usepackage[english]{babel} 
\PassOptionsToPackage{
    figuresep=colon,%
    hangfigurecaption=false,%
    hangsection=true,%
    hangsubsection=true,%
    sansserif=false,%
    configurelistings=true,%
    colorize=full,%
    colortheme=bluemagenta,%
    configurebiblatex=true,%
    bibsys=biber,%
    bibfile=bib-refs,%
    bibstyle=alphabetic,%
    bibsorting=nty,%
}{cleanthesis}
\usepackage{cleanthesis}

\hypersetup{
    pdftitle={\thesisTitle},    
    pdfsubject={\thesisSubject},
    pdfauthor={\thesisName},    
    plainpages=false,           
    colorlinks=false,           
    pdfborder={0 0 0},          
    breaklinks=true,            
    bookmarksnumbered=true,     %
    bookmarksopen=true          %
}

\begin{document}


\renewcaptionname{english}{\figurename}{Fig.}
\renewcaptionname{english}{\tablename}{Tab.}

\pagenumbering{roman}			
\pagestyle{empty}				
%
\begin{titlepage}
	\pdfbookmark[0]{Cover}{Cover}
	\flushright
	\hfill
	\vfill
	{\LARGE\thesisTitle \par}
	\rule[5pt]{\textwidth}{.4pt} \par
	{\Large\thesisName}
	\vfill
	\textit{\large\thesisDate} 
\end{titlepage}

\begin{titlepage}
	\pdfbookmark[0]{Titlepage}{Titlepage}
	\tgherosfont
	\centering

	\includegraphics[width=\linewidth]{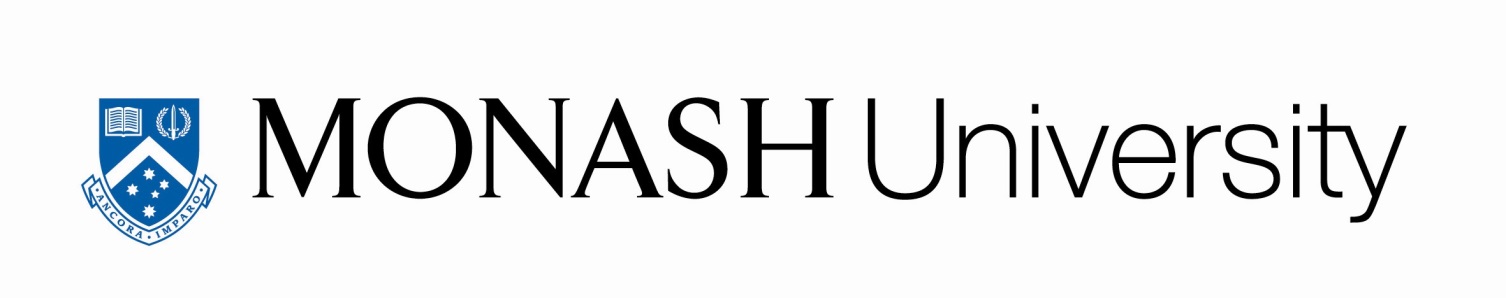} \\[2mm]

    \LARGE{\textbf{\textit{\thesisTitle}}}\\
    \vspace{30pt}
    \large\thesisName\\
    \textit{Bachelor and Master of Science}\\
    
    \vspace{34pt}
    A thesis submitted for the degree of \textit{Doctor of Philosophy} at \\
    Monash University in \textit{2022}\\
    \textit{Department of Human-Centred Computing}\\
    \textit{Faculty of Information Technology}


	\vfill
	\begin{minipage}[t]{.27\textwidth}
		\raggedleft
		\small\textit{Examiners}
	\end{minipage}
	\hspace*{15pt}
	\begin{minipage}[t]{.65\textwidth}
		{\small \thesisFirstReviewer} \\
	  	{\small \thesisFirstReviewerDepartment} \\[-1mm]
		{\small \thesisFirstReviewerUniversity}
	\end{minipage} \\[5mm]
	\begin{minipage}[t]{.27\textwidth}
		\raggedleft
		\textit{}
	\end{minipage}
	\hspace*{15pt}
	\begin{minipage}[t]{.65\textwidth}
		{\small \thesisSecondReviewer} \\
	  	{\small \thesisSecondReviewerDepartment} \\[-1mm]
		{\small \thesisSecondReviewerUniversity}
	\end{minipage} \\[10mm]
	\begin{minipage}[t]{.27\textwidth}
		\raggedleft
		\small\textit{Main Supervisor}
	\end{minipage}
	\hspace*{15pt}
	\begin{minipage}[t]{.65\textwidth}
	    \small\thesisFirstSupervisor
	\end{minipage} 
	\begin{minipage}[t]{.27\textwidth}
		\raggedleft
		\small\textit{Secondary Supervisor}
	\end{minipage}
	\hspace*{15pt}
	\begin{minipage}[t]{.65\textwidth}
		\small\thesisSecondSupervisor
	\end{minipage} 
	\begin{minipage}[t]{.27\textwidth}
		\raggedleft
		\small\textit{Associate Supervisor}
	\end{minipage}
	\hspace*{15pt}
	\begin{minipage}[t]{.65\textwidth}
		\small\thesisThirdSupervisor
	\end{minipage} \\[5mm]

	\small\thesisDate \\

\end{titlepage}







\hfill
\vfill
{
	\small
	\textbf{\thesisName} \\
	\textit{\thesisTitle} \\
	\thesisSubject, \thesisDate \\
	Supervisors: \thesisFirstSupervisor, \thesisSecondSupervisor\ and
	\thesisThirdSupervisor \\[1.5em]
	\textbf{\thesisUniversity} \\
	\thesisUniversityInstitute \\
	\thesisUniversityDepartment \\
	Faculty of Information Technology\\
	\thesisUniversityStreetAddress \\[1.5em]
	The style of this thesis is derived from Clean Thesis (\url{http://cleanthesis.der-ric.de/}).
}

\cleardoublepage
\pagestyle{empty}
\hfill
\begin{center}
\vspace*{0.5\textheight}
\textit{To my parents, Hong, and Elizabeth}
\end{center}


\pdfbookmark[0]{Copyright notice}{Copyright notice}
\addchap*{Copyright notice}
\label{sec:copyright}

© Kun-Ting Chen (2022) 

I certify that I have made all reasonable efforts to secure copyright permissions for third-party content included in this thesis and have not knowingly added copyright content to my work without the owner's permission.		

\pagestyle{plain}				
%
\pdfbookmark[0]{Abstract}{Abstract}
\addchap{Abstract}
\label{sec:abstract}

Visualisation has become increasingly important in helping people understand and gain insights from data; and, ultimately, to make better decisions. 
Traditional visualisations are designed to be shown on a flat surface (screen or page) but most data is not ``flat''.  For example, the surface of the earth exists on a sphere, however, when that surface is presented on a flat map, key information is hidden, such as geographic paths on the spherical surface being wrapped across the boundaries of the flat map.  Similarly, cyclical time-series data has no beginning or end. When such cyclical data is presented on a traditional linear chart, the viewer needs to perceive continuity of the visualisation across the chart's boundaries.  Mentally reconnecting the chart across such a boundary may induce additional cognitive load. More complex data such as a network diagram with hundreds or thousands of links between data points leads to a densely connected structure that is even less ``flat'' and needs to wrap around in multiple dimensions.

To improve the usability of these visualisations, this thesis explores a novel class of interactive wrapped data visualisations, i.e., visualisations that wrap around continuously when interactively panned on a two-dimensional (2D) projection of surfaces of 3D shapes, specifically, cylinder, torus, or sphere.
We start with a systematic exploration of the design space of interactive wrapped visualisations, characterising the visualisations that help people understand the relationship within the data and benefit from being understood as `cylindrical', `toroidal', or `spherical'. Subsequently, we investigate the design, development and implementation of a series of wrappable visualisations for cyclical time series, network, and geographic data. We show that these interactive visualisations better preserve the spatial relations in the case of geospatial data, and better reveal the data’s underlying structure in the case of abstract data such as networks and cyclical time series.
Furthermore, to assist future research and development, we contribute layout algorithms and toolkits to help create pannable wrapped visualisations.

\pdfbookmark[1]{Declaration}{Declaration}
\addchap{Declaration}
\label{sec:declaration}
\thispagestyle{empty}

I hereby declare that this thesis contains no material which has been accepted for the award of any other degree or diploma at any university or equivalent institution and that information derived from the published and unpublished work of others has been acknowledged in the text and a list of references is given.


\bigskip

\noindent\textit{\thesisDate}

\smallskip

\begin{flushright}
	\begin{minipage}{5cm}
		\rule{\textwidth}{1pt}
		\centering\thesisName
	\end{minipage}
\end{flushright}



%
\pdfbookmark[3]{Publications}{Publications}
\addchap{Publications} 
\label{sec:publications}

Some of the contents and ideas presented in this thesis have appeared previously in the following publications:

\textbf{Chapter \ref{sec:designspace}: Design Space for Pannable Wrapped Data Visualisations}

and

\textbf{Chapter \ref{sec:cylinder}: Cylindrical Wrapping of Cyclic Time Series}

[1] Kun-Ting Chen, Tim Dwyer, Benjamin Bach, and Kim Marriott. 2021. Rotate or Wrap? Interactive Visualisations of Cyclical Data on Cylindrical or Toroidal Topologies. IEEE Transactions on Visualisation and Computer Graphics 28, 1 (VIS 2021), 727–736

\textbf{Chapter \ref{sec:torus1}: Interactive Torus Wrapping for Network Visualisations - Small Networks}

[2] Kun-Ting Chen, Tim Dwyer, Kim Marriott, and Benjamin Bach. 2020. DoughNets: Visualising Networks Using Torus Wrapping. In Proceedings of the 2020 CHI Conference on Human Factors in Computing Systems (CHI 2020). 1–11.

\textbf{Chapter \ref{sec:torus2}: Interactive Torus Wrapping for Network Visualisations - Larger Networks}

[3] Kun-Ting Chen, Tim Dwyer, Benjamin Bach, and Kim Marriott. 2021. It’s a Wrap: Toroidal Wrapping of Network Visualisations Supports Cluster Understanding Tasks. In Proceedings of the 2021 CHI Conference on Human Factors in Computing Systems. (CHI 2021). 1–12.

\textbf{Chapter \ref{sec:spheremaps}: Spherical Wrapping for Geographical Data Visualisations}

and

\textbf{Chapter \ref{sec:spherevstorus}: Spherical and Toroidal Wrapping for Network Structured Data}

[4] Kun-Ting Chen, Tim Dwyer, Yalong Yang, Benjamin Bach, and Kim Marriott. 2022. GAN'SDA Wrap: Geographic And Network Structured DAta on surfaces thatWrap around. To appear in Proceedings of the 2022 CHI Conference on Human Factors in Computing Systems. (CHI 2022).
%
%
\pdfbookmark[3]{Acknowledgement}{Acknowledgement}
\addchap{Acknowledgement}
\label{sec:acknowledgement}


I am grateful to my supervisors: Prof. Tim Dwyer, Prof. Kim Marriott, and Dr. Benjamin Bach who always actively participated in my PhD meeting and offered professional support throughout my doctoral research and publications. I would like to thank my collaborator Dr. Yalong Yang who worked with us during the final year of my PhD and offered his expertise in geographic map projections and relevant empirical studies.

I am grateful to my thesis panel members Dr. Michael Wybrow, Dr. Sarah Goodwin, and Dr. Jarrod Knibbe for valuable feedback of possible future directions beyond the work presented in this thesis, A/Prof. Bernhard Jenny for chairing the review of my doctoral work and offered valuable advice and discussion on geographic map projections and practical applications, Dr. Arnaud Prouzeau and Dr. Lonni Besançon for useful advice of statistical analysis.

I would like to thank Prof. Stephen Kobourov and Prof. Melanie Tory who reviewed this thesis and offered many in-depth insights and comments into improving this manuscript.

I would like to thank for A/Prof. Rashina Hoda, Prof. John Grundy, and Dr. Chakkrit Tantithamthavorn for giving me an awesome opportunity to disseminate my research with audiences from different background in Department of Software Systems and Cybersecurity (SSC) lunch and learning seminar. I would like to thank Monash FIT academic language specialist Julie Holden who supported me for academic writing tips, structuring this thesis, and how to disseminate my research through Monash FIT Three Minute Thesis (3MT) competitions. I am grateful to Monash graduate research and FIT administrative team, including Ammie Julai, Sunny Singh, Sidalavy Chaing, and all the fellow colleagues in Data Visualisation and Immersive Analytics (DVIA) group who offered feedback and supports for my research and studies, with special thanks going to fellow students Sarah Schöttler, Kashumi Madampe, Stanislav Pozdniakov, Chunlei Chang, and Ishwari Bhade who actively participated in numerous pilot studies and provided their professional feedback. 

I would like to thank my parents, wife, and daughter for their wholehearted support during my PhD, in particular, my wife Elaine who pilots all of my studies and offers helpful advice on turning tough and boring user study materials into fun and easier to understand ones, and my daughter Elizabeth who is always a great supporter for her dad.

I am especially grateful to my ex-colleagues from LTA/MSI, including senior manager Sam Law, Shiyu Gao, Preetha Rajendran, Heng Kai Kung, Deborah Wong, and Doreen PG Hung who encouraged me to embark on this PhD research and offered their virtual support throughout, and other ex-colleagues who participated in the study and shared their opinions.

Finally, I would like to thank all the user study participants and paper reviewers. I would also like to thank Monash FIT for supporting my PhD research through Monash FIT postgraduate research scholarships. 
\cleardoublepage
\currentpdfbookmark{\contentsname}{toc}
\setcounter{tocdepth}{2}		
\tableofcontents				
\cleardoublepage

\listoffigures
\cleardoublepage

\pagenumbering{arabic}			
\setcounter{page}{1}			
\pagestyle{scrheadings}			

%
\chapter{Introduction}
\label{sec:intro}

\cleanchapterquote{You can’t do better design with a computer, but you can speed up your work enormously.}{Wim Crouwel}{(Graphic designer and typographer)}

Data analysis is essential to help make informed decisions in a variety of areas such as finance, market analysis, medical sciences, social sciences, and engineering. Data visualisation is commonly used to aid a user in the exploration and the comprehension of such data through visual representations in various forms such as statistical graphics, charts, networks and maps; to name but a few~\cite{munzner2014visualization,liu2014survey,ware2019information}.

Traditionally, these visualisations are laid out on a fixed-size flat rectangular surface designed for ease of reading on a printed page or standard screen, but many types of data are not ``flat'' in the sense that when the data is presented on a flat surface or screen some information must be lost. 
As a consequence, visualisations that are represented in this way could be hard to understand.

\begin{figure}
    \centering
    \subfigure[Flat map - shape distortion]{
    \includegraphics[width=0.6\textwidth]{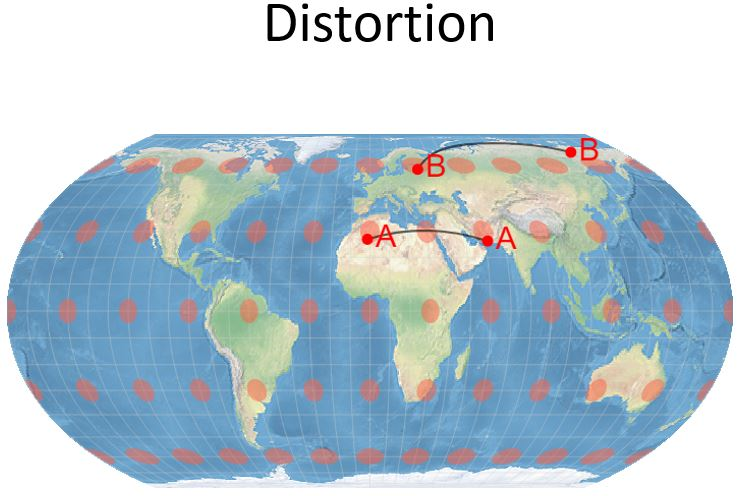}
    }
    \subfigure[Flat map - visual cut]{
    \includegraphics[width=0.6\textwidth]{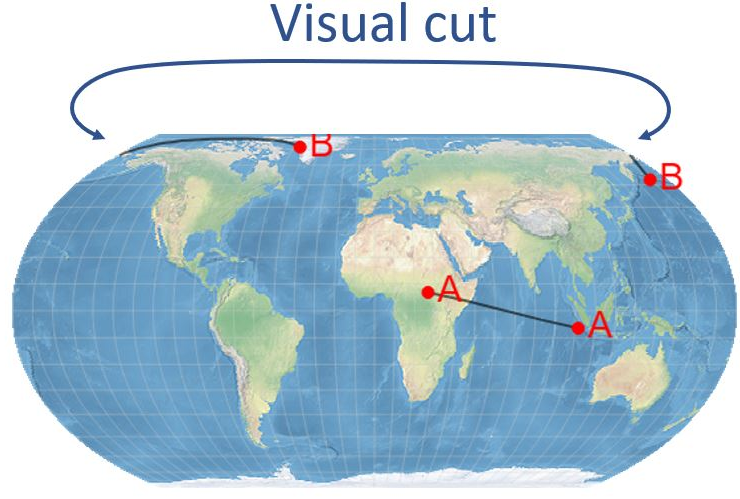}
    }
    \subfigure[Interactive spherical wrapping and rotation]{
    \includegraphics[width=0.6\textwidth]{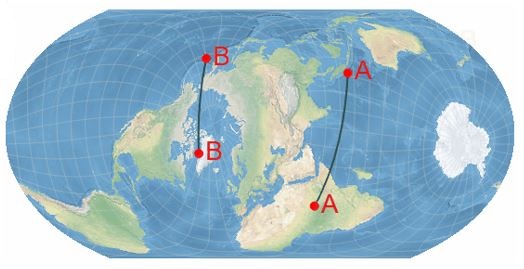}
    }
  \caption{Equal Earth map projection of pair of points (pair A and pair B). The black and curved line represents the geographical distance between each point pair on the surface of a globe. In such 2D flat maps, shapes are more distorted near the poles than the centre, as seen in the equal-sized orange dots (a). Path discontinuities (visual-cut) are inevitable in such maps (b). Interactive panning at the right allows a viewer to centre any region of interest for a less distorted view of the world. We discuss the effect of such interactive spherical wrapping on geographic comprehension tasks in~\autoref{sec:spheremaps}.}
  \label{fig:intro:visualcut_maps}
\end{figure}

For example, while most people agree that the world is spherical. To show maps of the entire earth’s surface on a screen we have to somehow cut, stretch and squash it. Cartographers and mathematicians have developed many methods to \textit{project} the surface of a sphere to a plane (i.e., techniques that connect the data points and graphical elements on a 2D plane by mapping their spatial relationship on a 3D spherical surface~\cite{snyder1997flattening}). However, due to the nature of a sphere, none of these map projection techniques can be considered optimal in depicting geographic information of a 3D globe on a 2D flat surface. Similarly, none of these projections are free of distortion and present the data without potentially introducing artefacts. Therefore, some projections cut, stretch pieces of maps, glue, and present the entire world in an interrupted view to preserve certain area or distance properties such as myriahedral projections~\cite{van2008unfolding}); with several other methods (described in~\autoref{sec:related:maps}) achieving different trade-offs between (e.g.) area, distance, or direction preservation and geographic path discontinuities~\cite{snyder1987map}.

As a consequence, geographic distances and areas closer to the north and south poles may appear larger and the shapes more distorted than those at the centre~\cite{jenny2017guide} (\autoref{fig:intro:visualcut_maps}(a)). Furthermore, distances between geographic locations (e.g., Tokyo and Greenland) or areas of continents or seas that are wrapped across the projection boundary, or the direction of aeroplane's route across the boundary are also easily misinterpreted~\cite{hennerdal2015beyond,hruby2016journey} (\autoref{fig:intro:visualcut_maps}(b)).


\begin{figure}
\centering
	\includegraphics[width=\textwidth]{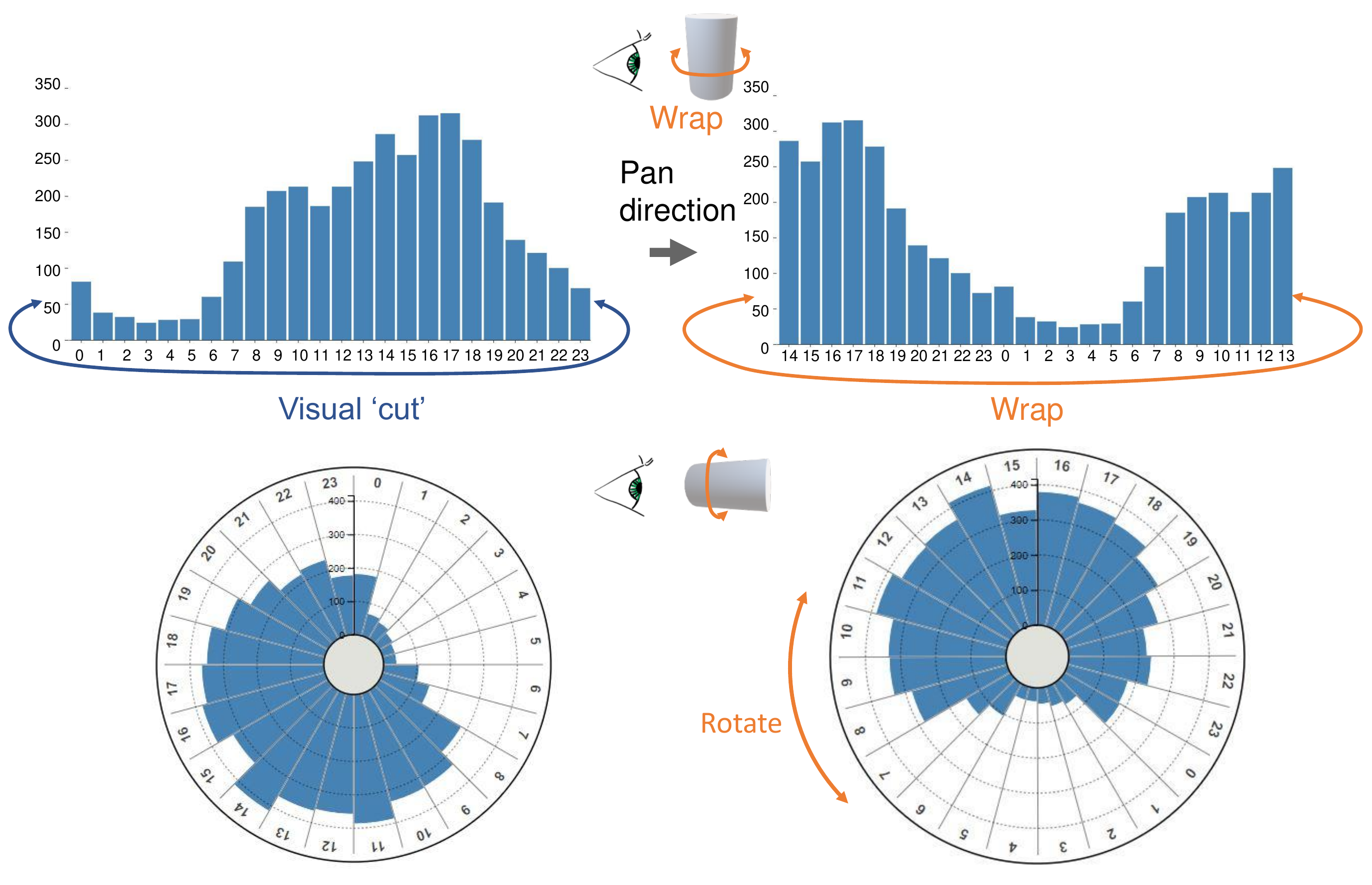}
	\caption{Bar chart and polar chart representation of hourly traffic flow over a 24-hour period in NYC in 2016~\cite{hourlytrafficaccidentsdataset}. To inspect whether the average hourly traffic accidents continue to increase or decrease from midnight till morning (upper-left), a user may need to mentally wrap the bars from 11pm at the right end across the page to 1am at the left end. Polar chart (lower-left) arranges the sequence of bars continuously, but certain sequences of bars (12pm to 5pm) are not aligned at the centre and thus a user may need to spin their head to make comparisons. Interactive wrapped and rotatable visualisations proposed in this thesis (upper-right, lower-right) allow a user to centre any group of bars while the visual cut is elsewhere in other hours (upper-right) and to have the same vertical baseline (lower-right) for comparing heights of bars. These images are created using our interactive tools described in~\autoref{sec:designspace}.
	We discuss the effect of such interactive cylindrical wrapping and rotation in~\autoref{sec:cylinder}. }
	\label{fig:intro:visualcut_cyclicaldata}
\end{figure}

Most of these projections were originally developed for \textit{static} display of the earth in printed maps and atlases. However, with modern computer graphics we can create interactive ``wrapped'' versions of these projections, such that they can be panned, reoriented and re-centred with simple mouse or touch drags~\cite{bostock_code_2013,Davies:2013ug}.  Such interactive panning allows the viewer to interactively rotate the 2D flat maps, i.e.\ changing the orientation of the projection, and to wrap the projection around the 2D plane to centre a region of interest (and therefore the visual ``cut'' (\autoref{fig:intro:visualcut_maps}(b)) is elsewhere in other parts of the map), minimising distortion and discontinuities (\autoref{fig:intro:visualcut_maps}(c)). As we show in \autoref{sec:spheremaps}, interactive panning of spherical maps provides promising results in terms of more accurate area, distance, and direction estimations than standard maps without such interaction.

Similarly, while often perceived as linear, many temporal phenomena have no beginning or end. Such data is intrinsically cyclical following cycles of day and night, seasons, or biorhythms. Common examples include time series of traffic flow over a twenty-four hour period; average temperature or birth rate over 12 months; or electrical and sound wave amplitude profiles. 
Non-temporal data can also be cyclical, such as average wind strength from different compass directions. 

Often, cyclical data is presented in traditional linear bar and line charts. However, this ignores the cyclical nature of the underlying data and so understanding of cyclical phenomena in these charts may be hindered by ``cuts'' in the visualisation; i.e. the analyst needs to mentally ``join'' the left and right sides of the visualisation together, as seen in~\autoref{fig:intro:visualcut_cyclicaldata}-Top-Left. This example also shows that an arbitrary cut could potentially make it hard to understand trends (e.g., what happened after 11pm? Does the number of hourly traffic accidents continue to decrease after 11pm till midnight?) or patterns that wrap across the boundary of the chart as well as to compare bars that are far apart in the chart but whose data is temporally close, e.g. comparing hour 23 with hour 0 on a 24-hour chart (~\autoref{fig:intro:visualcut_cyclicaldata}-Top-Left)~\cite{talbot2014four}.

One common solution to overcome this problem is to represent cyclical data in a polar (radial) representation. Like an analogue clock, polar visualisations show time in a circle, allowing for continuous representation of patterns and trends in any part of the cyclical data (\autoref{fig:intro:visualcut_cyclicaldata}-Bottom-Left). The disadvantage is that comparing lengths of bars in a polar visualisation is less effective than comparing bar heights in traditional linear bar charts where the bars are bottom-aligned~\cite{waldner2019comparison, adnan2016investigating, brehmer2018visualizing}.

To overcome the respective limitations of bar charts and polar charts, it is also possible for a ``flat'' or linear chart to be perceived as continuous when interactively connecting their left and right boundaries. That is, for linear charts, a novel ``wrapped'' panning interaction allows the viewer to pan the visualisations such that panning left causes marks which disappear from the left side to reappear on the right (and vice versa when panning right) (\autoref{fig:intro:visualcut_cyclicaldata}-Top-Right). Thus, bars that might otherwise be at opposite ends of the chart can be brought closer together for viewing. Similarly, a novel ``rotate'' gesture can be applied to polar charts, which allows the user to bring any bar to the top or centre bars for more aligned comparison (\autoref{fig:intro:visualcut_cyclicaldata}-Bottom-Right). 

The use of such wrapped panning interactions for polar and linear chart representations of cyclical data have not been previously studied. We investigate the effect of such novel wrapped panning on cyclical data analysis tasks in~\autoref{sec:cylinder}. Such one-dimensional wrapping or rotation described above can be considered to exist in a cylindrical topology\footnote{In this thesis, we use the term ``topology''' somewhat informally to refer to a small set of topological manifolds which correspond to the surfaces of shapes in $R^3$ (cylinder, sphere and torus).  Points on these surfaces can be mapped (projected) to bounded Euclidean spaces in $R^2$.  However, the different shapes (``topologies''), have different connectivity at the boundaries of these projections.  This connectivity is interesting to us, because it allows us to have interesting interactive ``wrapping'' behaviours in our projected visualisations.} (before being projected to the screen), since topologically data points continue to wrap at the sides or rotate along the circumference of a cylinder when panned.

\begin{figure}
\centering
	\includegraphics[width=0.85\textwidth]{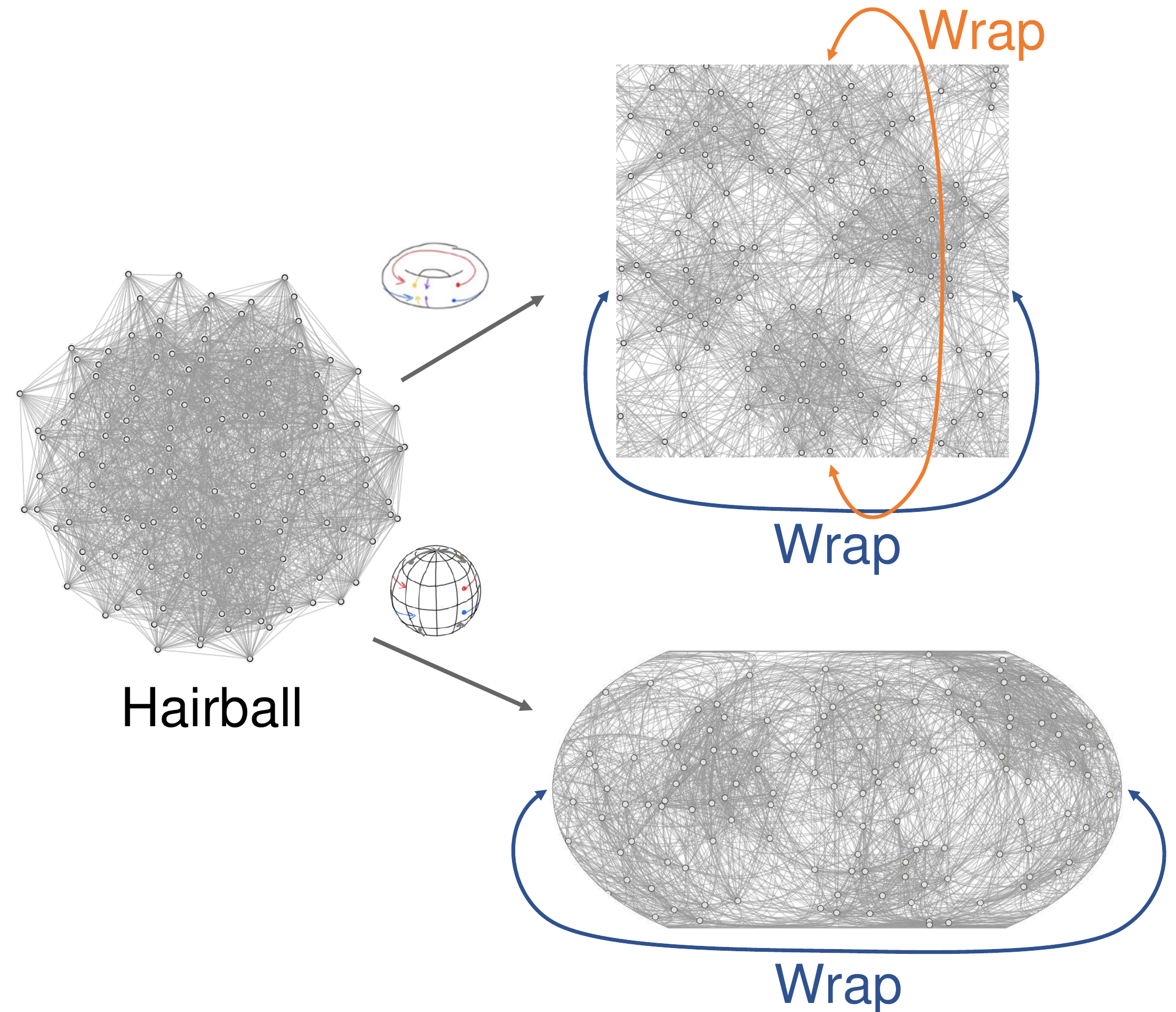}
	\caption{Node link representation of a densely connected network with a community structure. This network is created from public social network generators that simulate community structure. There are 128 nodes, 2090 links, and 6 clusters. The figures at upper-left, upper-right, bottom are laid out using a standard force-directed layout approach (discussed in~\autoref{sec:torus2}, torus-based (upper-right) and spherical-based layouts (bottom).}
	\label{fig:intro:hairball_networks}
\end{figure}

For more complex abstract data (data without inherent geometry, i.e., no information about mapping datasets to spatial position is provided), such as networks, highly connected networks with hundreds or thousands of links between data points lead to a densely connected structure that is even less ``flat'' when visualised in network diagrams (described below) and needs to wrap around in multiple dimensions.

Node-link diagrams are the most common way of visualising network data (or graphs) such as social and trade networks, software architecture, and biological networks across many domains~\cite{munzner2014visualization,yoghourdjian2020scalability,burch2020state}. They are traditionally laid out on a flat rectangular viewport corresponding to a printed page or computer display. Many networks are found to have a small-world property, i.e., the average shortest path length between nodes in a network is a few hops, defined by Watts and Strogatz~\cite{watts1998collective}, or a community structure, i.e., subsets of nodes are more tightly connected within the subsets than outside the subsets, defined by Newman~\cite{newman2006modularity}. 

However, in these networks, it could become too complex to understand or perceive the data's underlying structures or patterns for larger and denser datasets, because as the number of nodes and especially the number of links increases, their node-link diagrams become cluttered~\cite{van2008centrality,munzner2014visualization,saket2014node,wang2015ambiguityvis}. Highly connected network structures translate into diagrams with many links that tend to cross the diagram leading to a dense, cluttered ``hairball'' (\autoref{fig:intro:hairball_networks}-upper left)~\cite{yoghourdjian2018graph,zheng2018graph,roth2020socio,burch2020state,yoghourdjian2020scalability}.
\autoref{fig:intro:hairball_networks}-upper left shows a typical dense network (128 nodes, 2090 links) with a community structure, which becomes difficult both to disambiguate individual links but also to discern the high-level structure of the network, for example, to make out local clusters (defined shortly) of high connectivity. In order to better reveal and perceive such data's high-level network structures, its network layouts need to wrap around in multiple dimensions.


Discerning high-level structure, such as network clusters, is an important task in many domains, e.g., community structure in social networks\cite{roth2021socio,newman2006modularity}, or biological networks~\cite{girvan2002community}. Network clusters are loosely defined as subsets of nodes within the graph that are more highly connected to each other than would be expected from chance connection of edges within a graph of that density. More formally, a clustered graph has disjoint sets of nodes with positive modularity, a metric due to Newman which directly measures the connectivity of given clusters compared to overall connectivity~\cite{newman2006modularity}. Visual cluster analysis is an important application of network visualisation.  For example, an analyst may need to visually verify whether the output of an automatic cluster labelling algorithm makes sense with respect to the network structure.  For this task, they want a layout algorithm that displays the connectivity structure of the network as clearly as possible. 


However, visual clutter (due to the hairball effect described earlier) makes cluster disambiguation difficult with standard node-link diagrams.
As we explore in this thesis, there can be advantages to laying out networks on the surface of a 3D geometry, such as a torus or sphere, and then projecting the surface onto a 2D plane for reading on a paper or flat screen, such that in the case of torus, it has extra flexibility in avoiding or reducing crossings between links (edges) by wrapping some of them horizontally and vertically around the boundary of the display to reduce visual clutter (\autoref{fig:intro:torusnetwork}, detailed in~\autoref{sec:related:torus})~\cite{mohar2001graphs,kocay2016graphs}, and in the case of a sphere, it provides a layout that has no arbitrary edge to the display or privileged centre, as the network layout could be centred at any node of interest (\autoref{fig:intro:spherenetwork}, detailed in~\autoref{sec:related:sphere}), as demonstrated by Perry et al.~\cite{perry2020drawing} and Rodighiero\cite{rodighiero2020drawing}. 

In particular, node-link representations of highly connected network data (which is ubiquitous in the world around us, from protein-protein interaction (PPI) networks, to social, communication, trade or electrical networks) have no real ``inside'' or ``outside''. Arguably, they can be better distributed across the surface of a topology which ``wraps around''. Similar to geographic maps and cyclical data, the toroidal and spherical projections of networks onto a 2D plane result in network layouts that wrap around the plane's boundaries, that is, visualisations which when interactively panned (i.e. using simple mouse or touch drags) such that nodes or links disappear from one side of the display, these marks reappear on the opposite side.
In \autoref{fig:intro:hairball_networks} which we described earlier, toroidal (\autoref{fig:intro:hairball_networks}-upper right) and spherical (\autoref{fig:intro:hairball_networks}-bottom) layouts using  our algorithms presented in~\autoref{sec:torus2} and~\autoref{sec:spherevstorus} made clusters more distinct and easier to identify than the flat node-link layout (Hairball), and thus supports cluster identification tasks.

\section{Research Challenges}
\label{sec:intro:challenges}

To summarise, in all of the example data types shown above---including spherical maps, cyclical quantitative data, and networks---the visualisations can be wrapped around for better understanding about the patterns, structures or relationships within the data. Other examples of visualisations that wrap around will be explored in~\autoref{sec:designspace}.

However, as we discuss in detail in~\autoref{sec:related}, despite widespread use of map projections, spherical network layouts in 3D or immersive environments (such as Virtual Reality)~\cite{kwon2016study}, and work in graph theory and algorithms on how networks for certain classes of graphs, e.g., non-planar graphs which can be drawn on a toroidal surface without any link crossings~\cite{kocay2016graphs}, there has been no research exploring interactive wrapped visualisations on a more general and generalisable level, nor is there available software to create such visualisations for arbitrary data types. In addition, the evaluation of human readability studies of these types of displays has not been presented.

Thus, we argue that despite the multiple potential applications for interactive wrapped visualisations, prior to the work presented in this thesis there was a lack of research linking these concepts and evaluating this potential from a human-centred design point of view. This work aims to provide a unified view of interactive wrapped visualisations and proposes novel interactions (\autoref{fig:designspace:teaser}) that allow people to explore them and more effectively perform data analysis tasks.

\section{Research Goals}
\label{sec:intro:RGs}

The overarching design of this research follows five research goals (RG).
We outline the research goals. The specific research questions are described further in each respective chapter (\autoref{sec:cylinder}-\autoref{sec:spherevstorus}).

\subsection{RG1: Design space for interactive wrapped data visualisations}
\label{sec:intro:RGs:designspace}
We identify the types of real-world data whose visualisations can potentially benefit from being explored with one-dimensional and two-dimensional wrapping on a 2D plane of a 3D geometry such as a cylinder, torus or sphere. We describe a design space based on the concept of one-dimensional wrapping and two-dimensional wrapping, linked to 2D projections of three common 3D topologies: sphere, cylinder, and torus\footnote{While there are other shapes with interesting topological connectivity, we confine our exploration here to these three (cylinder, sphere and torus) because of their obvious wrapping properties when mapped to a 2D plane.  For example, Klein bottle or Möbius strips are not applicable to wrapped displays, as these surfaces are not orientable (the orientation is reversed when moved across the edge of the 2D plane unlike cylinder, torus, or sphere)}. This design space aims to consider a variety of real-world data types whose wrappable visualisations potentially benefit from being understood as `spherical', `cylindrical' or `toroidal' if it helps understand connectivity and patterns of the original data in these visualisations.

\textbf{RG1} is an exploration of the design space informed by a systematic literature review of related work as well as our own examples and exploration.

\subsection{RG2: Web tool for creating wrappable visualisations}
\label{sec:intro:RGs:tools}
We design and develop new web tools capable of creating wrapped visualisations for arbitrary data types, which can be interactively panned on a 2D projections of cylindrical and toroidal topologies. We apply cylindrical and toroidal topologies to a common range of data visualisations that allow a user to create wrappable visualisations for data types that make the most sense to be wrapped in this way. 


The research method for \textbf{RG2} is an experimental analysis method. We design and develop simple web tools that create arbitrary visualisations with wrapped interactive panning. We then analyse the types of real-world data wrappable visualisations that make the most sense. 

\subsection{RG3: Cylindrical wrapping of cyclical data}
\label{sec:intro:RGs:cylinder}
To test the feasibility of the proposed interactive wrapping, we first investigate whether interactive cylindrical wrapping and rotation of cyclical data, in particular, cyclical time series, represented in traditional linear bar charts and polar bar charts, provides benefits in terms of user task performance (e.g., task completion time and accuracy) and user reported feedback and preference, compared with standard non-wrapped linear bar charts and polar bar charts without such interaction. 


The research method for \textbf{RG3} is a mixed analysis method (i.e., combining quantitative and qualitative data analysis methods, described below). We pre-register our research questions, research design and hypotheses in Open Science Foundations (OFS). User studies of visualisation tasks, including time, accuracy, and subjective user rank, provide us with quantitative data. We collect participants’ post-study qualitative feedback. We then perform statistical tests such as parametric and non-parametric tests to evaluate the quantitative data and collate the qualitative feedback.

\subsection{RG4: Torus wrapping for network visualisations}
\label{sec:intro:RGs:torus}
To further explore the potential benefits of interactive wrapped visualisation, we explore and develop network layout algorithms that take advantage of the extra flexibility in a torus topology. We explore how to optimise node positions while wrapping certain links around the plane's edges horizontally and vertically to reduce crossing links (edges) and better capture the underlying data's structures and relationships for general networks (i.e.\ networks not just restricted to a certain class such as non-planar networks). We explore how best we can visualise the layout of a node-link diagram on the surface of a torus on a printed page or 2D flat screen. We explore whether toroidal network visualisation provides benefits in terms of graph layout aesthetics compared with standard non-wrapped layouts. We also evaluate whether toroidal layouts are more effective than standard non-wrapped visualisations in terms of network visualisation tasks. 

The research method for \textbf{RG4} includes an experimental analysis method for graph layout aesthetics and a mixed analysis method. For the former, we first use standard simulators to generate a variety of random abstract data (networks) that simulates real-world social and biological networks. We then evaluate the algorithms using performance indicators such as standard graph aesthetics that reflect readability of a given layout, as well as the algorithm's runtime performance. For the latter, the mixed analysis method is the same method as described in \textbf{RG3}. 


\subsection{RG5: Geographic and network structured data that wraps around}
\label{sec:intro:RGs:sphere}

We investigate the effectiveness of introducing interactive panning, combining wrapping and spherical rotation (described earlier) to a range of static map projections. We evaluate perceptual benefits of interactive spherical wrapping and rotation against static representations for geographic comprehension tasks.

To study whether interactive panning on 2D spherical projections are also useful for network data, we explore how to develop and adapt spherical network layout methods to find node positions that take advantage of the spherical surface. We explore the best interactive spherical projection suitable for node-link visualisations. 
We compare perceptual benefits of spherical network layouts with non-wrapped flat arrangements, and 2D projection of other 3D topologies, such as a torus for cluster understanding and network exploration tasks. 




The research method for \textbf{RG5} includes a mixed analysis method (as described in \textbf{RG4}) for maps and networks, and an experimental analysis method for evaluating the proposed method for improving spherical network layouts. 

\section{Contributions}
\label{sec:intro:results}

In this thesis, we contribute a novel class of interactive wrapped data visualisations, i.e., visualisations that wrap around two-dimensional projections of cylindrical, toroidal, or spherical surfaces. Such pannable wrapping is widely applicable to many data types, in particular, cyclical data, geospatial data, and networks. We offer the first web tool capable of creating pannable wrapped visualisation for arbitrary data types, and the first general-purpose torus network layout algorithms that afford better aesthetics and are more task-effective than standard non-wrapped representations. We also contribute to an adaptation of existing spherical network layout algorithms for spherical wrapping.

We show that these interactive visualisations better preserve the spatial relations in the case of geospatial data, and better reveal the data’s underlying structure in the case of abstract data such as networks and cyclical time series.

\subsection{Design space for interactive wrapped visualisations}
\label{sec:intro:results:designspace}

We contribute the first systematic exploration of the design space for interactive wrapped visualisations. Cylindrical topologies are well suited to data that is cyclical in one dimension. Toroidal topologies are useful for relational data and when the data has two cyclical dimensions. Spherical topologies are well suited to geographic data and relational data that can be arranged onto such a topology. 


We focus exclusively on the 2D projections that can be created from these views. This topological understanding of wrapping visualisations provides a unified view of wrapped panning interactions (using mouse or touch drags) and suggests new wrapped interaction designs (e.g.~\autoref{fig:designspace:teaser}, ~\autoref{fig:designspace:wrapchart_cylinder}, and ~\autoref{fig:designspace:wrapchart_torus}). Our design space can be a powerful tool to: (a) understand relations between visualisations (e.g., spherical maps, linear and polar bar charts, and networks), (b) apply interactive wrapping to a range of visualisations, and (c) to create new visualisations by mapping existing visualisations onto cylinder and torus visualisations to obtain wrappable projections.

The exploration of this design space was published in~\cite{chen2021rotate}.

\subsection{Web tool for creating wrappable visualisations}
To create wrappable visualisations, we contribute a small web tool that can present arbitrary 2D visualisations with one- or two-dimensional wrapped panning and allow for the selection of panning constraints in accord with the design dimensions, i.e., wrapping topology and projections. Web developers can prepare separate images such as their own static images of charts or other visuals for the visualisation, and optional axes-labels that remain fixed at the sides while the visualisation is being panned.


The web tool was published in~\cite{chen2021rotate}.

\subsection{Cylindrical wrapping of cyclic time series}
\label{sec:intro:results:cylinder}
We show that data that is cyclic in one dimension is suitable to be represented on a 2D plane with cylindrical wrapping. In a crowdsourced controlled experiment (defined in~\autoref{sec:related:evaluationmethods}) with 72 participants, our study offers the first evidence that pannable wrapped visualisation provides better task performance than standard non-wrapped representation for reading cyclic time series data.

Specifically, we find that bar charts with interactive cylindrical wrapping lead to less errors than standard bar charts for trend identification, and less errors than standard bar charts, standard polar charts, and interactive polar charts for pairwise value comparisons but slower than the standard charts.

The result of detailed evaluation was published in~\cite{chen2021rotate}.


\subsection{Torus wrapping for network visualisations}
\label{sec:intro:results:torus}
We offer the first general-purpose network layout method that we are aware of that is capable of laying out arbitrary networks on a 2D projected torus topology. Such topology permits additional spreading of node positions by wrapping certain links around the boundaries of the display to reduce visual clutter.
We use this to find layouts affording better graph layout aesthetics in terms of more equal link (edge) length, fewer link crossings, less deviation of incident link angles than standard non-wrapped representations. These aesthetics metrics have been shown relative to readability of network diagrams.
In two controlled user studies with 48 participants, we find that while torus layouts impose significant cost in accuracy and speed for understanding link wrapping across the plane's edges, adding either additional replicated context (to provide full network connectivity on the edge of the display, defined in~\autoref{sec:related:tiledisplayandpanning}), or interactive wrapping (panning interaction that allows for centring any region of interest) improve its performance to a point that it is comparable to standard non-wrapped layouts. The result of the algorithm and evaluation were published in~\cite{chen2020doughnets}.

To further study realistically larger networks, we offer two new algorithms for improving toroidal layout that is completely autonomous and automatic panning the viewport to minimise the number of link wrappings across the boundary. The resulting layouts afford fewer crossings, lower ``stress'', and greater visual cluster separation (i.e. less visual cluster overlapping). In a study of 32 participants comparing performance in cluster understanding tasks, we find that toroidal visualisation offers significant benefits over standard unwrapped visualisation in terms of improvement in error by 62.7\% and time by 32.3\%. The results of the layout algorithms and evaluation were published in~\cite{chen2021sa}.

\subsection{Geographic and Network Structured Data that Wraps Around}
\label{sec:intro:results:sphere}
To our knowledge, we offer the first systematic evaluation of the effect of introducing interactive spherical wrapping on different geographic map projection techniques. We find that interaction overwhelmingly improves accuracy and subjective user preference compared to static projections across all projection methods and tasks considered, at the cost of increased time due to interaction.
This suggests such interactive spherical wrapping should be routinely provided in online maps to alleviate misconceptions arising from distortions of map projection. The result of the evaluation was published in~\cite{chen2022gan}.

For networks, we present the first evaluation comparing spherical network projections against toroidal and non-wrapped flat layouts.
We contributed an adaptation of existing spherical network layout algorithms. To mitigate the link wrapping across the boundaries, we proposed algorithms to automatically centre the content of interest in the display. Our study confirms the benefits of topologically closed surfaces, such as the surface of a torus or sphere, when using node-link diagrams to investigate network structure. This finding suggests that interactive wrapping of networks arranged on 2D planes of 3D surfaces should be more commonly used for cluster analysis tasks and further, that toroidal layouts may be a good general solution, being not only more accurate than standard non-wrapped flat layout for cluster tasks but also at least as accurate for path following tasks. 

The result of the adapted layout algorithms and their evaluation were published in~\cite{chen2022gan}.


\section{Thesis Structure}
\label{sec:intro:strucutre}

We begin this thesis by reviewing the related research on data visualisation, revisiting existing data that wraps around topologies, network layout algorithms, and empirical evidence (\autoref{sec:related}). Afterwards, we present a design space exploration which identifies various data types that wrapping visualisations makes sense, and our interactive web tools that create novel applications (\autoref{sec:designspace}). We then discuss each of the wrapping topologies and effectiveness evaluation: cylindrical (\autoref{sec:cylinder}), toroidal (\autoref{sec:torus1} -- smaller networks and~\autoref{sec:torus2} -- larger networks), and spherical topologies (\autoref{sec:spheremaps} -- maps and~\autoref{sec:spherevstorus} -- networks). We conclude this dissertation with a discussion of the benefits and limitations of interactive wrapping and provide our view on the future of pannable wrapped visualisations and the potential challenges that may be explored by future researchers in this area (\autoref{sec:conclusion}).

%
\chapter{Background and Related Work}
\label{sec:related}

\cleanchapterquote{A picture is worth a thousand words. An interface is worth a thousand pictures.}{Ben Shneiderman}{Distinguished Professor of Computer Science (University of Maryland) and influential Human Computer Interaction researcher}


In this chapter, we examine traditional visualisation from the perspective of different data types that are laid out on a 2D rectangular plane, in particular, network visualisation (\autoref{sec:related:networks}), global map layout (\autoref{sec:related:maps}) and cyclical temporal data visualisation (\autoref{sec:related:tradition:temporal}). We also summarise relevant empirical evidence of the effectiveness of these types of data visualisations.
These are the data types that we argue can be interactively wrapped around a 2D display of 3D surfaces: cylinder, torus, or sphere, to better understand the relationships within the data. 
As we discuss each of these data types, where applicable we also look at existing work that has considered projections of these types of data from their arrangements on 3D sphere or toroidal topologies (defined in~\autoref{sec:intro}) and the very limited existing consideration of interaction with these projections.
In each case we discuss how this thesis relates and contributes to the existing literature.

\section{Network Visualisation}
\label{sec:related:networks}

Relationships between people or things can be modelled as a network of nodes (representing the things) and links between pairs of nodes (representing the relations between pairs of things).  Examples include social networks, electrical networks, software dependency networks~\cite{dwyer2013edge}, biological reaction networks~\cite{thomas2003structure}, and so on.
When it comes to visualising such networks, if the things represented by the nodes have no physical position in space (for example, software entities) or any physical position of those things is not relevant to the analysis, then we are free to place the nodes anywhere in the display.  That is, we are free to arrange the marks (e.g., nodes and links) corresponding to elements in the data in a way that is as readable as possible, in order to reveal important patterns or features. 

We consider data structures, such as networks, that have no inherent spatial structure to be ``abstract''. The most common way to visualise networks are node-link diagrams where the nodes are drawn as boxes with meaningful labels and the links are drawn as lines between the nodes.
If the aesthetic criteria for placing the nodes and routing the link lines can be precisely formulated then the problem of finding an aesthetically pleasing and readable layout can be considered as an optimisation problem.
A wide variety of automatic algorithms for finding readable and aesthetic layouts for these diagrams have been developed along with an extensive underlying mathematical theory~\cite{battista1998graph}, such as layered layout by Sugiyama et al.~\cite{sugiyama1981methods}, orthogonal layout by Batini et al.~\cite{batini1986layout,kieffer2015hola}, and many others~\cite{battista1998graph}.

\subsection{Force-directed network layouts} 
\label{sec:related:network:forcedirected}
Force-directed layout methods, which simulate a physical model of a network with springy links in the layouts, are the most widely used layout algorithm for general purposes in practical network visualisation applications~\cite{munzner2014visualization,liu2014survey,brandes2008experimental,dwyer2009scalable,devkota2019stress}. Such methods seek to optimise node-placement in node-link diagrams to support perception of clusters (as defined in~\autoref{sec:intro})~\cite{lu2020clustering, suh2019persistent, dwyer2008topology, dwyer2008exploration}, paths~\cite{zheng2018graph}, and spatial distances between nodes~\cite{devkota2019stress, brandes2008experimental}. 

\begin{figure}
    \centering
    \subfigure[Football-\nodelinklayout{}]{
    \includegraphics[width=0.4\textwidth]{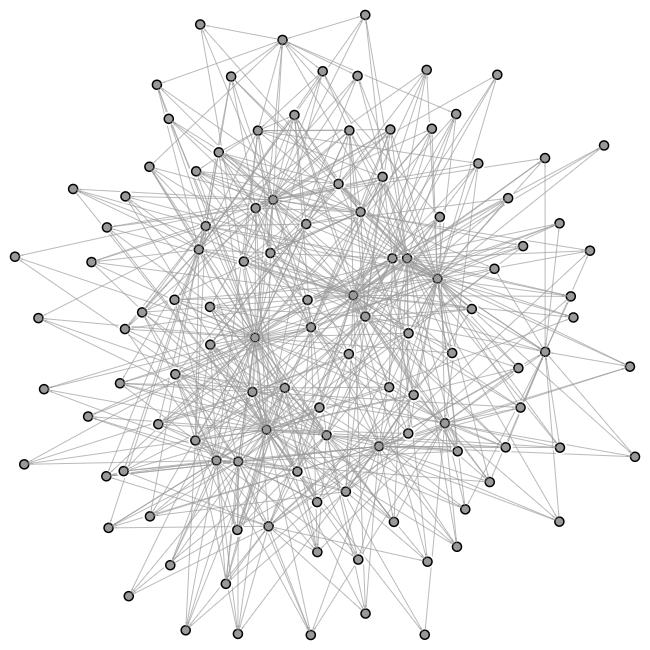}
    }
    \subfigure[Recipe-\nodelinklayout{}]{
    \includegraphics[width=0.4\textwidth]{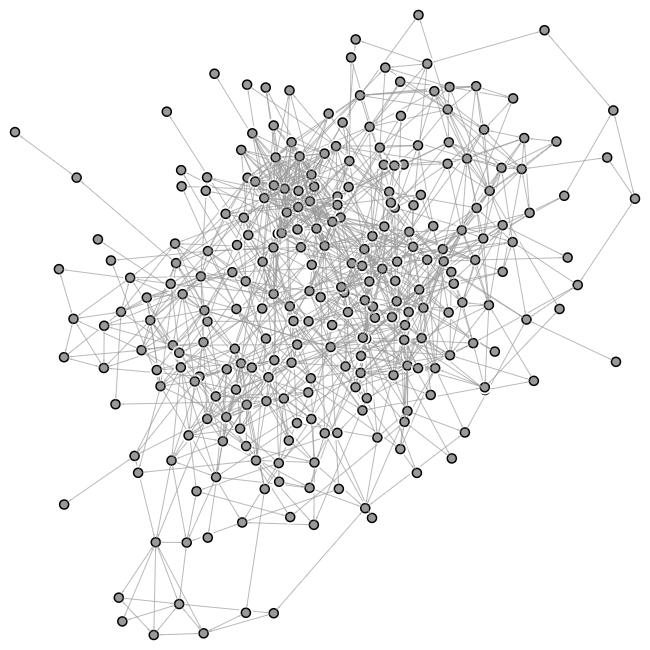}
    }
  \caption{Node-link representations of (a) an American College football network (115 nodes, 669 links) with links representing regular-season games between Division IA college teams (nodes) they connect during Fall 2000~\cite{girvan2002community}. (b) a recipe network (258 nodes, 1090 links) with links representing pairs of ingredients (nodes) frequently used together for cooking~\cite{ahn2011flavor}. The hairball in (b) makes it difficult to identify the main clusters in the network, which may impede cluster understanding tasks.}
  \label{fig:related:hairball}
\end{figure}

One of the most popular methods following this approach is by Kamada and Kawai~\cite{kamada1989algorithm} who simulate graph theoretic distance between nodes (computed by, e.g., all-pairs shortest path algorithms) in the layout by optimising energy functions where nodes with theoretic shorter path length are moved closer with a simulated spring systems and electrical forces, while nodes with larger shortest path length are placed further apart. 

The precise goal of the optimisation and the quality of its solution determine how well we can perceive patterns in the resulting visualisation and understand the data. Unfortunately, force-directed placement is a non-trivial optimisation problem, i.e., it is extremely difficult to find a global optimum with respect to the layout goals, and so imperfect heuristics must be used and the results are often sub-optimal. Such optimisations do not guarantee to deliver optimal solutions that are free of ambiguity and present the data without potentially introducing artefacts~\cite{binucci201910,munzner2014visualization}. 

While many of these methods can be applied to any type of networks and they typically improve the readability of network diagrams by reducing crossing links and node overlaps, there are certain types of dense network structures that can be particularly problematic for force-directed layout, producing messy ``hair-ball'' visualisations that are difficult to read~\cite{munzner2014visualization,burch2020state} (\autoref{fig:intro:hairball_networks}-Left and~\autoref{fig:related:hairball}).  Particularly problematic are:

\begin{itemize}
    \item scale-free networks (where connectivity of nodes
is distributed according to a power law~\cite{thomas2003structure}, i.e. most of the nodes have low connectivity but a few nodes have high degree of connectivity) as defined by Barabási and Albert~\cite{barabasi1999emergence};
\item small-world networks (most nodes are reachable by most other nodes with only a small number of hops) as defined by Watts and Strogatz~\cite{watts1998collective}; or
\item highly connected network structures such as tightly clustered nodes within subsets of nodes compared to their connectivity by chance in random networks, i.e., community structures defined by Newman~\cite{newman2006modularity},
\end{itemize}

The approach to layout considered in this thesis is a variant of force-directed layout, but extended to perform the unravelling of the network on the surface of a wrap-around topology such as the sphere (\autoref{sec:related:sphere}) and torus (\autoref{sec:related:torus}).  Performing layout on the surface of torus or sphere topologies potentially allows for better ``unfolding'' of dense networks, as seen in~\autoref{fig:intro:hairball_networks}-Right.

\subsection{Clustered network layout} 

Since force-directed layouts of highly connected dense network structures tend to become difficult to disambiguate individual links and to discern the high-level structure of the network such as clusters,
there are techniques which explicitly encode groups (subsets of nodes) or clustering information in visualisations, e.g.\ based on pre-identified clusters using community detection or clustering algorithms such as modularity clustering~\cite{newman2006modularity}. 
These include methods displaying clusters in map-like layouts that highlight the visual cluster boundary, such as GMap by Gansner et al.~\cite{gansner2009gmap}, Bubble Sets by Collins et al.~\cite{collins2009bubble}, and MapSets by Efrat et al.~\cite{efrat2014mapsets}. 
Similarly, to highlight common visual ambiguities such as visual overlap between community structures, an issue caused by the problematic hairball layout described in~\autoref{sec:related:network:forcedirected}, Wang et al.~\cite{wang2015ambiguityvis} encoded known clustering information in the visualisations to highlight visual cluster boundaries. Previous work also presented evaluations of task-performance for such group-level visualisations~\cite{saket2014node, okoe2018node, vehlow2017visualizing}. 

All of these algorithms rely on pre-identified cluster information---either from categorical variables within the data or pre-computed community detection---to highlight the cluster boundaries via layout or visual cues.
However, for the layout approach considered in this thesis, we do \textit{not} require knowledge of clusters and we do \textit{not} specifically optimise the network layout based on clustering information. 
Instead, our approach, like multidimensional scaling (defined in \autoref{sec:related:mds}) and force-directed network layout more generally, implicitly groups nodes by minimising a cost function over difference between ideal graph theoretic distances and actual node separation in the 2D wrap-around drawings of projected 3D torus and sphere topologies~\cite{gansner2004graph,munzner2014visualization}. 
This is different from a traditional non-wrapped node-link layouts. 
These topologies allow for extra flexibility in better distribution of node positions across the entire surface of a topology~\cite{rodighiero2020drawing,perry2020drawing}; and reducing link and node overlap (in terms of torus, \autoref{sec:related:torus}) and eventually optimise graph readability. 





\subsection{Network layout with data aggregation} 
To improve the usability of node-link diagrams for large or dense networks, many data aggregation layout methods reasonably reduce visual clutter and reveal high-level patterns by showing an aggregated view of nodes and edges.  Of course, they achieve this at the cost of reducing the level of detail present in the diagram, or requiring interaction to enable the user to recover such detail. Examples of such aggregated layout techniques include: link aggregation for creating high-quality layouts of network diagrams for software applications by Dwyer et al.~\cite{dwyer2013edge}; edge bundling which coalesced links with common direction (common start and end nodes similar to bundled train tracks) to reduce link clutter, such as confluent drawing by Bach et al.~\cite{bach2016towards}; an interactive lens approach where edge bundling is based on selective areas of interest while the edges outside the lens are less bundled and retain the original link information, i.e., Moleview by Hurter et al.~\cite{hurter2011moleview}; and work that provides a clean overview of large networks for identifying and comparing key structural information about a network, in GraphThumbnail layouts by Yoghourdjian et al.~\cite{yoghourdjian2018graph}. However, many of these methods reduce the amount of information perceived by a user. 

While such aggregation techniques may be compatible with the interactive wrapped visualisations introduced in~\autoref{sec:designspace}, the examples we explore present the full set of node-link data, taking advantage of the wrapping topology to reduce visual clutter. 

\subsection{Spherical network embedding}
\label{sec:related:sphere}
In previous subsections, we described how traditional network visualisations are laid out on a 2D flat screen to present a readable layout with meaningful patterns. Researchers have also explored visualisation of networks in three-dimensional views, with evidence that the third dimension can be used to improve readability of network clusters (Lu et al.~\cite{lu2020clustering}), or how to improve readability by removing link crossings for node-link diagrams~\cite{ware2008visualizing, greffard2011visual}. However, in such 3D representations occlusion is a significant problem~\cite{alper2013weighted}, because with introduction of the third dimension, some nodes and links can be occluded completely by others facing a viewer. 

\begin{figure}
	\includegraphics[width=\textwidth]{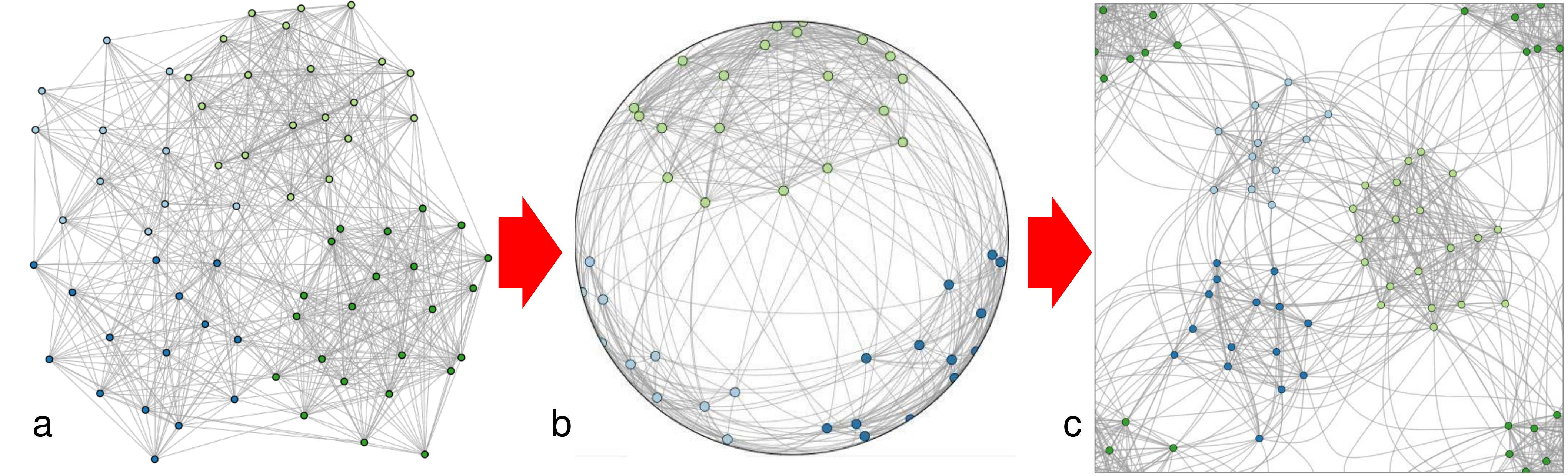}
	\caption{An example of a social network with community structures laid out using the traditional standard non-wrapped network layout (left); spherical network embedding on the surface of a 3D globe (middle) provides a better use of available screen space; the surface of a sphere projected onto a 2D plane to view the entire network using Charles Peirce’s Quincuncial projection (right)~\cite{peirce1879quincuncial}. All of the network layouts are created using the layout algorithms described in~\autoref{sec:spherevstorus}. The dataset is created using social network simulators by Fortunato et al.~\cite{fortunato2010community}}
	\label{fig:intro:spherenetwork}
\end{figure}

\begin{figure}
	\includegraphics[width=\textwidth]{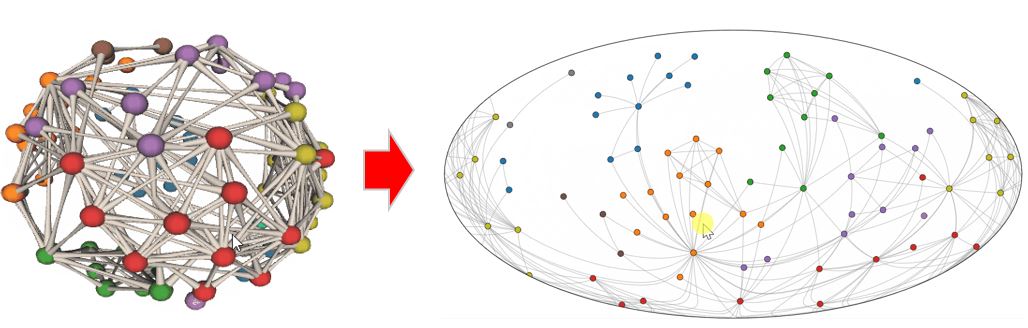}
	\caption{Spherical network layouts for a network of co-appearances of characters (nodes) in \textit{Les Misérables} by Victor
Hugo~\cite{knuth1993stanford}, where edges indicate that two characters coexist in one scene and colours indicate group affiliation of the characters (nodes) in the novel; Spherical layout (left) developed using the methods described in~\autoref{sec:spherevstorus}, and its 2D Hammer projection (right, described in~\autoref{sec:related:maps})}
	\label{fig:related:spherenetwork3d}
\end{figure}

\rev{Visualisation researchers have also explored the benefits of laying out networks on the surface of a sphere, or a torus (defined in~\autoref{sec:related:torus}). However, in the case of laying out a network over the surface of a 3D globe, for instance, one hemisphere is completely occluded by the other (\autoref{fig:intro:spherenetwork}(b) and~\autoref{fig:related:spherenetwork3d}-Left). A viewer needs to interactively turn to the opposite hemisphere to gain a full view of the entire network.}

The potential benefit of laying out networks on the surface of a sphere or torus is that they are topologically closed surfaces: there is no centre or border to the surface, as the surface of a sphere (or torus) allows the layouts to be centred at any data point of interest~\cite{brath2012sphere,perry2020drawing,rodighiero2020drawing} (\autoref{fig:intro:spherenetwork}-c and \autoref{fig:related:spherenetwork3d}-Right).  Thus, such spherical representation may allow the layout to better unravel the network and show its structure~\cite{brath2012sphere,perry2020drawing, rodighiero2020drawing}. For example, Kobourov and Wampler~\cite{kobourov2005non} have investigated generalisation of force-directed algorithms to the surfaces of non-Euclidean (non-flat) geometries, e.g., the sphere; Perry et al.~\cite{perry2020drawing} proposed a multidimensional scaling approach (defined in~\autoref{sec:related:mds}) to produce drawings of networks on the surfaces of a sphere.


Such spherical network layouts are most commonly viewed in immersive virtual reality (e.g. \cite{kwon2016study}) or as perspective projection of a 3D globe on a standard 2D monitor (e.g.~\cite{brath2012sphere,kobourov2008morphing}). The former provides the user with an immersive view from outside the sphere (\textit{exo}-centric) or \textit{within} the sphere (\textit{ego}-centric).  Ego-centric views are analogous to a planetarium, as users rotate their heads within the sphere, they see different parts of the network, effectively performing \textit{panning} across its surface. Compared to more traditional exo-centric 3D views, ego-centric views have been found to be less efficient~\cite{yang2018maps}. 

\rev{Note that in this thesis, we are \textbf{not considering direct 3D visualisation of surfaces of spherical and toroidal 3D shapes } (\autoref{sec:related:torus}).  While such 3D visualisation may be interesting in immersive environments (for example, headset virtual reality with stereopsis and true perspective views which respond to head position), it is much less common for network layouts to be  projected onto a 2D plane using a map projection due to the inherent problems in 3D with distortion and occlusion (detailed later, though some static map projections of graphs were demonstrated by~\cite{rodighiero2020drawing}). However, map projections have the great advantage over simple perspective rendering of a 3D globe, that the whole network can be seen at once (with Orthographic Hemisphere projection being the closest to a direct rendering of the 3D globe, but with both sides shown simultaneously). Furthermore, in this thesis, we focus on exploring the effectiveness and usability of projecting from such 3D shapes onto 2D wrapped display since we find that interactive 2D displays of such spherical layouts of data (whether geographic or network), and toroidal layouts of networks are not well studied.} 

\rev{Likewise, projections from \textit{inside} the 3D body~(e.g., \cite{yang2018maps, kwon2016study}) are not considered here. 
Nor do we consider other exotic projections of 3D surfaces such as \textit{Fish-eye projections}~\cite{du2017isphere} and \textit{Hyperbolic} visualisations~\cite{lamping1995focus,munzner1998exploring}, which require continuous morphing as they pan.}


\rev{Furthermore, despite the advantages of representing network layouts in 3D, visualisation design literature cautions against such use of 3D for visualisation if there are layout approaches better suited to the plane \cite[Ch.~6]{munzner2014visualization}.}


While one could simply use linear or perspective projection to project a 3D sphere (or torus, as described in~\autoref{sec:related:torus}) onto the 2D plane, occlusion (defined earlier) and distortion of the surface are significant problems~\cite{alper2013weighted}.
To minimise the problems of occlusion and perspective distortion when projecting the spherical surface of the earth onto 2D maps, cartographers have developed map projections (defined in~\autoref{sec:related:maps}) and visualisation researchers have applied such cartographic projection techniques to create 2D views of graphs on a sphere.
Spherical projections, i.e., techniques that connect the data points and graphical elements on a 2D plane by mapping their spatial relationship on a 3D spherical surface~\cite{snyder1997flattening}, have the great advantage over the aforementioned simple perspective rendering of a 3D globe, that the whole network can be seen at once while wrapping specific links around the plane's boundaries. For example, Orthographic Hemisphere projection (described in~\autoref{sec:related:maps}) is the closest to such direct rendering of the 3D globe, but with both sides shown simultaneously for viewing the entire network while the nodes and links across boundaries wrap around.

While it is much less common for networks to be projected onto a 2D plane using a map projection, some static map projections of graphs were demonstrated. Examples include map style network drawing on a sphere, projected onto a plane where nodes and links wrap around by Perry et al.~\cite{perry2020drawing}; spherical network embeddings with non-privileged views of social networks by Rodighiero~\cite{rodighiero2020drawing}  (\autoref{fig:intro:spherenetwork}(c)); and 2D projected sphere layouts using self-organising maps based approach (defined in~\autoref{sec:related:som}) by Wu and Takatsuka~\cite{wu2006visualizing}, visualising multivariate networks (a type of network with multiple attributes associated with nodes and links such as social, cultural and spatial relationships between people in a social network).

The downside of such 2D projected spherical layout and map projections is that the resulting visualisation introduces discontinuities in the layout which now wraps around the boundaries of the plane, potentially confusing the viewer. In the case of networks, the connectivity of wrapped links in such displays (such as~\autoref{fig:intro:spherenetwork}-right and~\autoref{fig:related:spherenetwork3d}-right) may not be obvious to all viewers, so the additional need for the viewer to understand link wrapping is the potential disadvantage of this style of visualisation~\cite{hennerdal2015beyond,hruby2016journey}. In this thesis, we explore the use of rotate-able spherical layouts and evaluate their readability for network visualisation tasks.
To the best of our knowledge, while interactive applications exist that allow the viewer to  interactively pan to rotate a geographic globe projection, we do not know of previous research on interactive panning of spherical network diagrams when they have been projected onto a 2D plane using a map projection. 
However, as we show in this thesis, interactive wrapping provides benefits for perception for visualisation tasks.
An exception is a web-based implementation of 2D rotate-able force-directed spherical layouts of networks by Manning~\cite{christophermanning:force}. However, in Manning's implementation the distance between nodes uses Euclidean distance not spherical great circle, which may not take full advantage of laying out a network on the surface of a sphere~\cite{perry2020drawing}).

Beyond node-link diagrams, a range of other information visualisations can potentially be projected onto a sphere to minimise artefacts that occur when trying to find an optimal embedding for a 2D plane, discussed in our design space in~\autoref{sec:designspace:sphere:subsec2}. 


\subsection{Toroidal network embedding}
\label{sec:related:torus}
\begin{figure}
	\includegraphics[width=\textwidth]{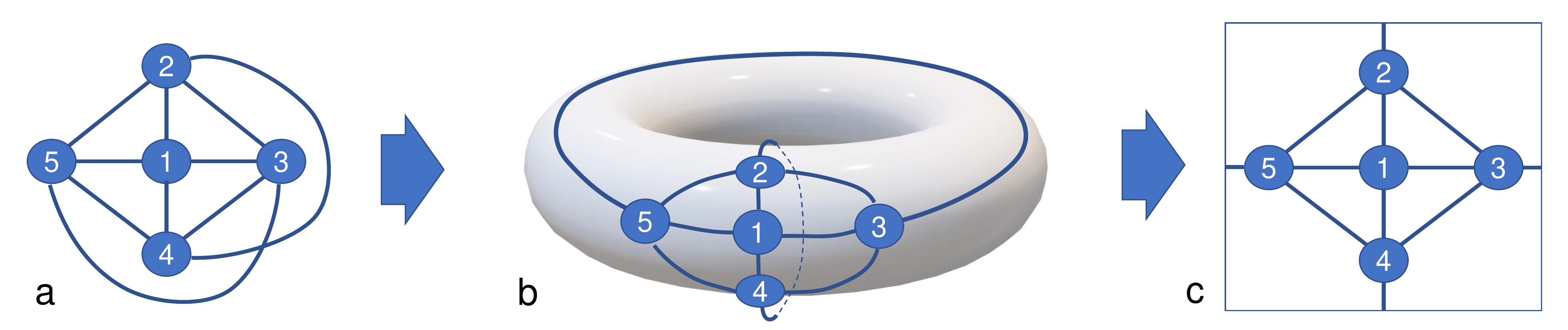}
	\caption{(a) A small complete graph K5,5 cannot be drawn on a 2D plane without line crossings. For example, there is an intersection of node pair (2, 4), and an intersection of (3, 5); (b) However, we can avoid such line crossings by routing the edge by connecting 2, 4 from top-bottom, and 3, 5 from left-right on a torus surface; (c) Cutting open the torus we can preserve the torus topology in a 2D plane. In~\autoref{sec:torus1} and~\autoref{sec:torus2} we propose torus-based network layout algorithms and evaluate their aesthetics compared with standard unwrapped representations.}
	\label{fig:intro:torusnetwork}
\end{figure}

\begin{figure}
    \centering
	\includegraphics[width=0.5\textwidth]{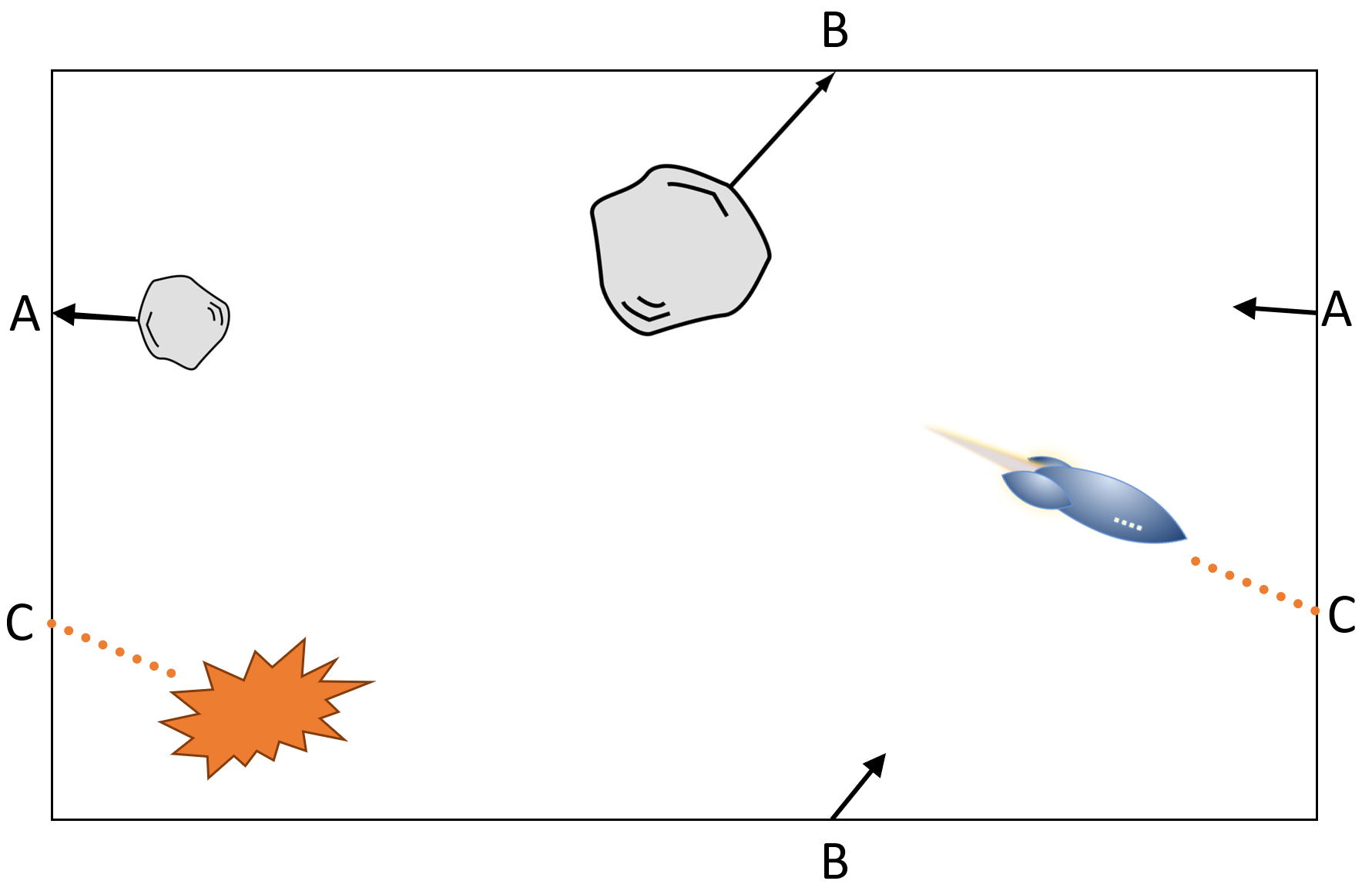}
	\caption{The video game ``asteroids'' is played on a flattened torus topology: here, we see that the path of the asteroid drifting to the left of the display will wrap around to the right at point A.  Similarly, the asteroid drifting to the top will wrap around at B and the bullets of the space ship will wrap right-to-left at point C.}
	\label{fig:intro:asteroids}
\end{figure}

In addition to embeddings of graphs on the sphere, mathematicians and graph theoreticians have also considered embeddings of graphs on other topologies, such as the torus.   For example, Mohar considers at length the theory of embedding graphs on various topological surfaces~\cite{mohar2001graphs}.   

Mathematicians have long been interested in the difficult combinatorial problems associated with identifying the smallest number of link (edge) crossings required to embed a graph in the plane~\cite{schaefer2013graph}.  Topological graph theory extends problems such as these into embeddings of graphs onto \emph{higher-genus} surfaces.  Higher-genus surfaces are essentially surfaces with holes which some of the links of a non-planar graph (i.e. the class of graphs that cannot be drawn in a plane without link crossings) can be routed through without intersecting other links~\cite{mohar2001graphs}. For example, a spherical surface has genus zero (no holes), while a torus surface has genus one (possessing a single ``hole''), and pretzels (double torus) have genus two, and so on. When a non-planar graph is drawn on the surface of a torus, it is possible to avoid or reduce crossings between links by routing some of them through the hole of the torus instead of around the outside, as shown in~\autoref{fig:intro:torusnetwork}(b).  This figure also shows that it is possible to then ``slice open'' and flatten the surface of the torus so that it can be rendered on a screen or printed on paper, as seen in~\autoref{fig:intro:torusnetwork}(c).  In such a flat rendering it is understood that edges which extend off the (for example) top, wrap around to the same horizontal position at the bottom of the drawing, and similarly for the left and right sides.

It would be fair to say that the majority of the work on higher-genus surface graph embedding has been theoretical or perhaps with application in areas such as circuit design.  While some of this work has appeared in the Graph Drawing literature, which has cross-over between theory and network visualisation applications, graph drawing and visualisation researchers have not seriously considered practical network visualisation on higher-genus surfaces.  Certainly, to the best of our knowledge, there was no available software that creates visualisations of arbitrary networks on higher-order surfaces.  Algorithms capable of computing embeddings of certain types of graphs do exist \cite{yu2014practical,duncan2011planar}, however, they are restricted to graphs that permit an embedding without crossings on the torus. 

While it seems they are rarely actually implemented, nor used in data visualisation, there are very notable examples of classic 2D video games using torus wrapping in a way that players intuitively understand the topology. Consider the classic 1980s video game \textit{Asteroids} \footnote{\url{https://en.wikipedia.org/wiki/Asteroids_(video_game)}} (\autoref{fig:intro:asteroids}).  In the Asteroids game, the video screen is like a box where players intuitively understand that an asteroid drifting off the left side of the TV screen box wraps around to the same vertical position at the right of the screen.  Similarly, vertical wrapping occurs where the asteroid drifts off the top of the boundary and wraps around to the same horizontal position at the bottom of the screen. Such two-dimensional wrapping (wrapping vertically and horizontally) can be described as toroidal, since topologically (as defined in~\autoref{sec:intro}) the data points are connected not only at the sides, but also at the top and bottom. In~\autoref{sec:torus1} we explore whether the intuitive understanding of such torus wrap-around topology is also applicable to network visualisation.

The closest to network visualisation on higher-genus surfaces that we are aware of in the literature is illustrations in the kind of theory papers described above.  That is, illustrations of the properties of embeddings on surfaces.  3D drawings of tori or higher-genus surfaces are sometimes used, i.e.\ doughnuts, pretzels and other holey pastries.
While Kobourov and Wampler~\cite{kobourov2005non} have presented a ``non-Euclidean spring embedder'', i.e., a generalisation of force-directed algorithms for non-Euclidean surfaces, which may be capable of producing 3D torus drawings, it has not actually been tested with a torus distance metric. Also, visualisation in their work is generally intended for 3D rendering and does not achieve a 2D embedding.  That is, it does not consider the question of how to project the torus surface back to a 2D display without occlusion.

An interesting research question may be whether such 3D representations might be usefully used in immersive environments, but this question is beyond the scope of this thesis.
Rather, torus embeddings have the interesting property that they can also be represented in a two-dimensional diagram or visualisation, where it is understood that the left and right sides of the drawing connect, as do the top and bottom - as per the asteroids game mentioned earlier. In this thesis, our central interest is whether these types of 2D toroidal drawings permit improved graph layout aesthetics (described in~\autoref{sec:related:tradition:empirical}) and readability of the connectivity of the network, such that they can usefully be used in real-world network visualisation applications.

Beyond node-link visualisations, other examples of information visualisations can potentially be arranged onto a torus topology, discussed in our design space in~\autoref{sec:designspace:torus}.

\subsubsection{Tile-display and interactive wrapped panning}
\label{sec:related:tiledisplayandpanning}

\begin{figure}
	\includegraphics[width=0.7\textwidth]{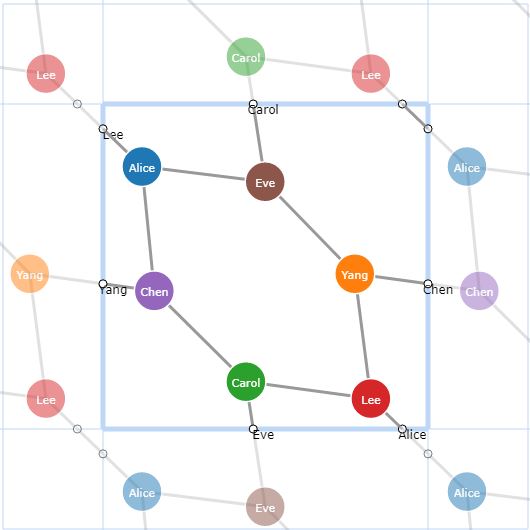}
	\centering
	\caption{Partial context showing replicated network layout along the 2D plane's edges where a node-link diagram of K3,3 graph is laid-out on a 2D torus topology (Kocay and Kreher 2016), image reproduced using our torus-based layout algorithm described in~\autoref{sec:torus1}, where small circles intersecting the boundary shows which node (name) the node connects to. The layout is created using our algorithm in~\autoref{sec:torus1:stressminimisation}.}
	\label{fig:related:tiledgraph}
\end{figure}

As already mentioned in relation to spherical wrapping~\autoref{sec:related:sphere}, one potential disadvantage of either toroidal or spherical wrapping is that this can break edges which now wrap horizontally and/or vertically  around the layouts. To aid comprehension of wrapping around the border, in some illustrations of toroidal graph embeddings the layout is replicated on the edges of the original layout to provide partial or full context to the static diagram, such as networks by Kocay and Kreher~\cite{kocay2016graphs} (\autoref{fig:related:tiledgraph}).
This means the edges or maps are no longer broken at the boundaries, but at the cost of replication consuming additional screen space. Another example of similar tile display illustration has been presented in self-organising maps (defined in~\autoref{sec:related:som}) display of high-dimensional data by Ultsch~\cite{ultsch2003maps}, which is explored in our design space in~\autoref{sec:designspace:torus}.

Another potential aid to understanding edge wrapping is interactive panning. In map projections (\autoref{sec:related:maps}), interactive panning allows a user to use simple mouse or touch drags to pan the diagram to move any region of interest to the centre.~\autoref{fig:designspace:dimensions}-top shows that when the data points are panned off the left side of the diagram, they reappear on the right side while the layout interactively  wrap around in other regions, and vice versa~\cite{Davies:2013ug}.

We might hypothesise that the user seeing the additional replicated wrapping context around the border in static diagrams aids comprehension. Similarly, we might also hypothesise that the user seeing the detail wrap around as they interactively pan will reinforce understanding. However, neither of these hypotheses have been tested before this thesis, which we explore in~\autoref{sec:torus1}. In~\autoref{sec:designspace}, we systematically explore tiled  representation and interaction as two wrapping approaches in our design space. This exploration leads us to create interesting wrapped visualisations for maps, cyclical data or relational data that may improve understanding of their data's relationships or connectivity.

\subsection{Aesthetics and other empirical evidence on network visualisation}
\label{sec:related:tradition:empirical}
This thesis explores networks, an important type of abstract data commonly used in social science, medical science, and engineering.
Effectiveness and usability are the key factors determining whether a given network layout and the patterns that it reveals can be of use for wider audiences and practitioners. 
The graph drawing community has a long history of establishing tools such as aesthetic metrics that measure the quality of network layout. These metrics give a quantitative model for aesthetic and readability requirements and have been correlated with user preference and performance in readability tasks. For example, fewer crossings between links and greater incidence angle between links where they connect to the same nodes have been shown to improve network path following tasks~\cite{huang2008effects, purchase2002empirical, purchase1997aesthetic, yoghourdjian2020scalability}. Furthermore, networks with more uniform link length have been found to be preferred by readers~\cite{dwyer2009comparison}. We demonstrate that such graph aesthetics measures can be improved by laying out networks on a 2D wrapped torus topology in~\autoref{sec:torus1} and~\autoref{sec:torus2}.

Furthermore, unlike drawings of networks in a traditional non-wrapped 2D plane, our graph layout approach is based on 2D drawings of projected 3D topologies. To evaluate the readability of wrapped graph layouts, we define new metrics that capture the amount of link wrappings across the boundaries, detailed in~\autoref{sec:torus1:aesthetics}, and the distance between clusters of nodes in the network, measuring how well a layout ``unfolds'' a network diagram with dense structures, detailed in ~\autoref{sec:torus2:qualitycomparison}.

Researchers have also explored readability of node-link diagrams across a variety of graph sizes and densities (e.g.\ Yoghourdjian et al.~\cite{yoghourdjian2020scalability}), and with a variety of network comprehension tasks (e.g.\ by Saket et al.~\cite{saket2014node} and Okoe et al.~\cite{okoe2018node}). For example, Yoghourdjian et al.\ found that reading a node-link diagram is error-prone (more than 50\% error rate) when graph density (ratio of edges to nodes) is equal to or greater than 6 for medium sized graphs (i.e., graphs with more than 50 nodes)~\cite{yoghourdjian2020scalability}. Therefore, improving network layout quality and its usability remains an open question. 

\subsection{Adjacency matrices} Since even the best layout algorithms must render very dense networks as hairballs (\autoref{sec:related:network:forcedirected}), alternatives to node-link diagrams such as a tabular or adjacency matrix format can be used~\cite{liu2014survey}. In an adjacency matrix, there is both a column and a row for each node, and the cells at the intersections of columns and rows of different nodes can hold a mark indicating the presence, absence, or weight of a link between those nodes.  Such representations are capable of completely eliminating occlusion of node-link views for large and dense networks since there is a distinct cell for each possible link~\cite{munzner2014visualization,liu2014survey}. 

\begin{figure}
    \centering
    \subfigure[Dense network]{
    \includegraphics[width=0.4\textwidth]{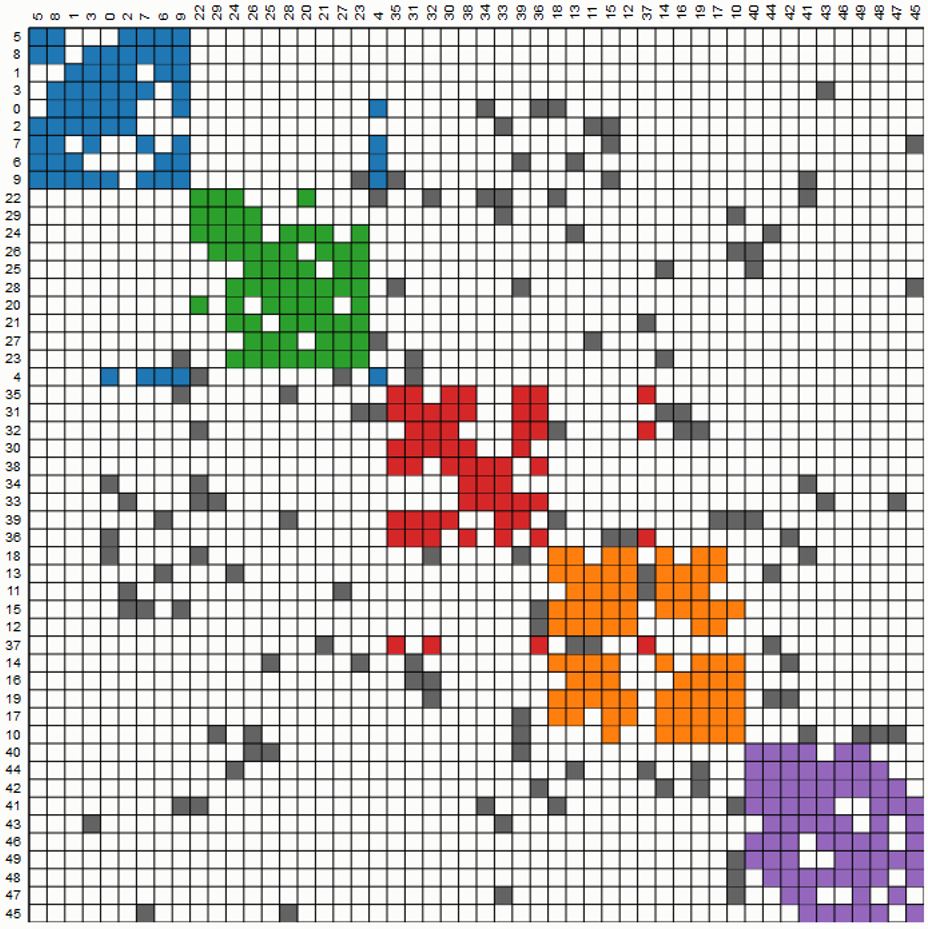}
    }
    \subfigure[Scale-free network]{
    \includegraphics[width=0.4\textwidth]{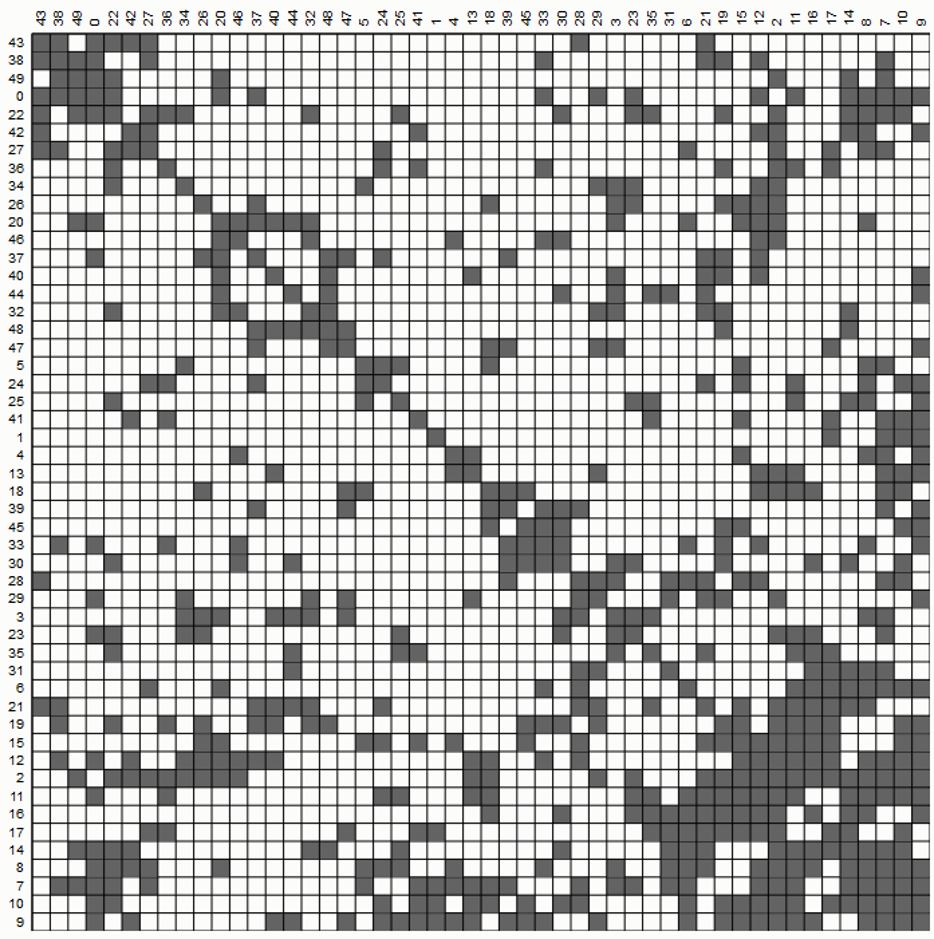}
    }
  \caption{Adjacency representations of (a) a dense network with community structures generated using Fortunato et al.'s community network model~\cite{fortunato2010community} (50 nodes, 211 links) and (b) a dense network generated using Barabási and Albert's scale-free network model~\cite{barabasi1999emergence} (50 nodes, 301 links); for a given dense network proper arrangement of rows and columns makes it easier to find that there are five main clusters (left) than a random order of rows/columns in the same network. Layouts were created by adapting existing library reorder.js by Fekete~\cite{fekete2015reorder}.}
  \label{fig:related:matrix}
\end{figure}

In such tabular and matrix representations, there is also a significant degree of layout freedom, in terms of how rows and columns may be ordered. 
For example, if the rows and columns can be ordered such that the cells representing links involved in network clusters can be grouped together, then the cluster structure may be more evident than a random ordering~\cite{behrisch2016matrix,burch2020state}. Examples include determining such aesthetic orderings of rows and columns that reveal visual cluster patterns using ``optimal leaf'' ordering (\autoref{fig:related:matrix}) or Cuthill Mckee ordering~\cite{fekete2015reorder}.


While adjacency matrices have been shown more effective than node-link diagrams for identifying connectivity~\cite{okoe2018node,burch2020state}, cluster identification tasks~\cite{okoe2018node}, and weighted graph comparisons~\cite{alper2013weighted} for large and dense networks, adjacency matrices suffer from the disadvantage that they are less familiar than node-link diagrams to most people and they can also be less intuitive to read~\cite{okoe2018node, munzner2014visualization}.


In~\autoref{sec:designspace}, we demonstrate novel use of interactive wrapping for matrices. That is, we consider interactive column and row reordering for matrices that wrap around -- again as if the matrix were projected onto 2D from the surface of either cylindrical or toroidal topologies.

\subsection{Multidimensional scaling}
\label{sec:related:mds}
Network layout algorithms have a strong connection with multidimensional scaling (MDS).

Many real-world data structures contain objects that have multiple attributes (e.g., a patient has temperature, weight, height, blood pressure, etc), however, when it comes to visualisation, people's ability to perceive such multidimensional data may be limited~\cite{borg2005modern,kirt2007comparison}. Therefore, in multidimensional data analysis, such data is commonly modelled using dimension reduction methods to preserve meaningful structures of a dataset when mapped to lower dimensions (e.g., 2D flat screens or 3D view) for ease of consumption by humans~\cite{ingram2008glimmer,cox2008multidimensional,bian2020implicit}. Examples include MDS which models high-dimensional data as data points (representing objects) and the distances between pairs of data points (representing similarity or dissimilarity between pairs of objects) in low-dimensional space, and self-organising maps (defined in~\autoref{sec:related:som}). These techniques have been used in a variety of fields such as social sciences and biological sciences~\cite{cox2008multidimensional,bian2020implicit}.

In MDS, the most common way of showing the resulting low-dimensional data points is scatterplots in 2D (or 3D) (i.e. X-Y or X-Y-Z dots plots). When visualising the resulting low-dimensional data points, similar to networks, if there is no physical positions that are relevant to the analysis of these data points, then it is considered ``abstract data'' and thus we have the freedom to lay out the data points to represent patterns or features corresponding to the high-dimensional data. One important goal of MDS is to preserve the high-dimensional data’s underlying structure in low dimension as well as possible in order to faithfully represent the structures, relationships and patterns of the data~\cite{borg2005modern,ingram2008glimmer,de2009multidimensional}. 

Examples of MDS plots that seek to arrange data to reveal structures, patterns or relations within the data include correlation between units sold and number of marketing promotions~\cite[Ch.~9]{Hsia-ChingChang2018AaKM}, voters' similarity ratings on political candidates, and embeddings of MDS plots for a better distribution of the datasets across the surface of a sphere by Lu et al.~\cite{lu2019doubly}.


Methods that seek to optimise the aforementioned arrangement of MDS plots typically solve a goal function called ``stress'', i.e. disparity of the summed distance between data points in low dimensions and the summed distance between original data points (defined more formally in~\autoref{sec:torus1}). These techniques solve exactly a variant of force-directed node-link layout problems (described in~\autoref{sec:related:network:forcedirected}) where the goal is to preserve the structure of the original data in the layouts. In fact, by specifying a complete distance matrix between all pairs of nodes of a network, the process of laying out node positions that simulates its graph theoretic shortest path distance is equivalent to the process of MDS finding suitable coordinates of data points in lower dimension that simulates its original high-dimensional distance~\cite{ingram2008glimmer,martins2012multidimensional,zheng2018graph}. Similarly, by mapping high-dimensional distance onto weighted edges between all pairs of objects, the process of MDS is the same as laying out a weighted graph. However, as mentioned in~\autoref{sec:related:network:forcedirected} and as we show in our evaluation (\autoref{sec:torus2}), for larger and denser networks, such traditional force-directed based approaches do not do a good job of untangling a network.

In this thesis, like Gansner et al.'s stress majorization (i.e. an optimisation method for minimising stress)~\cite{gansner2004graph}, Martins' social network layouts using MDS-based approach~\cite{martins2012multidimensional}, and Zheng et al.'s stochastic gradient descent (i.e.\ an optimisation method for minimising stress) for flat node-link layouts~\cite{zheng2018graph}, our layout approach (\autoref{sec:torus1}, \autoref{sec:torus2} and \autoref{sec:spherevstorus}) is also based on minimising this stress (defined earlier in this subsection) function, but on a more complex topology that provides us extra flexibility in laying out networks to reduce visual clutter. 

MDS has also been embedded on the surface of a sphere to achieve a cleaner view of the complex data structure and better reveal clusters, by Lu et al.~\cite{lu2019doubly}. However, the results were presented in a static 3D globe which may suffer from occlusion (defined in~\autoref{sec:related:sphere}) when viewed on a 2D flat screen or paper.  Such spherical layouts of high-dimensional data have not been evaluated by human readability studies. In~\autoref{sec:designspace}, we explore examples of interactive wrapped MDS plots on a torus topology using our network layout approach discussed in~\autoref{sec:torus2}.







\subsection{Self-organising maps}
\label{sec:related:som}
Self-organising maps (SOMs) are another example of arrangements of abstract data (defined in~\autoref{sec:related:networks}) that also seek to optimise the layouts to better understand their relationships or the structure.

Introduced by Kohonen~\cite{kohonen1982self}, SOMs have been popular for social sciences, web-document mining, and other domains. SOM refers to dimension reduction techniques commonly used in multidimensional data analysis (described in~\autoref{sec:related:mds}) in such a way that the original data's relationships and structure are preserved in lower dimensional projections (usually on a 2D hexagonal or rectangular grid) for ease of perception~\cite{kirt2007comparison}. SOM models high-dimensional data as map units (representing objects) arranged in a 2D grid, where each unit is associated with a weight vector (e.g., with its cardinality corresponding to the number of dimensions in the original data). At each layout iteration, SOM distributes the positions of the map units by first selecting a best matching unit where its weight vector is closest to an input training vector of the original high-dimensional data.  Then, the selected unit and its neighbouring units are moved such that the difference between the weight vector and the training vector decreases~\cite{squire2005visualization,wu2006spherical}. 

Discerning high-level structure, such as clusters, is an important task for SOMs. In the process described above, map units that are grouped spatially nearby (into clusters) have more similar values than those in distal clusters~\cite{kohonen1982self,wu2006spherical,kirt2007comparison}. Examples include studies showing SOMs revealing structural cluster changes and supporting visual comparisons tasks by Denny and Squire~\cite{squire2005visualization}, studies showing participants  distinguish clearly groups
within the data using SOMs for a real-world dataset by Kirt and Language~\cite{kirt2007comparison}, and a qualitative study showing superiority of the SOM for visual financial performance analysis by Sarlin~\cite{sarlin2015data}.

SOM has also been adapted as a general force-directed (defined in~\autoref{sec:related:networks}) layout approach for node-link visualisation, such as optimisation of the arrangements of data to reveal network clusters by Bonabeau, Eric and H{\'e}naux~\cite{bonabeau1998self}. 

SOMs have also been embedded on the surface of a sphere by Wu and Takatsuka~\cite{wu2006spherical,wu2006visualizing} or laid out onto a torus topology by Ultsch~\cite{ultsch2003maps}, as described in~\autoref{sec:designspace:torus}.

However, self-organising maps-based node-link layout approaches belong to a general class of force-directed methods~\cite{bonabeau1998self} and thus  still suffer from cluttered layouts for larger and denser networks. In~\autoref{sec:designspace}, we explore examples of SOMs that are represented on topologies that wrap around.






\section{Global Map Layout}
\label{sec:related:maps}
While there has been limited research into wrapped network visualisations that are projected from 3D shapes onto a 2D plane (\autoref{sec:related:sphere} and \autoref{sec:related:torus}), creating 2D-projections from sphere typologies has been a challenge for map makers for centuries. Geographic information visualisation is the most common example involving visualisation of the 2D surface of a 3D shape. While sometimes geographic data is presented on a physical globe, cartographers have long ago found ways to project the surface of the globe to a 2D plane, e.g. for presentation on printed paper or more recently on computer screens.
A plethora of methods for projecting a 3D globe onto a 2D map exist. Projecting the surface of a sphere onto a 2D plane is not novel: cartographers have been doing this for centuries. 

Cartographers and mathematicians have devised hundreds of projection techniques ~\cite{snyder1997flattening}, including rectangular projections such as Mercator projection, commonly used in Google Maps, Quincuncial, Miller, Patterson projections, and non-rectangular projections, such as Hammer, Mollweide, and unusual projections such as Lee's conformal projection (i.e., map projections that preserve every angle between two curves that cross each other on the surface of a globe~\cite{jenny2017guide}) in a tetrahedron, with different mappings between data points across the map's boundaries. The reason for this diversity is that the surface of a sphere is ``non-flat'' and thus when geographic data is presented on a 2D flat screen, none of existing cartographic projection techniques can be considered optimal in depicting the original geographic information on the surface of a sphere~\cite{schottler2021visualizing}. Rather, each projection is a trade-off between preserving shape, area, angle, distance or direction~\cite{snyder1997flattening} as it is not possible for a 2D projection to do all of these simultaneously. 
\begin{figure}
    \centering
	\includegraphics[width=0.8\textwidth]{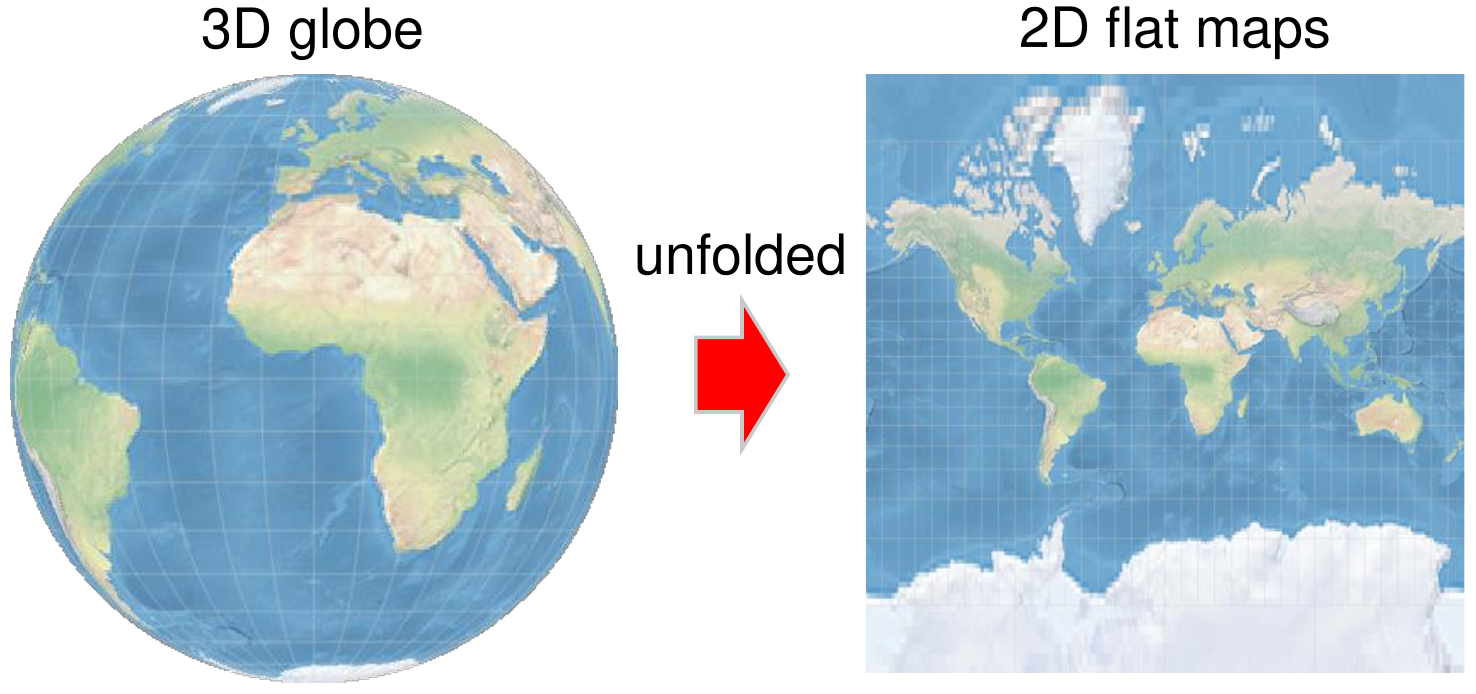}
	\caption{3D globe and common 2D flat maps (e.g., Mercator projection commonly used in Google Maps).}
	\label{fig:related:3dglobe_2dmap}
\end{figure}




As a consequence, cartographers have invented what are called \emph{equal area projections} that preserve the relative area of regions on the globe. These include \tequalearth{}, \thammer{} and \tmollweide{} (see \autoref{fig:related:map_projections})~\cite{vsavrivc2019equal, jenny2017guide}. They have also invented \emph{compromise projections} that do not preserve any of these criteria exactly but instead trade them off, creating a map that does not distort area, shape, distance or direction ``too much.''
These include \torthographic{} and \tequirectangular{} (see \autoref{fig:related:map_projections}). 

To alleviate the strong distortions at the poles (top and bottom), non-rectangular projections have been developed, including \textit{Fahey} projections (\autoref{fig:designspace:dimensions}-middle), Myriahedral projections~\cite{van2008unfolding}, i.e. the globe is fitted to a polyhedron which, when cut open and unfolded along its edges, results in maps with many interruption but that are conformal (preserving angles) and conserve areas. Other examples include Dymaxion projections and many others, summarised by existing literature~\cite{d3-map-collection,snyder1997flattening}. 

\begin{figure}
	\includegraphics[width=\textwidth]{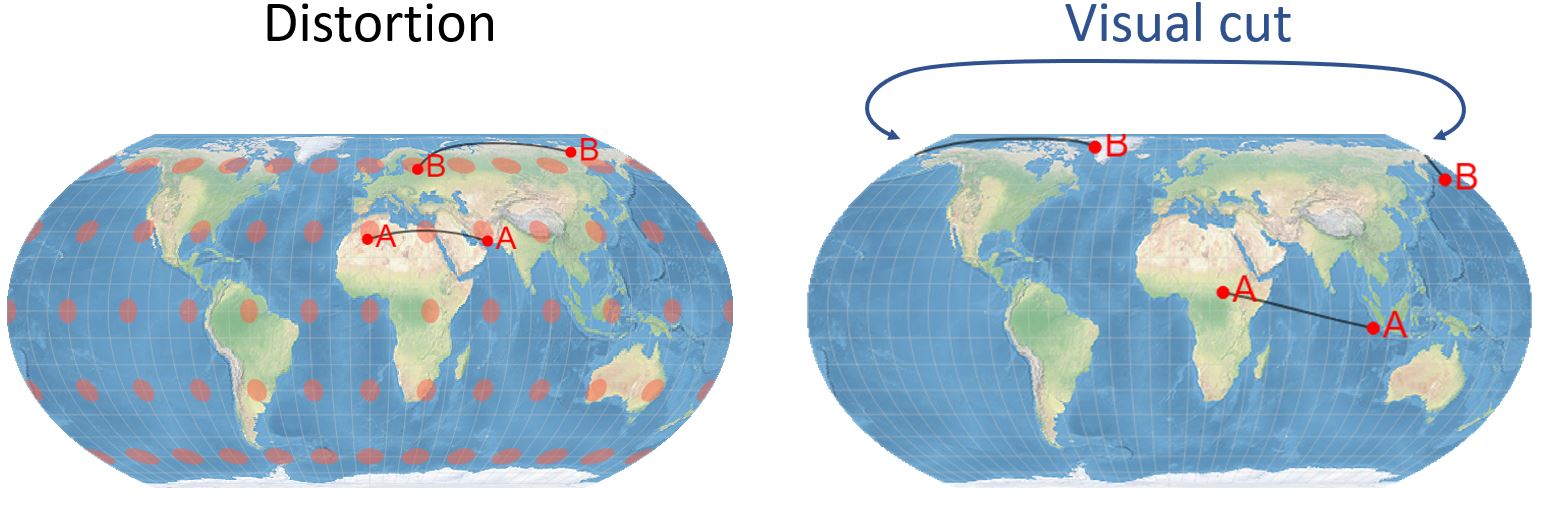}
	\caption{To judge which point pair represents the shorter true geographic distance on the surface of a globe, a user needs to take map distortion and discontinuities into account, which map may hinder geographic comprehension. Equal-sized dots on the left show the shapes are more distorted near the poles.}
	\label{fig:related:distortion_and_visual_cut}
\end{figure}

Globe projections also differ in the shape of the boundary of the projection. Some, such as \tequirectangular{} are rectangular, others, such as \tequalearth{} or \thammer{}, reduce distortion at the poles by projecting to a more ovoid shape. Some, such as the \tmollweide{} and the \torthographic{}, resemble the front and back views
of the 3D globe. Maps such as these in which the projection region is split are said to be \textit{interrupted}. The \torthographic{}, in particular, has a naturalistic appearance as it shows the Earth viewed from infinity~\cite{jenny2017guide}. 

There are several user studies on the readability and user preference of map projection visualisations~\cite{hennerdal2015beyond, hruby2016journey, avric2015user, carbon2010earth}. 
For example, Hennerdal~\cite{hennerdal2015beyond} evaluated with five rectangular and non-rectangular projections, and found that people may find it difficult to understand path continuity across the edges of the map. Hruby et al.~\cite{hruby2016journey} found that viewers find it difficult to understand the distance or direction between two points if this requires reasoning about the discontinuity in the projection and mentally wrapping the projection around a globe to understand their relative position. Avric et al. found users preferred continuous projection to interactive ones \cite{avric2015user}.

\begin{figure}
	\includegraphics[width=\textwidth]{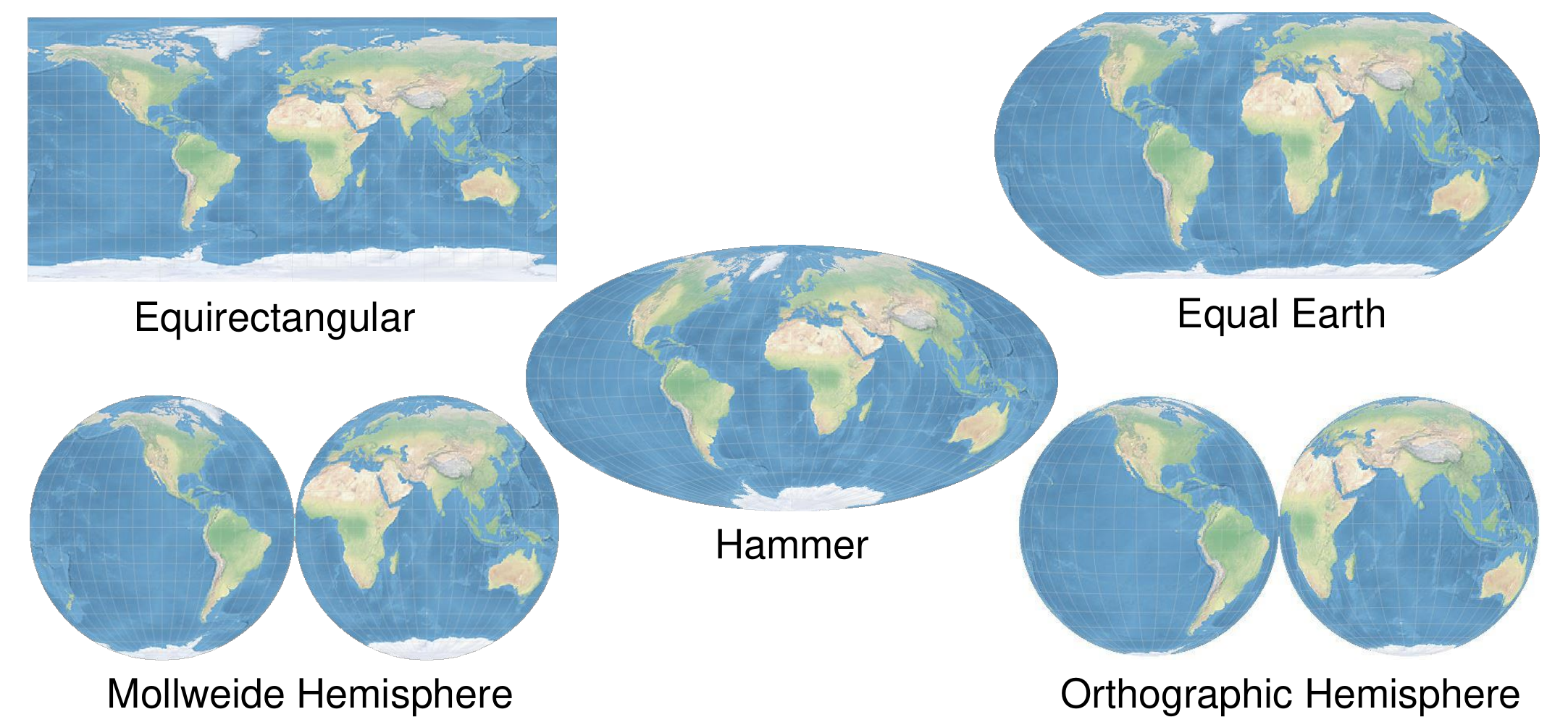}
	\caption{Five common map projections with a variety of area and/or distance preservation and shapes. The readability of these projections on geographic comprehension tasks are tested in~\autoref{sec:spheremaps}.}
	\label{fig:related:map_projections}
\end{figure}

This suggests that allowing the viewer to interactively pan the map projection so as to centre a region of interest may improve their understanding of the inherent distortion introduced by the projection and of the Earth's underlying geography. For instance, this allows them to reposition two points so that they are no longer separated by a discontinuity. Such interactive panning, also called spherical rotation~\cite{snyder1987map}, has been provided in many online maps for several years, e.g.\ by Bostock et al.~\cite{bostock_code_2013},  \textit{Rotate the World} by Davies et al.~\cite{Davies:2013ug}, and online interactive video maps showing how carbon dioxide travels around the globe over the course of one year by Jenny et al.~\cite{jenny_interactive_2016}.

One recent study of pannable terrain maps found that they perform more accurately than static maps but at the cost of additional time in task completion and concluded therefore that results of existing static map reading studies are likely not transferable to interactive maps~\cite{herman2018evaluation}.
However, surprisingly, as far as we are aware there has been no evaluation of whether interactive panning of globe projections improves performance on standard geographical tasks such as estimating the distance or direction between two points or the relative area of two regions. 

In~\autoref{sec:spheremaps}, we explore the effect of interactive panning on geographic comprehension tasks. 

The only direct user research of pannable globes that we know of are two studies in virtual reality (VR) investigating different visualisations for understanding origin-destination flow between locations on the Earth's surface by Yang et al.~\cite{yang_origin-destination_2019}. While it was not their main focus, the studies revealed that interactive panning improved task accuracy at the cost of task completion time when viewing flow shown using straight lines on a flat map. 
However, it is likely that this was not because panning was used to reduce geographical distortion but rather that it was used to separate the flow lines which were the focus of the tasks. Furthermore, the static and interactive conditions were across different studies so comparison was between groups. In~\autoref{sec:spheremaps}, we present a more systematic and direct study of interactive panning for geographic tasks.

\section{Visualising Cyclical Data}
\label{sec:related:tradition:temporal}

Cyclical data, such as: time series representations of average yearly rain-fall, monthly flu vaccines received by people,  and average birth rate over months of a year; non-temporal cyclical data such as average wind strength around the compass directions; and so on, have no beginning or end. Such data are often represented in traditional static linear bar and line charts where the series is assumed to wrap around from one side to the other (e.g. January to December). However, this traditional ``flat'' representation may hide important information as it hides the nature of the underlying data's structure, making it difficult to inspect trend or value comparisons across the left-and-right edge of the chart where the data is actually continuous~\cite{talbot2014four}. Compared to linear bar charts, polar (or radial) visualisations have the advantage that cyclical data can be presented continuously without mentally bridging the visual ‘cut’ across the left-and-right boundaries (to be explored in this chapter). However, as we discuss in this subsection, in general they have been shown to be less effective than a linear bar chart.

Despite extensive consideration from designers of visualisation techniques, temporal data visualisation continues to pose challenges for effective representation. 
Temporal data can be very rich, including discrete events, duration, arrays of quantitative values (time-series), as well as more complex combinations with other types of data such as networks and spatial data (i.e.~data combining spatial attributes such as locations), e.g. a geospatial map displaying hourly energy consumption in buildings within a city's downtown region by Zhang et al.~\cite{zhang2022timetables}. 
Various surveys of temporal data visualisation exist, citing general techniques by Aigner et al.~\cite{aigner2011visualization}, timelines by Brehmer et al.~\cite{brehmer2016timelines},  spatio-temporal data exploration techniques, for attributed trajectories by Tominski et al.~\cite{tominski2012stacking}, and more generally~\cite{andrienko2006exploratory}, and more abstract quantitative data that can be represented through space-time cubes (a three-dimensional (usually 3D Euclidean space consisting of a 2D geographical space and a time dimension)~\cite{bach2017descriptive}. Interactive techniques for exploring general temporal data (without any particular allowance for cycles) range from multilevel zooming to scale to large data or minute detail~\cite{zhao2011exploratory} to sophisticated visual organisation of sequences of state data (such as brain activity)~\cite{bach2015small}. 

Data for many temporal phenomena is characteristically cyclical, as nature follows the cycles of day and night, climate and seasonality, as well as biochemical processes that are continuous cycles. In the case of time series, polar visualisations have often been used either as simple cycles with radial bars or a line of varying distance from the centre.  Multiple layers of bars or lines have also been proposed, e.g., silhouette graphs~\cite{harris1999information}, overlaying multiple lines, e.g., one for every year, or a `timeline' spiralling outside the centre of the polar chart~\cite{tominski2008enhanced}. A particular way to visualise cycles and repetition in temporal data is to abstract temporal change through multidimensional reduction methods~\cite{bach2015time}. 

\begin{figure}
\centering
	\includegraphics[width=\textwidth]{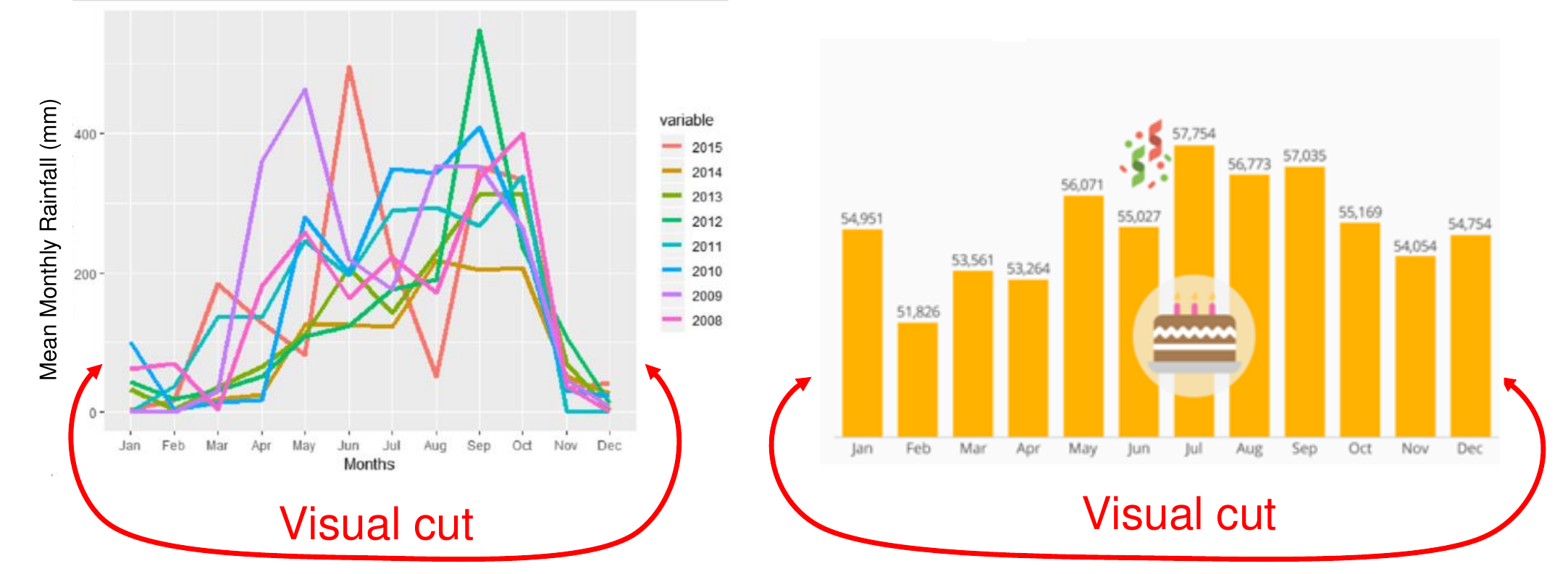}
	\caption{Average monthly rainfall over months of several years in Bangkok (image on the left adapted from Jain et al.\ \cite{jain2019prediction}) and credited to Jain et al.; average UK birth rate over 12 months of a year (image on the right adapted from \url{https://www.statista.com/chart/5814/the-months-of-the-year-with-the-most-births/}.}
	\label{fig:related:visualcut2}
\end{figure}

When comparing cyclical temporal data across 24 hours with static linear or polar bar charts, previous studies by Waldner et al.~\cite{waldner2019comparison}, Adnan et al.~\cite{adnan2016investigating}, and Brehmer et al.~\cite{brehmer2018visualizing} have shown that comparing lengths of bars in polar visualisation is less effective than  comparing bar heights in traditional linear bar charts. Likely (and in accordance with position encoding differences identified by Cleveland and McGill~\cite{cleveland1984graphical}), this is due to bars in the linear charts being parallel and aligned to a common baseline, as opposed to being at different angles and aligned to a circular base.

For example, Waldner et al.\ \cite{waldner2019comparison} found that for low-level tasks (e.g., locating extrema, reading values and comparing values at fixed 12-hour separation) a linear layout was significantly faster than a polar layout.
Other studies on time series visualisation also found that for positional and length judgements, people make less errors using linear bar charts than when using polar charts for tasks of finding trends~\cite{adnan2016investigating}, locating extrema~\cite{saket2018task, waldner2019comparison}, locating features at specific times~\cite{waldner2019comparison, brehmer2018visualizing}, comparing values~\cite{fuchs2013evaluation}, or proportion judgement~\cite{simkin1987information}. However, there is no empirical evidence comparing the readability of cyclical temporal intervals split across the `cut' on bar charts (\autoref{fig:intro:visualcut_cyclicaldata},~\autoref{fig:related:visualcut2}), compared with their corresponding polar bar charts showing the continuity. However, as we show in~\autoref{sec:cylinder}, interactive wrapping of bar charts to centre any region of bars provides promising results in terms of more accurate trend identification and pairwise value comparisons compared with standard bar charts or polar charts.

\section{Evaluation Methodology}
\label{sec:related:evaluationmethods}
In~\autoref{sec:related:tradition:empirical}, we described evaluation of network visualisations with established graph layout aesthetics metrics. This section describes other evaluation methods for visualisation performance discussed in this thesis.

User studies are the most commonly used evaluation method for data visualisations to gain insights from users, such as their preferences and reported qualitative experience on visual designs~\cite{liu2014survey,munzner2014visualization}. They are also used to understand the limit of visual perception, cognition of graphical encoding, and visualisation performance (in terms of effectiveness and usability) of visualisation techniques or interactive graphical tools~\cite{lam2011empirical,nobre2020evaluating}. The above data is often collected in multiple ways, including questionnaires, surveys (quantitative or qualitative), controlled experiments, and interviews. In this thesis, we used mixed evaluation methods including questionnaires and controlled experiments. Controlled experiments are widely used to systematically test the effect of controlled conditions (also known as independent variables) on dependent variables. Examples of dependent variables, also known as uncontrolled variables, include user’s task completion time, accuracy, number of mouse clicks, duration of mouse drags, and their gaze movement collected from stationary or mobile eye trackers. A comprehensive review of experimental designs and practical visual examples has been presented by Purchase~\cite{purchase2012experimental}.


There are two common ways for conducting controlled experiments: Laboratory-based or crowdsourced studies.
Lab-based supervised studies typically involve experimenters supervising user studies in a quiet room with participants operating dedicated equipment in a controlled environment. Experimenters have full control over the course of experiments, ensure participants closely follow experimental procedures, and alleviate other confounding factors such as disruption perceived by participants during the experiments~\cite{liu2014survey,borgo2018information}. In~\autoref{sec:torus:study1} and ~\autoref{sec:torus:study2} we present the user evaluation methods of lab-based studies. Supervised controlled experiments can also be done online, e.g., over a zoom session with participants sharing their computer screen operating dedicated experimental software. In~\autoref{sec:torus2:userstudy} we present an online supervised study. 

Controlled experiments are also commonly conducted through crowdsourcing. It collects samples with a wilder demographic range than a lab-based study which sometimes helps improve generalisation of empirical results~\cite{borgo2018information,okoe2018node,nobre2020evaluating}. However, in unsupervised studies, it often requires additional measures to ensure the quality of the data being collected, such as recruiting participants through established crowdsourcing platforms (Amazon Mechanical Turk~\cite{peer2017beyond}, Prolific~\cite{palan2018prolific}) and testing the attention of participants conducting the study, to ensure experimental procedures are followed appropriately and that the data collected could be reliable. In~\autoref{sec:cylinder:userstudy},~\autoref{sec:mapstudy}, and~\autoref{sec:networkstudy} we present user evaluation results through crowdsourcing experiments.
Gadiraju et al.~\cite{gadiraju2017crowdsourcing} compared lab-based and crowdsourced studies, while Borgo et al.~\cite{borgo2018information} surveyed visualisation studies in the past decade utilising crowdsourcing.  They identified that more details need to be reported (such as within-subjects, where all the subjects are given the same sets of conditions and trials being evaluated, or between-subjects, where the comparison of conditions is between two different groups using different trials) that may affect reproduction and rigorousness of this type of study. Meanwhile, Nobre et al.~\cite{nobre2020evaluating} mentioned that crowdsourcing was less used for evaluating complex interactive visualisation techniques due to potential challenges of sufficient training with remote novice participants, and thus they proposed improved study designs with extensive and careful training which makes crowdsourcing complex and interactive visualisation techniques possible with novice participants. 



In this thesis, we use crowdsourced studies in~\autoref{sec:cylinder},~\autoref{sec:spheremaps},~\autoref{sec:spherevstorus}; lab-based controlled experiments for~\autoref{sec:torus1}, and online supervised experiments for~\autoref{sec:torus2} to evaluate human readability of interactive wrapped data visualisations.
Other evaluation methodology such as experimental analysis (\autoref{sec:torus1} and \autoref{sec:torus2}) can be found in the respective chapters.

\section{Conclusion}
In summary, we reviewed traditional visualisations from the perspective of different data types that are laid out on a 2D rectangular plane. When presenting such data with complex relationships (whether networks, high-dimensional data, geographic data or other data with cyclical relations), it can be difficult to place elements in such a way that all similar or connected elements are close together without creating excessive overlaps of network clusters (\autoref{sec:related:network:forcedirected}) or loss of information when the data is split across the edges of the representation such as geographic maps (\autoref{sec:related:maps}) and cyclical time series data (\autoref{sec:related:tradition:temporal}). 


There can be advantages to 2D visualisation of 3D shapes. 
However, despite widespread use of projections in geovisualisation and the study of embeddings on spherical surfaces (\autoref{sec:related:sphere}) and torus topology in graph theory (\autoref{sec:related:networks}), before the work presented in this thesis, there has been no research exploring interactive wrapped visualisations and evaluation from the perspective of human-centred design or human-computer interaction. 


In this research, we aim to bridge this gap in literature by providing a unified view of interactive wrapped visualisations. In particular, we focus on 2D surfaces of 3D sphere, cylinder, and torus, because it allows us to have interesting interactive ``wrapping'' behaviours in our projected visualisations and improves data analysis tasks. To understand the concept of interactive wrapping and its potential applications we present a design space exploration in~\autoref{sec:designspace}, which unifies interactive wrapped visualisations and tools that allow a viewer to create them.


%
\chapter{Design Space for Interactive Wrapped Data Visualisation}
\label{sec:designspace}

\cleanchapterquote{Q: What is a topologist? A: Someone who cannot distinguish between a doughnut and a coffee cup.}{Paul Renteln and Alan Dundes}{Paul Renteln is emeritus professor of physics at California State University; Alan Dundes is distinguished professor of anthropology at
the University of California}

\begin{figure}
	\includegraphics[width=\textwidth]{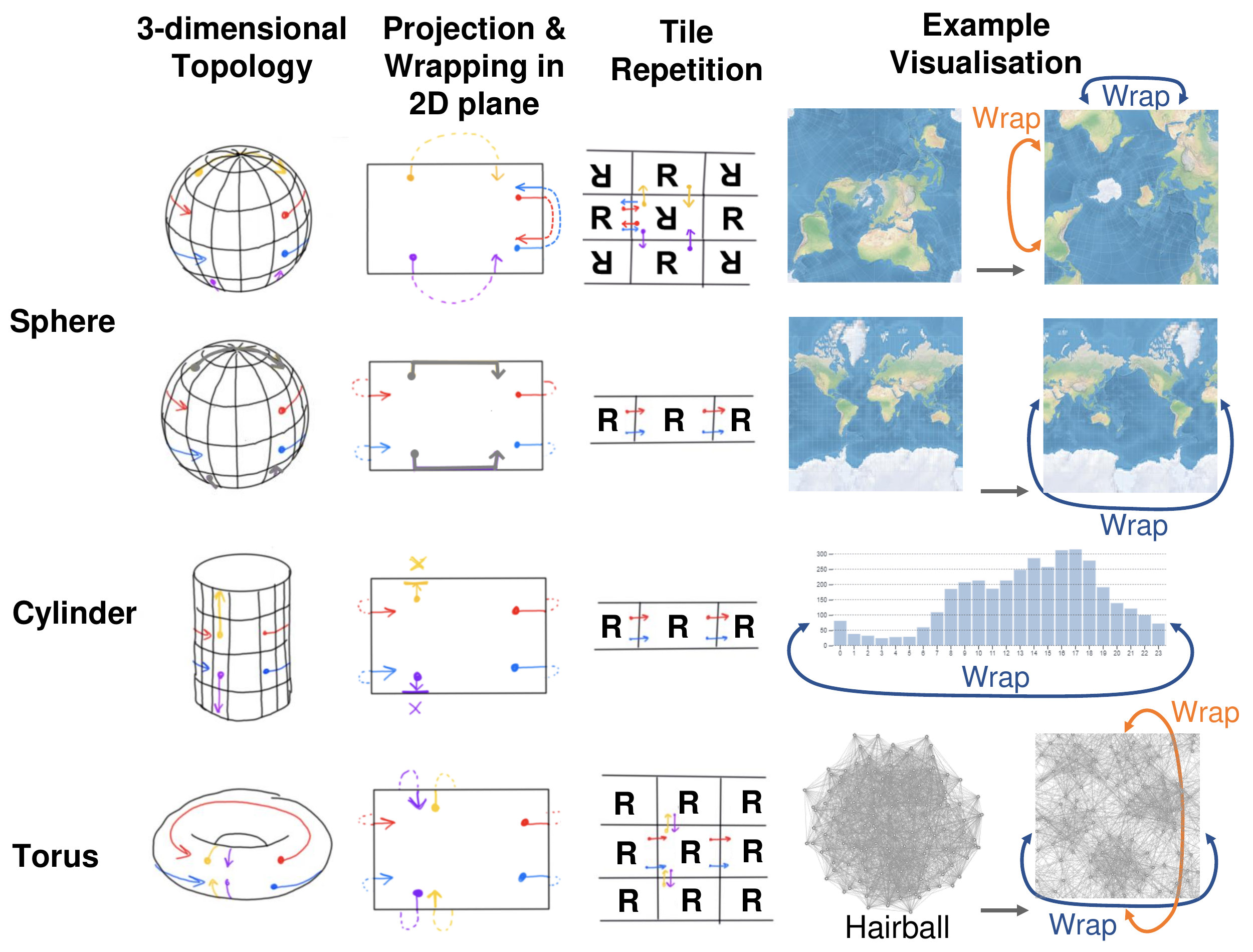}
	\caption{Design space of interactive wrapped data visualisations laid out on a 2D display of 3D cylinder, torus, and sphere geometries. The first column shows 3D topologies in our design space. The second column shows their resulting 2D rectangular projections. This also demonstrates how the points ``wrap'' around when panned. The third column shows tile replication which provides a continuous view of graphical elements wrapped across the boundary. The fourth column  \rev{illustrates example visualisations of these data types} applicable to each topology.}
	\label{fig:designspace:teaser}
\end{figure}

Most data types are not intrinsically ``flat'', including spherical maps, cyclical quantitative data, and networks.  By ``not flat'', we mean, informally, that visualisations of these data types can potentially be ``wrapped around'' the surface of different 3D shapes (cylinder, sphere, torus) and projected onto a 2D plane for better comprehension. By wrapping on a 2D plane, we refer to the fact that some visualisations can be perceived as continuous when connecting their left and right and/or their upper and lower boundaries (\autoref{fig:designspace:teaser}).



To better understand the concept of wrapped visualisations and their multiple potential applications, this chapter first presents a systematic exploration of the design space for wrappable visualisations. The design space is informed by related work in the literature as well as our own examples and exploration.

\rev{We examine data types whose visualisations can potentially benefit from being laid out on 2D representations of a 3D shape, in particular cylinder, torus or sphere shapes.
Naturally, spherical wrapping is well suited for geospatial data at a global scale (e.g., the world map) but may also be applicable to relational data types that have no spatial information that may be considered start and end points.  Such relational data types include network structured data and self-organising maps~\cite{ritter1999self,ito2000characteristics,ultsch2003maps,wu2006spherical} (\autoref{fig:designspace:teaser}-sphere, detailed in~\autoref{sec:designspace:sphere}). 
Cylindrical wrapping is well suited to data that
is cyclical in one dimension such that the representation benefits
from panning continuously along that spatial dimension. Examples of such data include cyclical time series, but also directed networks with cycles, which can be represented by Sankey diagrams\footnote{Sankey diagrams are a type of node-link diagram with directed links (arrows) used to depict flows between entities in which the width of the arrows is proportional to the flow rate} with cycles and metro maps with loops (\autoref{fig:designspace:teaser}-cylinder, detailed in~\autoref{sec:designspace:cylinder}).
Toroidal wrapping is applicable when the data has two cyclical dimensions and for relational data that can be arranged freely onto
such a topology for representation as node-link visualisations, adjacency matrices.  High-dimensional quantitative data can also be mapped to a torus topology using multidimensional scaling (\autoref{fig:designspace:teaser}-torus, detailed in~\autoref{sec:designspace:torus}). 
This design space can be a powerful tool to: \textit{(a)} understand relations between visualisations (e.g., linear and polar bar charts), \textit{(b)} apply interactive wrapping to a range of visualisations, and \textit{(c)} to create new visualisations by mapping existing visualisations onto sphere, cylinder and torus visualisations.} 

In~\autoref{sec:designspace:dimensions}, we describe design dimensions that characterise visualisations that can be laid out on the surface of a 3D cylinder, torus or sphere shape. Besides geographic maps and network structured data that are suitable for spherical wrapping, we propose novel uses of cylindrical and toroidal wrapping for a common range of data visualisation techniques, including: bar charts; line charts; polar (radial) charts; networks represented by adjacency matrices or node-link visualisations; high-dimensional data visualised by self-organising maps and multidimensional scaling.

In~\autoref{sec:designspace:tools} we present new tools capable of creating wrapped visualisations for arbitrary data types, which can be interactively panned on a 2D surface with cylindrical or toroidal wrapping for arbitrary data visualisations.

\section{Design Dimensions}
\label{sec:designspace:dimensions}
In this section, we introduce a design space which maps the possibilities for 3D cylindrical, toroidal and spherical \textit{topology} against the possibilities for mapping the topology to a two-dimensional visualisation (\textit{projection}) and the (\textit{wrapping}) affordances for showing how the visualisation wraps around.

Our design space is the composition of three dimensions:
\subsection{Topology} 
This dimension describes whether a visualisation (technique) is mapped onto the surface of a sphere (\autoref{fig:designspace:teaser}-top), cylinder (\autoref{fig:designspace:teaser}-centre) or torus (\autoref{fig:designspace:teaser}-bottom). The first column in \autoref{fig:designspace:teaser} shows three-dimensional topologies. Different topologies provide different connectivity across their surfaces, and hence different connectivity of those surfaces when they are projected to the Euclidean plane~\cite{gray2017modern,gardner1971martin}. We focus only on cylinder, sphere and torus topologies in this thesis because of their obvious ``wrapping'' behaviours when mapped to a 2D plane. They are called orientable surfaces because when data points are wrapped around the surface they maintain their orientation~\cite{gray2017modern,gardner1971martin}. With non-orientable topologies such as Klein bottle or Möbius strips the orientation reverses as their surfaces are traversed. Such topologies are not applicable to wrapped displays as they would have undesirable wrapping behaviour, such as inverting the graphical elements across the edges of the projected view.


\label{sec:designspace:dimensions:projection}
\subsection{Projection} 
This dimension describes the projection method used to obtain a 2-dimensional representation from the 3-dimensional topology. 
The most common examples of 2D-projections from 3D sphere topologies were developed for creating flat maps of the spherical surface of the world.  These include projections which result in rectangular 2D shapes, such as \textit{Mercator} (\autoref{fig:related:3dglobe_2dmap}-right), \textit{Equirectangular} projection (\autoref{fig:related:map_projections}) of a world map, \textit{Miller} projection (\autoref{fig:designspace:miller}-top, an example of a spherical network layout), Pierce\'s Quincuncial projection~\cite{peirce1879quincuncial} (\autoref{fig:designspace:teaser}-top row-example visualisation).  They can also result in non-rectangular shapes, such as \textit{Fahey} projection (\autoref{fig:designspace:dimensions}-middle), \textit{Hammer} and \textit{Orthographic Hemisphere} projections (\autoref{fig:related:map_projections}). In general, the different projection types strive to preserve different spatial aspects of the original sphere surface, such as area, distance or direction. For cylinder and torus topologies, we consider two projections motivated by our observation (detailed in~\autoref{sec:designspace:cylinder:motivation}): the \textit{side} and \textit{top} of the 3D surfaces, respectively, which results in a rectangular (side projection, detailed in~\autoref{sec:designspace:cylinder}) (\autoref{fig:designspace:wrapchart_cylinder}-centre) or a concentric radial (or ``polar'') (top projection, detailed in~\autoref{sec:designspace:cylinder}) (\autoref{fig:designspace:wrapchart_cylinder} and \autoref{fig:designspace:wrapchart_torus}).


 

\subsection{Wrapping} 
Wrapping is the technique used to show how the projected visualisation wraps (or connects up) around its boundaries. We consider two approaches: (1) interactive wrapped panning (using mouse or touch drags) affordances for visualisations (which can continuously cover each of these topologies) when projected back onto a 2D plane; (2) tile-display, i.e., by repeating the 2D rectangular plane partially (\autoref{fig:related:tiledgraph}) or in full (\autoref{fig:designspace:quincuncialtiles} and~\autoref{fig:designspace:mercatortiles}) while preserving the view of linear point movement wrapped across a plane's edges. 

The second column of~\autoref{fig:designspace:teaser} demonstrates how wrapped panning differs across the topologies, using a set of coloured points (red, blue, orange, purple). That is, it demonstrates how these points would wrap ``around'' the borders of a 2D-plane if panned, as a result of projecting from the surface of the underlying topology. 
While horizontal pan (red, blue points) wrap the 2D rectangular plane horizontally, vertical pan (yellow, purple) only wraps vertically in torus and spherical projections, e.g., Pierce\'s Quincuncial projection (\autoref{fig:designspace:teaser}-top row), as there are no data points wrapped (cut) across top or bottom boundaries in either cylindrical or other rectangular sphere projections, e.g., Mercator (\autoref{fig:designspace:teaser}-second row). The third column of~\autoref{fig:designspace:teaser} depicts how tile repetition differs across the topologies. 
While horizontal replication connects the graphical elements wrapped across the boundaries of horizontally adjacent tiles, vertical replication is only applicable to torus and spherical projections that are applicable to be tiled in vertical directions, e.g., Quincuncial projection (\autoref{fig:designspace:quincuncialtiles}).

\subsection{Constrained wrapping} 
The key problem caused by projecting a 3D shape's surface onto a 2D display is the distortion (as seen in~\autoref{fig:intro:visualcut_maps}) or discontinuity (i.e., visual ``cut'') created at the boundaries of the projection.  Following features of the surface across the boundary may be assisted by interactive panning.  That is, allowing the user of the visualisation to use a mouse or touch gesture to move (drag) and reorient centre projection such that features of interest are no longer split by the boundary.
Depending upon the underlying topology of the 3D surface and the style of projection used, such wrapped panning interaction may require limitations (or constraints) on the panning direction. For example, wrapping direction is constrained naturally in cylindrical and spherical projections (\autoref{fig:designspace:dimensions}-second row and~\autoref{fig:designspace:wrapchart_cylinder}-centre), while users can freely pan along both dimensions of the plane in torus topologies. To some extent, the constrained wrapping is also determined not only by the \textit{topology} and \textit{projection} dimensions but also by the type of data and conventions within the application domain.  We provide more details regarding the wrapping constraint in each respective topology later. 

In the following sections (\autoref{sec:designspace:sphere}, \autoref{sec:designspace:cylinder} and \autoref{sec:designspace:torus}), we provide galleries of wrapped visualisations that emerge from three design dimensions described above. We discuss each of the three wrapping topologies and their projections in detail. We show how each topology can be applied to create wrappable visualisations in information visualisation and identify the types of data for which wrappable visualisations may provide benefits. 

\section{Spherical Wrapping}
\label{sec:designspace:sphere}
\begin{figure}
    \centering
	\includegraphics[width=0.9\textwidth]{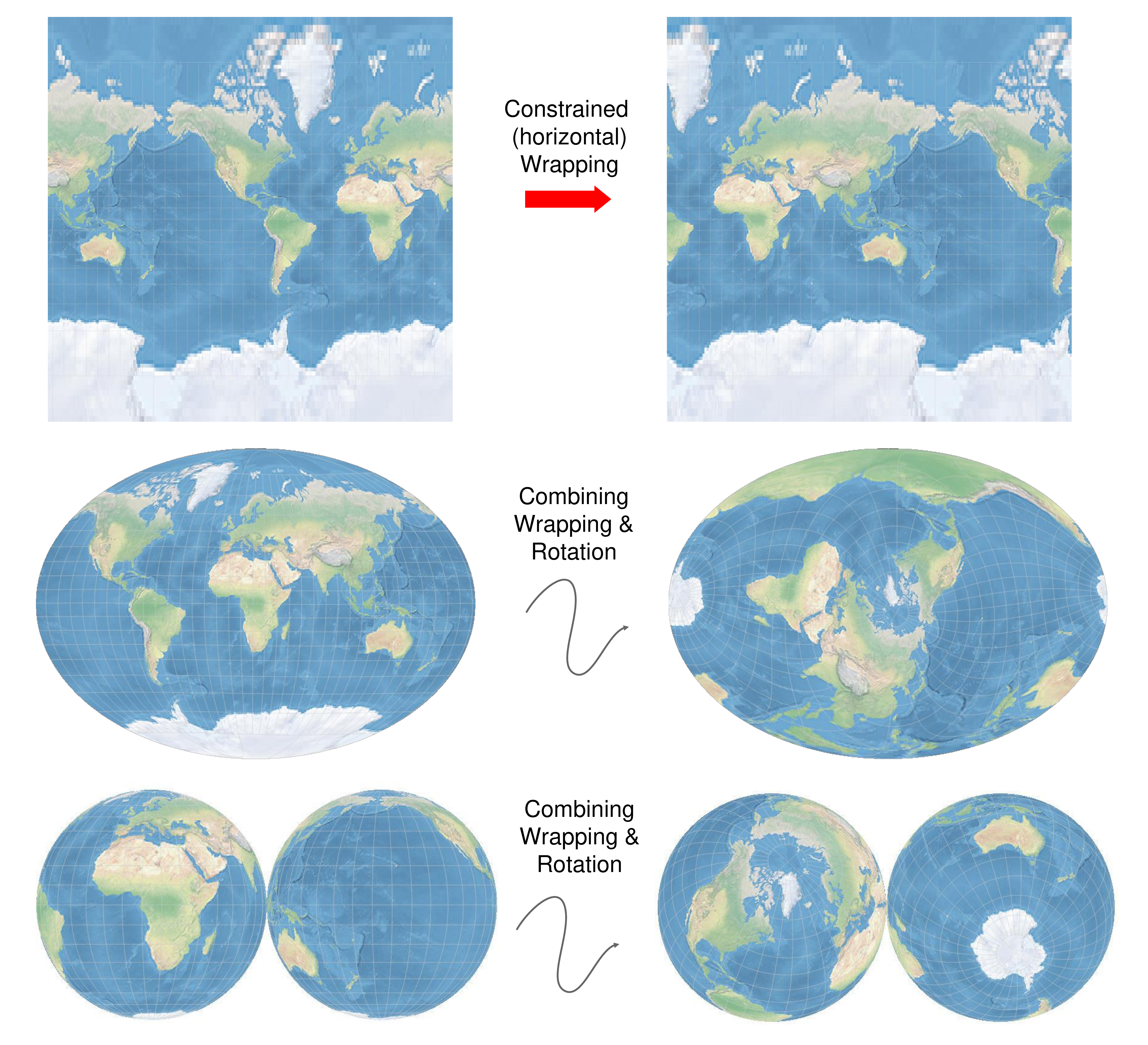}
	\caption{Examples of design space dimensions for sphere topology, including rectangular (top: Mercator), non-rectangular (middle: Fahey) spherical projections, and Orthographic Hemisphere (bottom). A constrained panning can be applied (top) such that the map is only wrappable in horizontal direction. A user can pan horizontally to explore the map. For Orthographic Hemisphere, we adapted D3's Orthographic Hemisphere projection to create two rotatable complementary Orthographic map projections, with one showing the western and the other showing the eastern hemisphere. They are placed close together. A user can drag one of the hemispheres, while the other hemisphere automatically adjusts rotation angles to show the opposite hemisphere. The evaluation of such interaction is presented in~\autoref{sec:spheremaps}.
}
	\label{fig:designspace:dimensions}
\end{figure}

Designing effective 2D-projections from sphere geometries has been a challenge for map makers for centuries. As discussed in~\autoref{sec:related:maps}, a plethora of projection techniques have been applied to \textbf{Spheres} by geographers, well summarised by Davies\footnote{\url{https://www.jasondavies.com/maps/transitions}}. Spherical projections can be rectangular (\autoref{fig:designspace:dimensions}-top) or non-rectangular (\autoref{fig:designspace:dimensions}-middle and \autoref{fig:designspace:dimensions}-bottom), referred to in the related work in~\autoref{sec:related}. 

In rectangular projections such as Pierce’s Quincuncial projection~\cite{peirce1879quincuncial}, data points can wrap around both horizontally and vertically. For example, in~\autoref{fig:designspace:teaser}-top row: example visualisation, the orange arrow shows that one tip of South America that wraps across the lower-left boundary is inverted at the corresponding upper-left position. Similarly, in the same visualisation, the blue arrow indicates that where one tip of Africa wraps around at the top-left boundary, it reappears and is inverted at the corresponding top-right position. The design of this inverted wrapping provides a seamless view in tiled repetitions  (\autoref{fig:designspace:mercatortiles} which reduces the discontinuities across the plane’s edges~\cite{peirce1879quincuncial,rodighiero2020drawing,hennerdal2015beyond}, discussed in \textbf{Tile-display}).

 
In other common rectangular spherical projections (e.g., Mercator), data points between the left and right boundaries are connected (red, blue in~\autoref{fig:designspace:teaser}-second row). Data points at the top (and bottom) of the 2D visualisation plane remain at the top while mirrored in a horizontal fashion (as shown in the grey arrow in~\autoref{fig:designspace:teaser}-second row). For example, in America centred world maps (\autoref{fig:designspace:dimensions}-top left), the continents of Asia wrap from the left end across the page to the right end. 

\begin{figure}
	\includegraphics[width=0.9\textwidth]{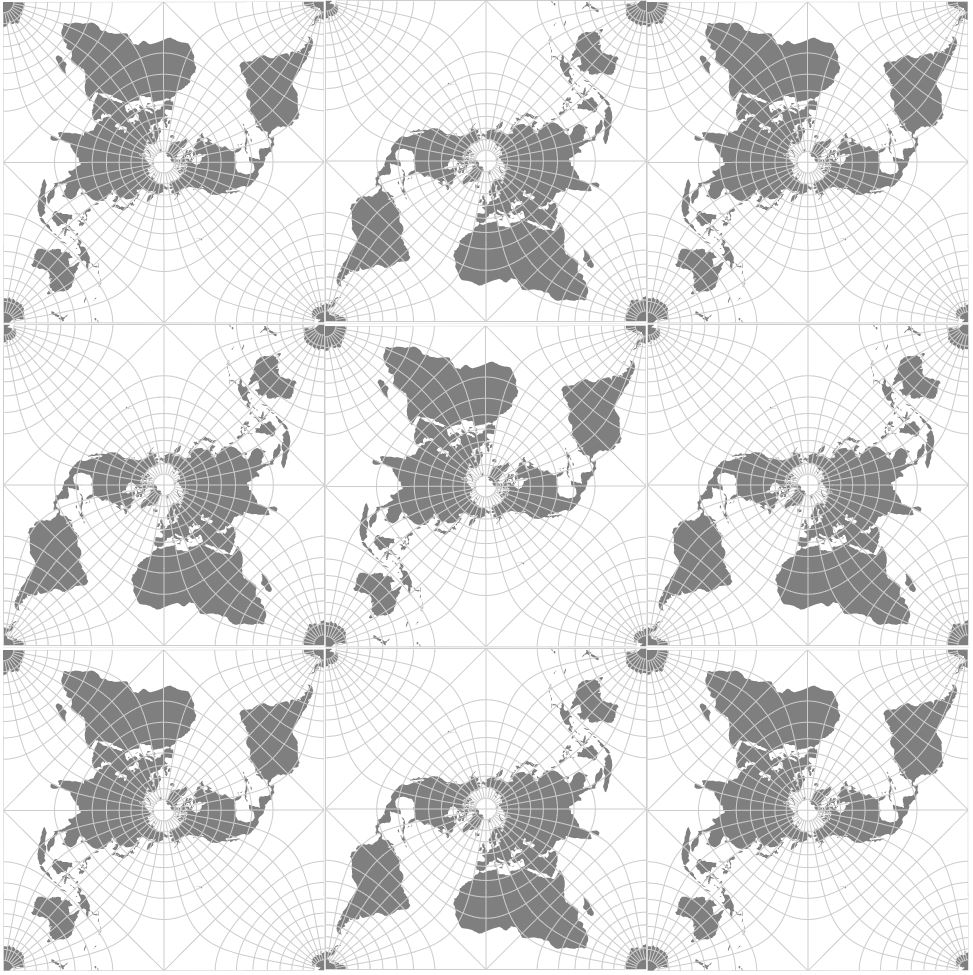}
	\centering
	\caption{Tile display in both horizontal and vertical directions. Since the continents are wrapped on the same side of the plane's edge, we develop a simple tool to rotate the tile 180 degrees in (counting from the upper-left to lower-right) tile 1, 3, 5, 7, 9. The repeated tiles show continuity of data points at the boundaries which preserves the data point movement wrapped across the boundaries.}
	\label{fig:designspace:quincuncialtiles}
\end{figure}

In a non-rectangular projection, the boundary could be curved (e.g., Equal Earth), oval (e.g., Hammer), and the projections could be interrupted (e.g., Orthographic Hemispheric), as discussed previously in \autoref{sec:related:maps}. There exist a range of different mappings between points across the boundary. Such non-rectangular visualisations still provide for wrapping. For example, in Orthographic Hemisphere projections, there is wrapping between inner sides of two circles in addition to the outer sides of the circles (\autoref{fig:designspace:dimensions}-bottom). 


\textbf{Constrained wrapping} In common rectangular map projections, e.g., Mercator or Hobo-Dyer projections, the poles remain at the north and south, so the wrapped panning is usually constrained to the horizontal axis (\autoref{fig:designspace:dimensions} - top row). In other words, the map can be panned horizontally with points on the right and left being close. This results in a cylindrical wrapping which is only pannable in the horizontal direction.

\textbf{Combining wrapping with rotation} Many spherical projections can be moved in different directions. For example, in Fahey or Orthographic Hemisphere projections, wrapping can be combined with rotating the projection. The rotation allows the user to reorient and re-centre the projected visualisation with simple mouse or touch drags while wrapping gives the ``cut'' of the projection, as seen in Fahey projection (\autoref{fig:designspace:dimensions}-middle). In the case of Orthographic Hemisphere projection, when one hemisphere is dragged, the Orthographic projection of the other hemisphere automatically adjusts three-axis rotation angles such that it shows the correct opposite hemisphere (\autoref{fig:designspace:dimensions}-bottom).



\textbf{Tile-display} If such a spherical projection were repeated to create a seamless tile display, it would show a simple repetition of the projection. In some projections, methods to reduce the discontinuities across the plane’s edges exist, such as the 3 by 3 tile display of Pierce’s Quincuncial projection~\cite{peirce1879quincuncial} (\autoref{fig:designspace:quincuncialtiles}). 
In common rectangular map projections, e.g., Mercator, Miller, Equirectangular,  horizontal tile display preserves the view of the discontinuities across the left-and-right boundaries (\autoref{fig:designspace:mercatortiles}). 

\begin{figure}
	\includegraphics[width=\textwidth]{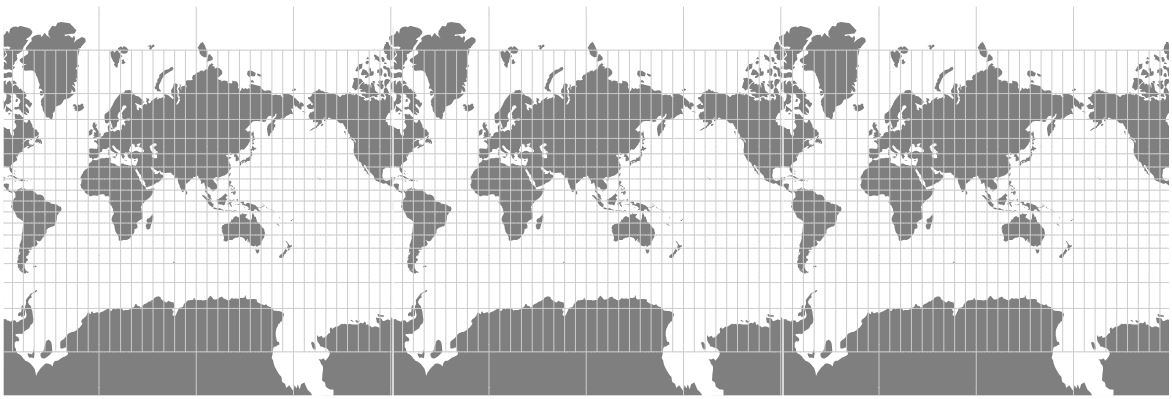}
	\centering
	\caption{``Wrapping'' dimension in horizontally repeated tiles. The repeated tiles show continuity of data points at the boundaries which preserves the data point movement wrapped across the boundaries.}
	\label{fig:designspace:mercatortiles}
\end{figure}

In~\autoref{sec:spheremaps}, we explore the usability and effectiveness of a range of spherical projections described above on geospatial data analysis tasks such as area, distance, and direction estimation tasks. We compare different map projection techniques with and without interactive spherical wrapping and rotation as described in our design space.

\subsection{Non-geographical examples}
\label{sec:designspace:sphere:subsec2}

\begin{figure}
	\includegraphics[width=\textwidth]{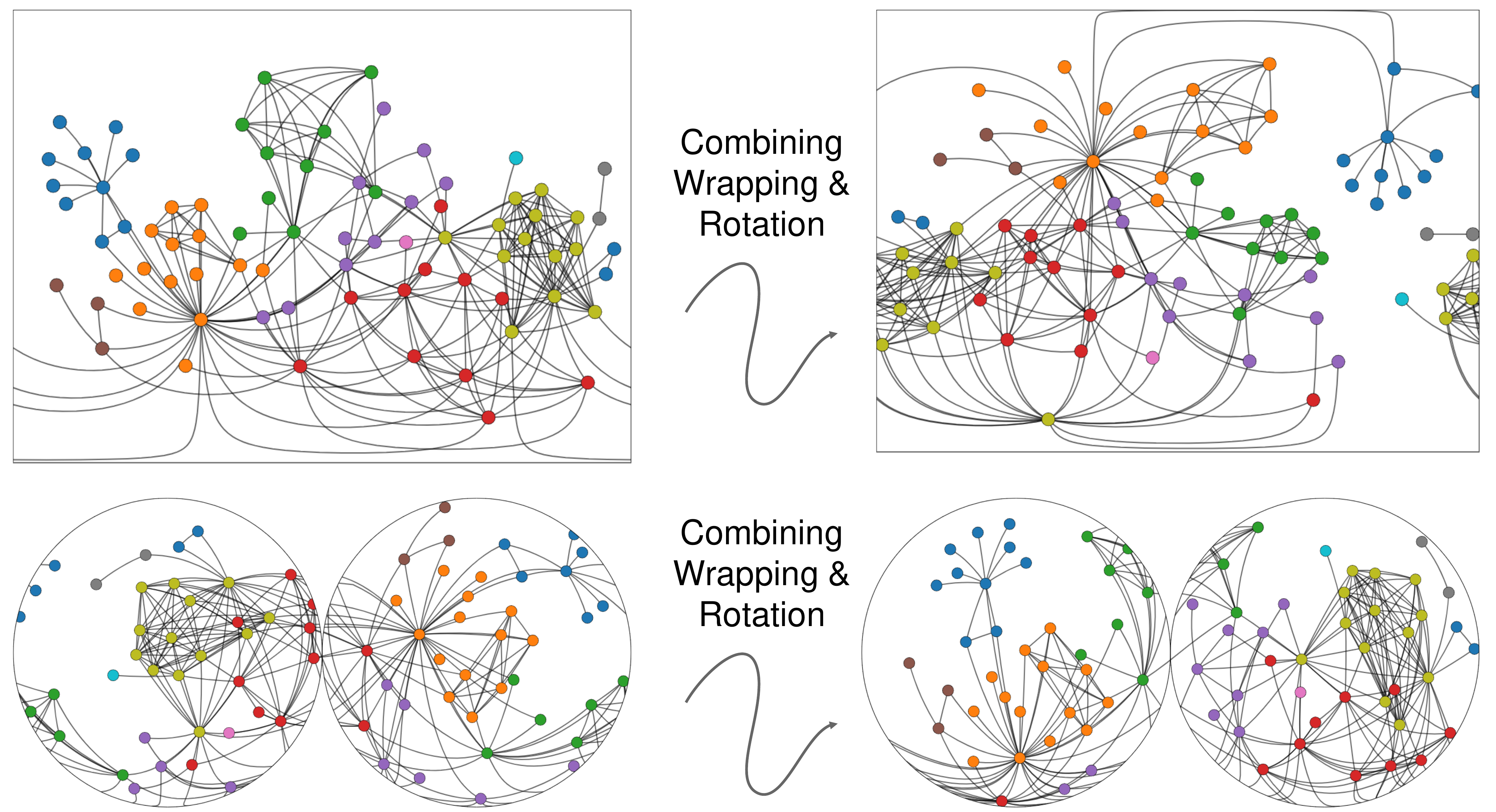}
	\caption{A rectangular (top: Miller) and non-rectangular (bottom: Mollweide Hemisphere) spherical projection for a network layout of co-appearances of characters (nodes) in \textit{Les Misérables} by Victor
Hugo~\cite{knuth1993stanford}, where edges indicate that two characters coexist in one scene and colours indicate group affiliation of the characters (nodes) in the novel. The network is laid out using our layout algorithm described in~\autoref{sec:spherevstorus}.}
	\label{fig:designspace:miller}
\end{figure}

\begin{figure}
	\includegraphics[width=0.9\textwidth]{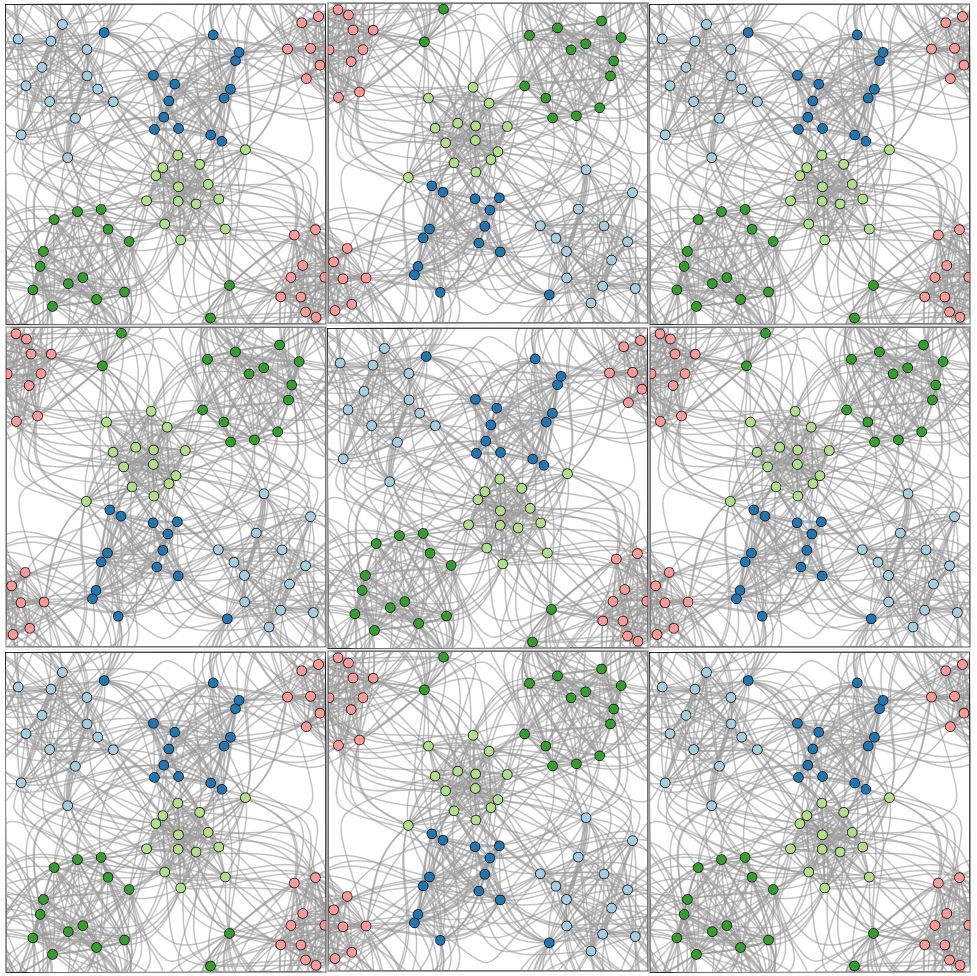}
	\centering
	\caption{Pierce\'s Quincuncial projection of ``Wrapping'' dimension in both spatial directions: horizontally and vertically repeated tiles. The repeated tiles show continuity of data points at the boundaries which preserves the data point movement wrapped across the boundaries.}
	\label{fig:designspace:quincuncialnetworktiles}
\end{figure}
Beyond geographic visualisations, spherical geometries have been used to create embeddings for node-link graph visualisations. This includes: generalisation of force-directed network layout methods over non-Euclidean surfaces (e.g., sphere) by Kobourov et al.~\cite{kobourov2008morphing}; laying out social networks on the surface of a sphere to provide better use of available space, using multidimensional scaling by Martins et al.~\cite{martins2012multidimensional} and Perry et al.~\cite{perry2020drawing}; 2D projections of spherical embeddings of networks without more centred or privileged nodes, compared to traditional flat node-link layout by Rodighiero~\cite{rodighiero2020drawing}; and force-directed implementation of 2D projected spherical network layouts by Manning~\cite{christophermanning:force}. 


Despite the work described above, spherical projections of networks are less common than their 3D representation in immersive environments, as described in~\autoref{sec:related}. When presenting node-link layouts in a 2D rectangular projection, e.g., Miller or Mercator, using our adapted spherical network layout algorithm (detailed in~\autoref{sec:spherevstorus}), links between nodes wrap across the boundary with data points between the left and right meet up, while the links at the top or bottom remain at top or bottom being horizontally mirrored. Those links appear greatly distorted (\autoref{fig:designspace:miller}-top). In non-rectangular projections (e.g., Mollweide Hemisphere, introduced in~\autoref{sec:related}), the links wrap around the circumference between the inner (and outer) side of the circles.

Compared to geographic data, a network has no inherent spatial structure. Therefore, wrapped panning for a network has no inherent constraint. The wrapping and rotation interaction could be applied freely in both horizontal and vertical directions. This allows the viewer to move any region of interest to the centre, as seen in~\autoref{fig:designspace:miller}.

To show the link continuity wrapped across the boundaries of a 2D sphere layout, a network can be represented with tile-display. For example, tile repetition of Quincuncial projection of a network shows link connectivity wrapped across the horizontal and vertical boundaries~\autoref{fig:designspace:spherenetworktiles}, while in Equirectangular (introduced in~\autoref{sec:related}) projected network layout it shows the link continuity wrapped horizontally (\autoref{fig:designspace:spherenetworktiles}).




\begin{figure}
	\includegraphics[width=\textwidth]{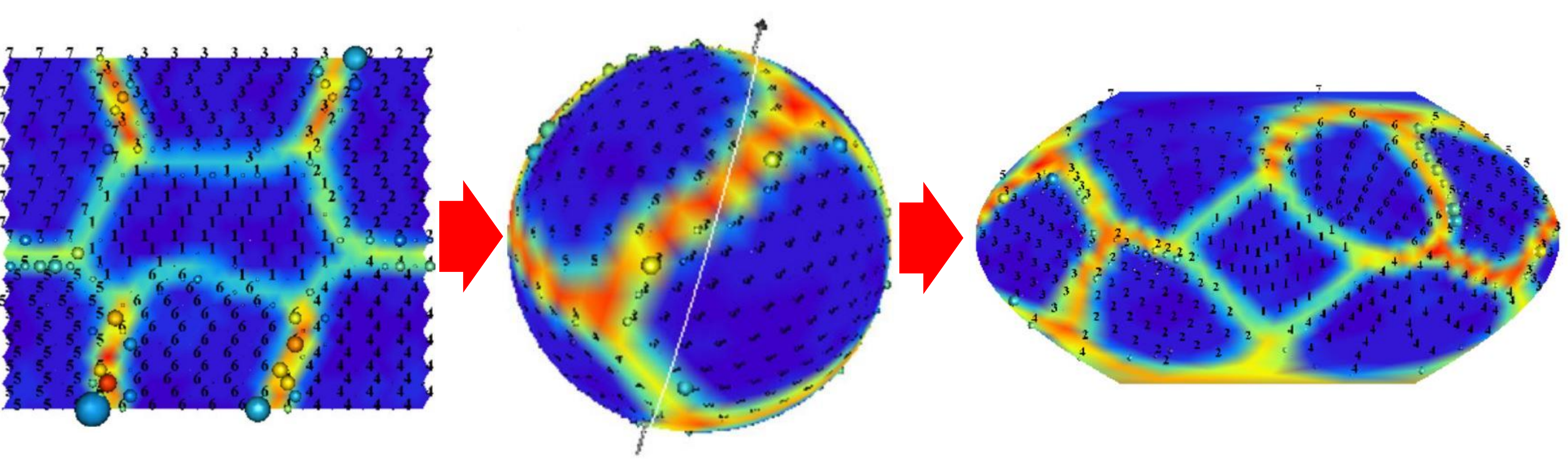}
	\caption{Traditional arrangement of SOMs on a 2D plane (left) are known to have a privileged centre (or ``border effect''). This can be alleviated by rendering on a continuous surface -- sphere (middle). The spherical SOM is then projected onto a 2D plane (right) where regions between the left-and-right boundaries meet up. The images are adapted from and credited to Wu et al.~\cite{wu2006spherical}.}
	\label{fig:designspace:som}
\end{figure}

Similar to node-link embeddings, spheres have been used to embed multidimensional scaling (MDS) plots for better capturing the data's underlying topology on the surface of a sphere such as international trade data between cities by Leeuw and Mair~\cite{de2009multidimensional}, and by Lu et al.~\cite{lu2019doubly}. Spheres have also been used to embed self-organising maps (SOMs) with the advantage that there is no privileged centre (i.e. as shown in~\autoref{fig:designspace:som}-left, the regions at the centre have more neighbourhood relationships and thus are updated more often) than those at the centre by Wu et al.~\cite{wu2006spherical}, shown on both 3D representations (\autoref{fig:designspace:som}-middle), and projected onto a 2D-plane (\autoref{fig:designspace:som}-right). 

\autoref{fig:designspace:som} shows a 3D globe representation of a SOM and its 2D pseudo-cylindrical projection showing that clusters between the left-and-right boundary wrap around~\cite{wu2006spherical}. Beyond node-link diagrams and SOMs, other information visualisations can potentially be projected onto a sphere such that there is no arbitrary edge to the display or privileged centre: multidimensional scaling by Papazoglou et al.~\cite{papazoglou2017examination} and Lu et al.~\cite{lu2019doubly} and their extension time curves~\cite{bach2015time}. However, we did not find any implementations or visualisations and their implementation might be non-trivial and thus to be studied in future. 


In~\autoref{sec:spheremaps} and \autoref{sec:spherevstorus}, we investigate the effectiveness of interactive spherical wrapping on maps and networks.

\begin{figure}
	\includegraphics[width=\textwidth]{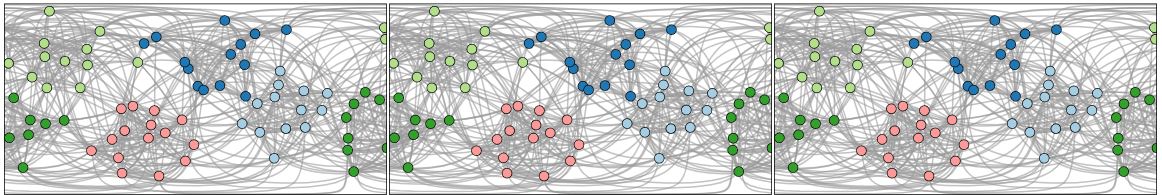}
	\centering
	\caption{Tile repetition of an Equirectangular projected spherical network layout.}
	\label{fig:designspace:spherenetworktiles}
\end{figure}

\section{Cylindrical Wrapping}
\label{sec:designspace:cylinder}
In this section, we motivate our design space of cylindrical wrapping by considering data types which are intrinsically cyclical (e.g., cyclical time series) which have no beginning or end but are often represented in standard linear bar charts or polar charts. 



\subsection{Motivation: wrap and rotate panning interaction}
\label{sec:designspace:cylinder:motivation}
\begin{figure}
	\includegraphics[width=\textwidth]{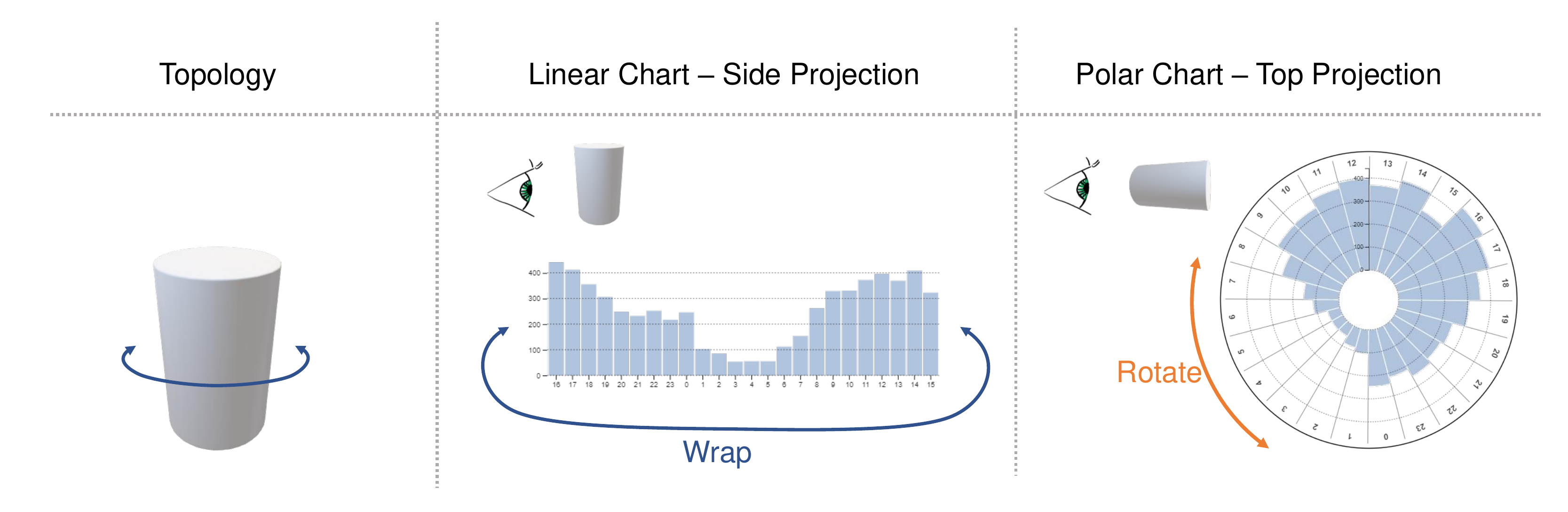}
	\caption{Rotate or Wrap: average traffic accidents per hour on Thursdays in Manhattan in 2016 (Troshenkov 2021), represented in a 2D projected cylinder with wrap (linear bar chart) and rotate (polar chart) interactions.}
	\label{fig:designspace:wrapchart_cylinder}
\end{figure}

As described in~\autoref{sec:intro}, the arrangement of cyclical data, such as time series in traditional linear bar charts, ignores the cyclic nature of the underlying data and thus may hinder identification of trends or value comparison of bars split across the left-and-right boundary (cut). Existing polar representation provides a common solution for showing time in a continuous manner without introducing the cut, but at the cost of reduced efficiency of data analysis performance.

In order to overcome respective limitations of linear charts and polar charts when representing cyclical time series, we propose bar charts with novel interactive wrapping and polar charts with interactive rotation.

An \textbf{interactive bar chart} is a linear bar chart (\autoref{fig:designspace:wrapchart_cylinder}-centre) where a user can drag horizontally using mouse or touch interaction to pan the visualisation, such that the chart wraps around from the left to the right and vice versa. This wrap interaction is intended to avoid the issue of bars at the extreme left of the chart being difficult to compare against bars at the extreme right, and making it difficult, for example, to detect trends that continue across this arbitrary cut. The user can simply bring such bars back together by re-centring.

An \textbf{interactive polar chart} is a polar bar chart like a static polar (\autoref{fig:designspace:wrapchart_cylinder}-right) while a user can drag on any part of the polar bar chart to rotate the chart. The chart spins around the centre in a clockwise or counterclockwise fashion as the user pans accordingly. This rotation interaction was motivated from the arrangement of bars in a polar chart that seemed easier to compare when they could be centred around the vertical centre line. When a pair of bars is centred in this way, the two bars have the same vertical baseline, meaning their heights can be compared directly. When they are off-centre, they are at an awkward angle. Rotation allows the user to centre the two bars they are interested in.

\subsection{Cylindrical projections}
The interaction techniques described in \autoref{sec:designspace:cylinder:motivation} can be thought of as stemming from the connectivity in cyclical data being mapped to a 3D topology, as per~\autoref{fig:designspace:wrapchart_cylinder}. For the data with a single cyclical domain, the underlying topology we consider for both linear and polar charts is a cylinder. Both the wrap and rotate interactions can be thought of as a rotation of the 3D cylindrical surface before projecting back to a 2D visualisation along one of two possible view angles of the cylinder, as follows. 

\noindent\textbf{Side projection: wrap interaction for bar charts} A cyclical bar chart can be wrapped around the cylinder such that the full time period covers the circumference of the cylinder, and the range of the bars representing values will extend to the height of the cylinder.  We can rotate the cylinder to choose the time interval that is centred in the field of view.  To provide a 2D view of the whole data domain, we can ``slice'' the cylinder on the side farthest from the viewer and flatten the resulting sheet, as per Fig.~\ref{fig:designspace:wrapchart_cylinder}-centre.  

\noindent\textbf{Top projection: rotate interaction for polar charts} For a polar chart, imagine shrinking the base of the cylinder such that it forms a cone, with the bars extending from the narrow base, outward to the wider top.  The polar chart is then simply a view of the cone with the point facing the viewer.  Rotation of the cone now changes which bar is centred at the top.  

\subsection{Cylindrical wrapping}
We now describe cylindrical wrapping that links to our design dimensions. In \textbf{Cylinders}, there is no movement from points from the bottom to the top, or vice versa, while points on the right and left of a visualisation plane wrap (Fig \ref{fig:designspace:teaser}-centre). 
Cylinder wrapping can be applied in any visualisations that can benefit from panning continuously along one spatial dimension: vertically or horizontally. 

\noindent\textbf{Constrained wrapping} For side projected cylinder (as described in~\autoref{sec:designspace:dimensions:projection}), wrapping direction is constrained naturally to only one dimension, but the orientation may be horizontal or vertical.
Most interactive Mercator projections, such as Google's 2D map, are in fact cylindrical wrappings as panning is allowed only horizontally (infinitely) (\autoref{fig:designspace:dimensions}-top).

\noindent\textbf{Rotational panning} For a top-projected cylinder, a user rotates the resulting circular visualisation, so the panning is rotational. 

\textbf{Tile-display} Similar to sphere topology, a visualisation can be shown with repeated tiles in horizontal way (\autoref{fig:designspace:mercatortiles} and~\autoref{fig:designspace:spherenetworktiles}).

In the following, we give further examples. We show how cylindrical topology leads to wrappable or rotatable visualisations and for which kinds of data wrappable visualisations make most sense.

\subsection{Cyclical time series}
We have seen examples of cyclical time series on bar charts and polar charts in~\autoref{sec:intro}. Such cyclical data with one periodic dimension (e.g., 24 hours over a day, or 12 months over a year) is suitable to be shown on a cylinder. Their interactive wrapping (for bar charts) or rotation (for polar charts) can be mapped to a cylinder with side or top projection.

\autoref{fig:intro:visualcut_cyclicaldata} shows hourly traffic accidents. Panning horizontally allows focusing on specific sections on the cycle, such as temporally (e.g., \texttt{Jan}-\texttt{Dec}, (left), or based on min and max values in the data (right). In~\autoref{fig:intro:visualcut_cyclicaldata}-upper left, having 0am on the left and 11pm on the right is completely arbitrary. If we want to look at the midnight (11pm, 0am and 1am) together, we can pan them so the wrap is elsewhere in the day (\autoref{fig:intro:visualcut_cyclicaldata}-upper right). 

\begin{figure}
    \centering
    \includegraphics[width=1\columnwidth]{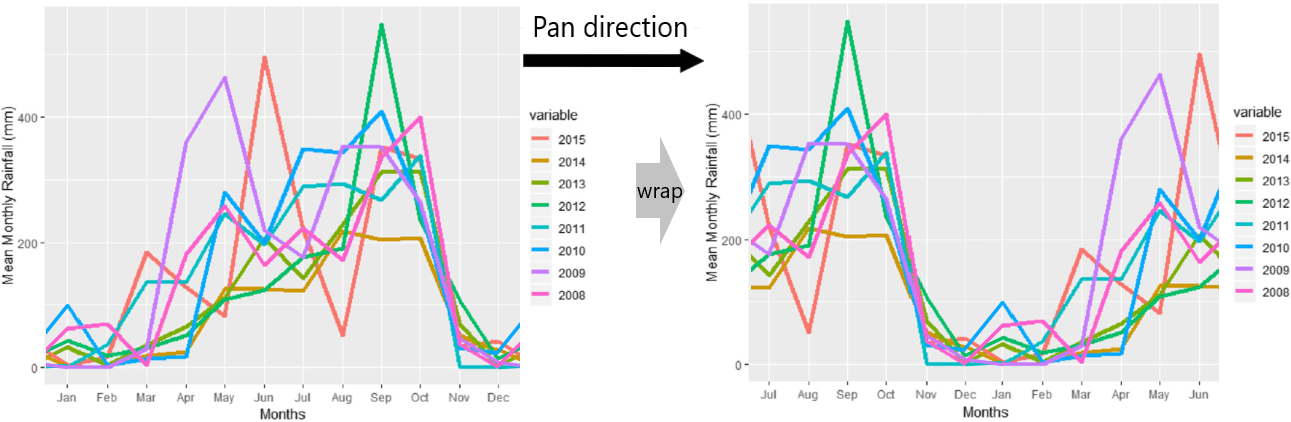}
    \label{fig:timeseries}
    \caption{Cylindrical interaction for standard visualisations of Bangkok's average monthly rainfall from 2008 to 2015: a time series wrapping horizontally and allowing to perceive patterns within individual time periods (e.g,. year.) Image is adapted from Jain et al.~cite{jain2019prediction} and credited to Jain et al.}
    \label{fig:designspace:cylinder:linechart}
\end{figure}

Apart from bar charts, cylindrical wrapping can be applied to other types of visuals in a straightforward manner. \autoref{fig:designspace:cylinder:linechart} shows a set of average yearly rainfall data shown as lines plotted across 12 months.

\subsection{Parallel coordinate plots}
\begin{figure}
    \centering
    \subfigure[PCP side-projection]{
        \includegraphics[width=0.5\columnwidth,trim=3cm 9cm 19cm 2cm,clip]{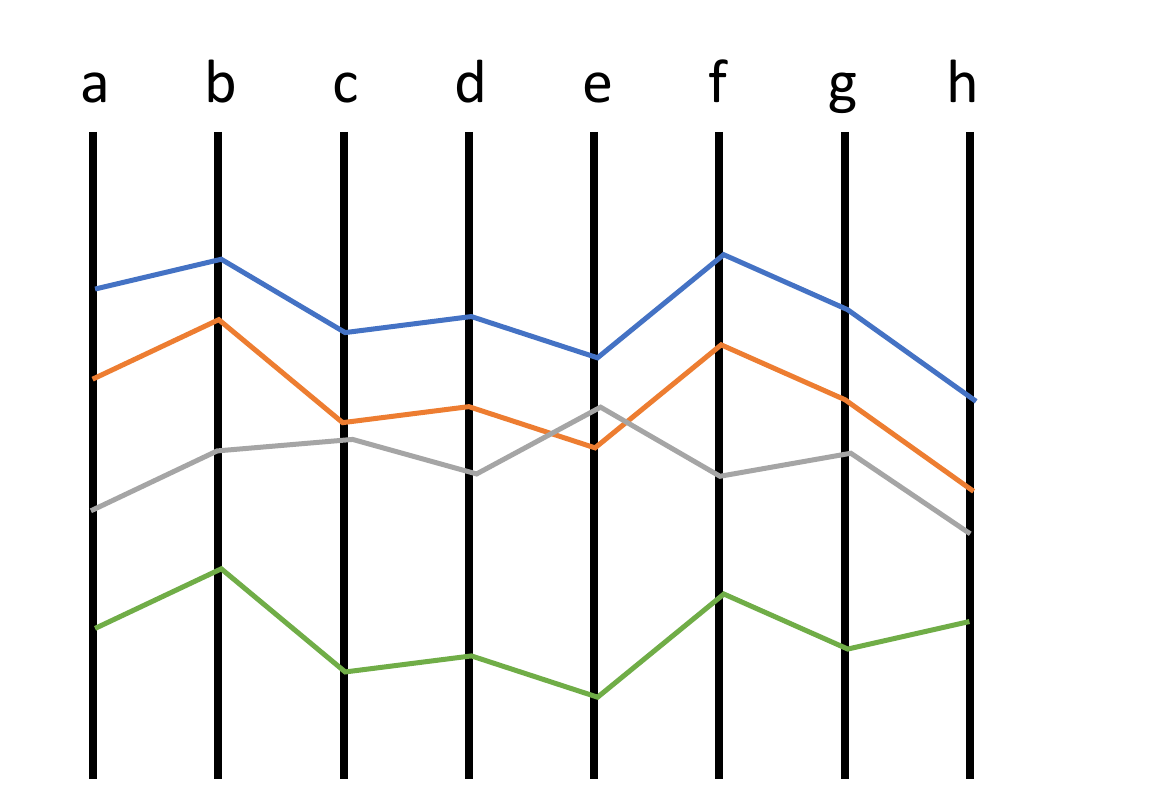}
    }
    \subfigure[PCP top-projection]{
        \includegraphics[width=0.4\columnwidth,trim=16cm 7cm 5cm 0.5cm,clip]{gfx/design_space/cylinder/radialpcps.pdf}
    }
    \caption{Cylindrical interaction for standard parallel coordinates plot (PCP) with cyclical dimensions in (a) side-projection; (b) the PCP in top-projection radial view.}
    \label{fig:pcps}
\end{figure}
Parallel coordinates plots are a common type of visualisation used to represent multidimensional data (i.e., mapping n-dimensional relations into 2D patterns), where the dimensions have a sensible cyclical ordering, can be mapped to the cylinder and be either side-projected (\autoref{fig:pcps}a), or top-projected in a radial form (\autoref{fig:pcps}b).

\subsection{Metro maps}
\begin{figure}
    \centering
    \subfigure[Two connected lines from the Singapore metro map, causing a cycle.]{
    \includegraphics[width=\columnwidth]{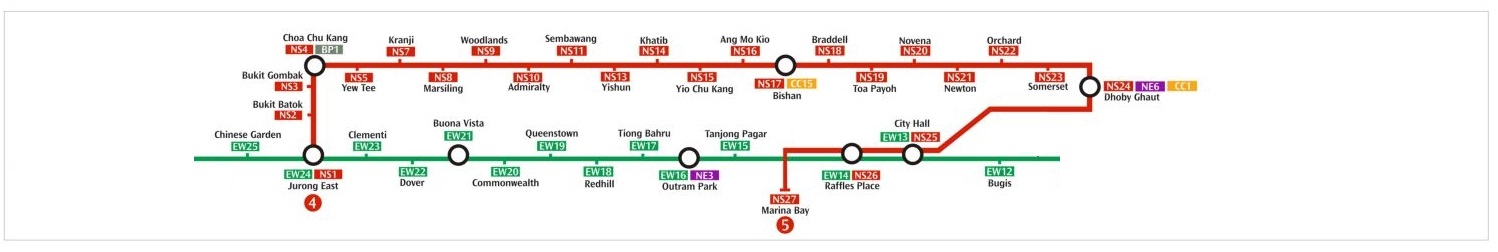}
    }
    \subfigure[The map from above, wrapped around a cylinder topology and side-projected.]{
    \includegraphics[width=\columnwidth]{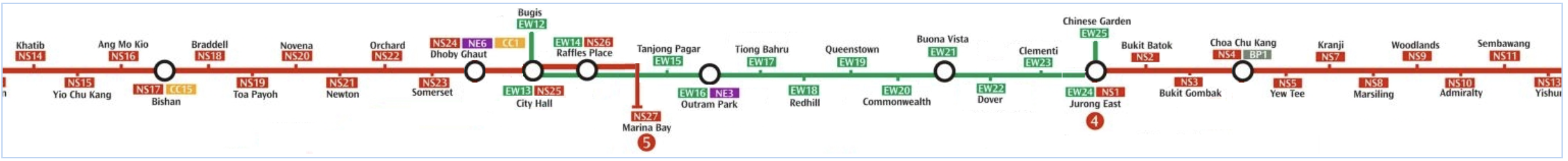}
    }
    \subfigure[Result of panning the map above to the left.]{
    \includegraphics[width=\columnwidth]{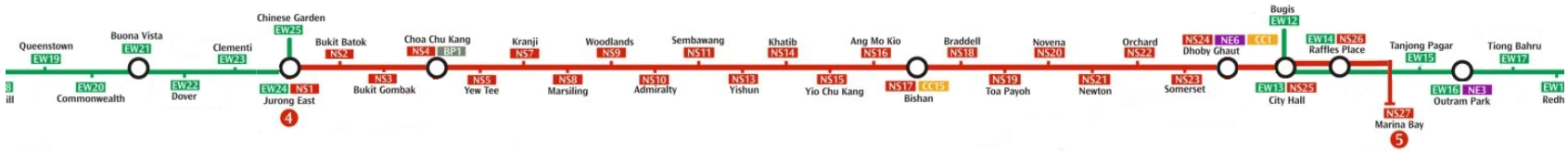}
    }
    \caption{Example of cylindrical wrapping: transportation map; image adapted from Singapore metro maps~\cite{singaporemetro}}
    \label{fig:transportline}
\end{figure}
The concept of interactive wrapping can equally be applied to other types of data.  For example, ~\autoref{fig:transportline}a shows a transportation map (two intersecting lines from the Singapore Metro) with a cycle. The transport map above shows a city loop interconnecting the red and green lines. Such a map often shows a cycle interconnecting individual lines. We show that by redrawing the map on a cylinder such that the loop is routed around the circumference, and then by a side-projecting, we can unwrap the cycle (\autoref{fig:transportline}b).  The projected view can then be endlessly panned left or right (\autoref{fig:transportline}c). If we want to look at the remaining stations of the red line before a transit, we can pan them so the red line starts from the left and wraps around. Furthermore, such a representation could be used on (e.g.) a circle line train to continuously show the stations ahead in the order they will be visited.
Such narrower maps can be shown more space-efficiently on, e.g., narrow static or digital displays above doors inside trains.

\subsection{Cyclical Sankey diagram}
\begin{figure}
    \subfigure[Cyclical links are circular.]{
        \includegraphics[width=.5\columnwidth]{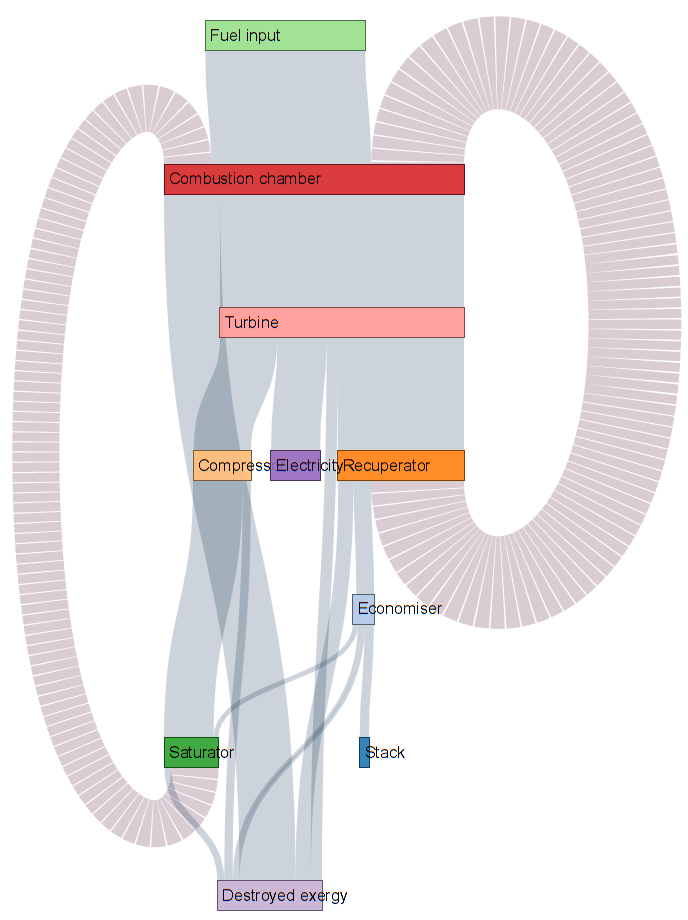}
        }
    \subfigure[Cylindrical projection to allow continuous vertical panning.]{
        \includegraphics[width=.4\columnwidth]{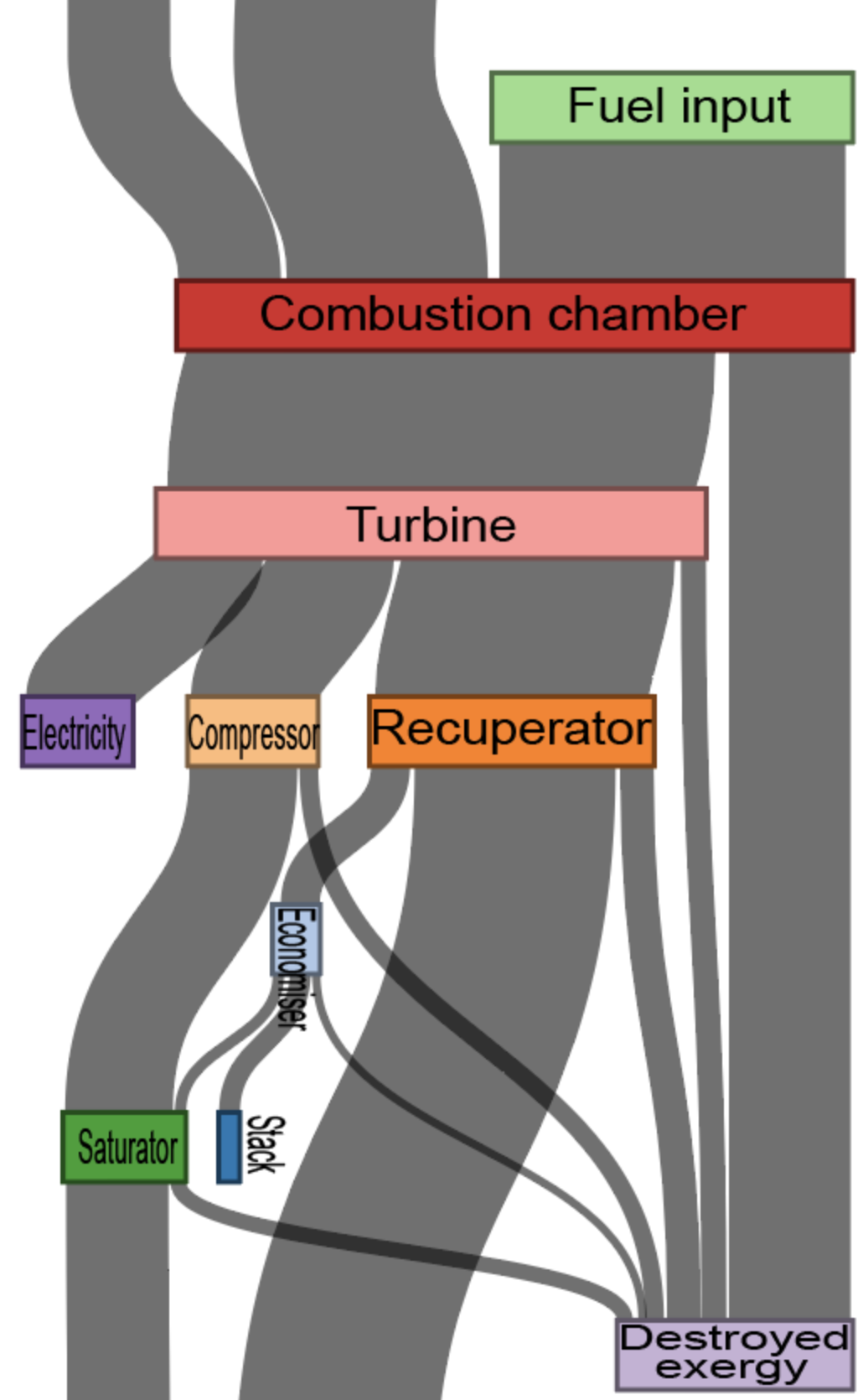}
    }    
    \caption{ Examples of cylindrical wrapping: a Sankey diagram that includes a cycle (left) can wrap vertically to highlight the continuous nature of flows (right).}
    \label{fig:sankey}
\end{figure}

In a similar (cyclical network) vein,~\autoref{fig:sankey} shows a Sankey diagram depicting thermodynamic analysis of water injection in a micro gas turbine\cite{CARRERO20171414}. 
Sankey diagrams are a type of node-link diagram with directed links (arrows) used to depict flows between entities in which the width of the arrows is proportional to the flow rate.
Originally developed by Matthew Henry Phineas Riall Sankey to demonstrate flow of energy in steam engines in 1898, Sankey diagrams have become popular in information visualisation for showing movements of data elements between different groupings.  For example, Google Analytics uses them to show click-through behaviour of users of web pages.  However, in such abstract information visualisation they are rarely depicted with cycles, possibly because they start to look messy, as in ~\autoref{fig:sankey}a\footnote{generated with a fork of the d3-Sankey software: \url{http://bl.ocks.org/soxofaan/bb6f91d57dc4b6afe91d}}.  In ~\autoref{fig:sankey}b, we redraw the same Sankey diagram on a cylinder topology, and project from the side to afford vertical panning.

\section{Toroidal Wrapping}
\label{sec:designspace:torus}
In this section, we extend the concept of one-dimensional wrap and rotate to consider 2-dimensions for visualisations that can be wrapped both horizontally and vertically. We first describe toroidal interaction techniques. Next, we propose novel uses of toroidal wrapping for cyclical time-series data that have multiple levels of periodicity respectively (e.g.\ weekly as well as daily, as shown in~\autoref{fig:designspace:wrapchart_torus}). Subsequently, we demonstrate examples of toroidal wrapping for networks represented by node-link diagrams or adjacency matrices~\cite{ghoniem2005readability} and high-dimensional data visualised by self-organising maps~\cite{ito2000characteristics,ultsch2003maps} or multidimensional scaling (as defined in~\autoref{sec:related}).


\subsection{Toroidal interaction techniques: combining wrapping and rotation}
\label{sec:designspace:torus:torustopology}
\begin{figure}
	\includegraphics[width=\textwidth]{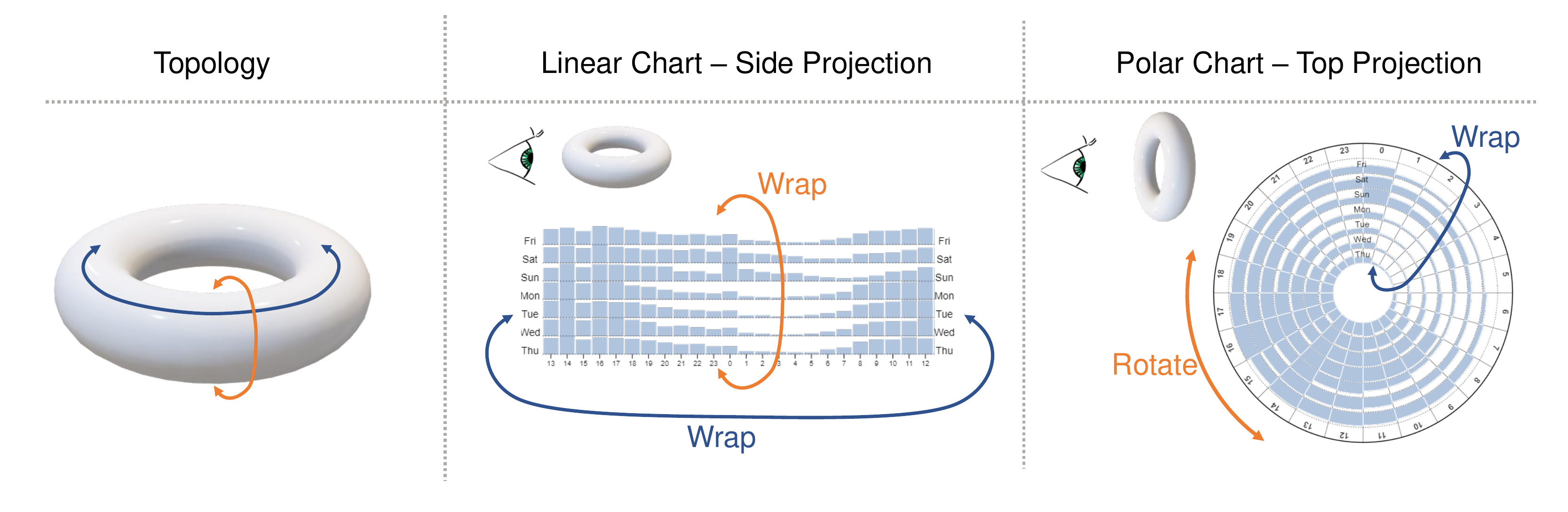}
	\caption{Rotate or Wrap: average traffic accidents per hour across the week in Manhattan in 2016; dataset taken from Troshenkov~\cite{hourlytrafficaccidentsdataset}.}
	\label{fig:designspace:wrapchart_torus}
\end{figure}
\textbf{Side projection: 2D wrapped bar chart arrays}
The concept of a visualisation that can wrap in two-dimensions is best understood, initially, by extending our earlier charts.  In ~\autoref{fig:designspace:wrapchart_torus}-centre, we further break down the hourly traffic accident data by day of the week.  Each bar still represents one hour, but there are now seven small multiple visualisations, one for each day.  Like the 24-hour day, the seven-day week is also cyclical.  It's conventional to display the week from Sunday to Saturday, with the weekdays centred and consecutive.  But what happens if an analyst particularly wants to compare Saturday activity against Sunday?  Here is where we can introduce a second (vertical) dimension of interactive wrapping, such that vertical mouse or touch drags cause the day order to wrap, while still allowing horizontal drags to wrap the hours.
Topologically, this can be seen as a side view of the torus, where now the hour dimension spans the long circumference of the torus, while the days progress around the circumference of a torus segment (as per~\autoref{fig:designspace:wrapchart_torus}-centre).

\textbf{Top projection: polar chart arrays that wrap and rotate}
In ~\autoref{fig:designspace:wrapchart_torus}-right, we introduce a new type of visual inspired by a top-down view of the torus.  The result is a polar visualisation where, as before, the disc can be rotated through a mouse or touch drag tangential to the disc to change the hour that is centred at the top.  In addition, a mouse drag outward from, or towards, the centre of the disc changes the day order. A top projection of a torus results in a polar visualisation, similar to top-down views on cylinder topologies. The difference is that panning can now happen in two ways: using rotation along one dimension and using wrapping to move elements from the \textit{inside} of the polar visualisation to its outside (\autoref{fig:torus-topdown}). This kind of panning can also overcome the common problem of polar visualisations that visual information at its centre is rendered smaller than on its outer boundary.

\begin{figure}
    \centering
    \includegraphics[width=1\columnwidth]{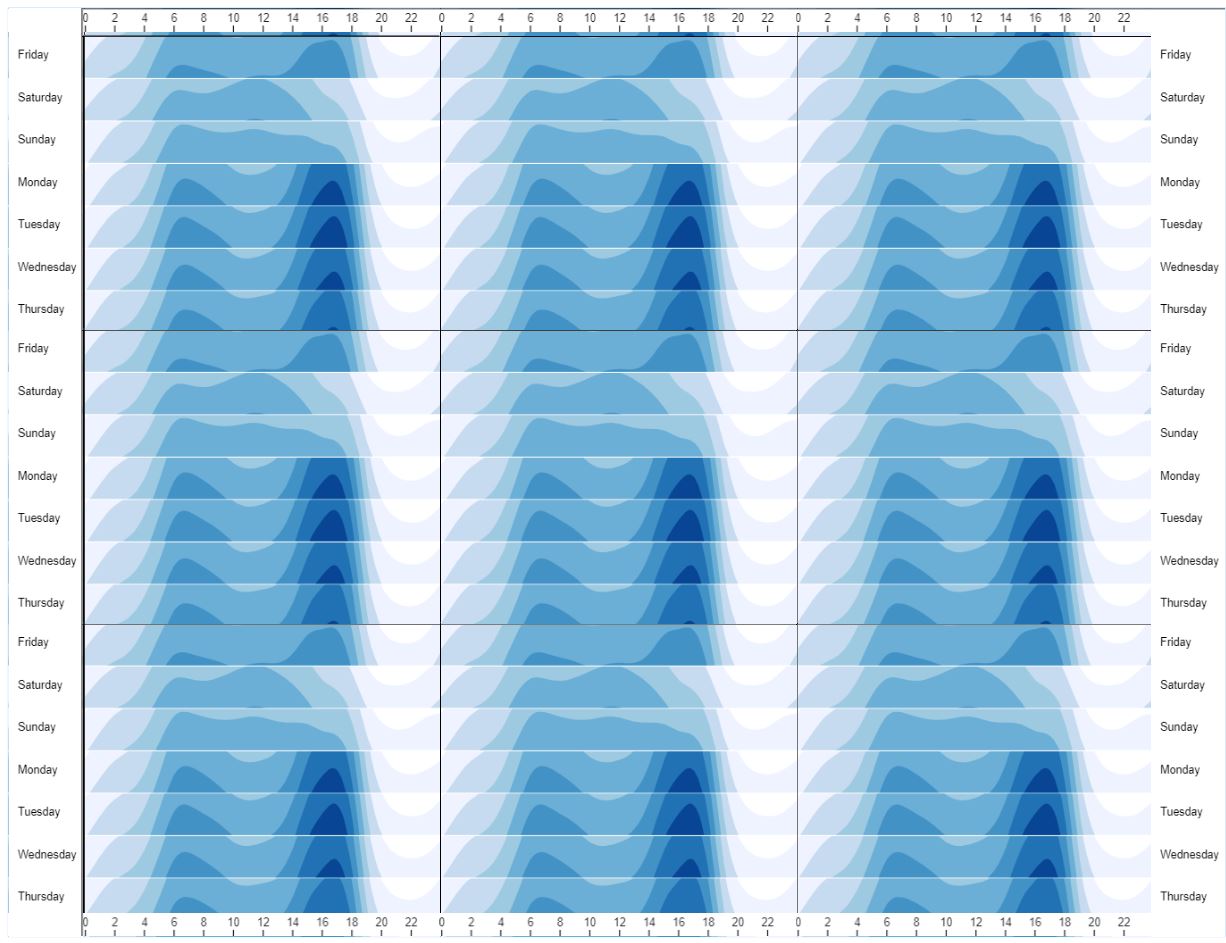}
    \caption{Example of cyclical time series of New South Wales traffic accidents (dataset taken from Roads and Maritime Services~\cite{nswtrafficaccidentsdataset}) represented on a 2D projected torus topology with $3\times 3$ tile display. Two periodic dimensions such as 24 hours of a day across 7 days a week repeat across the top, bottom, left, and right edges of the original chart (shown in the central tile) without repetition. It allows a viewer to observe continuous ranges of both dimensions across the chart's boundaries with any break in the temporal dimensions.} 
    \label{fig:torustiles}
\end{figure}

\subsection{Toroidal wrapping}
We now describe toroidal wrapping and link visualisations to our design dimensions. In \textbf{Toruses}, a 2D visualisation can be panned (wrapped) both horizontally and vertically (\autoref{fig:designspace:teaser}). 
In user interface design, the earliest example of such torus wrapping may be the classic Asteroids game, described in~\autoref{sec:related:torus}. Torus topologies have also been explored for mouse pointers on computer screens by Huot et al.~\cite{huot2011torusdesktop}.
However, in Information Visualisation torus topologies have not been systematically explored. Some algorithmic work has been applied to create self-organising maps (SOMs)~\cite{ito2000characteristics,ultsch2003maps} on a torus, but visualisation has always happens in a static plane. Our simple web tool (described in~\autoref{sec:designspace:tools}) allows us to create torus topologies for any static visualisation, given that axis labels and visualisation ``content'' are uploaded separately. 

\noindent\textbf{Constrained wrapping} While panning direction is constrained naturally in cylinder topologies to only one dimension (e.g.\ horizontally or vertically), on toroidal topologies, users can freely interact with two-dimensional panning, in both horizontal and vertical for a side projection. Such free panning allows for rapid navigation, however, in many visualisations, spatial dimensions have meaning and constraining interactions to one dimension at a time facilitates exploration. In addition, for some toroidal projections of certain data types, it makes sense to further constrain the pannability. For example, a symmetric matrix could be panned both horizontally and vertically, but it usually makes sense to keep the matrix diagonal centred.

\noindent\textbf{Wrapping combining rotation} For toroidal topologies, users can freely interact with rotational and radial panning for a top projection. 

\noindent\textbf{Tile-display}
The (static) tile replication for a torus layout provides a continuous view of graphical elements that are wrapped across the boundaries, as seen in~\autoref{fig:torustiles}. This allows a viewer to see the continuity of cyclical time series with two periodic dimensions. 


In the following, we give further examples. We show how toroidal topology leads to interactive wrappable visualisations, static tiled display, and for which kinds of data wrappable visualisations make most sense. 

\subsection{Cyclical time series}
\begin{figure}
    \subfigure{
    \includegraphics[width=1\columnwidth]{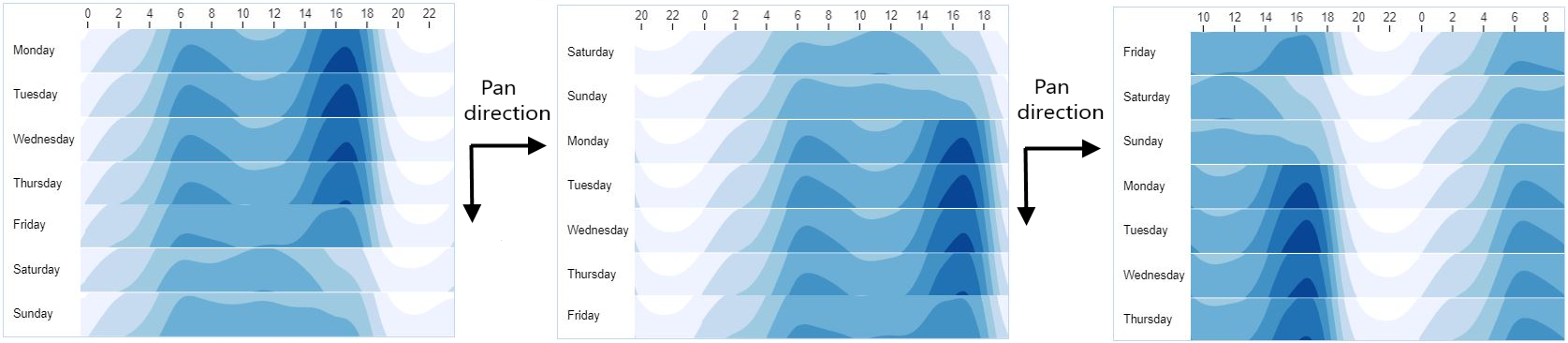}
    }
\caption{Example for torus horizontal and vertical torus wrapping, individually.}
\label{fig:horizon}
\end{figure}
\begin{figure}
    \centering
  \includegraphics[width=0.7\columnwidth]{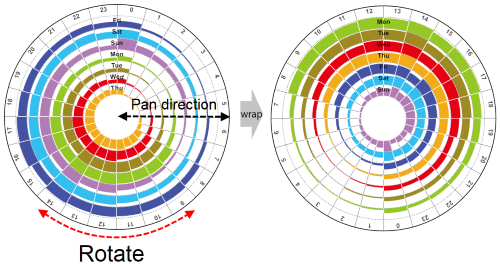}
    \caption{Example of a top-down view onto a torus topology, allowing to move elements from the centre of the visualisation to its outer boundary. Dataset taken from Troshenkov~\cite{hourlytrafficaccidentsdataset}}
    \label{fig:torus-topdown}
\end{figure}

The first example that shows temporal dimensions have meaning and that constraining interactions to one dimension at a time facilitates exploration is cyclical time series. For cyclical data with two periodic dimensions, the charts can be designed to be arranged in a 2D bar chart arrays described in~\autoref{sec:designspace:torus:torustopology}, which can be interactively wrapped around horizontally and vertically in a side-projected torus, or arranged in a 2D polar chart arrays that combine rotating in one dimension while wrapping in another, in a top-projected torus, as described in~\autoref{sec:designspace:torus:torustopology}.

\autoref{fig:horizon} shows an horizon graph~\cite{heer2009sizing} for aggregated time series data over days (vertically) and hours (horizontally). When represented on a side-projected torus, panning horizontally will cycle through the hours of the day (\texttt{0}-\texttt{24h}) and panning vertically will cycle through the days of the week (\texttt{Mon}-\texttt{Sun}). Panning in this example is restricted to one of these dimensions at a time to allow for exploring values across either days or hours without accidentally changing the context of the other dimension. For example, an analyst might be interested in exploring the exact time of peaks across days, and therefore requires the hour-dimension (horizontal) to remain fixed, while panning vertically through the days. The pannable version of this visualisation is created from our web tool. Detailed configuration is discussed in~\autoref{sec:designspace:tools}.

Take another example. \autoref{fig:torus-topdown} shows hourly traffic flow over seven days a week in New York City~\cite{hourlytrafficaccidentsdataset}, represented using our 2D  polar chart arrays, introduced in~\autoref{sec:designspace:torus:torustopology}. Such a dataset has been evaluated by visualisation literature on a static 2D plane by Waldner et al.~\cite{waldner2019comparison}. In a typical arrangement the hours are arranged analogous to a natural clock. However, to compare values between two bars or a range of bars in a day period (e.g., 7am to 6pm) one may need to turn the head around to mentally centre the bars (readability of polar charts compared to linear charts is discussed in~\autoref{sec:related} and evaluated in~\autoref{sec:cylinder}). In a top-projected torus, panning interaction is supported in two ways. A viewer can use rotation along the hour dimension to centre on the day period from 7am to 6pm for a more aligned comparison, while panning (wrapping) allows for moving, e.g., Monday from the inside of the polar visualisation to its outside, as seen in~\autoref{fig:torus-topdown}-Right, to provide an ordering starting from the first day of a week beginning, as opposed to a typical arrangement shown in~\autoref{fig:torus-topdown}-left. 

\subsection{Heatmaps}
\begin{figure}
    \centering
	\includegraphics[width=\textwidth]{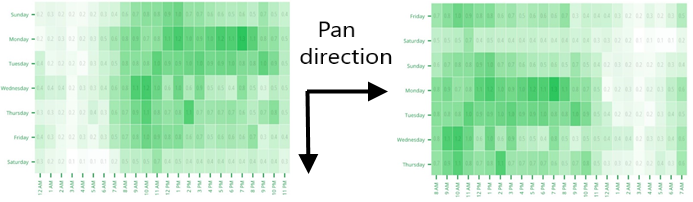}
	\caption{Average Reddit submissions by hour (horizontal) and day-of-week (vertical) represented on a heatmap can be laid out on a 2D side-projected torus. Images are adapted from average Reddit submissions~\url{https://www.reddit.com/r/dataisbeautiful/comments/1pe4vm/heatmap_of_all_link_submissions_to_reddit_which/}}
	\label{fig:designspace:torus:heatmaps}
\end{figure}

Similar to cyclical time series, heatmaps can also be used to represent cyclical time series with two periodic dimensions. The example in~\autoref{fig:designspace:torus:heatmaps} shows average Reddit submissions by hour (horizontal) and day-of-week (vertical). Since the data has two periodic dimensions, it is suitable to be laid out on a 2D side-projected torus. We specify images for bottom and left axes and no pan constraint.

\subsection{Adjacency matrices}
\begin{figure}
    \centering
    \includegraphics[width=1\columnwidth]{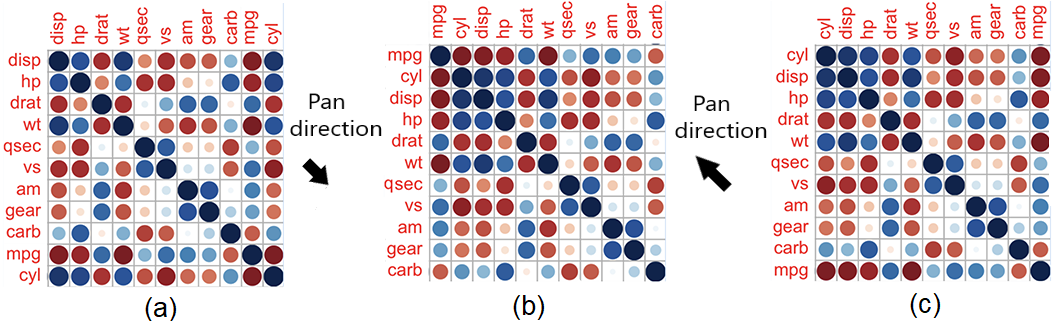}
    \caption{Effects of diagonal panning on a torus-topology matrix. Using the same row/column ordering: 
    (a) random start and end row/column shows two clusters (blocks along the diagonal), 
    (b) horizontal panning now shows a single cluster (top-left), and 
    (c) panning again highlights strongly correlated \texttt{mpg} row/column. Taken from ~\url{http://www.sthda.com/english/wiki/visualize-correlation-matrix-using-correlogram}}
    \label{fig:matrix}
\end{figure}

An example for a torus topology that uses constrained panning are adjacency matrices for network visualisation (\autoref{fig:matrix}). Some matrix orderings are in fact \textit{cycles} such as those based on the travelling salesman problem~\cite{behrisch2016matrix}. These matrix orderings make the matrix a torus since their start and end point have to be taken randomly from the permutation and matrices can be panned along the two spatial dimensions. Which of the elements becomes the first and last row is usually determined using external heuristics such as the highest-degree node, however, there is no given starting point. Choosing a random starting point from a permutation can lead to arbitrary patterns in a matrix. For example, the matrix in \autoref{fig:matrix}a seems to show two strong clusters, one on the top-left and one on the bottom-right. 
The torus panning for matrices therefore can serve two purposes: 
\textit{(a)} learning and communicating the idea of ordering rows and columns and 
\textit{(b)} exploring a specific ordering to avoid, e.g., overlooking cells at the margins of the matrix or misinterpreting clusters cut in half by the ordering (\autoref{fig:matrix}a).


For undirected networks, matrices are symmetric to the diagonal (usually top-left to bottom-right diagonal, if both rows and columns run left-right and top-down respectively). To preserve this important feature of symmetry and to keep viewers' mental map of the matrix preserved, we can constrain the panning so that user can pan only along the diagonal, i.e., the diagonal is fixed while vertical and horizontal panning happen at the \textit{same} time. 

Panning the matrix diagonally (\autoref{fig:matrix}b) reveals that the two clusters visible in \autoref{fig:matrix}a are in fact one large cluster situated at the top-left of the matrix. Panning further along the diagonal for just \textit{one} single row (\autoref{fig:matrix}c) highlights the \texttt{mpg} row/column by emphasising its strong connections (correlations) with any of the other nodes in the network. The same ``pan-configuration'' also highlights the red cells in the \texttt{mpg}-row/column, which depict negative correlations. In contrast, the configuration in  \autoref{fig:matrix}b visually emphasises the \textit{strength} of the correlation, rather than their \textit{type}. In summary, torus wrapping can help exploration and scrutinising data as well as finding the views most appropriate for a given task or message. 

\begin{figure}
    \centering
	\includegraphics[width=0.9\textwidth]{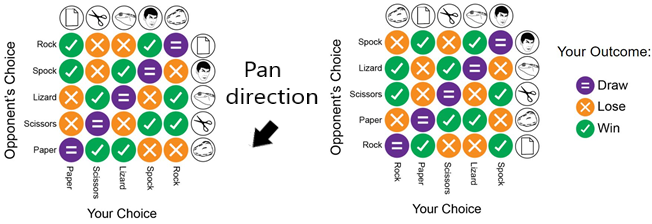}
	\caption{A directed graph with the special self edges (the draws) mapped to the anti-diagonal. An ``antidiagonal'' pan constraint keeps those draws where we can easily track them. Adapted from A Rock, Paper, Scissors, Lizard, Spock Chart In R. Images Wikimedia: Creative Commons~\url{https://upload.wikimedia.org/wikipedia/commons/8/89/Rock_paper_scissors_lizard_spock.svg}}
	\label{fig:designspace:torus:antidiagonal}
\end{figure}

A matrix can also be panned along the anti-diagonal direction. \autoref{fig:designspace:torus:antidiagonal} shows a directed graph with the special self edges (the draws) mapped to the anti-diagonal. An ``antidiagonal'' pan constraint keeps those draws where we can easily track them.

\subsection{Node-link diagrams}
\begin{figure}
    \centering
    \includegraphics[width=1\columnwidth]{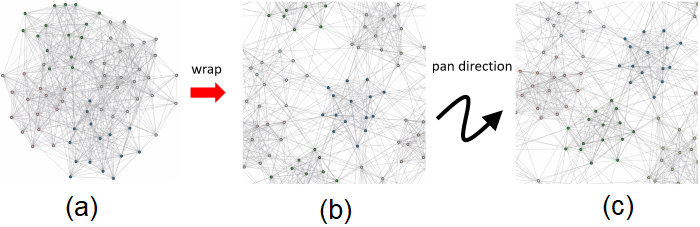}
    \caption{Examples of torus panning on a network using the random partition network data set~\cite{fortunato2010community} with already known clustering information:
    (a) traditional (unwrapped) network on a 2D-plane; (b, c) two configurations wrapped network laid-out on a 2D torus, using the algorithm presented in~\autoref{sec:torus2}.} 
    \label{fig:torusnetworks}
\end{figure}
\begin{figure}
    \centering
    \includegraphics[width=0.9\columnwidth]{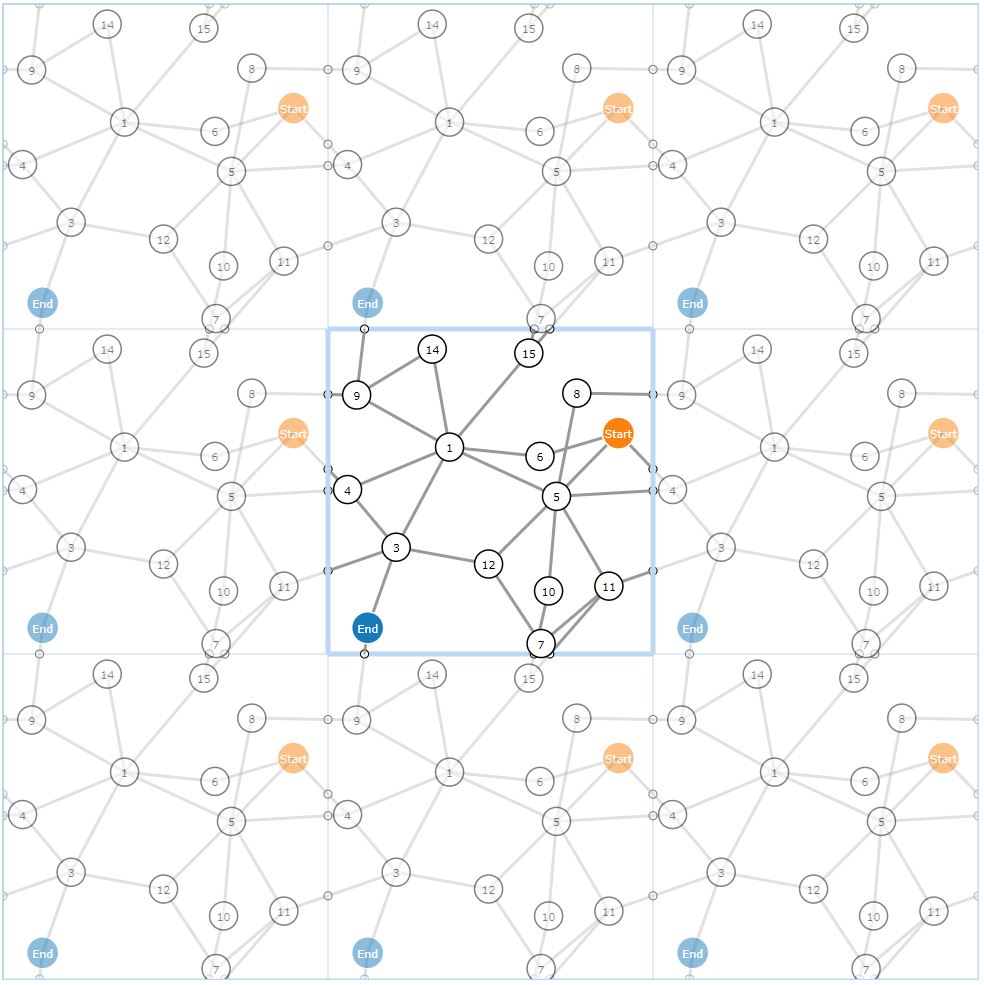}
    \caption{Example of tile repetition in a torus embedding of a network, laid out using our layout algorithms described in~\autoref{sec:torus2}.} 
    \label{fig:torusnetworktiles}
\end{figure}
A node-link diagram of a network can be embedded onto a torus and side-projected onto a 2D plane, while wrapping specific links around the borders of the display. An illustrative example is shown in~\autoref{fig:intro:torusnetwork}, discussed in~\autoref{sec:related:torus}.
While wrapping some 2D visualisations, such as charts, heatmaps, matrices, or horizon graphs is straightforward, wrapping a network layout on a torus needs to optimise their node positions in a torus topology to better show the patterns or relationships within the data.

\autoref{fig:torusnetworks} shows a larger and denser network example with community structures (defined in~\autoref{sec:related:networks}) laid out in node-link diagrams. In (a) it shows that a traditional non-wrapped node-link representation of the network appears more tangled with many links crossing the diagram, making it difficult to inspect the data's underlying structure (investigated in~\autoref{sec:torus2}. In (b) the same network is laid out in a 2D side-projected torus using our torus wrapping algorithm detailed in~\autoref{sec:torus2}. It shows that torus layouts make clusters appear more distinct. However, some clusters may be split across the edges of the diagram. In (c) it shows such torus-based network layout is freely pannable in both horizontal and vertical directions without any panning constraint. When a cluster is panned off on the left side of the diagram, it reappears on the horizontally right position, and vice versa. When a node or link is panned off on the top boundary of the diagram, it reappears on the vertically bottom position, and vice versa. This allows a viewer to interactively wrap the network to bring any region of interest to the centre (e.g., to see full clusters).

A torus-based network can also be illustrated with a tile display, with replicated network layouts shown on the edges of the central highlighted diagram to show the link continuity wrapped across the boundaries. For example, \autoref{fig:torusnetworktiles} shows tile repetition of a network which shows link connectivity wrapped across the left-and-right and top-and-bottom sides of the original layout.

In~\autoref{sec:torus1} and~\autoref{sec:torus2}, we explore layout algorithms that can produce the drawings of above examples, and investigate whether such tiled display (\autoref{sec:torus1}) and the aforementioned interactive wrapping (\autoref{sec:torus1} and~\autoref{sec:torus2}) improve the usability of torus-based network layouts.

\subsection{Self-organising maps}
\begin{figure}
    \centering
	\includegraphics[width=0.5\textwidth]{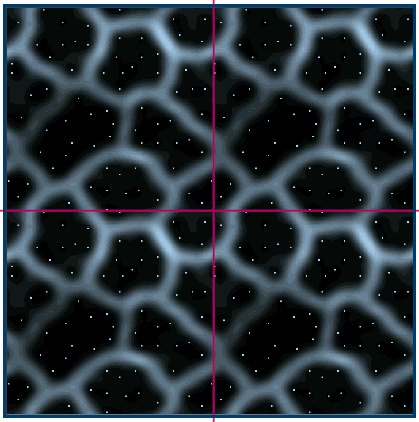}
	\caption{SOM rendered on a tiled 2D toroidal layout by Ultsch~\cite{ultsch2003maps}. Image is credited to Ultsch.}
	\label{fig:related:tiledsom}
\end{figure}
An illustrative example of self-organising maps is shown in~\autoref{fig:related:tiledsom} where tiled repetition is shown that preserves the connectivity of the graphical elements split across the left-and-right, or top-and-bottom boundaries by Ultsch~\cite{ultsch2003maps}. SOM is also used for preserving the topological structure of multidimensional data in lower dimensions by Kohonen~\cite{kohonen1982self}.

\subsection{Multidimensional scaling}
\begin{figure}
    \centering
    \includegraphics[width=1\columnwidth]{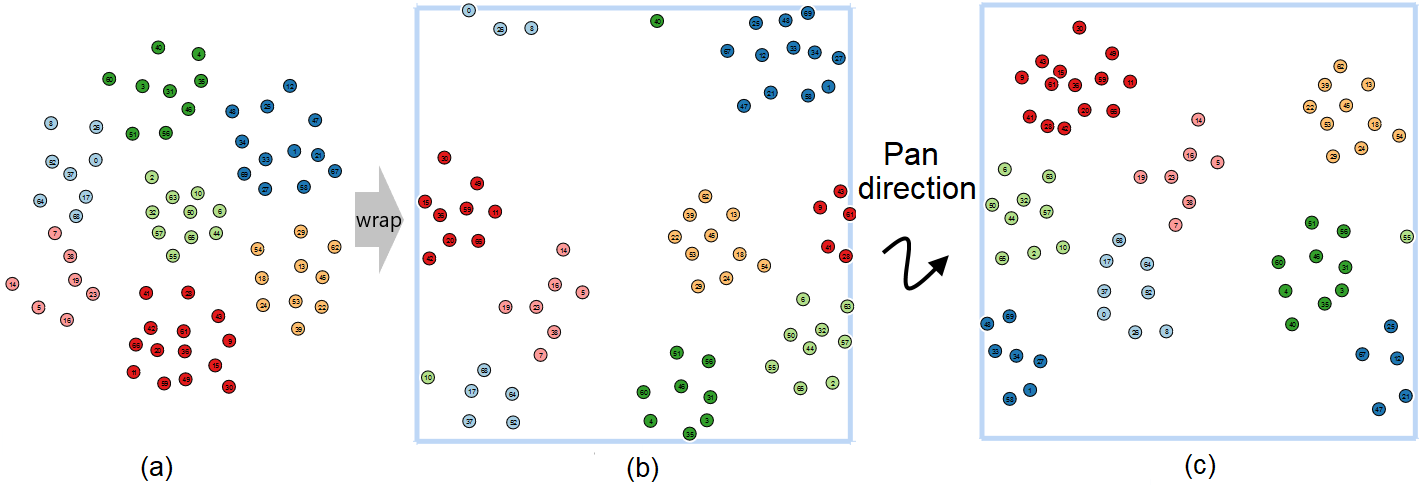}
    \caption{Examples of torus panning on an MDS using the Hepta data set~\cite{thrun2020clustering} with already known clustering information:
    (a) traditional (unwrapped) MDS plot on a 2D-plane; (b, c) two configurations wrapped MDS plot laid-out on a 2D torus, using the algorithm presented in~\autoref{sec:torus2}. Datasets was adapted from Benchmark data set from Fundamental Cluster Problem Suite (FCPS).}
    \label{fig:mds}
\end{figure} 

A MDS plot can be laid-out on a torus topology. This example (\autoref{fig:mds}) shows a MDS plot based on the hepta data set. The 3-dimension attributes can be converted into a 2D similarity matrix between all pairs of nodes. The diagrams in the left column show an unwrapped 2D MDS plot of 50-70 points, 7 classifications using a standard force-directed layout method. The diagrams in the right column show the same plot laid-out on a torus, using the same algorithm we developed in~\autoref{sec:torus2}. \autoref{fig:mds} shows a multidimensional scaling (MDS) calculated on a torus topology. ~\autoref{fig:mds}(b) and~\autoref{fig:mds}(c) show our torus layouts in two different panning configurations (free vertical and horizontal panning, no panning constraints). 

\section{Web Tools for Creating Wrappable Visualisations}
\label{sec:designspace:tools}
\begin{figure}
    \centering
	\includegraphics[width=0.7\textwidth]{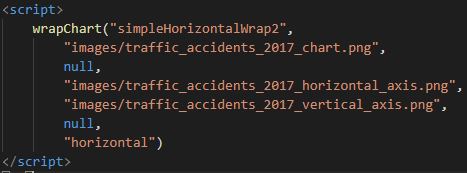}
	\caption{Interactive tool library for one-dimension (cylindrical) wrapping or two-dimensional toroidal) wrapping of visualisations}
	\label{fig:designspace:wrapchart_api}
\end{figure}

To create the wrappable visualisations described and figured in previous sections, we implemented a small web tool\footnote{\url{https://github.com/Kun-Ting/WrappingChart}} that can present arbitrary 2D visualisations with one- or two-dimensional wrapped panning and allows for the selection of panning constraints in accord with the topology (cylinder or torus). 
The tool, called \texttt{WrappingChart} is a small JavaScript library which allows web developers to easily create interactive wrap-around visualisations like those depicted in this chapter from their own static images of charts or other visuals.
The user can prepare separate images for the visualisation and optional axes-labels that remain fixed at the sides while the visualisation is being panned. Example is shown~\autoref{fig:designspace:wrapchart_api}.


The API and parameters are described below.
The user can upload separate images for the visualisation and optional axes-labels that remain fixed at the sides while the visualisation is being panned.
The tool was developed using Typescript.
\begin{itemize}
    \item HTML element that hosts the interactive visualisation (\texttt{targetElementSelector})
    \item chart body (\texttt{bodyImageURL})
    \item image URL of x-axis (e.g., tick marks, bars, labels) at the top boundary (\texttt{xAxisTopImageURL})
    \item image URL of x-axis at the bottom boundary (\texttt{xAxisBottomImageURL})
    \item image URL of y-axis at the left boundary (\texttt{yAxisLeftImageURL})
    \item image URL of y-axis at the right boundary (\texttt{yAxisRightImageURL})
    \item constrained panning: \texttt{panConstraint} (possible value: horizontal, vertical, diagonal, antidiagonal)
\end{itemize}

\begin{figure}
    \centering
	\includegraphics[width=0.5\textwidth]{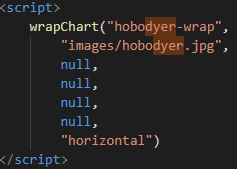}
	\caption{Cylindrical wrapping of Hobo-Dyer projection}
	\label{fig:designspace:wrapchartexample_maps}
\end{figure}

\textbf{Geographic visualisation} For example, in Mercator projection of the world map (\autoref{fig:designspace:dimensions}-top), to have poles remain at the north and south and to keep adjacencies of continents on the earth's surface correct, only horizontal panning makes sense. We call our \textit{wrapChart} library (with a ``horizontal'' pan constraint to limit the panning (\autoref{fig:designspace:wrapchartexample_maps}).

\textbf{Hourly traffic flow over 24-hour period} 
\begin{figure}
    \centering
	\includegraphics[width=0.7\textwidth]{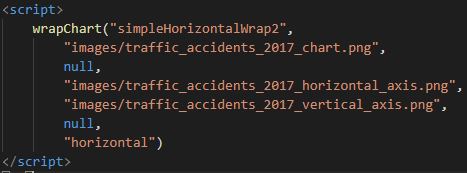}
	\caption{Interactive cylindrical wrapping configuration of cyclical time series data using our tool}
	\label{fig:designspace:wrapchart_timeseries}
\end{figure}

To create wrappable visualisations as shown in~\autoref{fig:horizon}, we can upload the images of the bar charts and x-axis, y-axis images of hourly traffic flows over months or years. We call wrapChart with a ``horizontal'' pan constraint to limit the panning (\autoref{fig:designspace:wrapchart_timeseries}).

Our small library supports interactive panning of data visualisations. While we develop separate tools for creating tile-display and interactive rotation for our cylindrical and toroidal design space, in future, we consider integrating them as options: “rotate”, “wrapandrotate”, “sphericalrotate” (“sphericalrotate” needs to specify the projection type as wrapping differs accordingly). We will also add a new boolean variable to turn on/off tile-display. The tile repetition is automatically computed based on Constraint.

For cylindrical topology, we also implement a similarly novel `rotate' gesture for polar charts, which allows the user to bring any bar to the top or centre bars for more aligned comparison. 



\section{Conclusion}
\label{sec:designspace:conclusion}

In this chapter, we present a design space exploration for a novel class of interactive wrapped visualisations. The outcomes of this exploration characterise \rev{data types whose visualisations can potentially benefit from being} wrapped around 2D surfaces of 3D shapes for projection back to a 2D interactive wrap-around display.  They may benefit from being understood as `cylindrical’, `toroidal’, or `spherical’, addressing \textbf{RG1} (\autoref{sec:intro:RGs:designspace}). This design space exploration leads us to develop novel interaction techniques for not only geographic data, but also cyclical time series, multidimensional and network structured data. The design space also contributes to a number of caveats and design guidelines for future applications. 



We showed that the design space can be a powerful tool to: \textit{(a)} understand relations between visualisations (e.g., linear and polar bar charts), \textit{(b)} apply interactive wrapping to a range of visualisations, and \textit{(c)} to create new visualisations by mapping existing visualisations onto sphere, cylinder and torus visualisations. 


We presented a simple interactive web tool that is capable of wrapping arbitrary data visualisations around 2D projected cylindrical and toroidal topologies with support for constrained panning, addressing \textbf{RG2} (\autoref{sec:intro:RGs:tools}).  The JavaScript source code for our library API is available to web developers to easily create their own wrap-around data visualisations.

In the following chapters, we present design, development and evaluation of cyclical, toroidal, spherical wrapping for cyclical time series, networks, geographic maps, showing the benefits of interactive wrapped visualisations compared with standard unwrapped representations.


%
\chapter{Cylindrical Wrapping of Cyclical Data}
\label{sec:cylinder}

\cleanchapterquote{The interactive charts were significantly easier than the standard charts. Being able to move the chart and line up the areas instead of having them go off screen was a great advantage. By far, the hardest was the standard circle chart. I found myself having to really turn my neck to get a good look at some of them.}{Anonymised user study participant, P2}{Trend identification tasks}

Inspired by the observations in our design space (\autoref{sec:designspace}), this chapter demonstrates the perceptual benefits of interactive (pannable) wrapped visualisation over standard visualisations. We start with cylindrical topology that is suitable for data that is cyclic in one temporal or spatial dimension. Examples include average rainfall over 12 months, hourly traffic accidents rate, or average wind strength over different compass directions.


This chapter investigates the trade-off between the continuity provided by polar visualisations versus the ease of height comparison on linear charts, and---most noteworthy---the use of novel (cylindrical) interactions for polar and linear chart representations of cyclical data to work around their respective limitations.




\textbf{Specifically, we address the following open questions (linked to RG3 in \autoref{sec:intro:RGs:cylinder})}: 
\begin{itemize}[noitemsep,leftmargin=*]
    \item \textbf{[RQ3.1]}: Does adding interactive wrapping to linear bar charts improve their effectiveness?
    \item \textbf{[RQ3.2]}: How effective is interactive wrapping for linear bar charts compared with static polar bar charts?
    \item \textbf{[RQ3.3]}: Does adding interactive rotation to polar bar charts improve their effectiveness?
\end{itemize}

To investigate these questions, we report on a controlled user study with 72 participants comparing four different visualisations of cyclic data: 
(a) linear bar chart (without interactive wrapping); 
(b) linear bar chart with interactive wrapping;
(c) polar bar chart (without interactive rotation); and
(d) polar bar chart with interactive rotation. 



\section{Empirical Evidence}
\label{sec:cylinder:empirical}
Several studies  have investigated error performance with bar charts, finding that people are more accurate at comparing \textit{adjacent} bars than bars that are further apart by Cleveland et al.~\cite{cleveland1984graphical}, Heer et al.~\cite{heer2010crowdsourcing}, and the reproduced version of Cleveland et al.'s in four experiments by Talbot et al.~\cite{talbot2014four}. 
Other studies of comparison of cyclical data, in particular, temporal data (hourly, weekly, or monthly) with static linear or polar bar charts have been discussed in~\autoref{sec:related:tradition:temporal}. 

To cope with the complexity of temporal data, interaction has been investigated. For example, pan and zoom along a linear timeline by Schwab et al.~\cite{schwab2019evaluating}, mouse hover and showing hints on a linear chart or circular chart by Adnan et al.~\cite{adnan2016investigating}, and a study of interaction and single-scale selection techniques on a linear timeline by Schwab et al.~\cite{schwab2021evaluation} have been presented.  However, we are not aware of wrapped panning (as defined in Sec.\ \ref{sec:techniques}), being evaluated with cyclical time-dependent data.

In our work, we investigate interactive panning as one possibility to wrap a linear visualisation. This notion is different from other notions of wrapping in visualisations, such as ``\textit{DuBois Wrapping}'' by Karduni et al.~\cite{karduni2020bois}, which wraps oversized bars in bar charts into zig-zag lines.

In summary, while visualisation tasks described in this chapter have been previously tested in literature (described in~\autoref{sec:related:tradition:temporal}), such as trend detection~\cite{adnan2016investigating, zacks1999bars}, pairwise group comparison~\cite{adnan2016investigating}, and pairwise single value comparison~\cite{waldner2019comparison}, no study has particularly focused on bars far apart but which could be brought closer through wrapping. Our contribution to this corpus of knowledge is the particular study of interaction for wrapping in linear bar charts and rotation in radial charts. This task suggests looking at intervals across both ends of a linear bar chart as well as bars that are `far apart' in both chart types that could be brought `closer together' in interactive linear bar charts.  






\section{User Study 1: Wrapped Time Series Readability}
\label{sec:cylinder:userstudy}
\begin{figure}
\centering
	\includegraphics[width=\textwidth]{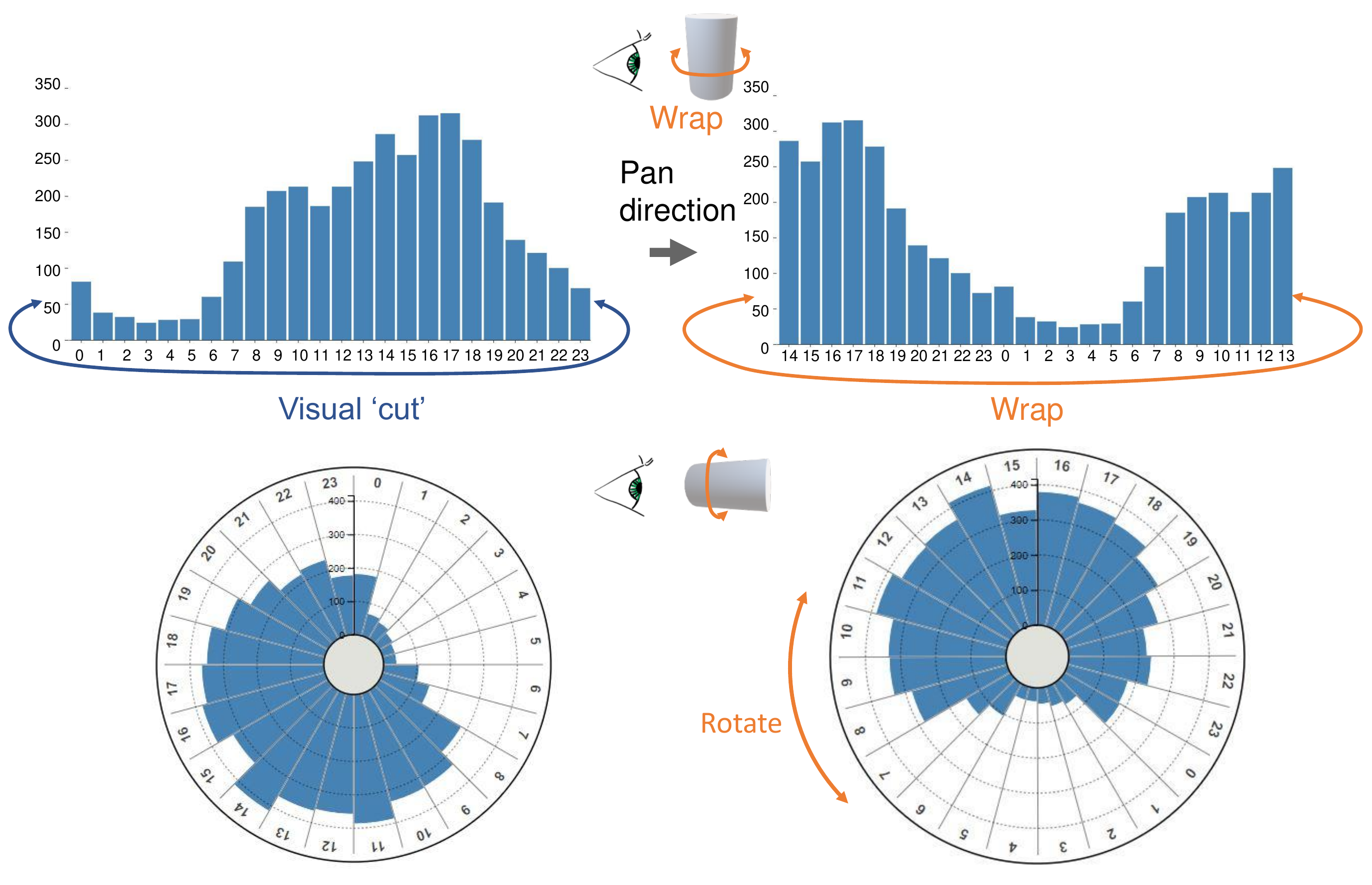}
	\caption{Bar chart and polar chart representation of hourly traffic flow over a 24-hour period in NYC in 2016~\cite{hourlytrafficaccidentsdataset}. To inspect whether the average hourly traffic accidents continue to increase or decrease from midnight till morning (upper-left), a user may need to mentally wrap the bars from 11pm at the right end across the page to 1am at the left end. Polar chart (lower-left) arranges the sequence of bars continuously, but certain sequences of bars (12pm to 5pm) are not aligned at the centre and thus a user may need to spin their head to make comparisons. Interactive wrapped and rotatable visualisations proposed in this thesis (upper-right, lower-right) allow a user to centre any group of bars while the visual cut is elsewhere in other hours (upper-right) and to have the same vertical baseline (lower-right) for comparing heights of bars.}
	\label{fig:cylinder:cylinder:timeseries}
\end{figure}

This section reports on a controlled user study with 72 participants investigating the performance of linear and interactive wrapped charts to understand time series data. The study aims to answer our questions Q1 to Q3 stated earlier. Apart from the interactive wrapping, our study differs from past work by focusing on tasks that require considering multiple bars: pair comparisons and aggregation across intervals (trend identification and average value estimation), as we expect these to be the tasks most affected by wrapping in bar charts and rotational centring in polar charts.

The full study material, illustrative examples of interactive wrapping or rotation are available in the Open Science Foundation: \url{https://osf.io/r8cw4/}.

\subsection{Techniques}
\label{sec:cylinder_techniques}
In our study, we compare four techniques for the visualisation of cyclical data: \textit{static bar}, \textit{interactive bar}, \textit{static polar} and \textit{interactive polar}. 

\begin{itemize}[noitemsep,leftmargin=*]
    \item \textbf{\tstaticbar} represents a traditional static bar chart, as in~\autoref{fig:designspace:wrapchart_cylinder}a. 
    \item \textbf{\tstaticpolar} represents a traditional static polar bar chart, arranging bars in a radial fashion, as in~\autoref{fig:designspace:wrapchart_cylinder}b.
    \item \textbf{\tinteractivebar} is a linear bar chart (\autoref{fig:designspace:wrapchart_cylinder}a) with interactive cylindrical wrapping, \rev{using mouse or touch drags}, as described in~\autoref{sec:designspace:cylinder:motivation}.
    \item \textbf{\tinteractivepolar} is a polar bar chart like \tstaticpolar{} (\autoref{fig:designspace:wrapchart_cylinder}b) with interactive cylindrical rotation, \rev{using mouse or touch drags}, as described in~\autoref{sec:designspace:cylinder:motivation}. 
\end{itemize}

\subsection{Hypotheses}
\label{sec:hypotheses}

We group our hypotheses by our research questions RQ4.1-RQ4.3. Hypotheses have been pre-registered with the Open Science Foundation: \url{https://osf.io/9k5bm}.

\subsubsection{[RQ3.1] Does adding interactive wrapping to linear bar charts improve their effectiveness?}
 
\begin{itemize}[noitemsep,leftmargin=*]
    \item H3.1-error: \tinteractivebar{} has less \merror{} than \tstaticbar{}. This is observed by the existing study results that people perform more accurately when comparing adjacent bars than bars that are far apart by Talbot et al.~\cite{talbot2014four}. With our interactive technique, it allows them to bring bars that are far apart closer.
    \item H3.1-time: \tstaticbar{} has less \mtime{} than \tinteractivebar{} as the latter requires panning.
    \item H3.2-error: \tinteractivebar{} has less \merror{} than \tinteractivepolar{}.
    \item H3.2-time: \tinteractivebar{} has less \mtime{} than \tinteractivepolar{}.
\end{itemize}

\subsubsection{[RQ3.2] How effective is interactive wrapping for linear bar charts compared with static polar bar charts?}

\begin{itemize}[noitemsep,leftmargin=*]
    \item H3.3-error: \tinteractivebar{}{} has less \merror{} than \tstaticpolar{}.
\end{itemize}

\subsubsection{[RQ3.3] Does adding interactive rotation to polar bar charts improve their effectiveness?}
\begin{itemize}[noitemsep,leftmargin=*]
    \item H3.4-error: \tinteractivepolar{} has less \merror{} than \tstaticpolar{} since participants are able to rotate the chart to best solve the task.
    \item H3.4-time: \tstaticpolar{} has less \mtime{} than \tinteractivepolar{} since the latter potentially involves panning.
\end{itemize}

Our last hypothesis is about user preference:
\begin{itemize}[noitemsep,leftmargin=*]
    \item P4.1: user prefers \tinteractivebar{} to \tstaticbar,
\tstaticpolar, or \tinteractivepolar{} since \tinteractivebar{} facilitates comparisons across the cut while allowing to best compare bar heights. 
\end{itemize}

\subsection{Tasks}
\label{sec:cylinder:tasks}

\begin{figure*}
    \centering
    \includegraphics[width=\textwidth]{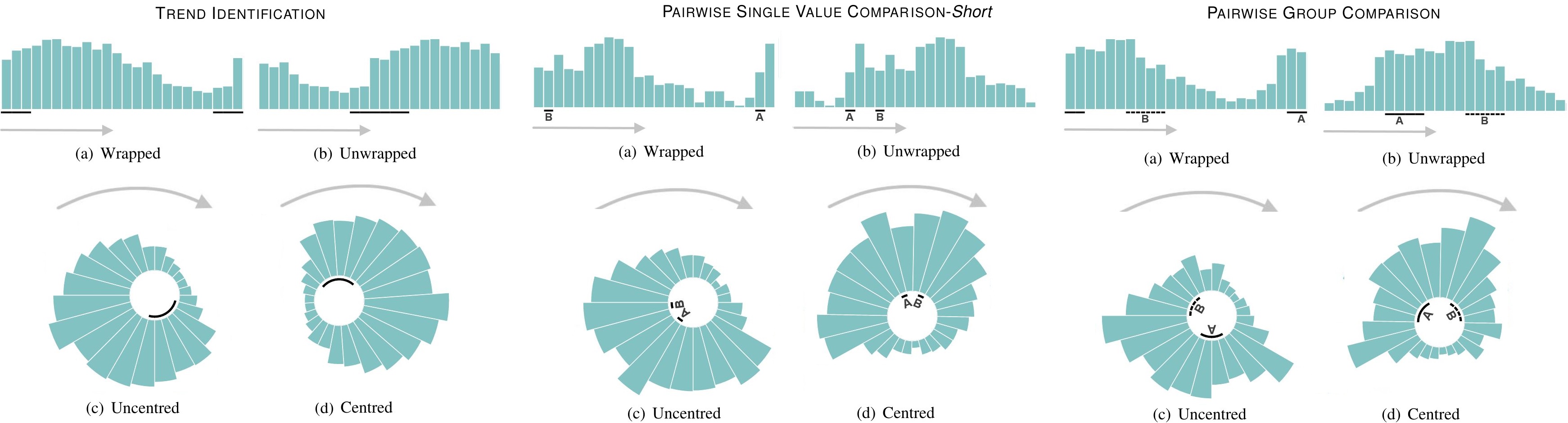}
        \caption{Examples of data sets and visual stimuli for all three tasks used in the study: \tintervaltask (left), \tmultiplebars (centre) and \tsinglebar (right). Sub figures (a)-(d) show the individual conditions for wrapped and unwrapped as well as centred and uncentred across both visualisation techniques.}
    \label{fig:tasks}
\end{figure*}

We designed three tasks to compare our four techniques. The tasks are motivated by existing time series visualisation research~\cite{waldner2019comparison, brehmer2018visualizing, adnan2016investigating, fuchs2013evaluation, correll2012comparing, albers2014task}, and graphical perception task typology by Brehmer and Munzner~\cite{brehmer2013multi}. However, they did not look at the effect of interactive panning nor did they look at a group of bars that span across the cuts. We focus on the particular effect of analysing data across the cuts in linear static and interactive bar charts. To that end, we created two levels of difficulty for each task: \textit{unwrapped} (for bar charts) or \textit{centred} (for polar charts) and \textit{wrapped} (for bar charts) or \textit{uncentred} (for polar bar charts). In the \textit{wrapped} condition, the user is required to analyse the data across the cuts, compared to the \textit{unwrapped, centred, and uncentred} conditions. Examples for all tasks and conditions are shown in~\autoref{fig:tasks}. 
\begin{itemize}[noitemsep,leftmargin=*]

\item \textbf{\tintervaltask:} \textit{``What describes the highlighted sequence of bars best, e.g., continuously increasing/continuously decreasing/neither?''} (\autoref{fig:tasks}-\tintervaltask{}) The participants were asked to identify the trend of a highlighted set of adjacent bars. The timeout for this task was 5 seconds, inspired by our pilot studies. We recorded participants’ responses with multiple-choice questions with 4 options with \textit{continuous decreasing, continuous increasing, neither}, or \textit{unsure}. We created 10 trials for this task per technique, 3 of them were monotonically increasing, 3 were monotonically decreasing intervals, 4 were neither. We fixed the length of all intervals to be 6 bars. 

\item \textbf{\tmultiplebars:} \textit{``Which group of bars, \textit{A} or \textit{B}, has the higher 
total values?''} (\autoref{fig:tasks}-\tmultiplebars{}) The timeout for this task was 10 seconds. We gave this task more time, as it required mental aggregation of a group of bar height. We recorded participants’ responses with multiple-choice questions with 3 options with \textit{A}, \textit{B}, and \textit{unsure}.  We created 10 trials for this task per technique, balancing the answers of \textit{A} and \textit{B}.
 
\item \textbf{\tsinglebar:} \textit{Which bar, \textit{A} or \textit{B}, has the higher value?} (\autoref{fig:tasks}-\tsinglebar{}) 
The timeout for this task was 5 seconds. We created 12 trials for this task per technique. We controlled for the distance between targeted bars, i.e., 6 trials per technique for \textit{Short} (2 bars apart) and 6 trials per technique for \textit{Long} (6 bars apart). We balanced the answers of \textit{A} and \textit{B}.  We recorded participants’ responses with multiple-choice questions with 3 options with \textit{A}, \textit{B}, and \textit{unsure}.

\end{itemize}

\subsection{Visual configuration}

The width of the bar chart was fixed to 628 pixels, which was the same as the median grid line of the circumference of a polar chart with 200-pixels radius, as seen in~\autoref{fig:designspace:wrapchart_cylinder}-middle and~\autoref{fig:designspace:wrapchart_cylinder}-right. The y-axis scale of the bar chart was the same as the y-axis scale in the polar chart, i.e., 200 pixels. This setting was similar to prior visualisation studies comparing polar and bar charts by Waldner et al.~\cite{waldner2019comparison}. 
For the study, we removed all tick marks, grid lines and labels, such that visual comparison is based purely on bar height. We used a monochrome colour for the bars, as Adnan et al.~\cite{adnan2016investigating} found colour visual encoding makes the effect of linear or polar chart negligible for time series visualisation. For the polar chart, we used an inner circle with a radius of 50 pixels for showing a curve. This marks the bars in each task, as shown in ~\autoref{fig:tasks}. The corresponding visual encoding was an underline for bar charts.
To support \tintervaltask, each chart was accompanied by a grey arrow indicating the direction of time inside the visualisation. For the linear bar charts, the arrow was shown from the left to the right on the bottom of the chart. We implemented each technique using D3.js~\cite{d3js:2020}.

\subsection{Data}
\label{sec:cylinder:dataset}
Our stimuli were generated from New York hourly traffic accident datasets between 2013 and 2016~\cite{hourlytrafficaccidentsdataset}. We selected Brooklyn, Bronx, Manhattan, Queens, and Staten Island borough data. These data show a non-smooth distribution of the aggregated number of traffic accidents over 24 hours, which we felt would provide a suitable and ecologically-valid data set for our study. The same data has been used in prior time series visualisation studies comparing linear and polar charts by Waldner et al.~\cite{waldner2019comparison}. To generate cyclical temporal data for the study tasks, we first aggregated the hourly traffic accidents on a monthly basis for each borough. We then obtained average hourly traffic accident datasets in 24 hours (i.e., 24 data points) for a given borough, month and year to obtain hundreds of candidate sample data sets.  We selected from these candidates for our task stimuli, controlling for difficulty based on pilot testing, as follows:

\begin{itemize}[noitemsep,leftmargin=*]
\item \textbf{\tintervaltask:} Across all samples we restricted the candidates to instances with a monotonically increasing or decreasing set of six adjacent bars, with a minimum height difference of 1-2\% (i.e., small but  apparent on the displays tested). For neither monotonically increasing nor decreasing instances, there are no other monotonic sequences of more than two bars. 
Furthermore, we added one quality control (obvious) trial per technique with a much larger minimum height difference of greater than 20\%.

\item \textbf{\tmultiplebars:} 
The difference of total values between two groups was 5-15\% of bar chart height. For the quality control trial, the difference is set to be greater than or equal to 20\% of bar chart height.
  
\item \textbf{\tsinglebar:} 
The difference of heights between the two bars was 1-1.5\% of bar chart height. For the quality control trial, the difference is set to be greater than or equal to 20\% of the bar chart height.

\end{itemize}

\subsection{Experimental design}

Our experimental design follows the practical guideline for experimental human-computer interaction by Purchase~\cite{purchase2012experimental}. Similar to many existing time series studies that crowdsourced (explained in~\autoref{sec:related:evaluationmethods}) their studies~\cite{schwab2019evaluating,waldner2019comparison,brehmer2018visualizing}, and to evaluate the applicability of our interactive wrapped visualisation based on how a time series visualisation would be accessed on the Internet today, we opt for a crowdsourced study.

We decided on a within-subject design (explained in~\autoref{sec:related:evaluationmethods}), where each participant performed all of the 4 techniques $\times$ 3 tasks. We used 10 repetitions for \tintervaltask{}, \tmultiplebars{}, and 6 repetitions for \tsinglebar{}-\textit{Short} and 6 repetitions for \tsinglebar{}-\textit{Long}. Following existing unsupervised time series studies by Brehmer et al.~\cite{brehmer2018visualizing}, we inserted a quality control trial for each technique per task to check participants' attention. We used 4 practice trials for each technique per task. This gives a total of 47 trials $\times$ 4 techniques = 188 trials per participant. We used a full-factorial design to counterbalance the learning effect of four techniques (which results in a total of 24 different orderings of techniques). Each recorded trail has a timer associated with the task type, as described in~\autoref{sec:cylinder:tasks}. We expected the study to complete within 40 minutes. The order of \textit{wrapped}, \textit{unwrapped}, \textit{centred, and uncentred} questions for each task was randomised across tasks per technique. Each trial used a randomly selected dataset from 5 borough data that satisfies the constraints as described in~\autoref{sec:cylinder:dataset}. Therefore, none of the same graphics appear twice throughout the study.

We used equal numbers of trials for each of these four groups. We recorded \textit{task-completion time} (\mtime{}), \textit{task-error} (\merror{}), and \textit{subjective user preference} (\mpref{}) as dependent variables across all the tasks.

\subsection{Participants and procedure}

We crowdsourced the study via both convenience sampling and the Prolific platform~\cite{palan2018prolific}. 
The participants on the Prolific platform have been reported producing data quality comparable to the dominant Amazon Mechanical Turk by Peer et al.~\cite{peer2017beyond}. Many time series visualisation studies were in fact crowdsourced~\cite{schwab2019evaluating, waldner2019comparison, brehmer2018visualizing, di2020evaluating, satriadi2021quantitative, albers2014task, correll2012comparing}, with increasing number of visualisation studies deployed over the Prolific platform, e.g., by Satriadi et al. ~\cite{satriadi2021quantitative}. 

We hosted the study on our web server application. We set a pre-screening criterion on performance, i.e., minimum approval rate of 95\%, and minimum number of previous submissions of 10. We also limited our study to desktop users with larger screens. We provided a payment of \pounds5 (i.e.\ a rate of \pounds7.5/h) to Prolific participants. This is considered good payment according to the Prolific platform. We recorded 72 participants who passed quality control tasks and completed the study. This comprised 3 full counterbalanced blocks of participants. There were 48 participants from the Prolific group and 24 participants from local sampling (recruited via email). Out of all participants, there were 23 identified as females, 47 identified as males and 2 participants preferred not to disclose their gender. The age of participants was between 18 and 50. Each participant went through all of the 4 techniques with the order assigned by the software with the following procedure. 
First, they completed a tutorial explaining the technique and task. For \tintervaltask{}, participants were instructed to read the trend from a set of adjacent bars highlighted by either an underline (for bar charts) or a curve (for polar charts). Examples of monotonically increasing, monotonically decreasing, and neither were given during the training as well as in practice trials. An example of neither is shown in~\autoref{fig:tasks}-\tintervaltask{}(a-d). During the training, when the participant's answer was incorrect they would be shown the same practice trial again until a correct answer was given.

For \tinteractivebar{} and \tinteractivepolar, an animated image demonstrating the wrapping or rotation interaction was shown for the task. Participants were encouraged to try the interaction themselves with the same example as the one in the animation. Following that, participants were required to successfully complete 4 practice trials before proceeding to the recorded trials. For recorded trials, each trial was first loaded on a participant’s browser before the software started the timer. Each participant went through the same task order, i.e., \tintervaltask{}$\rightarrow$\tmultiplebars{}$\rightarrow$\tsinglebar{}-\textit{Short}$\rightarrow$ \tsinglebar{}-\textit{Long}.

\subsection{Results}
\label{sec:cylinder:results}
\begin{figure}
    \centering
    \vspace{-2em}
    \includegraphics[width=.9\textwidth]{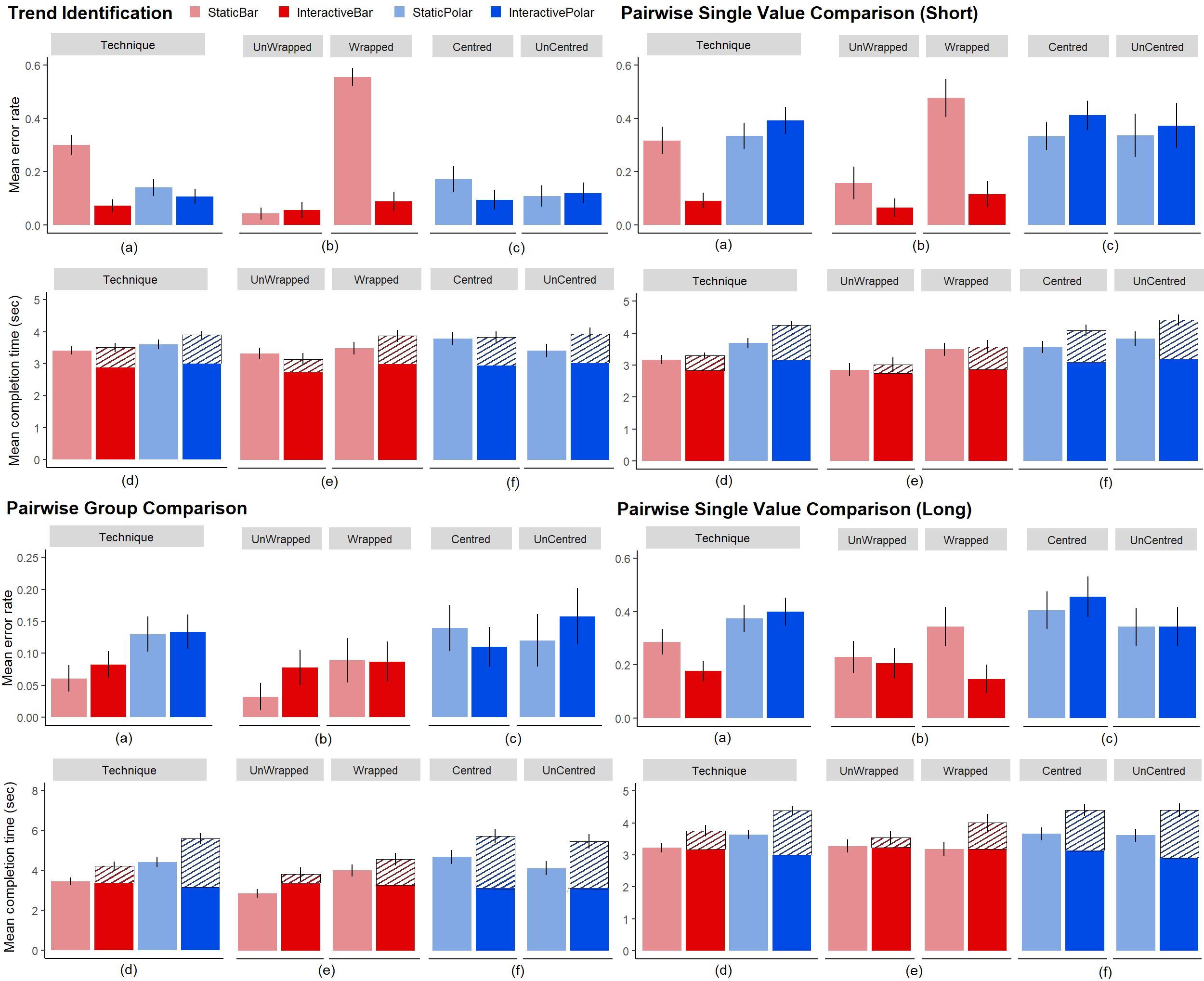}
    \vspace{-1em}
    \caption{Quantitative results for \merror{} and \mtime{} for all tasks and techniques: bar charts (red) and polar charts (blue). Static techniques are indicated in a lighter colour. Textured parts of some bars indicate the fraction of time used for panning (\tinteractivebar) and rotating (\tinteractivepolar). The bars show the results of average performance in terms of error rate and task completion time. Error bars indicate 95\% confidence interval.}
    \label{fig:cylinder:results}
\end{figure}
We report on the results of 72 participants from both the Prolific group and the convenience sampling group. All of these participants passed the attention check trials, completed the training and recorded trials. This resulted in 9,216 trials. We excluded attention check trials in the analysis, as they used a much looser constraint than recorded trials, as described in~\autoref{sec:cylinder:dataset}. Therefore, we have equal numbers of \textit{wrapped}, \textit{unwrapped} trials for bar charts, and equal numbers of \textit{centred} and \textit{uncentred} trials for polar charts for each task in the analysis. 

Since the distribution of \merror{} and \mpref{} of each technique did \textit{not} follow a normal distribution, to compare error rates between different techniques we used standard non-parametric statistics described by Field et al.~\cite{field2012discovering}. To compare multiple (four) conditions, we used Friedman's non-parametric test and Tukey's post hoc pairwise comparison with Bonferroni correction~\cite{field2012discovering} to identify significant differences between \tstaticbar{}, \tinteractivebar{}, \tstaticpolar{}, and \tinteractivepolar{}. The confidence interval is 95\%. Since \mtime{} of each technique was normally distributed, tested with Shapiro-Wilk's normality test~\cite{shaphiro1965analysis} and visually checked by Q-Q plot, we used standard parametric statistics methods, described by Field et al.~\cite{field2012discovering}. To compare multiple (four) techniques, we used analysis of variance (ANOVA) repeated measures (i.e. hypothesis testing for within-subject analysis) and Tukey's post hoc pairwise comparison with Bonferroni correction~\cite{field2012discovering} to test significance. 
The results of error bar graphs are shown in~\autoref{fig:cylinder:results} and~\autoref{fig:userrank}. 
We report on the most significant findings by tasks as follows. 


\subsubsection{Trend identification} 
\begin{figure}
    \centering
    \includegraphics[width=0.85\columnwidth]{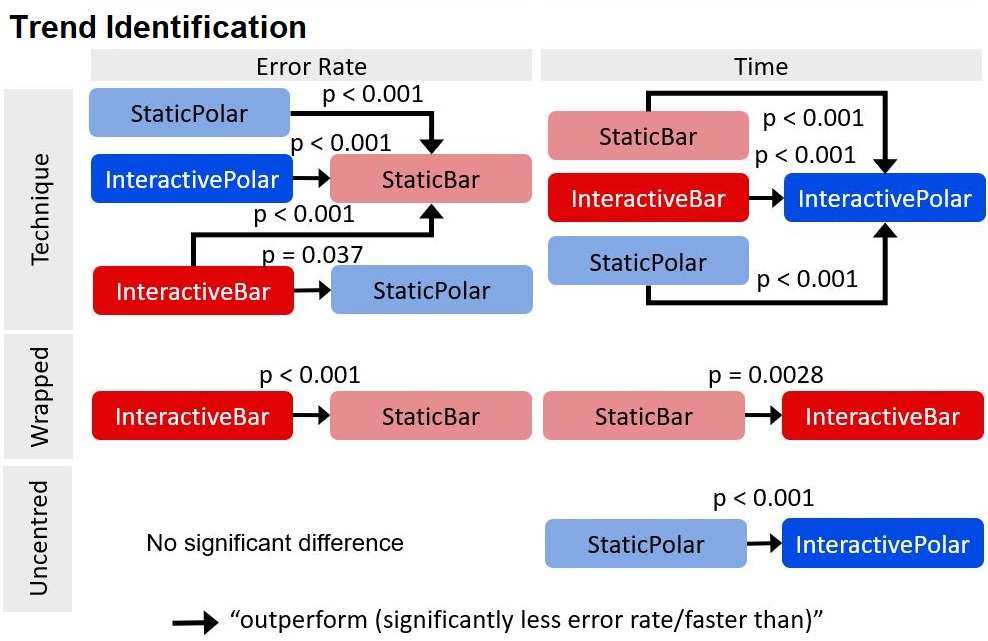}
    \caption{Statistically significant results of \tintervaltask}
    \label{fig:intervalStat}
\end{figure}

Friedman's test revealed a statistically significant effect of techniques on ERROR ($X^2(3) = 142.6; p < .001$). Post-hoc tests showed that \tinteractivebar{} ($.07, SD = .19$), \tstaticbar{} ($.14, SD = .26$) and \tinteractivepolar{} ($.1, SD = .22$) resulted in significantly less errors ($p < .001$) than \tstaticbar{} ($.3, SD = .3$), as seen in~\autoref{fig:cylinder:results}-\tintervaltask{}(a) and~\autoref{fig:intervalStat}-\merror{}. We also found that with interactive wrapping, \tinteractivebar{} resulted in significantly less errors than \tstaticpolar{} ($p = .037$), as seen in~\autoref{fig:cylinder:results}-\tintervaltask{}(a) and~\autoref{fig:intervalStat}-\merror{}. For wrapped trials, \tinteractivebar{} ($.08, SD = .21$) significantly outperformed \tstaticbar{} ($.56, SD = .19$) in terms of error ($p < .001$), as seen in~\autoref{fig:cylinder:results}-\tintervaltask{}(b) and~\autoref{fig:intervalStat}-\merror{}. The results of repeated measure ANOVA showed a statistically significant effect of techniques on \mtime{} ($F = 8.54, p < .001$). Post-hoc tests showed that \tinteractivebar{} ($3.95s, SD = 1.08$) was significantly ($p < .001$) slower than \tstaticbar{} ($3.45s, SD = 1.09$), \tstaticpolar{} ($3.65s, SD = 1.24$) and \tinteractivebar{} ($3.57s, SD = 1.17$), as seen in~\autoref{fig:cylinder:results}-\tintervaltask{}(d) and~\autoref{fig:intervalStat}-\mtime{}. \tinteractivebar{} ($3.93s, SD = 1.07$) was significantly ($p = .0028$) slower than \tstaticbar{} ($3.54s, SD = 1.13$) for wrapped trials, as seen in~\autoref{fig:cylinder:results}-\tintervaltask{}(e) and~\autoref{fig:intervalStat}-\mtime{}. For Uncentred trials, \tinteractivepolar{} ($4s, SD = 1.11$) was significantly ($p < .001$) slower than \tstaticpolar{} ($3.46s, SD = 1.22$), as shown in~\autoref{fig:cylinder:results}-\tintervaltask{}(f) and~\autoref{fig:intervalStat}-\mtime{}.

\subsubsection{Pairwise single value comparison-Short}
\begin{figure}
    \centering
    \includegraphics[width=0.85\columnwidth]{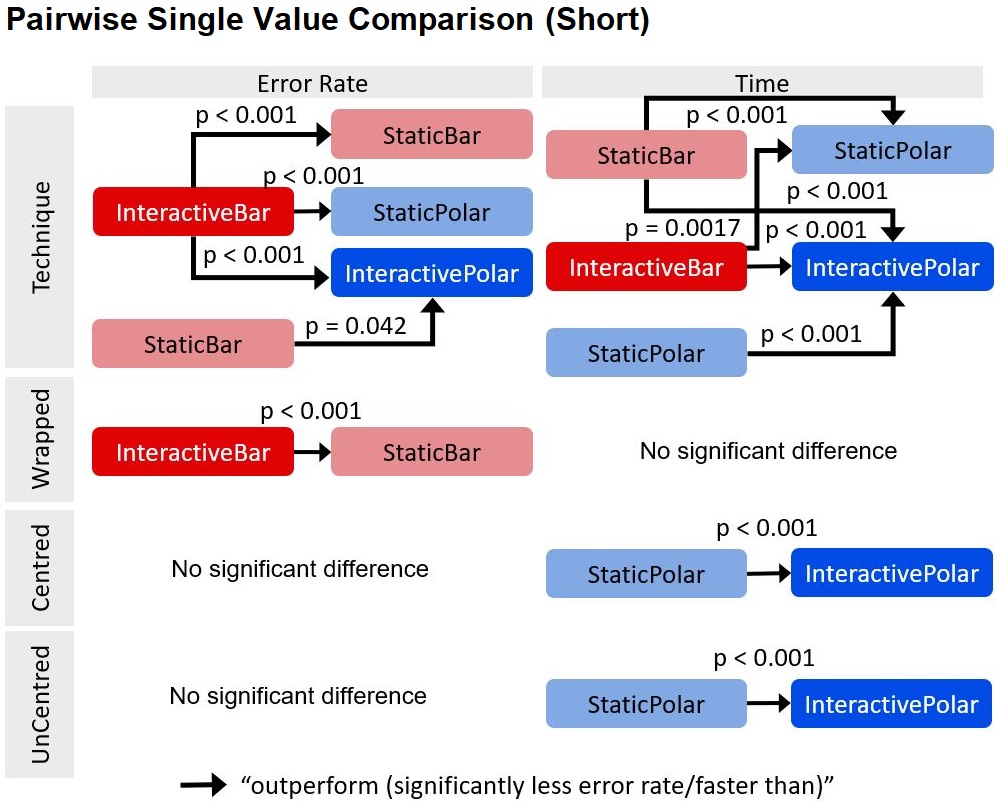}
    \caption{Statistically significant results of \tsinglebar{} (Short distance - 2 bars apart)}
    \label{fig:compareSingleShortStat}
\end{figure}
Friedman's test revealed a statistically significant effect of techniques on \merror{} ($X^2(3) = 118.88; p < .001$). Post-hoc tests showed that \tinteractivebar{} ($.09, SD = .24$) resulted in significantly ($p < .001$) less error than \tstaticbar{} ($.32, SD = .41$), \tstaticpolar{} ($.33, SD = .39$), and \tinteractivepolar{} ($.39, SD = .41$), as seen in~\autoref{fig:cylinder:results}-\tsinglebar{}-Short(a) and~\autoref{fig:compareSingleShortStat}-\merror{}. \tstaticbar{} significantly ($p = .042$) outperformed \tinteractivepolar{} in terms of error. For wrapped trials, \tinteractivebar{} ($.11, SD = .28$) resulted in significantly ($p < .001$) less errors than \tstaticbar{} ($.48, SD = .41$), as shown in~\autoref{fig:cylinder:results}-\tsinglebar{}-Short(b) and~\autoref{fig:compareSingleShortStat}-\merror{}. The results of repeated measure ANOVA showed a statistically significant effect of techniques on time ($F = 38.2, p < .001$). Post-hoc tests showed that \tinteractivepolar{} ($4.31s, SD = 1.03$) was significantly slower than \tstaticbar{} ($3.21s, SD = 1.23$), \tstaticpolar{} ($3.74s, SD = 1.22$) and \tinteractivebar{} ($3.34s, SD = 1.29$), as shown in~\autoref{fig:cylinder:results}-\tsinglebar{}-Short(d) and~\autoref{fig:compareSingleShortStat}-\mtime{}. \tstaticpolar{} was significantly ($p < .001$) slower than \tstaticbar{} and \tinteractivebar{}. There was no significant difference between \tinteractivebar{} and \tstaticbar{}. For centred trials, \tinteractivepolar{} ($.4.15s, SD = 1.04$) was significantly ($p < .001$) slower than \tstaticpolar{} ($3.61s, SD = 1.11$). For uncentred trials, \tinteractivepolar{} ($4.46s, SD = 1$) was significantly ($p < .001$) slower than \tstaticpolar{} ($3.87s, SD = 1.31$).

\subsubsection{Pairwise single value comparison-Long}
\begin{figure}
    \centering
    \includegraphics[width=0.85\columnwidth]{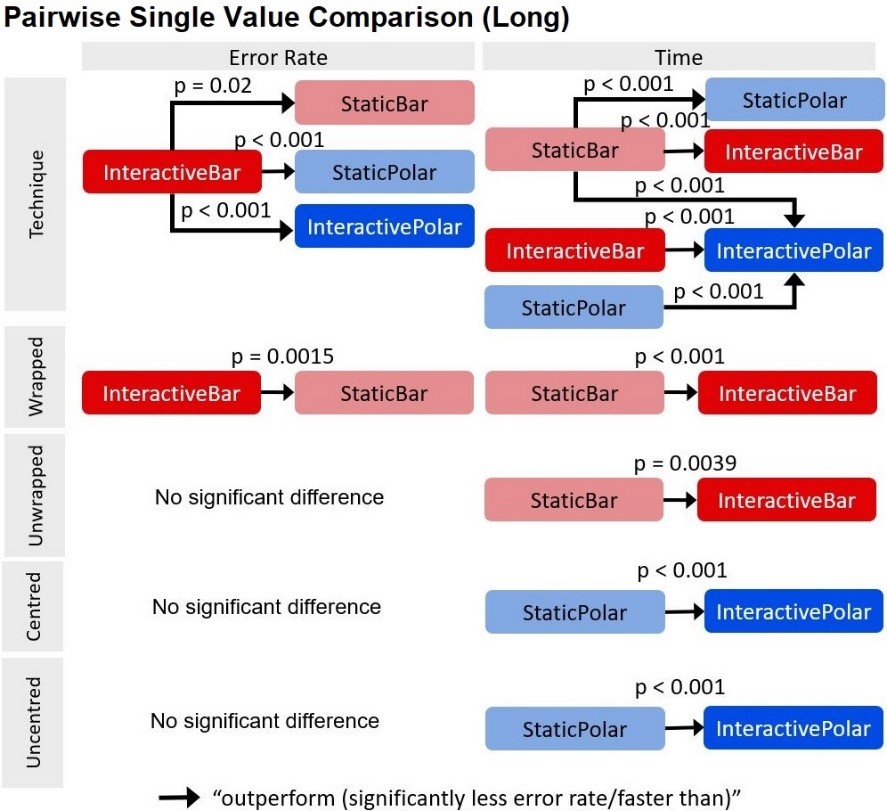}
    \caption{Statistically significant results of \tsinglebar{} (Long distance - 6 bars apart)}
    \label{fig:compareSingleLongStat}
\end{figure}
The Friedman's test revealed a statistically significant effect of techniques on \merror{} ($X^2(3) = 61.3; p < .001$). Post-hoc tests showed that \tinteractivebar{} ($.17, SD = .32$) resulted in significantly less errors ($p < .001$) than \tstaticpolar{} ($.37, SD = .41$) and \tinteractivepolar{} ($.4, SD = .43$). \tinteractivebar{} resulted in significantly less errors ($p = .02$) than \tstaticbar{} ($.29, SD = .38$), as seen in~\autoref{fig:cylinder:results}-\tsinglebar{}-Long(a) and~\autoref{fig:compareSingleLongStat}-\merror{}. For wrapped trials, \tinteractivebar{} ($.14, SD = .3$) significantly outperformed \tstaticbar{} ($.34, SD = .42$) in terms of error ($p = .0015$), as seen in~\autoref{fig:cylinder:results}-\tsinglebar{}-Long(b) and~\autoref{fig:compareSingleLongStat}-\merror{}. The results of repeated measure ANOVA showed a statistically significant effect of techniques on time ($F = 33.03, p < .001$). Post-hoc tests showed that \tinteractivepolar{} ($4.45s, SD = 1.15$) was significantly ($p < .001$) slower than \tstaticbar{} ($3.27s, SD = 1.21$), \tstaticpolar{} ($3.68s, SD = 1.16$) and \tinteractivebar{} ($3.83s, SD = 1.48$), as seen in~\autoref{fig:cylinder:results}-\tsinglebar{}-Long(d) and~\autoref{fig:compareSingleLongStat}-\mtime{}. \tstaticpolar{} was significantly ($p < .001$) slower than \tstaticbar{}, as seen in~\autoref{fig:cylinder:results}-\tsinglebar{}-Long(d) and~\autoref{fig:compareSingleLongStat}-\mtime{}. \tinteractivebar{} was significantly ($p < .001$) slower than \tstaticbar{}.
We also found that \tinteractivebar{} ($4.08, SD = 1.6$) was significantly slower than \tstaticbar{} ($3.23, SD = 1.23$) for wrapped ($p < .001$) and unwrapped ($p = .0039$) trials, as seen in~\autoref{fig:cylinder:results}- \tsinglebar{}-Long(e) and~\autoref{fig:compareSingleLongStat}-\mtime{}. For centred/uncentred trials, \tinteractivepolar{} ($4s, SD = 1.11$) was significantly ($p < .001$) slower than \tstaticpolar{} ($3.46s, SD = 1.22$), as shown in~\autoref{fig:cylinder:results}-\tsinglebar{}-Long(f) and~\autoref{fig:compareSingleLongStat}-\mtime{}.

\subsubsection{Pairwise group comparison}
\begin{figure}
    \centering
    \includegraphics[width=0.85\columnwidth]{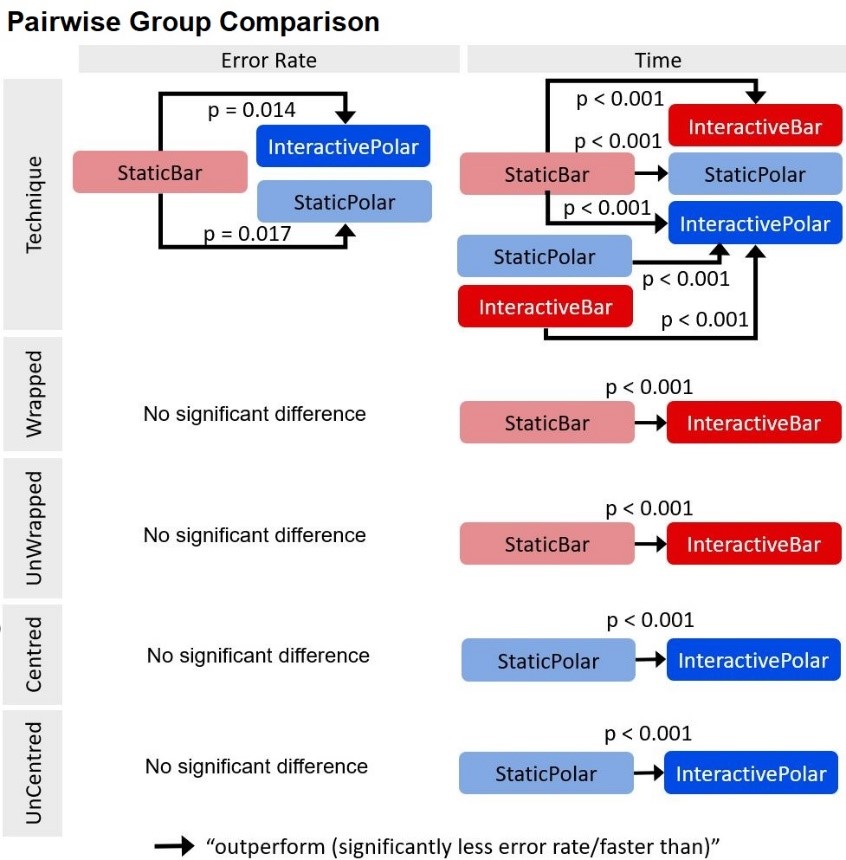}
    \caption{Statistically significant results of \tmultiplebars}
    \label{fig:compareAvgStat}
\end{figure}
Friedman's test revealed a statistically significant effect of techniques on \merror{} ($X^2(3) = 32.9; p < .001$). Post-hoc tests showed that \tstaticbar{} ($.06, SD = .16$) resulted in significantly less errors ($p = .017$) than \tstaticpolar{} ($.131, SD = .22$). \tstaticbar{} ($.06, SD = .16$) resulted in significantly less errors ($p = .014$) than \tinteractivepolar{} ($.135, SD = .22$), as seen in~\autoref{fig:cylinder:results}-\tmultiplebars{}(a) and~\autoref{fig:compareAvgStat}-\merror{}. The reults of repeated measure ANOVA showed a statistically significant effect of techniques on \mtime{} ($F = 33.03, p < .001$). Post-hoc tests showed that \tinteractivepolar{} ($4.45s, SD = 1.15$) was significantly ($p < .001$) slower than \tstaticbar{} ($3.27s, SD = 1.21$), \tstaticpolar{} ($3.68s, SD = 1.16$) and \tinteractivebar{} ($3.83s, SD = 1.48$), as seen in~\autoref{fig:cylinder:results}-\tmultiplebars{}(d) and~\autoref{fig:compareAvgStat}-\mtime{}. \tstaticpolar{} was significantly ($p < .001$) slower than \tstaticbar{}, as seen in~\autoref{fig:cylinder:results}- \tmultiplebars{}(d) and~\autoref{fig:compareAvgStat}-\mtime{}. \tinteractivebar{} was significantly ($p < .001$) slower than \tstaticbar{}, as seen in~\autoref{fig:cylinder:results}-\tmultiplebars{}(d) and~\autoref{fig:compareAvgStat}-\mtime{}. For either wrapped or unwrapped trials, \tinteractivebar{} was significantly ($p < .001$) slower than \tstaticbar{}, as seen in~\autoref{fig:cylinder:results}- \tmultiplebars{}(e) and~\autoref{fig:compareAvgStat}-\mtime{}.

\subsubsection{User preference}
\begin{figure}
    \centering
    \includegraphics[width=0.9\columnwidth]{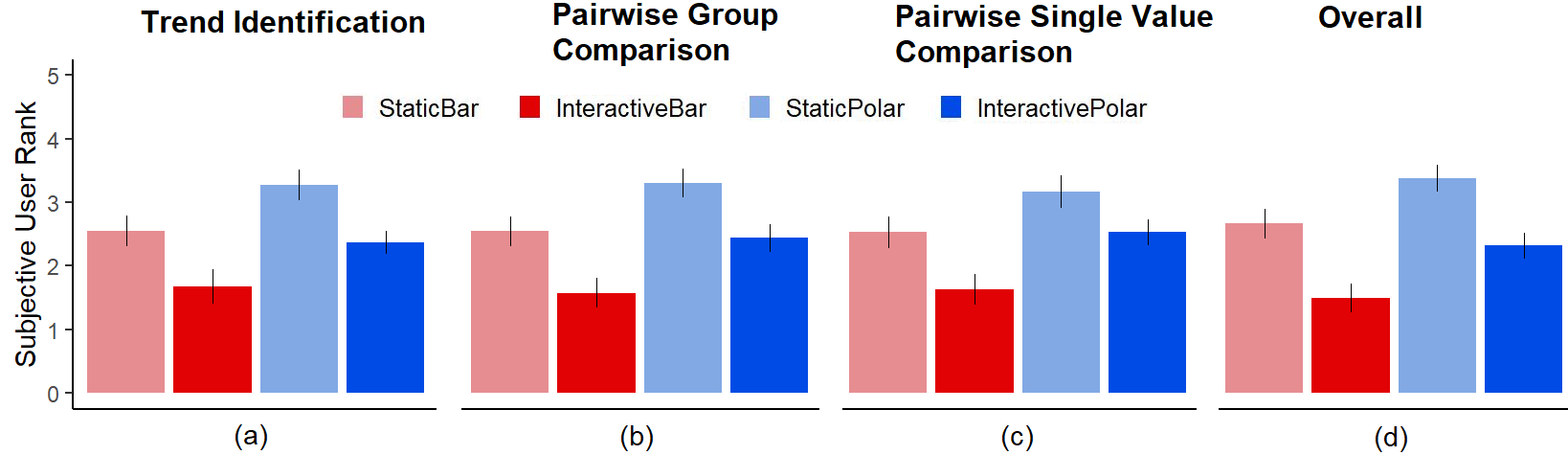}
    \caption{Subjective User ranking for all tasks and techniques (lower is better).}
    \label{fig:userrank}
\end{figure}
For \tinteractivebar{}: Friedman's test revealed a statistically significant effect of techniques on \mpref{} ($X^2(3) = 57.7; p < .001$). Post-hoc tests showed that \tinteractivebar{} was significantly preferred over \tstaticbar{} ($p < .001$), \tstaticpolar{} ($p < .001$) and \tinteractivepolar{} ($p = .0054$). \tstaticbar{} was significantly preferred over \tstaticpolar{} ($p = .0034$). \tinteractivepolar{} was significantly preferred over \tstaticpolar{} ($p < .001$).

For \tsinglebar{}: Friedman's test revealed a statistically significant effect of techniques on \mpref{} ($X^2(3) = 53.1; p < .001$). Post-hoc tests showed that \tinteractivebar{} was significantly preferred over \tstaticbar{} ($p < .001$), \tstaticpolar{} ($p < .001$) and \tinteractivepolar{} ($p < .001$). \tstaticbar{} was significantly preferred over \tstaticpolar{} ($p = .015$). \tinteractivepolar{} was significantly preferred over \tstaticpolar{} ($p = .015$).

For \tmultiplebars{}: Friedman's test revealed a statistically significant effect of techniques on \mpref{} ($X^2(3) = 66.5; p < .001$). Post-hoc tests showed that \tinteractivebar{} was significantly preferred over \tstaticbar{} ($p < .001$), \tstaticpolar{} ($p < .001$) and \tinteractivepolar{} ($p < .001$). \tstaticbar{} was significantly preferred over \tstaticpolar{} ($p = .0021$). \tinteractivepolar{} was significantly preferred over \tstaticpolar{} ($p < .001$).

Overall: Friedman's test revealed a statistically significant effect of techniques on \mpref{} ($X^2(3) = 79.7; p < .001$). Post-hoc tests showed that \tinteractivebar{} was significantly preferred over \tstaticbar{} ($p < .001$), \tstaticpolar{} ($p < .001$) and \tinteractivepolar{} ($p < .001$). \tstaticbar{} was significantly preferred over \tstaticpolar{} ($p = .005$). \tinteractivepolar{} was significantly preferred over \tstaticpolar{} ($p < .001$).

\subsection{Qualitative user feedback}

The majority of participants reported more confidence in using \tinteractivebar{} across all tasks. Some participants mentioned that panning the chart brings bars closer to one another, making it significantly easier to inspect bars and come to a decision. For example, 
\emph{``Interactive bar chart was the easiest because I had a clear vision of whether the bars were getting smaller or larger.''} (P4, \tintervaltask{}), \emph{``with the interactive bar chart, it was possible to change the position of the bars to either not have an interruption in a group of bars or compare the groups/single bars better.''} (P33, \tintervaltask{}), \emph{``The interactive bar chart makes it easier by allowing me to more closely align the 2 bars to work out the difference between them.''} (P22, \tsinglebar{}).

Some participants reported more confidence in \tstaticbar{} than \tstaticpolar, citing that the \tstaticpolar{} looked more confusing and was harder to read. This confirms existing studies that polar charts are generally less effective than linear bar charts~\cite{waldner2019comparison,adnan2016investigating,brehmer2018visualizing}. For example, participants mentioned they had to turn their head or neck to spot the answer and static polar charts caused confusion about comparing the bars, and therefore they ranked it the worst, e.g., \emph{``Bar charts are easier in [\tmultiplebars{}] than circular charts. For both [\dstatic{} and \dinteractive] circular charts, I found myself turning my head to the side to get a better look. The standard bar chart, although it went off the side was still easier than both circular.''} (P2, \tmultiplebars{}), and \emph{``It's easier to see it in the bar charts because all the bars are at the same level unlike in circular charts''} (P40, \tsinglebar{}).

Furthermore, some participants mentioned the ability to rotate allows them to see in different angles and make them feel more confident in the analysis, e.g., \emph{``It is easier for me to compare the lengths when the bars are close together (even if they are not straight lines). It is also easier when the bars are growing upwards (they do not point downwards).''} (P37, \tsinglebar{}), and \emph{``Overall the interactive circular chart is more helpful in working out the patterns and difference in height of the bars because you can move it so the highlighted bars are at the top and facing clockwise plus the style of the bars makes it easier to notice differences.''} (P22, Overall).
Some other participants mentioned the polar charts provided a more panoramic view. \tstaticbar{} made it hard to focus when seeing broken ranges or bars that are located far apart.
 

\section{Discussion and future work}
\label{sec:cylinder:discussion}

Overall, the results of our study indicate that interactive cylindrical wrapping leads to significant improvements in error over static representations for \tintervaltask{} and \tsinglebar{}, despite the additional time required to pan. 

Returning to our questions from the Introduction, to answer \textbf{[RQ3.1]} \textit{``Does adding interactive wrapping to linear bar charts improve their effectiveness?''}, we begin by noting that
contrary to findings from past studies~\cite{waldner2019comparison, brehmer2018visualizing, adnan2016investigating} (that static linear bar charts outperform static polar bar charts), we found that for  \tintervaltask{} across the `cut' in linear bar charts, the polar representation is actually significantly better in terms of error rate.  However, introducing interaction to linear bar charts reverses this result. We can conclude that the \tinteractivebar{} clearly outperforme \tstaticbar{} for \tintervaltask{} and \tsinglebar{} tasks in terms of error, especially for trials where the bars being compared are separated by the `cut'. These results allow us to reject the null-hypothesis for H3.1-error for some tasks. We found that \tinteractivebar{} results in less errors than \tinteractivepolar{} for some tasks, rejecting the null-hypothesis for H3.2-error and sometimes is significantly faster, rejecting the null-hypothesis for H3.2-time.

While we found that \tstaticbar{} was faster than \tinteractivebar{} across all tasks except \tsinglebar{}-Short (rejecting the null-hypothesis for H3.1-time for some tasks), the time spent actually moving the charts was a notable fraction of the trial time. We see from the textured parts of bars in~\autoref{fig:cylinder:results}-\tintervaltask{} and~\autoref{fig:cylinder:results}-\tsinglebar{}-\textit{Short} that for \tinteractivebar{} the interaction time in \tintervaltask{} and \tsinglebar{}-short tasks were greater than the time difference compared to \tstaticbar{}.
We observed that most people use the interaction once, moving \tinteractivebar{} or \tstaticpolar{} so that they could comfortably solve the task.
While it is tempting to assume the time difference is due to the extra time spent performing the interaction, it is impossible to know whether people are able to actively reason about the visualisation during interaction, suggesting an interesting direction for future study.

We also found that \tinteractivebar{} significantly outperformed \tstaticpolar{} in terms of error for \tintervaltask{} and \tsinglebar{} \textbf{[RQ3.2]}. We can thus reject the null-hypothesis for H2-error for these tasks. We conclude that adding interactive wrapping to bar charts---while incurring a cost in terms of the time spent interacting---decreases errors compared to static polar versions which technically do not require rotation to avoid the cut-problem. This trend is similar across all tasks but significant only for two.

Eventually, while we did not find any significant result that interaction reduces errors in polar charts \textbf{[RQ3.3]}, we found \tstaticpolar{} was significantly faster than \tinteractivepolar{} across all tasks, we therefore accept the null-hypothesis for H3.4-error and reject the null-hypothesis for H3.4-time. We also found that \tinteractivebar{} was significantly preferred to \tstaticbar{}, \tstaticpolar{}, and \tinteractivepolar{}. We therefore reject the null-hypothesis for P4.1. 

In our study, we exclusively focus on evaluating cyclic temporal data in one spatial dimension to keep the crowdsourced study from becoming overly complicated. Based on the study results, we might hypothesise that, for a more complicated case of 2D temporal data exploration in a 2D torus topology, interactive wrapping with two spatial (or temporal) dimensions may still outperform the rotational and wraparound polar chart. In future, we intend to further investigate this hypothesis, which we discussed further in~\autoref{sec:conclusion:cyclicaltimeseries}. Another direction is to investigate if the performance benefits afforded by interactive wrapping across the boundaries also applies to other non-temporal cyclic data types, such as geographic maps or flow diagrams with cycles.
We would also like to further investigate the aspects of the wrapping method that drive better performance (\autoref{sec:conclusion:others}). For example, we cannot say whether the interactive panning we provided is better than a passive animation of wrapping affording different views.  Our feeling, however, is that the interaction gives users a better understanding of the paradigm as well as a sense of control.

In summary, we take these results as strongly encouraging interactive wrapping for linear bar charts, especially since the upfront cost of adding an interactive wrapping technique is small and the interaction is \textit{not} required to read the bar charts.

\section{Conclusion}
\label{sec:cylinder:conclusion}
We have presented a study comparing the effect of interactive wrapping of bar charts and rotation of polar charts on the readability of real-world cyclical data.
Our study confirms the benefits of one-dimensional interactive wrapping on a 2D projected cylindrical topology for cyclical time series, as described in our design space (\autoref{sec:designspace:cylinder}). This is the first study demonstrating that a pannable wrapped visualisation offers significant benefits in terms of error rate over the equivalent static visualisation for reading ranges and comparing values that are split across the edges of the chart, which addresses \textbf{RG3} (\autoref{sec:intro:RGs:cylinder}). Specifically, our study indicates that standard bar charts with interactive wrap panning offers significant benefits over standard unwrapped bar charts in accuracy for reading ranges or value comparison tasks and overall user preference. However, an interactive bar chart is significantly slower than standard bar charts. For polar charts, interactive panning makes them significantly slower and less accurate than static polar charts or bar charts, except for tasks of reading monotonic ranges. In the latter, the standard bar chart is least accurate due to the need to mentally connect discontinued ranges across two ends of the charts.

In the next chapter, we extend the concept of wrapping to two-dimensional and explore a more complex data type -- network, with toroidal wrapping.
%
\chapter{Interactive Torus Wrapping for Network Visualisations - Smaller Networks}
\label{sec:torus1}

\cleanchapterquote{Clutter and confusion are failures of design, not attributes of information.}{Edward Tufte}{(American statistician and emeritus professor of political science, statistics, and computer science at Yale University.)}

Some types of visualisations are fairly straightforward to either wrap or rotate.  Examples include bar charts, polar charts, or horizon graphs (described in~\autoref{sec:designspace}). 
However, arranging a network on the surface of a torus such that it can be projected to a 2D panel that wraps around horizontally or vertically requires optimisation of node positions on that surface to better show the patterns or relationships within the data.

In this chapter, we investigate visualisations of networks on a two-dimensional torus topology, like an opened-up and flattened doughnut (\autoref{fig:intro:torusnetwork}). That is, the network is drawn on a rectangular area such that certain links may be ``wrapped'' across the boundary, allowing for additional spreading of node positions to reduce visual clutter. 
Apart from crossing reduction as discussed in~\autoref{sec:related:torus}, toroidal layouts may afford other benefits: e.g., greater angular resolution between links connected to the same node and more uniform link (edge) lengths. We find it interesting that torus drawings of graphs, like those in~\autoref{fig:intro:torusnetwork}(c), have not (to our knowledge), been seriously considered as a practical means of visualising networks.  

A well known example for torus wrapping is the game Asteroids (\autoref{fig:intro:asteroids}) where players intuitively understand the torus topology.
We consider whether this intuitive understanding is also applicable to network visualisation.
However, as we described in \autoref{fig:intro:torusnetwork}(c), one possible disadvantage of such toroidal drawing is that it requires users to follow wrapped links from one side of the display to the other.  
We are inspired by illustrations of tile-display of networks and SOMs (\autoref{sec:related:tiledisplayandpanning}) that show connectivity of graphic elements wrapped across the boundary, and the significant improvement of introducing interactive cylindrical wrapping of bar charts in understanding cyclical data (\autoref{sec:cylinder}).
Therefore, this chapter investigates if these wrapping approaches aid comprehension of torus wrapping. Three fundamental research issues that link to \textbf{RG4} (\autoref{sec:intro:RGs:torus}) must be addressed if torus layout is to become more widely used.

\begin{itemize}[noitemsep,leftmargin=*]
    \item \textbf{[RQ4.1]}: How to develop layout algorithms that take advantage of the extra flexibility of graph layout on the torus?
    \item \textbf{[RQ4.2]}: How can we best visualise the layout of a node-link diagram on the surface of a torus on a piece of paper or 2D computer monitor?
    \item \textbf{[RQ4.3]}: What, if any, are the perceptual benefits graph layout on a torus has over standard layout on a 2D plane?
\end{itemize}

To answer these questions, we first present a modified version of a stress minimisation algorithm that supports interactive human-guided layout (\autoref{sec:torus:interactivelayout}) of graphs on the torus (\textbf{RQ4.1}). 
In~\autoref{sec:torus1:aesthetics:subsec1}, we describe graph aesthetic metrics~\cite{purchase2002metrics}, that have been correlated with user preference and performance in readability tasks \cite{ware2002cognitive}.  
Number of link-link crossings, the first metric that we can demonstrate can be improved by toroidal layout, is certainly known to affect readability.  Another metric that may benefit from a toroidal topology is the minimum angle between links incident on nodes. 

In two controlled experiments with 48 participants, we explore if these improvements in metrics permitted by toroidal embeddings lead to an improvement in readability greater than any detrimental effect of, for example, the challenge of following links that wrap around the boundaries of the drawing, and whether providing partial, full context, or interactive panning aide understanding of toroidal graphs (\textbf{RQ4.2}), compared with their standard non-wrapped representations (\textbf{RQ4.3}).




\section{\twebcola{} Stress Minimising Torus Layout}
\label{sec:torus1:stressminimisation}
As described in~\autoref{sec:related:network:forcedirected}, for very dense, scale-free, or small-world graphs, standard force-directed methods generate node-link layouts that are difficult to read.
\begin{figure}
    \centering
    \includegraphics[width=0.6\columnwidth]{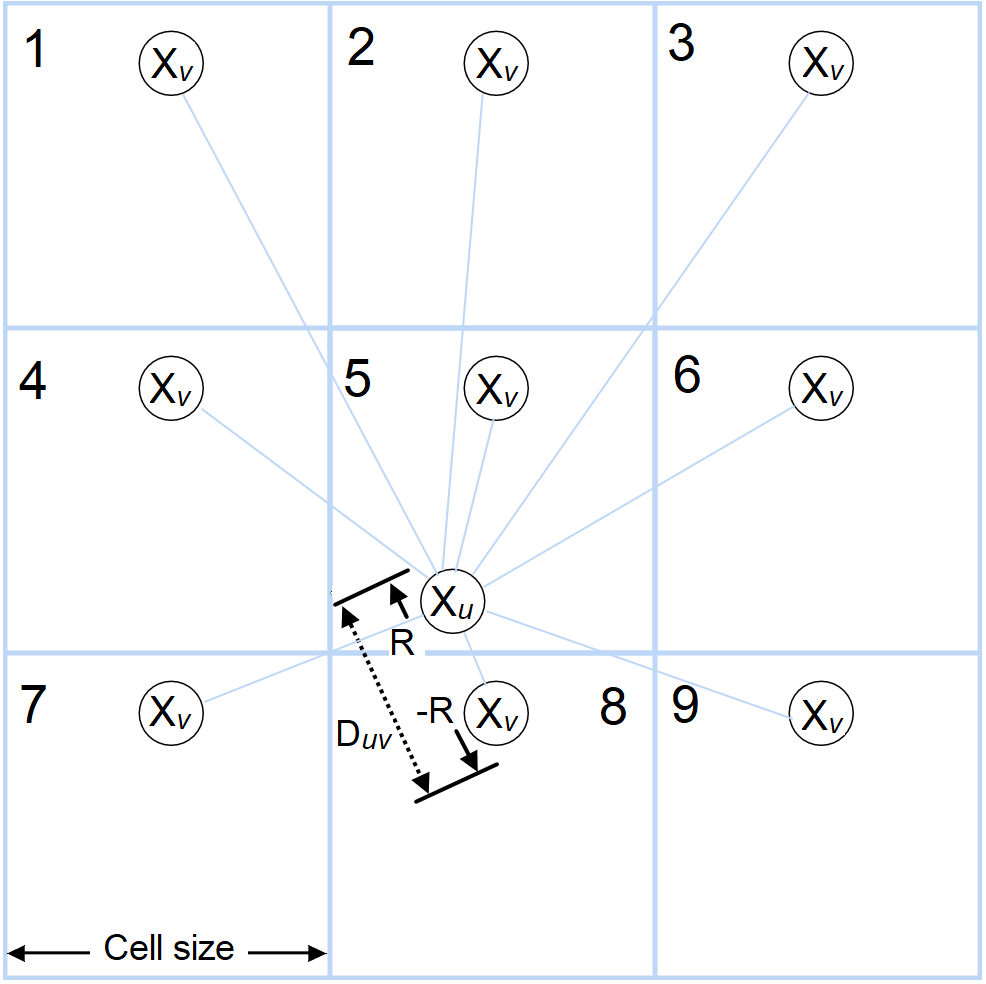}
    \caption{The stress minimising algorithm finds the wrapping that is closest to its graph theoretic distance out of 9 possible placement of nodes for each node pair. For example, the Euclidean distance between a node $X_u$ with respect to another node $X_v$ is closest to its graph theoretic distance $D_{uv}$. Therefore the adjacency between cell 5 and 8 is chosen, resulting in a vertical wrapping.}
    \label{fig:threebythreetile}
\end{figure}

Force-directed methods are the most commonly-used layout algorithms for general purpose graph visualisation.
These methods find a layout by minimising an objective function, such as the standard \textit{stress} function, based on differences between ideal and the actual distances in a low-dimensional coordinate system~\cite{brandes2008experimental,dwyer2009scalable,zheng2018graph,devkota2019stress}.
We adapt the stress function for 2D toroidally-wrapped topology.
We consider the positions of each pair of nodes in a 3 $\times$ 3 repeated tiling, with equal square \textit{cell size} as shown in~\autoref{fig:threebythreetile}, which is the length of each square tile. 
Each node is considered to have nine positions, with the same offset position within every cell.  Like \twebcola{}, for each pair of nodes, we compute the gradient information across the nine possible ways to consider their adjacency.  
Thus, we have the following definitions for stress for conventional unwrapped graphs \nodelinklayout{}, and then for toroidal wrapped graphs \toruslayout{}.
Given a graph $G$ with nodes $V$, we define:
\begin{equation}
    \label{eqn:stress}
    \mathit{stress} = \left\{ \begin{array}{rl}
    \sum_{(u,v) \in V\times V, u\ne v} \frac{(L\times D_{uv} - d_{uv} )^2}{(L\times D_{uv})^2} & \nodelinklayout{} \\ 
    \sum_{(u,v) \in V\times V, u\ne v}\ min_{w \in\ W} \frac{(L\times D_{uv} - d_{uvw})^2}{(L\times D_{uv})^2} & \toruslayout{} \\
    \end{array}\right.
\end{equation}
For a pair of nodes $(u,v)$, $d_{uv}$ is the Euclidean distance between $u$ and $v$, $D_{uv}$ is the shortest graph-theoretic path length between them. 
In \toruslayout{}, $w \in \{1\ldots 9\}$ selects one of the nine possible adjacencies which informs the wrapping as described previously, such that each $d_{uvw}$ is the actual (Euclidean) distance of $(u,v)$ between the centre cell and adjacent cell $w$. For both \nodelinklayout{} and \toruslayout{}, the term $\frac{1}{L\times D_{uv}^2}$\ is used to penalise long-range attraction.
The constant $L$ indicates the desired link length in the resulting layout. 

Various methods for minimising \emph{stress} are used, detailed in~\autoref{sec:related:mds}.  We follow Dwyer et al.~\cite{dwyer2008topology} in using a gradient descent approach.  Our approach is implemented as a modification of the \emph{WebCoLa}~\cite{webcola} implementation of the algorithm from Dwyer et al.~\cite{dwyer2008topology}.
Note, that this technique moves all pairs of nodes at a time by taking the partial derivatives across all pairs of nodes, summed to give a gradient vector across all node positions, and thus the placement of nodes occurs for all pairs of nodes. 

In this thesis, we name this approach \twebcola{}\footnote{We name our simple interactive layout approach \twebcola{}, given that it minimises the stress function by moving all pairs of nodes at a time using gradient descent (an optimisation method for finding a local minimum of a differentiable function), in contrast to a different approach, named \textit{\tpairwise{}}, which we explore in~\autoref{sec:torus2} that minimises the stress function by moving a single pair of nodes at a time using gradient descent.}. Meanwhile, the \twebcola{} approach using \emph{WebCoLa} also allows us to add constraints to avoid overlapping node labels.


We take a straightforward approach to adapting the goal function (1) to the torus.
At each iteration, we compute gradient contributions of each node, not just with respect to each other node, but with respect to the positions of each other node in the eight cells adjacent to the central cell in a $3\times 3$ tiling corresponding to the possible adjacencies in a torus topology, as seen in~\autoref{fig:threebythreetile}.  The positions of each node within each cell are computed by simple offset, i.e.\ in~\autoref{fig:threebythreetile} the position of node $X_{v}$ in the top-left cell is simply $X_{v}$'s $x,y$ position in the central cell, minus the cell size, and so on for the other cells.  This results in nine different sets of gradient contributions.  

From these nine choices of gradient components for each node, we compute the one that results in the largest reduction of stress, and take this to be the entry in the gradient vector for that node.  The resultant descent vector (computed from the gradient and a step size from the corresponding second derivative information, as per \cite{dwyer2008topology}) has horizontal and vertical components for each node with contributions from all other nodes e.g.\ for node $X_{u}$ in~\autoref{fig:threebythreetile}. If the descent vector takes a node beyond the cell boundaries, then the position for that node wrap around, as per asteroids rules.

\subsection{Interactive layout}
\label{sec:torus:interactivelayout}
As with all gradient descent approaches to optimisation of non-convex functions, it can happen that the layout can converge to a configuration corresponding to a poor local minimum of the stress function.  In future, new approaches for initialising the layout to a better state may autonomously give a high-quality trade-off between aesthetic measures.  For the layouts used in our study stimuli, however, we achieve this with an interactive, human-guided approach, made possible by our iterative stress minimisation approach.
A user can drag nodes while layout is proceeding which can be useful to guide the layout away from poor local minima, but also to explore different configurations of torus layout and the different symmetries that they reveal, as seen in ~\autoref{fig:k3_3_four_conditions}. This interactivity was useful in preparing stimuli for our controlled study (\autoref{sec:torus1:aesthetics}), which is not intended as an evaluation of the automatic layout algorithm, but rather an evaluation of the torus drawing paradigm in general as per \textbf{[RQ4.3]}, and for which we needed to control for the various aesthetic measures. 

In summary, a human-guided stress minimising approach to generate a torus layout is as follows\footnote{The detailed pseudocode to our method is available from \url{https://github.com/Kun-Ting/WebCola}}.

\begin{enumerate}[noitemsep,leftmargin=*]
\item Automatically obtain an initial layout, starting with all nodes at the centre of the centre cell in a 3$\times 3$ grid.

\item Of the nine possible wrappings for each pair of nodes, choose the wrapping such that the torus distance is closest to the ideal (graph-theoretic related) distance, \autoref{fig:threebythreetile}.  

\item 
For each pair of nodes, compute the partial derivative of the stress function for the chosen wrapping for that pair.   

\item As per \cite{dwyer2008topology}, we sum the partial derivatives across all pairs to give a gradient vector across all node positions.
\item Also per  \cite{dwyer2008topology} we use second derivative information to compute an optimal descent vector in the gradient direction.
\item After moving all nodes according to the descent vector some may have moved outside of the centre cell.  These are translated back to the centre cell (e.g., by moving $X_{v}$ in Figure \ref{fig:threebythreetile} to its corresponding position in cell $5$).
\item The above steps are repeated until convergence to a local minimum, i.e.\ movement falls below a predefined threshold.
\item User drags a node to a desired position.
\item The layout algorithm automatically adjusts the positions of all other nodes using aforementioned steps 2-7.
\end{enumerate}

\section{Comparing Torus and non-Torus Drawings}
\label{sec:torus1:aesthetics}
In this section, we compare different ways to render torus drawings of graphs with standard non-toroidal node-link diagram representations. 
We do this using both 
\textit{i)} established metrics for assessing network layout aesthetic quality, and 
\textit{ii)} some new metrics specific to torus drawings, as described below.
We then prepared a corpus of 72 graphs using standard graph generation techniques that simulate real-world graphs as described in Section \textit{\nameref{sec:torus1:graph_corpus}}.  These were then laid out to balance the aesthetic measures using our interactive layout for both toroidal and non-toroidal conditions.
Finally, we prepared three different styles of renderings of these graph layouts for use as stimuli in our two controlled studies.

\subsection{Standard layout aesthetics measures}
\label{sec:torus1:aesthetics:subsec1}
The diagrams in our graph corpus were arranged to find a balance of a number of aesthetic measures.
For a graph with node set $V$ and edge (link) set $E$ we follow past work in considering the following graph layout aesthetics:

\noindent\textbf{Link Length Variance}---Graphs with more uniform link length have been found to be preferred by readers \cite{dwyer2009comparison}.  
We compute link length variance by first scaling the graph layout such that the average link length is 1.  Thereafter, we take the Link Length Variance as the average squared deviation from 1 across all links $E$.
\begin{equation}
    \frac{1}{|E|}\sum_{e \in E} ( 1 - l_e)^2
\end{equation}
where $l_e$ is the length of link $e$ in the scaled drawing.

\noindent\textbf{Minimum Angle}---For a node $v$ we take the ideal angle between adjacent links incident to $v$ to be $\theta_v = \frac{360^{\circ}}{\mathit{deg}(v)}$, where $\mathit{deg}(v)$ is the number of incident links on $v$.  Then, our Minimum Angle metric, is the average difference between this ideal angle of incidence and the actual minimum angle of incidence for links incident to $v$ in the drawing:
\begin{equation}
    \frac{1}{|V|}\sum_{v \in V} \frac{|\theta_v - \mathit{min}\theta_{v}|}{\theta_v}
\end{equation}

\noindent\textbf{Link-link Crossings}---The adverse effect of link-link crossings on readability of graphs is well studied \cite{huang2009measuring}. Since we consider only straight-line drawings link-link crossings is a straightforward count of the number of times in each diagram that the line segments for pairs of links intersect.
In all of the diagrams in our corpus we generally tried to keep crossing counts low by steering the stress minimisation algorithm, but truly crossing minimal drawings with straight-line links are difficult; first, to identify, especially in larger graphs, and second---especially for the non-torus condition---it is difficult to nudge the stress layout into a state that leads to poor link length variance.  In general, we did not want to overly prioritise link-link crossing minimisation over the other metrics, as studies such as \cite{kieffer2015hola} indicate a balance of metrics is required.

Past work has found node-node overlaps and link-node crossings significantly worsen graph readability \cite{bennett2007aesthetics}.  As per \cite{dwyer2006ipsep}, we include constraints in late iterations of the stress minimisation algorithm to prevent node overlaps.  There are automatic processes for removing link-node crossings \cite{simonetto2011impred, dwyer2008topology}.  For the layouts in our diagram corpus, we remove node-link crossings through a simple manual nudging, taking care not to overly disturb the other metrics.

\subsection{Torus layout aesthetics measures}
\label{sec:torus1:aesthetics:subsec2}
We define the following two measures for torus layouts, specifically.

\noindent\textbf{Left-right and Top-bottom Wrapping}---Straight-line links lead the readers' eye in an undemanding way from one node to its neighbours.  However, where the links wrap around the links of a torus drawing, it seems reasonable to assume that finding the continuation of the link on the other side of the diagram is a more demanding task.  We report left-right and top-bottom wrapping separately, as we have controlled the stimuli in our study such that their numbers are similar overall.

\begin{figure}
    \centering
	\includegraphics[width=4cm]{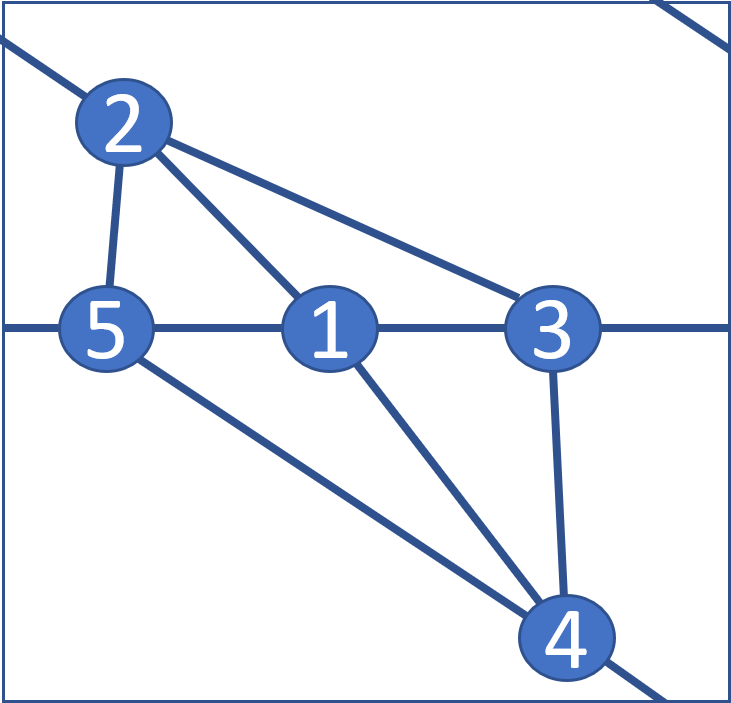}
	\caption{Corner wrapping}
	\label{fig:torus1:cornerwrap}
\end{figure}

\noindent\textbf{Corner Wrapping}---When diagonal links wrap near the corners of a torus diagram the link must be split into three segments, as per the example to the right (\autoref{fig:torus1:cornerwrap}).  In pilots for our first study we quickly realised that such corner-wrapping is especially confusing to readers.  As described in Study 1, we therefore introduced additional training material to prepare participants for this case and controlled for corner-wrapping while generating our stimuli.

\section{Graph Corpus}
\label{sec:torus1:graph_corpus}

\begin{figure}
    \centering
	\includegraphics[width=\textwidth]{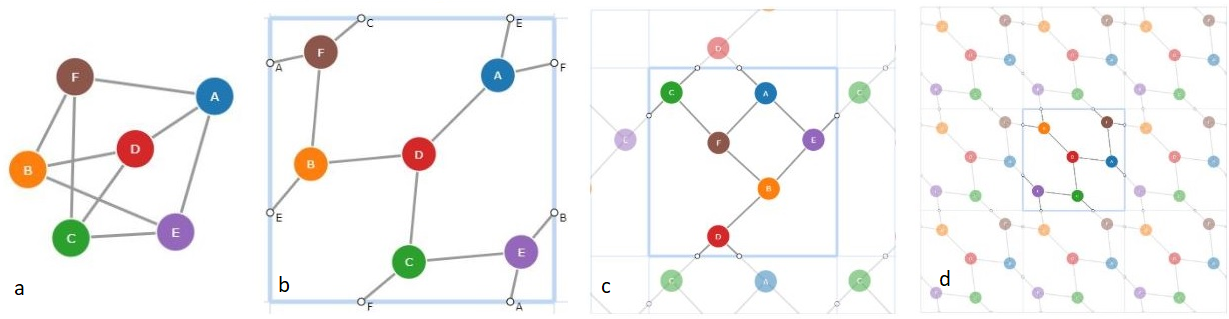}
	\caption{The complete bi-partite k3,3 graph drawn using the four different techniques considered in this chapter: (a) \nodelinklayout; (b) \tnocontext; (c) \tpartialcontext{}; (d) \tfullcontext}
	\label{fig:k3_3_four_conditions}
\end{figure}

\begin{figure}
    \centering
	\includegraphics[width=\textwidth]{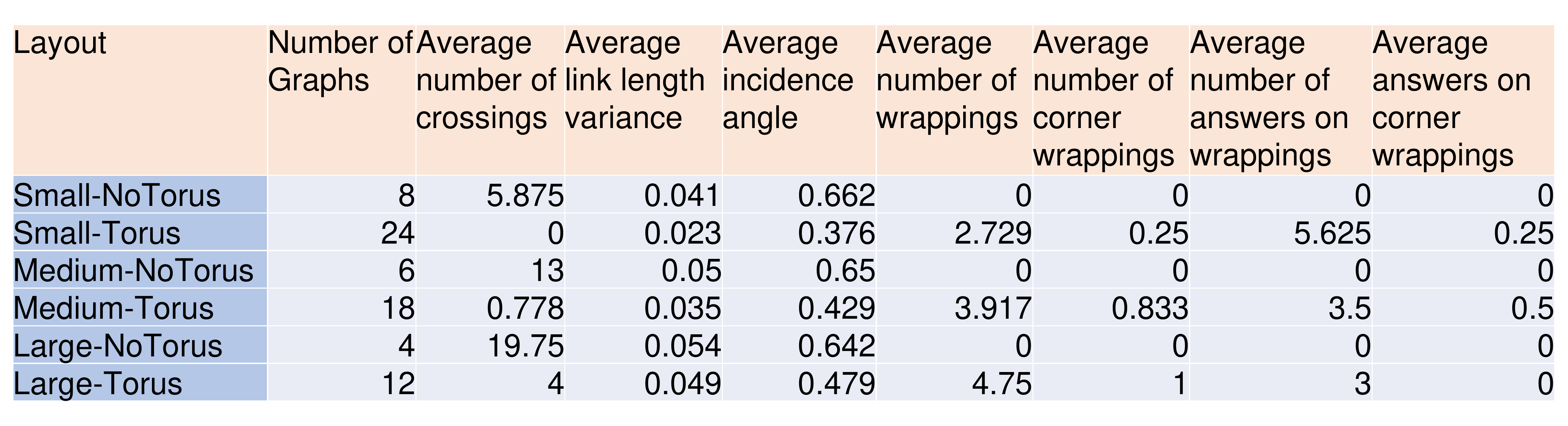}
	\caption{Average number of crossings, link length variance, and incidence angle. Total number of torus wrappings and number of answers requiring a wrapping.}
	\label{fig:torus1:graphstats}
\end{figure}

In this section we describe the test corpus of graphs, generated as described below, that we used in our studies.   The layouts were generated using the human-guided stress minimising approach as described above and \emph{WebCoLa}~\cite{webcola} layout for non-torus graphs, also with some manual adjustment.  In general we tried to find layouts that balanced the crossings, minimum angle, and link length variance statistics.  However, there is a tradeoff.  For example, in some cases fewer crossings may be possible (as in~\autoref{fig:k3_3_four_conditions}(a)), but it would come at a great cost to the other measures.  
As can be seen from the table, the torus embedding permitted significantly better layouts in terms of all three of these measures.  However, it came at the new cost of wrapping.
Table \ref{fig:torus1:graphstats} provides summary statistics for these aesthetic measures.

\begin{figure}
    \centering
	\includegraphics[width=\textwidth]{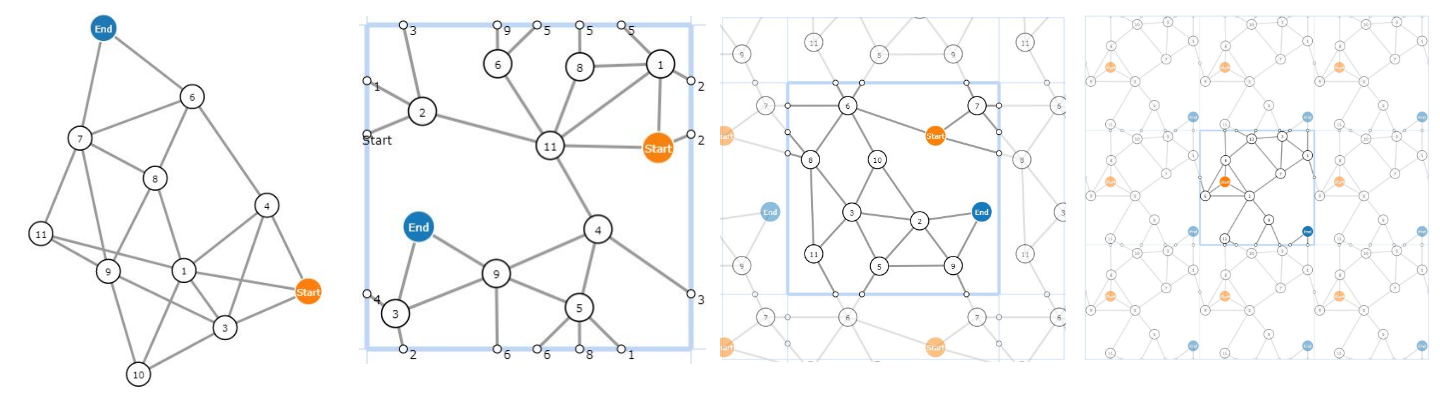}
	\caption{Example graph for our studies for \dmedium{}. Techniques left-to-right are \nodelinklayout, \tnocontext, \tpartialcontext{} and \tfullcontext.}
	\label{fig:torus1:four_layouts_low_degree_line_crossings}
\end{figure}

\begin{figure}
    \centering
	\includegraphics[width=\textwidth]{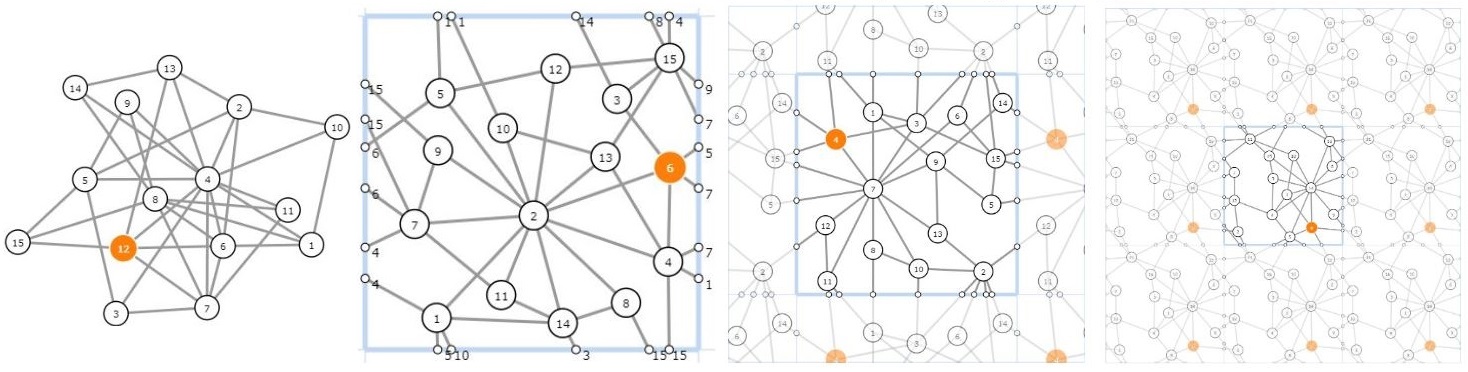}
	\caption{Example graph for our studies for \dlarge{}.  Techniques left-to-right are \nodelinklayout, \tnocontext, \tpartialcontext{} and \tfullcontext.}
	\label{fig:torus1:four_layouts_high_degree_line_crossings}
\end{figure}

We generated a sample corpus of 18 graphs rendered using the four different techniques for a total of 72 torus and non-torus drawings. 
The graphs were generated using algorithms designed to simulate real-world social networks and biological networks, using generators from NetworkX~\cite{networkx}. The majority of the graphs we used are Watts-Strogatz's small-world and Barabasi-Albert's scale-free graphs. We also used Erdos-Renyi's binomial network model to generate networks that balance the total number of wrappings. Therefore, we used one scale-free, small-world, and binomial model generator for creating graph instances for \dsmall{} class. For \dmedium{} class, we used two scale-free generators and one small-world generator , an example is shown in~ \autoref{fig:torus1:four_layouts_low_degree_line_crossings}.  For \dlarge{} class, we used two scale-free generators and one small-world generator, an example is shown in~ \autoref{fig:torus1:four_layouts_high_degree_line_crossings}. 

Each class (\dsmall{}, \dmedium{}, \dlarge{}) had a constant number of nodes (8, 11, and 15, respectively) and number of links in a tight range (12--18, 18--28, and 26--36).  

\section{Torus Rendering Techniques}
\label{sec:torus1:rendering}
In this section we comparatively evaluate four different visual representations for node-link diagrams.  An example of each representation applied to the complete bi-partite graph known as \textit{k3,3} is shown in Figure \ref{fig:k3_3_four_conditions}. 

\begin{itemize}[noitemsep,leftmargin=*]
\item\nodelinklayout{} is a conventional non-toroidal node-link diagram as laid out by a force-directed technique (WebCoLa \cite{webcola}) with manual refinements  to balance metrics, Fig.\  \ref{fig:k3_3_four_conditions}(a).  

\item\tnocontext{} is a torus drawing without any contextual tiling, as described below, Fig.\  \ref{fig:k3_3_four_conditions}(b).  Note that the torus drawings used in our studies include labels at the boundaries indicating to which nodes wrapped links are connected.  The need for such labelling in torus drawings without additional context became clear from the pilots for our first study.  

\item\tpartialcontext{} shows only partial context, that is, we see part of a repeated $3\times3$ tiling to show some torus adjacencies, Fig.\  \ref{fig:k3_3_four_conditions}(c). 

\item\tfullcontext{} shows the full $3\times3$ layout tiles (context) of a torus drawing, Fig.\  \ref{fig:k3_3_four_conditions}(d).  

\end{itemize}

Note that the diagrams including context (\tpartialcontext{} and \tfullcontext) are scaled to the same size in the figures in this paper, but in our studies they were shown to participants so that the central cell is the same size as the \tnocontext{} drawing; i.e., the area of \tfullcontext{} drawings shown to study participants were nine times that of \tnocontext{} drawings.


In Study 1 we evaluate static versions of these four representations.  In Study 2 we introduce a simple panning interaction for the torus rendering techniques, giving us three more conditions: \tnocontextpan{}, \tpartialcontextpan{}, and \tfullcontextpan{}.

\section{User Study 2: Static Torus Drawing Readability}
\label{sec:torus:study1}
In this study, we investigate if our torus drawings are effective to solve low-level network analysis tasks and if they are more efficient than baseline drawings of node-link diagrams without torus link wrapping. Thus the techniques in our study were exactly the same as described in the previous section. We were also interested in subjective user feedback and if the amount of visual \textit{context} shown has an impact on effectiveness and efficiency of torus drawings.

\subsection{Tasks}
There are many tasks that users typically perform while analysing network visualisations \cite{lee2006tasktaxonomy}.  For our study, we selected representative link and path following tasks for network analysis \cite{lee2006tasktaxonomy} whose performance we believed would be influenced by torus link wrapping. 
\begin{itemize}[noitemsep,leftmargin=*]
	\item \tshortestpath: \textbf{What is the shortest path in terms of number of links that need to be traversed between the nodes labelled \emph{Start} and \textit{End}?} Participants had to count the number of links between the marked nodes. We recorded participants' responses with multiple-choice questions with 8 options with answers of similar length but varying length.
	\item \tneighbours: \textbf{Identify all the friends (neighbouring nodes) of the orange node.} We recorded participants' responses with multiple-choice questions with 8 options with similar ordered lists of nodes of various length. 
	\item \tnodecount: \textbf{Identify the total number of people (nodes) in the network}. To answer, we provided a slider with a range of 1 to 20. Error rate was obtained by calculating the absolute difference between participant's answers and actual answers divided by the actual answers.
	\item \tlinkcount: \textbf{Identify the total number of relationships (links) of the network}. To answer, we provided a slider with a range of 1 to 20. Error rate was calculated the same way as for \tnodecount.
\end{itemize}

\subsection{Data sets}
For all tasks, we chose graphs from one of the 18 graphs generated as described in Section \textit{\nameref{sec:torus1:graph_corpus}}. 
Out of 18 trials in each condition in our study, there were 3 \dsmall{}, 3 \dmedium{}, 2 \dlarge{} graphs (difficulty levels as defined above) used in \tshortestpath{} tasks, 3 \dsmall{}, 3 \dmedium{}, 2 \dlarge{} in \tneighbours{} tasks, 1 \dsmall{} graph in \tnodecount{}, and 1 \dsmall{} graph in \tlinkcount. While the graph size (up to 15 nodes, 36 links) is relatively small, as pointed out in \cite{dwyer2009comparison, dwyer2008exploration, kieffer2015hola}, such small graphs already present a significant challenge to readability and present a suitable level of difficulty for path-following tasks in a study. 

\subsection{Hypotheses}
Our predictions were pre-registered with the Open Science Foundation: \url{https://osf.io/2e6bm}.

\textbf{Effect of layout}
\begin{itemize}[noitemsep,leftmargin=*]
\item\textbf{L4.1}: \tfullcontext{} has better task effectiveness (in terms of participant speed and error) than \tnocontext{}, and \tpartialcontext{} across all task difficulties ([RQ4.2]).
\item\textbf{L4.2}: \tfullcontext{} has better task effectiveness than \nodelinklayout{} (involves too many link crossings), \tnocontext{} (requires too much mental wrapping), and \tpartialcontext{} (requires certain mental wrapping) for difficult tasks ([RQ4.2][RQ4.3]).
\item\textbf{L4.3}: \tpartialcontext{} has better task effectiveness than \tnocontext{} across all task difficulties ([RQ4.2]).
\item\textbf{L4.4}: \nodelinklayout{} has better (participant reported) task effectiveness than \tfullcontext{} for \dsmall{} and \dmedium{} graphs (\textbf{RQ4.3}).
\end{itemize}

\textbf{Effect of tasks}
\begin{itemize}[noitemsep,leftmargin=*]
\item\textbf{T4.1}: Participants will perform better (in terms of participant speed and error) using \tfullcontext{} than \tnocontext{} and \tpartialcontext{} on \tshortestpath{} and \tneighbours{} (\textbf{RQ4.2}).
\end{itemize}


\textbf{Effect of size of graphs}
\begin{itemize}[noitemsep,leftmargin=*]
\item\textbf{D4.1}: Participants will perform better using \tfullcontext{} across all task difficulties than \tnocontext{} and \tpartialcontext{} (\textbf{RQ4.2}).
\end{itemize}


\textbf{Participant Preference}
\begin{itemize}[noitemsep,leftmargin=*]
\item\textbf{P4.1}: Participants will prefer \tfullcontext{} over either \tnocontext{} or \tpartialcontext{} ([RQ4.2]).
\item\textbf{P4.2}: Participants will prefer \tpartialcontext{} over \tnocontext{} ([RQ4.2]).
\item\textbf{P4.3}: Participants will report more confidence in using \tfullcontext{} than \tnocontext{} and \tpartialcontext{} ([RQ4.2]).
\item\textbf{P4.4}: Participants will report more confidence in using \nodelinklayout{} than \tfullcontext{} ([RQ4.3]).
\end{itemize}

\subsection{Participants}
We recruited 24 participants through email and snowball from our institute. 6 people were female while 18 were male. There were 2 participants aged below 20, 16 between 20-30, 5 between 30-40, and 1 greater than 50. While 5 participants never see social network diagrams, the rest 19 responded they either seldom (15) or often (4) see network diagrams from a course, textbook, or Internet. \rev{In pilot studies, we found participants were confused about where the data points and link connect when they were split across the boundaries, especially for low-level path following tasks. Therefore, we opt for a controlled lab-based study where the experimenter went through the tutorials with participants and supervised the study.} 

\subsection{Design and procedure}
We opted for a within-subject design study with 3 factors: 4 techniques $\times$ 3 difficulty levels $\times$ 4 tasks. We used a full-factorial design to balance for the 4 techniques (24 orderings). We used 18 trials with the number of graphs in each difficulty level (Table~\autoref{fig:torus1:graphstats}) + 10 training trials = 28 trials in each technique. For each technique, participants progressed through the same order of graph sizes from \dsmall{}, \dmedium{} to \dlarge{}.  
Participants sat in front of a Dell 22-inch LCD screen. To collect visual focus of each representation, we equipped the Tobii-pro X3-120 eye-tracking system to record visual focus of participants doing trials. We used a laptop with Intel I5 8350U (1.7GHz), Intel UHD Graphics 620 to run the website and control experiment. 

\subsection{Results}
\begin{figure}
    \centering
	\includegraphics[width=\textwidth]{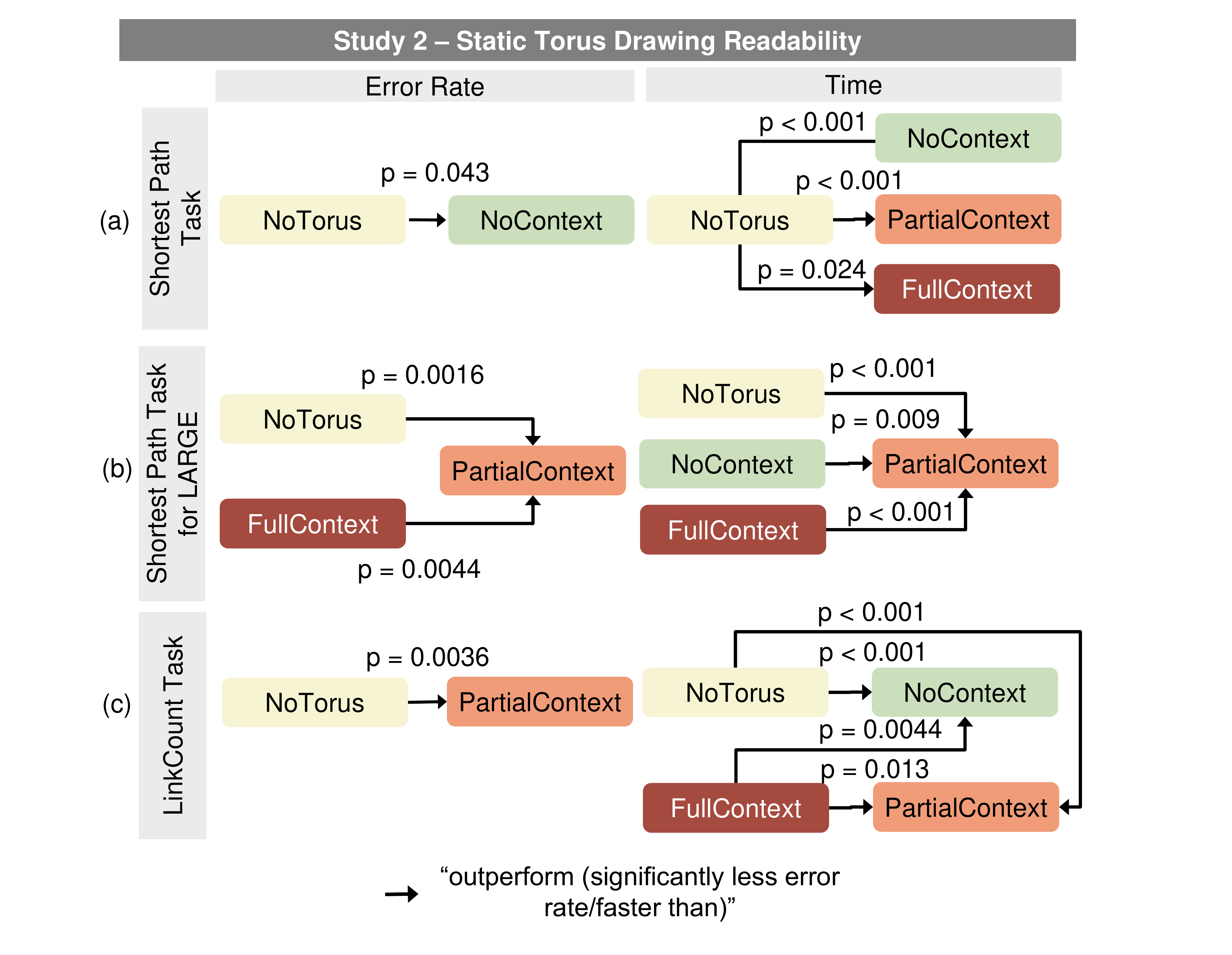}
	\caption{Study 2: Statistical results of performance comparisons between \nodelinklayout{} and Torus drawings under 95\% confidence level.}
	\label{fig:torus1_study2_performancesummary}
\end{figure}

\begin{figure}
    \centering
	\includegraphics[width=0.65\textwidth]{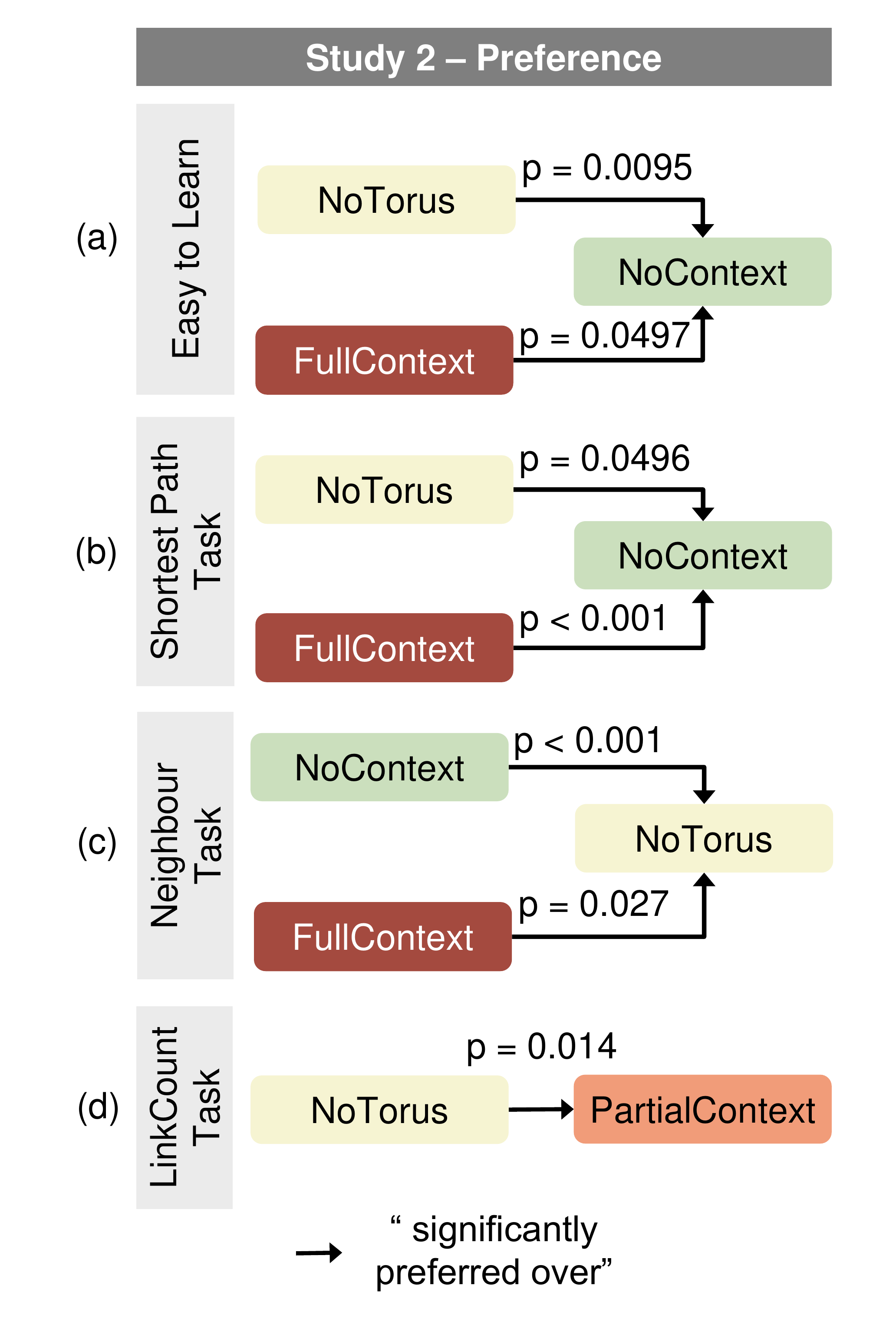}
	\caption{Statistical results of preference differences of Study 2 (95\% confidence level).}
	\label{fig:studypreferencesummary}
\end{figure}

\begin{figure}
    \centering
	\includegraphics[width=10cm]{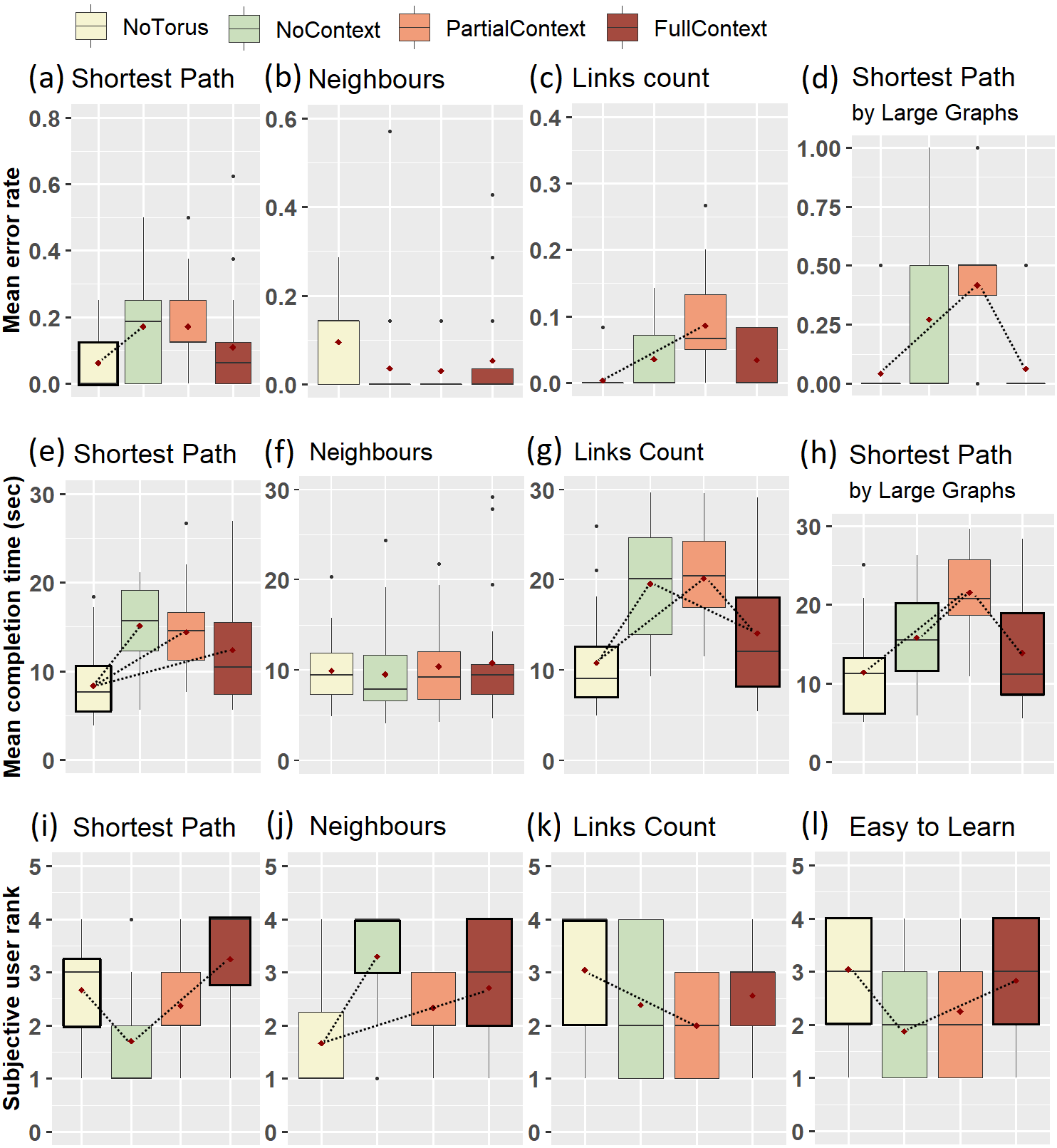}
	\caption{Study 2: Results for error, time, and subjective user rank by task. Higher rank indicates stronger preference. Dotted lines indicate significant differences for $p < .05$. The best significant results are highlighted in the border of bars.}
	\label{fig:study1results}
\end{figure}

All of the 24 participants completed 72 trials in the real (non-training) question set. They answered all the questions in tutorial and training sections correctly before entering the real question set. Therefore, we recorded the performance of 1,728 trials. There were 29 significant differences in Study 2 as summarised below, based on 95\% confidence levels. Since the distribution of the logarithm of completion time of each condition followed normal distribution, we used ANOVA repeated measures and Tukey's post-hoc pairwise comparison to test significance. Since the distribution of error rate and subjective rank of each condition did \textit{not} follow normal distribution, we used Friedman's non-parametric test to test null-hypothesis and Nemenyi's post-hoc pairwise comparison to test significant difference. Significant differences are summarised in \autoref{fig:torus1_study2_performancesummary}(error and time), and \ref{fig:studypreferencesummary}(participant preferences). Graphics with detailed results are found in \autoref{fig:study1results}.

For \textbf{\tshortestpath} and \textbf{\tlinkcount}, our most significant results showed the following:
\begin{itemize}[noitemsep,leftmargin=*]
    \item \nodelinklayout{} significantly outperformed \tnocontext{} in error and time independent of size of graphs (as seen in~\autoref{fig:torus1_study2_performancesummary}(a), \autoref{fig:torus1_study2_performancesummary}(c), box plots in ~\autoref{fig:study1results}(a), \autoref{fig:study1results}(e), \autoref{fig:study1results}(g)) and outperformed \tpartialcontext{} (as seen in ~\autoref{fig:torus1_study2_performancesummary}(a), \autoref{fig:torus1_study2_performancesummary}(c), with box plots in \autoref{fig:study1results}(c), \autoref{fig:study1results}(e), \autoref{fig:study1results}(g)); 
    \item \tfullcontext{}{} significantly outperformed \tpartialcontext{} in time independent of size of graphs (as seen in ~\autoref{fig:torus1_study2_performancesummary}(c), with its box plot in ~ \autoref{fig:study1results}(g));
    \item For \tshortestpath{} by \dlarge{} graphs, \nodelinklayout{} and \tfullcontext{} both significantly outperformed \tpartialcontext{} in error and time (as seen in ~\autoref{fig:torus1_study2_performancesummary}(b), with box plots in ~\autoref{fig:study1results}(d), \autoref{fig:study1results}(h));  
    \item Compared to \tnocontext{}, participants found it easier and reported greater confidence using both \nodelinklayout{} and \tfullcontext{} (as seen in~\autoref{fig:studypreferencesummary}(a), \autoref{fig:studypreferencesummary}(b), with box plots in~\autoref{fig:study1results}(i), \autoref{fig:study1results}(l)). \nodelinklayout{} was significantly preferred over \tpartialcontext{} (as seen in~\autoref{fig:studypreferencesummary}(d) with its box plot in~\autoref{fig:study1results}(k));
\end{itemize}


For \textbf{\tneighbours}, we found
\begin{itemize}[noitemsep,leftmargin=*]
    \item \tnocontext{} and \tfullcontext{} were significantly preferred over \nodelinklayout~(\autoref{fig:studypreferencesummary}(c), \ref{fig:study1results}(j)).
    \item This correlates with a weak trend for \tnocontext\ to outperform \nodelinklayout\ in error rate ($p = 0.23$) independent of size of graph, as seen in \autoref{fig:study1results}(b). The results of performance by graph size are omitted for space limitations and can be found in the study material web link.
\end{itemize}
Based on participants' qualitative feedback, they used \tnocontext\ with labelled display to quickly identify neighbours of a chosen node. \tnocontext{} drawing was (surprisingly to us) the most preferred torus display.

For \textbf{\tnodecount}, results did \textit{not} show any significant difference between techniques and most participants could correctly identify the number of
nodes in all conditions, which suggests that most participants were not confused by the torus wrapping. The result is omitted and can be found in the study material web link.

Based on these results, we \textit{rejected} the following hypotheses for [RQ4.2]: L4.3, T4.1 (for \tneighbours{}), P4.2, P4.4, and for [RQ4.3]: L4.2, L4.4 (for \dsmall{}), P4.4. We accepted hypotheses for [RQ4.2]: L4.1, L4.2, T4.1 (for \tshortestpath), D4.1, P4.1, P4.3 and for [RQ4.3]: L4.4 (for \dmedium{}).

\subsection{Qualitative user feedback}

The majority of participants mentioned that their preferences were dominated by the technique, independent from task and graph size. Participants favoured \tfullcontext{} for \tshortestpath{} over \tnocontext{} because it helped understanding link wrapping and provided a great overview over the network. At the same time, they liked the cleanness of \tnocontext{} which has no repetition and extra information, which allowed them to concentrate on the task. 

\subsection{Summary}
In our first study, we generally found an improvement to torus drawing by adding full context to torus drawing. Our results can be summarised as follows. For \textbf{[RQ4.3]} (What, if any, are the perceptual benefits graph layout on a torus has over standard layout on a 2D plane?), \tfullcontext{} is as good as \nodelinklayout{} for \tshortestpath{}-\dlarge{}, \tnodecount{} and \tlinkcount{}. Static \tnocontext{} or \tpartialcontext{} torus is clearly worse than \nodelinklayout{}. For \textbf{[RQ4.2]} (How can we best visualise the layout of a node-link diagram on the surface of a torus on a piece of paper or 2D computer monitor?), \tfullcontext{} is the best static torus layout. Static \tnocontext{} or \tpartialcontext{} is the worst torus layout. \tnocontext{} had a tendency (confidence intervals $\le$ 77\%) to perform better in accuracy than \nodelinklayout, independently of graph size for \tneighbours{}. For \tneighbours, \tnocontext{} and \tfullcontext{} were both significantly preferred over \nodelinklayout{} in subjective user ranking (\textbf{[RQ4.3]}). Participants indicated that the \tnocontext{} technique appeared cleaner and would be better suited to show larger graphs.


The role of context is crucial yet not entirely conclusive from our results. Result 3 above, regarding preference for \tnocontext{} over torus drawings with context, implies that the redundant information disturbs users, even as it assists them (as per Result 1 and 2).  Moreover, context requires screen space.  We therefore designed a 2nd study (\autoref{sec:torus:study2}) to investigate the effect of interactive panning across the three torus representations.
\section{User Study 3: Torus drawings + Panning}
\label{sec:torus:study2}

For our second study, we repeat the evaluation of the three torus conditions from Study 2 (\tnocontext, \tpartialcontext, \tfullcontext): the visual representation in each of these techniques was exactly the same (e.g.~\autoref{fig:torus1:four_layouts_low_degree_line_crossings} and~\autoref{fig:torus1:four_layouts_high_degree_line_crossings}). The sole interaction was panning, using the mouse \rev{or touch drags}. Since link wrapping is not applicable to the node link representation we do not repeat trials for \nodelinklayout, but instead perform a between-groups analysis of the Study 2 results for this condition, with those in Study 3. Tasks and graphs were also the same as in Study 2. 
\subsection{Hypotheses}
Our predictions were pre-registered with the Open Science Foundation: \url{https://osf.io/v3756}.
\begin{itemize}[noitemsep,leftmargin=*]
    \item \textbf{I4.1}: \tnocontextpan\ has better task effectiveness than \tnocontext{} (\textbf{RQ4.2}).
    \item \textbf{I4.2}: \tpartialcontextpan{} has better task effectiveness than \tpartialcontext{} (\textbf{RQ4.2}).
    \item \textbf{P4.5}: Participants will prefer \tnocontextpan{} to \tfullcontextpan{} (\textbf{RQ4.2}).
\end{itemize}
\subsection{Participants, design, and procedure}
We recruited a new set of 24 participants from a local institute through email and snowball, none of which had participated in Study 2. One person was aged below 20, 17 between 20-30, 4 between 30-40, and 2 greater than 40. All of the participants never or seldom see social network diagrams. Similar to Study 2, we used a within-subject design study with 3 factors: 3 techniques, 18 graphs, and 4 tasks. We performed a between-subject study comparing Interactive panning as in this study to the results of no-panning in Study 2. We use the same trials as previous user study and add interactive panning to the trials, using the mouse. 

\begin{figure}
    \centering
    \includegraphics[width=\textwidth]{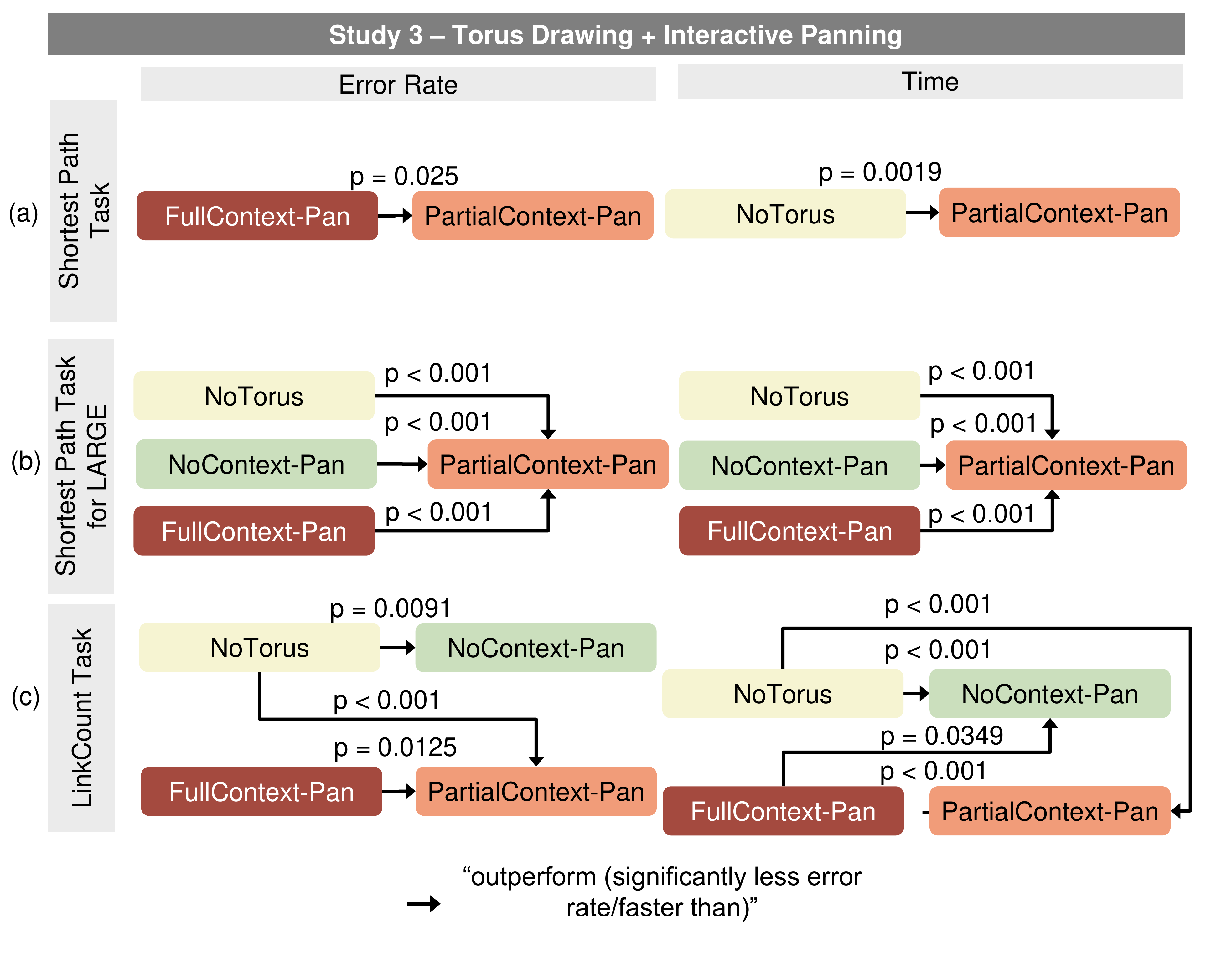}
    \caption{Study 3: Statistical results for performance of \nodelinklayout{} and the torus+panning conditions, with 95\% confidence level.}
    \label{fig:study2performancesummary}
\end{figure}
\begin{figure}
    \centering
	\includegraphics[width=0.65\textwidth]{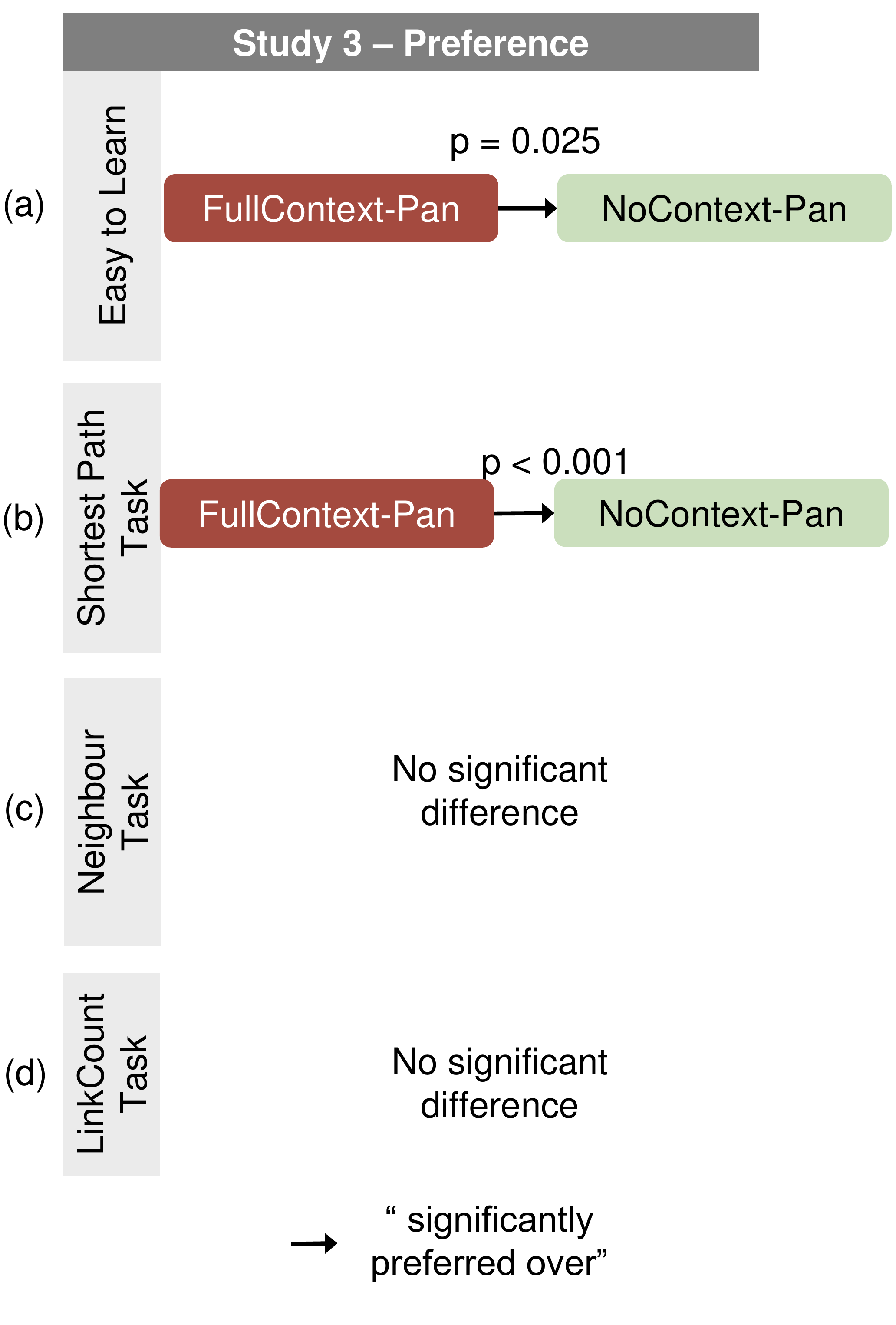}
	\caption{Statistical results of preference differences of Study 3 (95\% confidence level).}
	\label{fig:study3preferencesummary}
\end{figure}
\begin{figure}
    \centering
    \includegraphics[width=10cm]{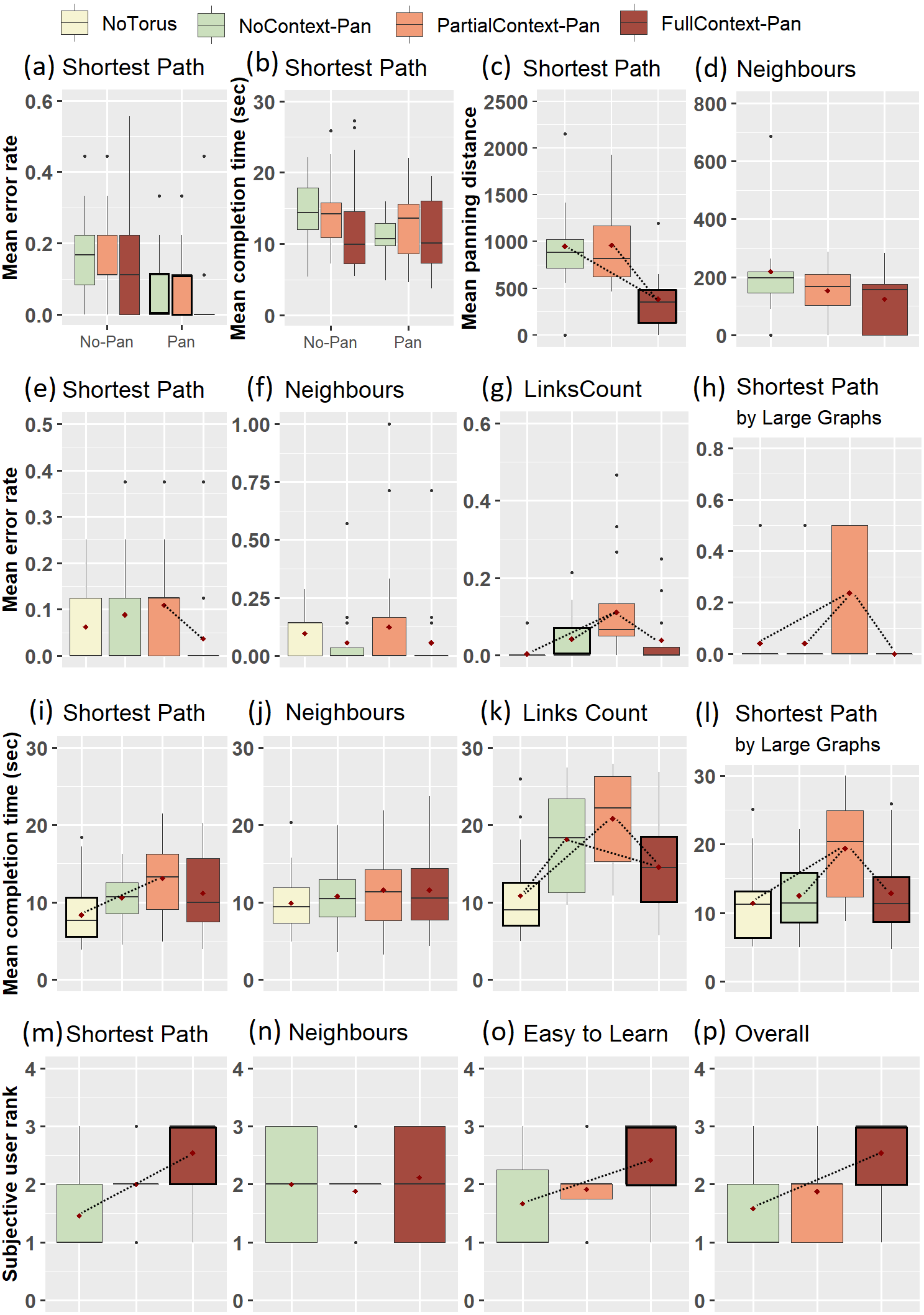}
\caption{Study 3: Results for error, time, panning distance, and subjective user rank by task. Higher rank indicates stronger preference. Dotted lines indicate significant differences for p < 0.05. The best significant results are highlighted in the border of bars.}
    \label{fig:study2results}
\end{figure}

\subsection{Results}
All 24 participants successfully went through 54 trials in the real question set. They answered all the questions in the tutorial and training sections correctly before entering the real question set. Therefore we recorded a performance of 1,296 trials. 

\begin{figure}
    \centering
    \includegraphics[width=\textwidth]{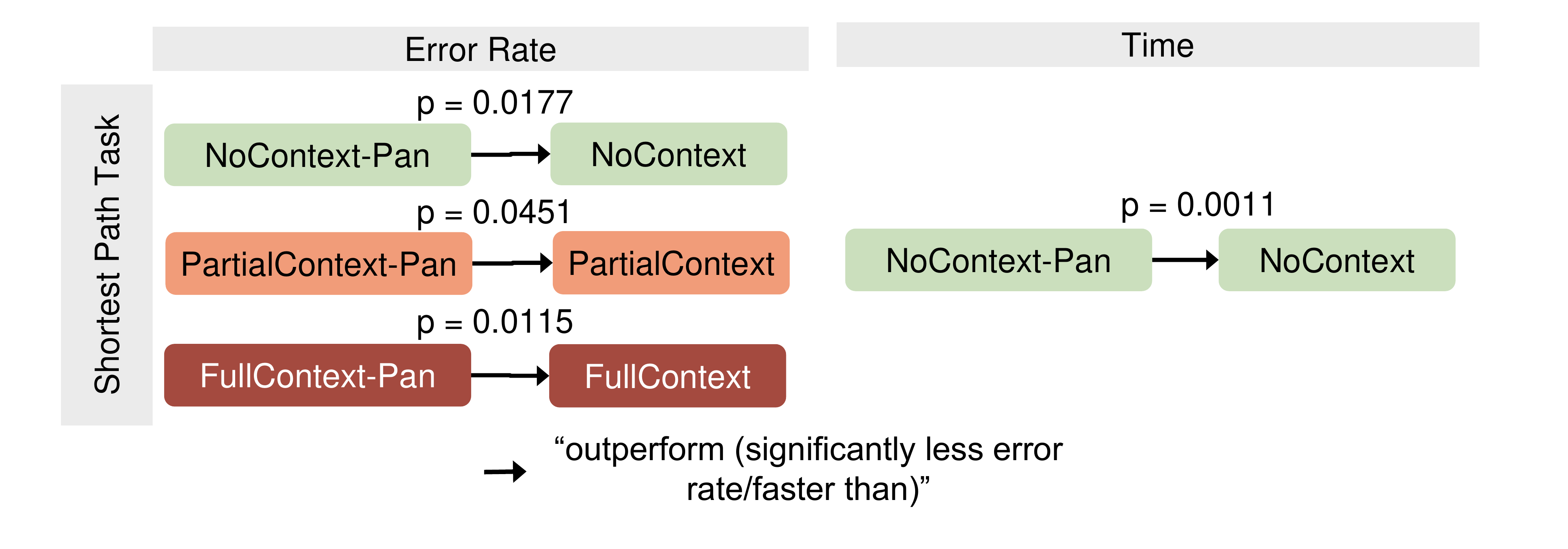}
    \caption{Statistical results of performance comparisons between groups of static torus and interactive panning, under 95\% confidence level.}
    \label{fig:study2betweengroupperformancesummary}
\end{figure}

To test for significance, of panning on performance between panning and no-panning conditions, we used paired t-test. We used Wilcoxon's signed rank test to test significance difference in error rate between 2 groups. To test differences in performance between \nodelinklayout{} in Study 3 and torus with panning, we opted for ANOVA independent measures and Tukey's post-hoc pairwise comparison, as the logarithm of completion time followed normal distribution. We used Kruskal-Wallis's non-parametric test and Wilcoxon multiple pairwise comparisons to test error. To test panning distance between layout conditions we used ANOVA RM and Tukey's post-hoc pairwise comparison. Significant differences are shown in~\autoref{fig:study2betweengroupperformancesummary} (between groups of static and interactive panning), \autoref{fig:study2performancesummary}(for indiviual techniques) and~\autoref{fig:study3preferencesummary} (subjective participant rankings). Graphics with detailed perfomance results are found in~\autoref{fig:study2results}. 

In contrast to Study 2, where there were significant differences between no-context and full-context torus or standard layout representations, with panning, we did not find any significant differences between no-context and full-context or standard layout, independent of graph size for \tshortestpath{} task. This suggests that panning makes no-context torus techniques equally performant to full-context or standard layout, as participants pan more in the conditions where less context is given.

We found significant differences in the panning conditions for \textbf{\tshortestpath}:
\begin{itemize}[noitemsep,leftmargin=*]
    \item Independently of graph size, all torus panning significantly outperformed their non-panning counterpart in error (as shown in~\autoref{fig:study2betweengroupperformancesummary},  \autoref{fig:study2results}(a)). \tnocontextpan{} significantly outperformed \tnocontext{} without panning in time (as shown in~\autoref{fig:study2betweengroupperformancesummary},  \autoref{fig:study2results}(b));
    \item \tnocontextpan{} and \tpartialcontextpan{} were panned significantly more than \tfullcontextpan{}, as seen in~\autoref{fig:study2results}(c);
    \item \nodelinklayout{} significantly outperformed \tpartialcontextpan{} in  time (as shown in~\autoref{fig:study2performancesummary}(a), \autoref{fig:study2results}(i)), and \tfullcontextpan{} significantly outperformed \tpartialcontextpan{} in  error (as shown in~\autoref{fig:study2performancesummary}(a), \autoref{fig:study2results}(e)) independently of size of graphs.
    \item With panning, \tnocontextpan{}, \tpartialcontextpan{} and \tfullcontextpan{} all significantly outperformed \tpartialcontextpan{} in error and time by \dlarge{} (as seen in~\autoref{fig:study2performancesummary}(c), \autoref{fig:study2results}(h), \autoref{fig:study2results}(l));
    \item Compared to \tnocontextpan{}, participants found it easier, and reported greater confidence in using \tfullcontextpan{} (as seen~\autoref{fig:study3preferencesummary}(a), \autoref{fig:study3preferencesummary}(b), \autoref{fig:study2results}(m), \autoref{fig:study2results}(o)). Overall, \tfullcontextpan{} was significantly ($p=0.0026$) preferred over \tnocontextpan{}, as seen in~\autoref{fig:study2results}(p);
\end{itemize}

For \textbf{\tneighbours}, we did not find significant differences between any of the torus techniques, 
but for \textbf{\tlinkcount}, both \nodelinklayout{} and \tfullcontextpan{} significantly outperformed \tnocontextpan{} and \tpartialcontextpan{} in error and time (as shown in~\autoref{fig:study2performancesummary}(b),  \autoref{fig:study2results}(g), \autoref{fig:study2results}(k)), but \tfullcontextpan{} outperformed \tnocontextpan{} only with significant support in terms of time.

Based on these results, we rejected P4.5 and accepted I4.1 and partially accepted I4.2 (with significant support in terms of Error).

\section{Discussion and Limitation}
\label{sec:torus1:discussion}
With respect to \textbf{[RQ4.3]}, our most significant results are 1) full-context or pannable full-context and no-context torus is as good as (not demonstrably different from) standard layout; 2) static or pannable partial-context torus is, despite being a drawing style most commonly used in the literature, e.g. by Kocay et al.~\cite{kocay2016graphs} (\autoref{fig:related:tiledgraph}), the worst torus layout and clearly worse than standard non-wrapped layout.
For \textbf{[RQ4.2]}, 1) full-context is the best static torus layout; 2) no-context + pan is as good as (not demonstrably different from) full-context. 



Overall, we found that while torus topology allowed for significant improvements in the standard graph layout aesthetic measures (as per~\autoref{fig:torus1:graphstats}), the wrapping of links at the sides of the diagram imposes a significant cost in terms of speed and accuracy in tasks requiring users to follow links (as demonstrated in~\autoref{sec:torus:study1}).  However, this cost is mitigated by providing either more wrapped context in static diagrams (\autoref{sec:torus:study1}), or by allowing interactive panning (\autoref{sec:torus:study2}).  Of the three torus representations shown in~\autoref{sec:torus:study2}, the \tnocontextpan{} representation seems to have shown the greatest benefit from the introduction of panning, and indeed panning was used more in the \tnocontextpan{} condition than \tfullcontextpan{}. It is interesting that different levels of replicated context (no, partial, full) have significant differences on torus readability. As a majority of users commented that they like the overview provided by full-context, it may be interesting to explore gaze distribution in different levels of context in future. 


There are also limitations. First, while we focus on precise graph readability and path/link following, the small graphs tested ($\le 15$ nodes, $\le$ 36 links) are not really representative of real-world networks of interest in many domains, such as social and biological networks. We recognise that testing other types of tasks (such as cluster identification) on realistically larger graphs than those in our studies is an important next step - but it is beyond the scope of this initial study which focuses on precise graph readability and path/link following. In the next chapter, we investigate if torus drawing works well for graphs larger than those in our studies. Based on the improved graph aesthetics such as crossing reduction, such torus topology with link wrapping has a potential to better untangle a network than non-wrapped layouts. We expect that the reduced clutter torus diagrams may work well for tasks particularly important to large networks, and those with complex structures that can benefit from a relaxation of the structure and less link crossings - an important network exploration task - visual cluster identification, which is explored in the next chapter. 

\rev{Another limitation is that we only tested one static force-directed layout algorithm. There are many different layout algorithms for 2D node-link visualisations, such as orthogonal link layout by Batini et al.~\cite{batini1986layout,kieffer2015hola}, layered layout by Sugiyama et al.~\cite{sugiyama1981methods}, network layout with data aggregation by Yoghourdjian et al.~\cite{yoghourdjian2018graph}, as described in~\autoref{sec:related:networks}. We leave detailed discussion to~\autoref{sec:conclusion:discussion}.}

Second, the automatic layout algorithm based on \twebcola{} stress-minimisation approach is known to easily become stuck in local minima of the stress function (as described in~\autoref{sec:torus:interactivelayout}), corresponding to a poor choice of link wrapping across the torus surface.  Therefore, the layouts tested in both studies involved human-guidance of the algorithm. The interactive nature of the algorithm means that a user can guide it to a quite reasonable layout, certainly layouts that were good enough for our study. However, such an interaction would not scale to larger, real-world networks. In the next chapter, we explore a different algorithm that is more robust and completely autonomous. We further evaluate the new algorithm with realistically larger networks that addresses the limitations stated in this section. 



\section{Conclusion and Future Work}
\label{sec:torus1:conclusion}
We have investigated a technique using a force-directed approach for mapping arbitrary network structures onto a two-dimensional toroidal topology with links wrapping around the boundaries for relaxing node positions to reduce visual clutter.


Our studies indicate that torus-based network layout could be a practical technique (not worse than the standard force-directed graph visualisation technique tested) for link-following tasks (identify neighbours/find the shortest path) and feature identification tasks (estimate the number of nodes or links), but that either redundant context or interactive panning are necessary. The graphs tested were automatically generated using algorithms designed to simulate naturally occurring graphs, but we would like to further evaluate the torus representations in a real-world application and see if it is usable by domain experts. This chapter addressed  \textbf{RG4} (\autoref{sec:intro:RGs:torus}). The study results confirmed the feasibility of two wrapping approaches: tile-display and interactive wrapping described in~\autoref{sec:designspace:torus} that aid understanding of torus wrapping. 

Technically speaking, our \twebcola{} stress-minimising torus layout method is the first method we are aware of that is able to layout all graphs (not just graphs limited to a particular genus) on a torus topology.  While we can force it to converge we cannot make strong guarantees that it converges to a local optimum.  However, we are interested to see if the combinatorial techniques for layout of restricted classes of graphs on the torus, can be adapted to help us find good starting conditions for our stress-minimising torus layout in order to create a robust and completely autonomous high-quality torus layout. 




%
\chapter{Interactive Torus Wrapping for Network Visualisations - Larger Networks}
\label{sec:torus2}

\cleanchapterquote{It is not how much empty space there is, but rather how it is used. It is not how much information there is, but rather how effectively it is arranged.}{Edward Tufte}{(American statistician and emeritus professor of political science, statistics, and computer science at Yale University.)}



Building upon the work presented in~\autoref{sec:torus1}, this chapter addresses \textbf{RG4} (\autoref{sec:intro:RGs:torus}). We introduce and evaluate two new algorithms for improving toroidal wrapped layouts.
We address limitations of our previous investigation of toroidal network layouts (described in~\autoref{sec:torus1:discussion}) by considering larger networks, and investigating whether torus-based layout might also better display high-level network structure like clusters.

First, we present a new algorithm for computing toroidal node-link diagram layout that is completely autonomous and considerably more robust than the previous \twebcola{} method described in~\autoref{sec:torus1:stressminimisation}, and consistently producing high-quality layouts (\autoref{sec:torus2:empiricalevaluation}). Like \twebcola{} stress minimising torus layout method, this algorithm is a general-purpose toroidal layout algorithm based on a variant of a force-directed placement, and therefore it is capable of laying out arbitrary networks. We extend the standard non-wrapped pairwise gradient descent algorithm of Zheng et al.~\cite{zheng2018graph} to handle the more complex case of layout on a torus.

\begin{figure}
    \centering
    \subfigure[Unwrapped]{
        \includegraphics[width=0.3\textwidth]{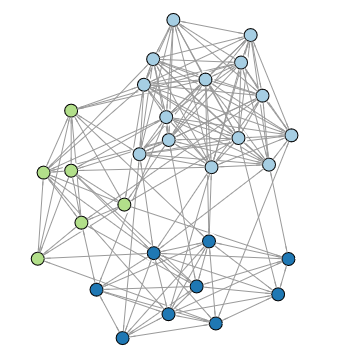}
    }
    \subfigure[Wrapped]{
        \includegraphics[height=0.3\textwidth]{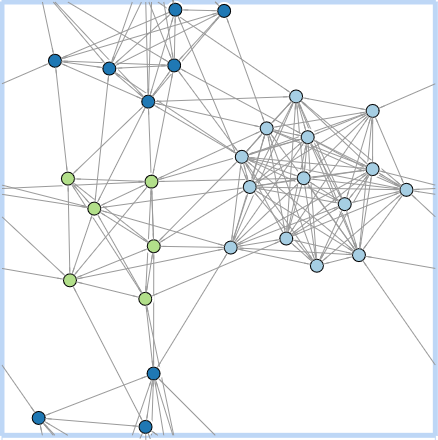}
    }
    \subfigure[Best pan]{
        \includegraphics[height=0.3\textwidth]{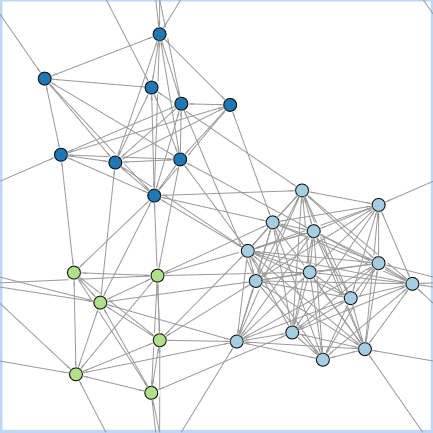}
    }
    \caption{Example of unwrapped and wrapped networks; Colours are used to illustrate already known clustering information: (a) standard force-directed layout; (b) torus layout wraps the links around the boundaries vertically top-to-bottom, or horizontally left-to-right using our toroidal layout algorithm (\autoref{sec:torus2:algorithm}); (c) wrapping links are minimised by our automatic pan algorithm (\autoref{sec:torus2:autopanalgorithm})}
    \label{fig:torus2:wrapgraph}
\end{figure}

\begin{figure}
    \centering
    \includegraphics[width=\textwidth]{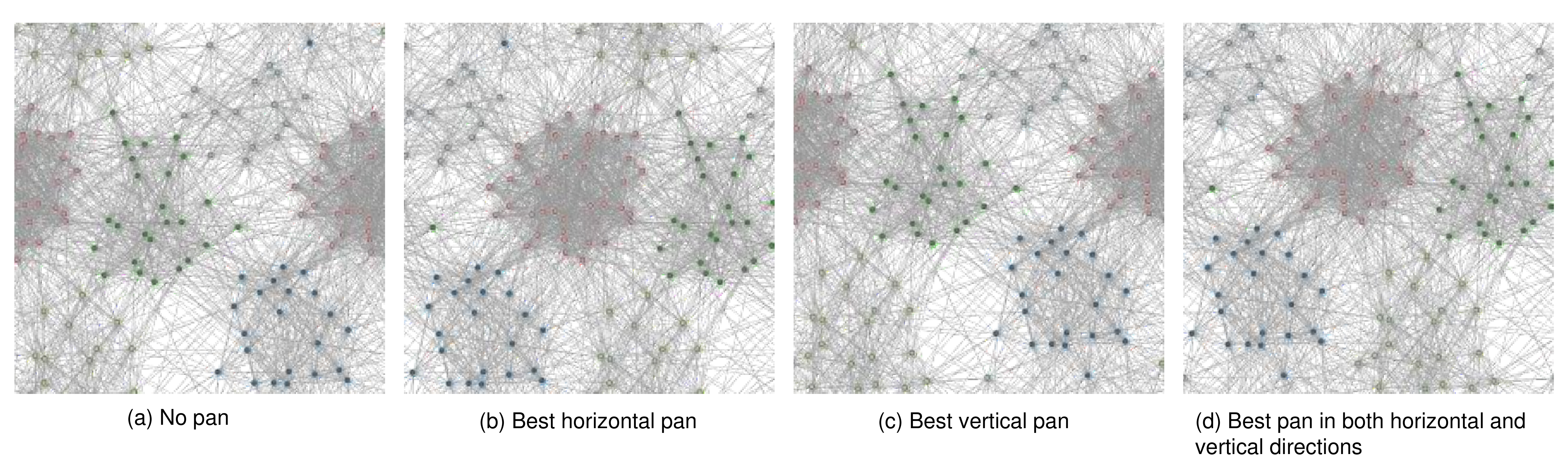}
    \caption{Example of automatic panning of toroidal layouts of a network with 126 nodes and 2496 links from \dsmallhard{} (described in~\autoref{sec:torus2:graphcorpus} and~\autoref{fig:torus2:graphicaesthetics}): (a) Original toroidal layout without auto pan requires a user to pan to navigate the network; (b-c) Either  horizontal or vertical pan reduces link wrappings across the edges of the display; (d) Best pan in both horizontal and vertical directions minimises the number of link wrappings and better reveals 5 main clusters.}
    \label{fig:torus2:autopan}
\end{figure}

Second, to better discriminate clusters, we explore a completely novel issue identified through our testing (\autoref{sec:torus2:userstudy}) as well as our previous study results (\autoref{sec:torus1:discussion}) which indicate that link wrapping across the plane's edges impedes network understanding tasks when neither tile-display nor interactive panning (\autoref{sec:torus1:discussion}) is introduced. We present a new algorithm for computing how best to ``cut'' a toroidal network layout to the surface of a torus, i.e., to automatically pan the viewport to reduce the number of links wrapped. To the best of our knowledge this has not been previously considered. As shown in~\autoref{fig:torus2:wrapgraph}(c) and~\autoref{fig:torus2:autopan}(d), toroidal layouts with automatic panning better reveal the structure of the networks such as clusters and makes a considerable difference to the quality of the final layout. 





Third, we compare \twebcola{} and our new automatic toroidal layout method with non-wrapped flat layout algorithms, using a large corpus of 200 networks with community structures (defined in~\autoref{sec:torus2:empiricalevaluation}). We explore whether our improved torus-based layout algorithm is able to find node positions affording better graph aesthetics. 

Fourth, we conduct a new study (\autoref{sec:torus2:userstudy}) with 32 participants, evaluating (a) a new and optimised layout algorithm with (b) automatic wrapping, using (c) larger networks ($\le 134$ nodes, $\le$ 2590 links), and (d) focus on cluster discrimination tasks. We show that torus layout significantly outperformed non-torus for cluster identification tasks in terms of improvement in error by 62.7\% and time by 32.3\%.

\section{Improved Algorithm for Torus Layouts}
\label{sec:torus2:algorithm}

We introduce a toroidal layout algorithm that solves a key limitation of the \twebcola{} gradient descent algorithm proposed in~\autoref{sec:torus1:stressminimisation} which could become stuck in local minima (as shown in~\autoref{fig:lowmodularityexample}(a)) and thus which often requires manual intervention to guide the algorithm to find a reasonable layout.
We use this algorithm to find layouts that consistently yield better graph aesthetics: fewer crossings, less stress, greater incidence angle, and greater cluster distance (\autoref{sec:torus2:empiricalevaluation}).


Following \twebcola{} gradient descent (\autoref{sec:torus1:stressminimisation}), our layout algorithm finds node positions of each pair of nodes in a 3×3 repeated tiling. We develop an iterative algorithm which seeks to minimise \emph{stress} of the layout across the surface of the torus. However, while in \twebcola{} gradient descent the layout algorithm moves all pairs of vertices at each iteration and also needs a good initialisation to find a suitable torus wrapping, our new approach randomly moves a single pair of vertices at a time.  In our empirical testing (see~\autoref{sec:torus2:runtimecomparison}), we find that this stochastic approach avoids the algorithm getting stuck prematurely in local minima of the stress function and leads autonomously to a high-quality torus layout without the need of a good initial state.

We name this new approach \tpairwise{}, given that it minimises the stress function by moving a single pair of nodes at a time using gradient descent, as opposed to our previous layout approach, \twebcola{}.  
The idea to randomly select and move a single pair of nodes at a time is inspired by an approach suggested for general (non-toroidal) stress-minimising graph layout by Zheng et al.~\cite{zheng2018graph}.  

The key component to adapting a gradient-descent layout approach (such as stress minimisation) to a 2D torus layout, is to consider the nine different possible choices for wrapping each link, as shown in~\autoref{fig:threebythreetile}.  We find that the \tpairwise{} approach works particularly well for toroidal layout because at each iteration, as well as optimising stress for a single link, we can choose the optimal link wrapping configuration for that link.  In the \twebcola{} gradient descent approach using \twebcola, selecting the wrapping configuration for all links at once was the source of significant instability.

\subsection{Pairwise Gradient Descent}
\label{sec:torus2:pairwisealgorithm}

Following \twebcola{}, our toroidal layout approach is based on minimising the stress function defined in~\autoref{eqn:stress}). We consider a non-wrapped node-link layout, named \nodelinklayout{}, and a wrapped layout on a torus topology, named \toruslayout{}. For each pair of nodes, we compute the gradient information across the nine possible ways to consider their adjacency.  

The constant $L$ is selected proportionally to \textit{cell size} and the graph diameter, i.e., the longest of the shortest-paths of $G$, as defined in Algorithm \autoref{algorithm:pairwise}. 
Note that it is possible to change the choice of $L$ to produce a layout that has a desired level of torus wrapping. Unlike \twebcola{} where the choice of $L$ was not clearly shown and relied on a good initialisation, the method presented in this section produces deterministic layouts.

A gradient-descent approach to reducing the stress function uses the gradient information for this function in a given graph configuration to choose descent vectors, or directions by which to move the nodes to reduce the overall stress function, as per Dwyer et al.~\cite{dwyer2008topology}.
For \toruslayout{}, for a given pair of nodes we choose the adjacency across cells which contributes to the greatest reduction in stress as the descent vector by which the nodes will be moved.  
In \twebcola{}, such descent vectors were computed simultaneously across all pairs of nodes, before moving all nodes according to the computed vectors.  In this paper, however, we move just a pair of nodes at a time and follow an annealing schedule to enforce convergence.

A summary of our annealing schedule is shown in \autoref{eqn:annealingschedule}, where $\eta(t)$ is a time-dependent scale factor applied to descent vectors before moving nodes accordingly. 

\begin{equation}
    \label{eqn:annealingschedule}
    \mathit{\eta(t)} = \left\{ \begin{array}{rcl}
    min\ (1, \frac{D_{max}^2}{D_{uv}^2}e^{-\lambda t}) & \mbox{for} & t\leq\ \tau \\ 
    min\ (1, \frac{D_{min}^2}{D_{uv}^2}\frac{1}{1+\lambda t}) & \mbox{for} & \tau\ < t \leq \tau_{max} \\
    \end{array}\right.
\end{equation}
We follow Zheng et al.~\cite{zheng2018graph} in choosing an exponential decay schedule for a fixed number of iterations $\tau$. 
Starting from the next iteration after $\tau$, the algorithm uses a $\frac{1}{t}$ schedule to converge to a stable configuration. 
The annealing schedule begins with a maximum step size, 1 at first iteration $t = 0$. This avoids local minima, as reported by Zheng et al.~\cite{zheng2018graph}.
$\lambda$ is a decay constant determined by a given parameter $\epsilon$ such that the schedule's step size at iteration $\tau$ is constrained from above by $\epsilon$, i.e., $D_{max}^2e^{-\lambda \tau}=D_{min}^2\epsilon$.

\begin{algorithm}
 \KwData{$graph\ G=(V, E),\ wrapping\ W=3\times 3\ tiles$}
 \KwResult{$Graph\ embeddings\ on\ a\ 2D\ plane$}
 $L\leftarrow \frac{Cell\ Size}{min(Graph\ Diameter\ ,\ 2) +\ 1}$\;
 $X\leftarrow |V| \times |V| \times 2$ node position matrix\;
 $D\leftarrow \emph{ShortestPathsMatrix(G)}$\;
 \For{$\eta\ in\ each\ annealing\ schedule\ in\ \autoref{eqn:annealingschedule}$}{
    \For{each $u, v\in V$\ in\ random\ order}{
        \eIf{wrapped}{
          $X^\prime_{uv} \leftarrow$ set of 9 possible vectors from $u$ to $v$ across the cells of $W$\;
          $d_{uv} \leftarrow $ euclidean lengths of each $X^\prime_{uv}$\;
      $R\leftarrow (min_{w \in\ W} \frac{(L\times D_{uv} - (d_{uvw})^2}{2})\frac{\overrightarrow{X^\prime_{uvw}}}{d_{uvw}}$\;
      }{
      $R\leftarrow \frac{(L\times D_{uv} - d_{uv})^2}{2}\frac{\overrightarrow{X_{uv}}}{d_{uv}}$\;
      }
        $X_{u}\leftarrow X_{u} - \eta R$\;
        $X_{v}\leftarrow X_{v} + \eta R$\;
        $\textbf{if}\ wrapped\ \textbf{then}\ translate\ X_{u},X_{v}\ back\ to\ centre\ cell$\;
    }
  }
 \caption{\tpairwise{} gradient descent layout algorithm, which provides a \toruslayout{} layout when \emph{wrapped} is true, or a \nodelinklayout{} layout when \emph{wrapped} is false.
 \label{algorithm:pairwise}}
\end{algorithm}

\begin{figure}
    \centering
    \includegraphics[width=\textwidth]{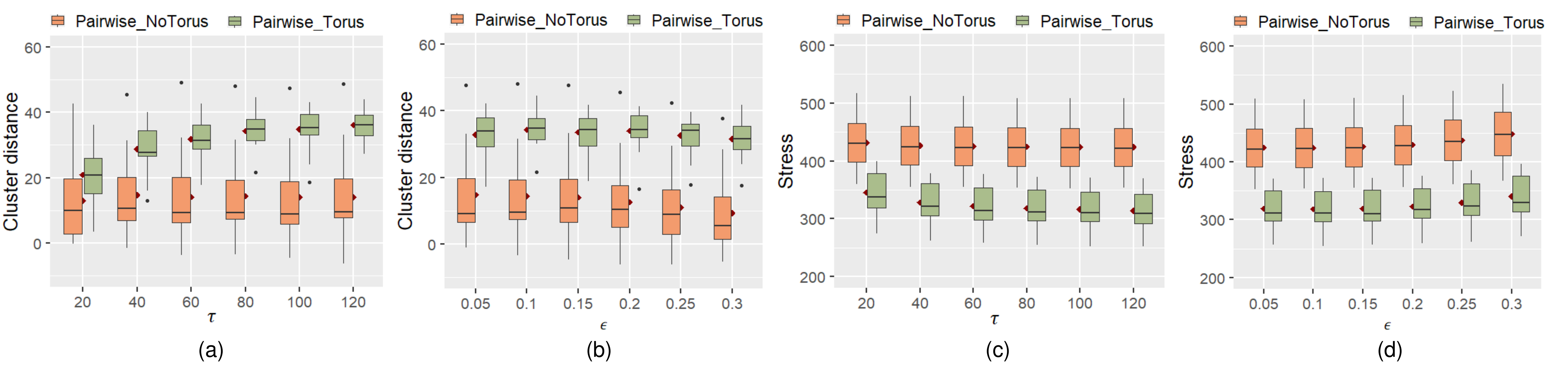}
    \caption{The cluster distance and stress of networks from \dsmallhard{} (\autoref{sec:torus2:graphcorpus}) with 20 runs when varying  parameters $\tau$ and $\epsilon$ on \autoref{eqn:annealingschedule}. A larger value of $\tau$ gave greater cluster distance (a) and less stress (c) with $\epsilon=0.1$. There was not much improvement for either $\tau > 80$ (a,c) or $\epsilon > 0.1$ (b,d). Therefore we chose $\tau=80$ and $\epsilon=0.1$ for our \tpairwise{} layout algorithm.}
    \label{fig:torus2:parametersettings_on_cluster_distance}
\end{figure}

\begin{figure}
    \centering
    \subfigure{
    \includegraphics[width=0.4\textwidth]{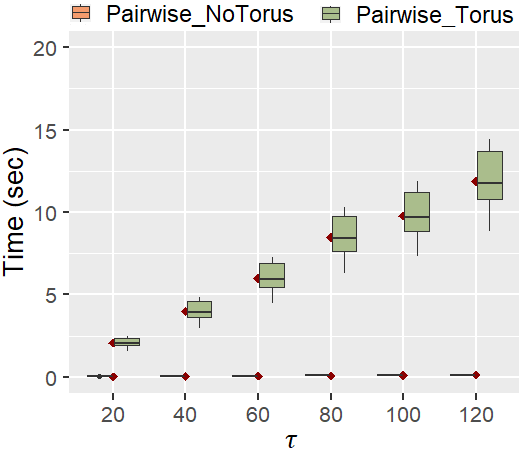}
    }
    \subfigure{
    \includegraphics[width=0.4\textwidth]{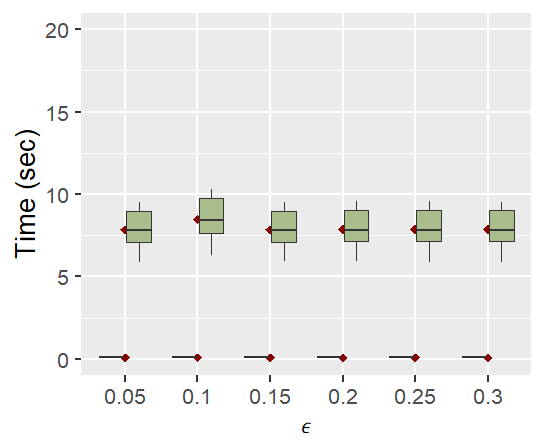}
    }
\caption{The running time of layout algorithms from \dsmallhard{} (\autoref{sec:torus2:graphcorpus}) with 20 runs when varying  parameters $\tau$ and $\epsilon$ on \autoref{eqn:annealingschedule}.}
\label{fig:torus2:parametersettings_on_running_time}
\end{figure}

The parameter that determines how well a toroidal layout spreads out clusters of a network is (1) $\tau$ which controls when to switch from exponential schedule to the $\frac{1}{t}$ schedule (\autoref{eqn:annealingschedule}), (2) $\epsilon$ which determines the constraint of $\eta(t)$. We experimented with a variety of parameter $\tau$ and $\epsilon$ for the step size schedule $\eta(t)$ as shown in~\autoref{fig:torus2:parametersettings_on_cluster_distance} and \autoref{fig:torus2:parametersettings_on_running_time}. For $\tau$, we fixed $\epsilon=0.1$ while varying $\tau$ in a range of 20 and 120. We used the graphs from \dsmall-\dhard~graph corpus described in~\autoref{sec:torus2:graphcorpus} and~\autoref{fig:torus2:graphicaesthetics}. We find increasing $\tau$ led to better separation of clusters (defined more formally in~\autoref{sec:torus2:clusterdistance}) and less stress, as shown in~\autoref{fig:torus2:parametersettings_on_cluster_distance}(a, c). For $\epsilon$, we then fixed $\tau=80$ while varying $\epsilon$ from 0.05 to 0.3. The results indicate that $\tau$ greater than 0.2 is a poor choice and thus it has worse cluster separation and stress, as shown in~\autoref{fig:torus2:parametersettings_on_cluster_distance}(b, d). We therefore set $\tau=80$ and $\epsilon=0.1$ for the exponential schedule. 
For $\frac{1}{t}$ schedule we set $\epsilon=0.001$. This gives a smaller step size and thus the layout algorithm always converges.

The stopping criterion is set to a maximum pairwise movement $\delta<0.03$ or a maximum of $\tau_{max}=200$ iterations , whichever comes first. 
 Algorithm 1 gives pseudocode for this process \footnote{The detailed pseudocode to our method is available from \url{https://github.com/Kun-Ting/its-a-wrap}}.~\autoref{fig:lowmodularityexample} shows layout examples from \dlargehard{} graphs of \twebcola{} and \tpairwise{} for \nodelinklayout{} and \toruslayout{} at convergence. Similarly, examples of \dlargeeasy{} graphs using these layout algorithms are shown in~\autoref{fig:highmodularityexample}.


\subsection{Computational Complexity}
The time complexity of each iteration of \tpairwise{} and \twebcola\ stress minimisation is O($|V|^2$) where $V$ is a set of vertices. There is a constant factor $W$ due to torus adjacencies which does make our reported run-times uniformly slower than non-torus layout.
The space complexity of all the methods are the same, i.e., O($|V|^2$)
\tpairwise{}-\toruslayout{}’s asymptotic computation time complexity is the same as traditional force-directed layout so we would expect similar scalability (subject to the constant factor $W$).  Further improvement may be possible using multilevel spatial decomposition methods, e.g.\ by Quigley and Eades~\cite{quigley2000fade} and Walshaw~\cite{walshaw2000multilevel}.

\section{Automatic Panning Algorithm}
\label{sec:torus2:autopanalgorithm}
As investigated in~\autoref{sec:torus:study1}, additional context and interactive panning have been shown to improve understanding of wrapped visualisations. The qualitative feedback in that study revealed that repetition of networks in adjacent cells is not practical for displaying large graphs. Rather, interactive panning was found to be highly-beneficial in helping study participants to pan, such that links of interest are not wrapped.
However, we can greatly support this process by automatically panning to minimise the number of links that are split across viewport boundaries.

To perform automatic panning of a given layout (e.g.\ as determined by the algorithm given in~\autoref{sec:torus2:pairwisealgorithm}, we take a sweepline approach by Shamos and Hoey~\cite{shamos1976geometric} to search for pan positions which lead to the least number and severity of split links, horizontally and vertically.
A given layout provides a fixed set of node positions relative to a viewport.  A ``pan'' of a \toruslayout{} layout, involves translating all node positions uniformly, except where a node would move outside the viewport, it is wrapped to the other side of the viewport, top-to-bottom or left-to-right.  For a given layout of a graph with $|V|$ nodes, the node positions left-to-right define an ordering over the nodes, and the positions top-to-bottom provide a second ordering.  The set of links that are wrapped across viewport boundaries $E_{wr} \subseteq E$ is constant under translation of the nodes, until the translation is sufficiently large that a node must be wrapped around.  Thus, in each axis (horizontal and vertical) there are precisely $|V|$ distinct translations that must be considered in order to examine all sets of possible wrapped links for a given layout.  We can examine each of these sets to determine which induces the lowest \textit{wrapping cost}, defined as~\autoref{eqn:edgelengthpenalty}, where $d_{uv}$ is the Euclidean distance between $u$ and $v$ connected by a link $e$:

\begin{equation}
    \label{eqn:edgelengthpenalty}
    \mathit{wrapcost}(E_{wr}) = \sum_{e \in E_{wr}}\frac{1}{ d_{uv}}
\end{equation}

Algorithm \ref{alg:autopan} details the steps of this procedure.  The runtime complexity of Automatic Panning is O(|V|log|V|+|E|) due to the need to sort node positions and examine all links.

\begin{algorithm}
\small
\KwData{Node position vector $X$ with position $(x,y)\in X$ for each node, for a torus layout of a graph $G$ from Algorithm\ \autoref{algorithm:pairwise}.~\autoref{fig:torus2:autopan}(a) shows an example.}
\KwResult{Torus layout with minimum link wrappings on the boundary and less disconnected clusters centred within the viewport}
\textbf{Horizontal Sweep (result shown in~\autoref{fig:torus2:autopan}(b))}: 

(1) Sort nodes by increasing $(x,\_)\in X$ and initialise sweep-line positions $S$ with the mid-points $s_i$ of all adjacent $x_i, x_{(i+1)}$;

(2) For each sweep line position $s_i \in S$, maintain a set of open links. Assuming we sweep left-to-right, at each $s_i$ we add to $E_{wr}$ the links outgoing from the right side of node $i$ and remove any links incoming to the left-side of $i$.   If \textit{wrapcost}$(E_{wr})$ (\autoref{eqn:edgelengthpenalty}) of open links is smaller than $minCost$, take this as the new $minCost$ and set $minX$ to the current sweep line position $s_i$\;

\textbf{Vertical Sweep (result shown in~\autoref{fig:torus2:autopan}(c))}: 

(3) Repeat step 1 and 2 for all $y$-positions $(\_,y) \in X$. Therefore, $minY$ is the sweep line position with minimum cost.

\textbf{Apply optimal pan and centre within viewport \\
(result shown in \autoref{fig:torus2:autopan}(d))}: 

(4) Translate all node positions based on $minX$, $minY$.

(5) Centre the layout such that the centre $x$-position of the left-most and the right-most nodes is at the centre of the cell; and similar to centre vertically.
\caption{Auto Panning Procedure
\label{alg:autopan}
}
\end{algorithm}

\section{Algorithm Evaluation}
\label{sec:torus2:empiricalevaluation}


In this section, we compare two layout conditions: torus (\toruslayout) and traditional 2D planes (\nodelinklayout) for both our proposed algorithm \tpairwise{} (Algorithm \ref{algorithm:pairwise}) and the \twebcola{} algorithm (\autoref{sec:torus1:stressminimisation}) against a large corpus of 200 graphs.
We show that \tpairwise{} has convergence and run-time performance benefits over its predecessor \twebcola{}.
We assess layout quality using established graph aesthetics measures and a novel cluster readability metric, \textit{cluster distance}, measuring visual distance between boundaries of clusters in a given layout, finding that \tpairwise{} toroidal layout algorithm outperforms either \tpairwise{} non-torus or \twebcola{} layout algorithms.

In order to evaluate how well a layout method separates nodes and clusters, we look at graphs with community structures~\cite{mishra2007clustering, saket2014group}. 
We control for \textit{modularity}, a metric for graph theoretic community structure, defined by Newman~\cite{newman2006modularity}.
A graph with high modularity indicates that the cluster structure is more distinct, as there are more links within each cluster than between clusters \cite{newman2006modularity, fortunato2010community}. 

We compared means of runtime, stress, crossings, incidence angle, and cluster distance of four different layout methods, i.e., \tpairwise{}-\toruslayout{}, 
\tpairwise{}-\nodelinklayout{}, 
\twebcola{}-\toruslayout{} and 
\twebcola{}-\nodelinklayout{}, using Friedman's non-parametric test and Nemenyi's post-hoc pairwise comparison, as they did \textit{not} follow normal distribution. We report significant differences under 95\% confidence.  

\subsection{Graph Corpus}
\label{sec:torus2:graphcorpus}

\begin{figure*}
    \centering
    \subfigure[Stress Against Time] {
        \includegraphics[width=0.6\textwidth]{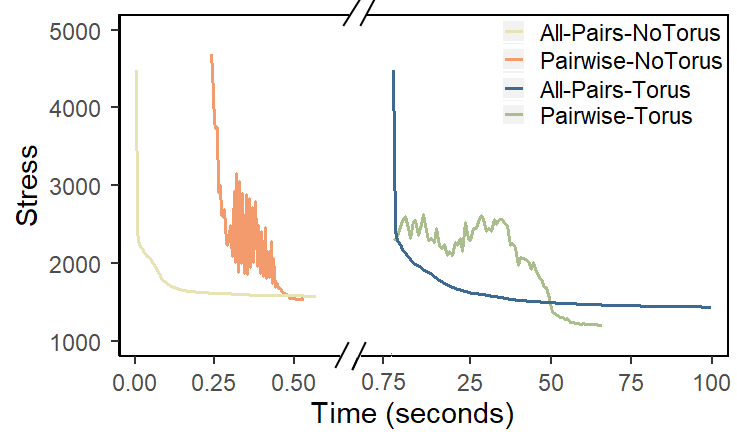}
    }
    \subfigure[\twebcola{} \nodelinklayout{}]{
        \includegraphics[width=0.4\textwidth]{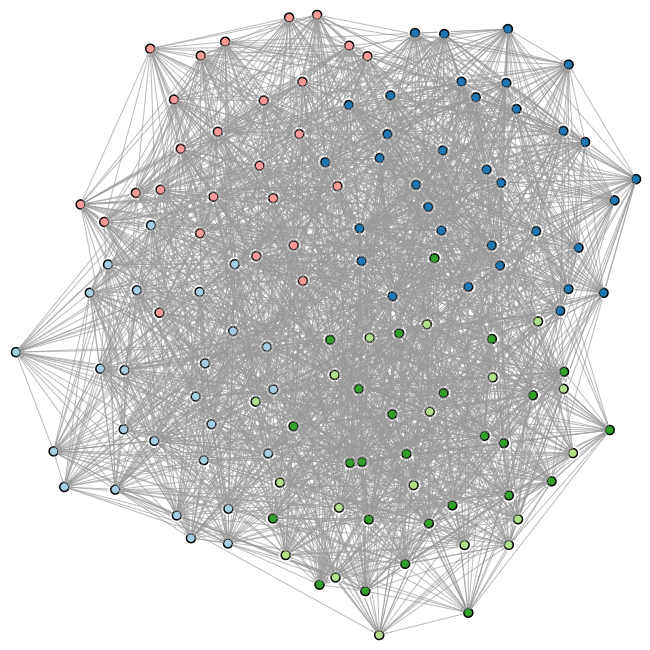}
    }
    \subfigure[\twebcola{} \toruslayout{}]{
        \includegraphics[width=0.4\textwidth]{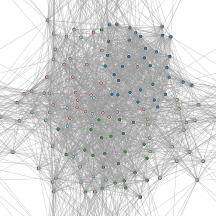}
    }
    \subfigure[\tpairwise{} \nodelinklayout{}]{
        \includegraphics[width=0.4\textwidth]{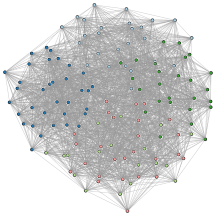}
    }
    \subfigure[\tpairwise{} \toruslayout{}]{
        \includegraphics[width=0.4\textwidth]{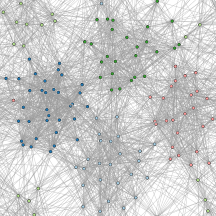}
    }
    \caption{(a) Mean stress over time of 20 runs of torus and standard node-link for both \tpairwise{} and \twebcola{} algorithms for a network with 130 nodes, 2504 links from \dlargehard{} (\autoref{sec:torus2:graphcorpus}). (b-e) show the network layouts at convergence. \tpairwise{}-\toruslayout{} reached a lower stress level, was faster to converge, and better revealed network clusters than  \twebcola{}-\toruslayout{}.}
    \label{fig:lowmodularityexample}
\end{figure*}

\begin{figure*}
    \centering
    \subfigure[Stress Against Time] {
        \includegraphics[width=0.6\textwidth]{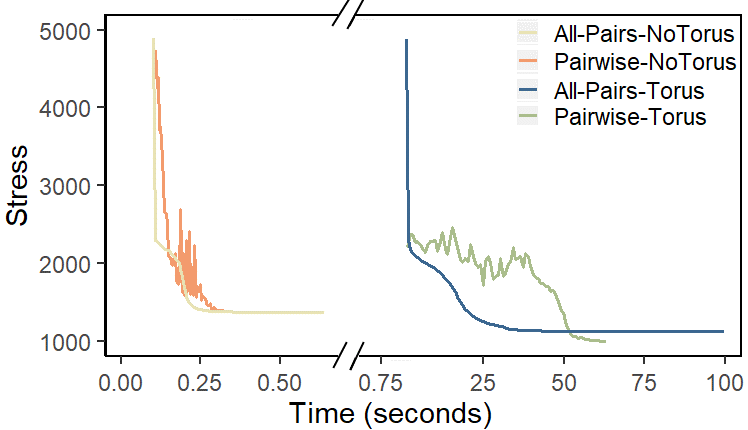}
    }
    \subfigure[\twebcola{} \nodelinklayout{}]{
        \includegraphics[width=0.4\textwidth]{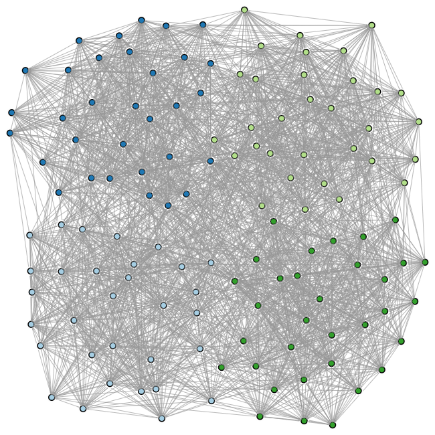}
    }
    \subfigure[\twebcola{} \toruslayout{}]{
        \includegraphics[width=0.4\textwidth]{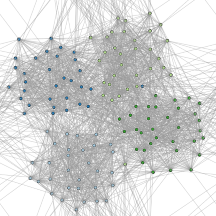}
    }
    \subfigure[\tpairwise{} \nodelinklayout{}]{
        \includegraphics[width=0.4\textwidth]{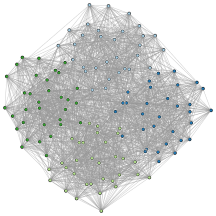}
    }
    \subfigure[\tpairwise{} \toruslayout{}]{
        \includegraphics[width=0.4\textwidth]{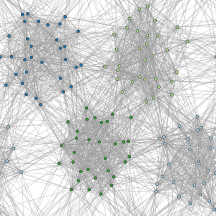}
    }
    \caption{(a) Mean stress over time of 20 runs of torus and standard node-link for both \tpairwise{} and \twebcola{} algorithms for a network with 130 nodes, 2504 links from \dlargeeasy{} (\autoref{sec:torus2:graphcorpus}). (b-e) show the network layouts at convergence. \tpairwise{}-\toruslayout{} reached a lower stress level, was faster to converge, and better revealed network clusters than  \twebcola{}-\toruslayout{}.}
    \label{fig:highmodularityexample}
\end{figure*}

\begin{figure}
    \centering
    \includegraphics[width=\textwidth]{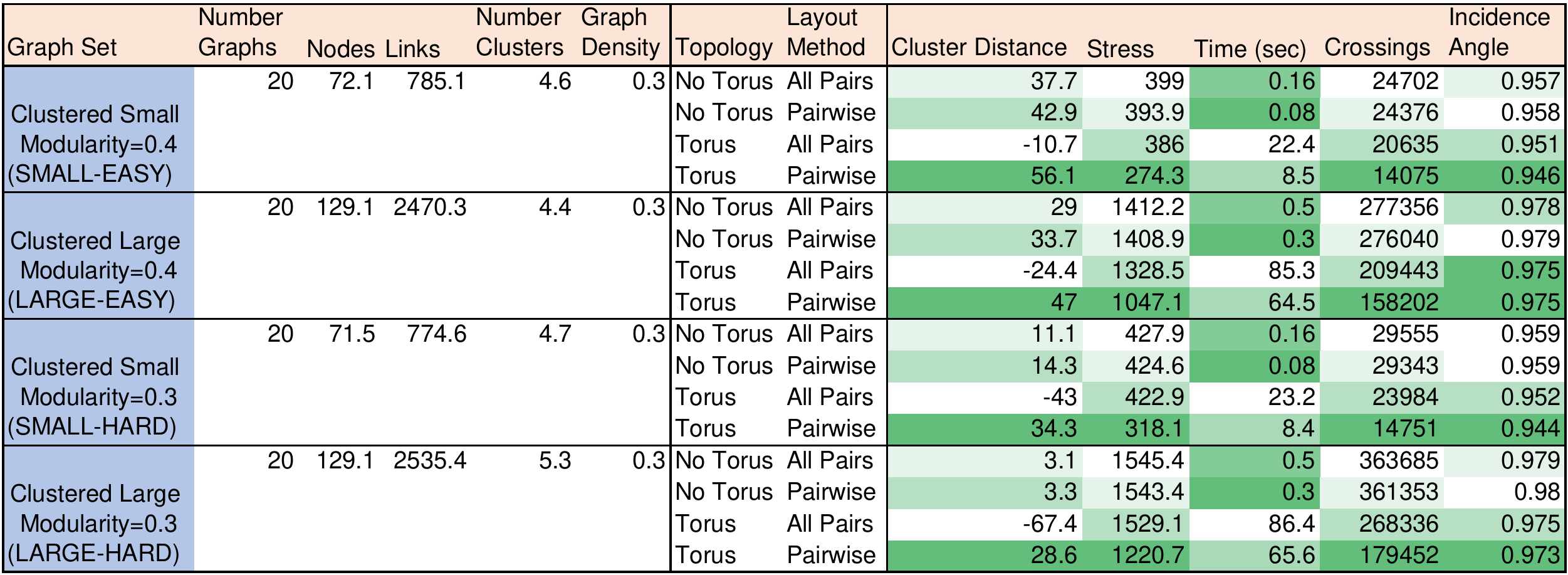}
    \caption{Average properties of 80 random partition networks and graph aesthetics results of 20 runs of each network rendered using \toruslayout{}, \nodelinklayout{} for both \tpairwise{} and \twebcola{} algorithms when varying modularity and size.}
    \label{fig:torus2:graphicaesthetics}
\end{figure}

\begin{figure}
    \centering 
    \includegraphics[width=\textwidth]{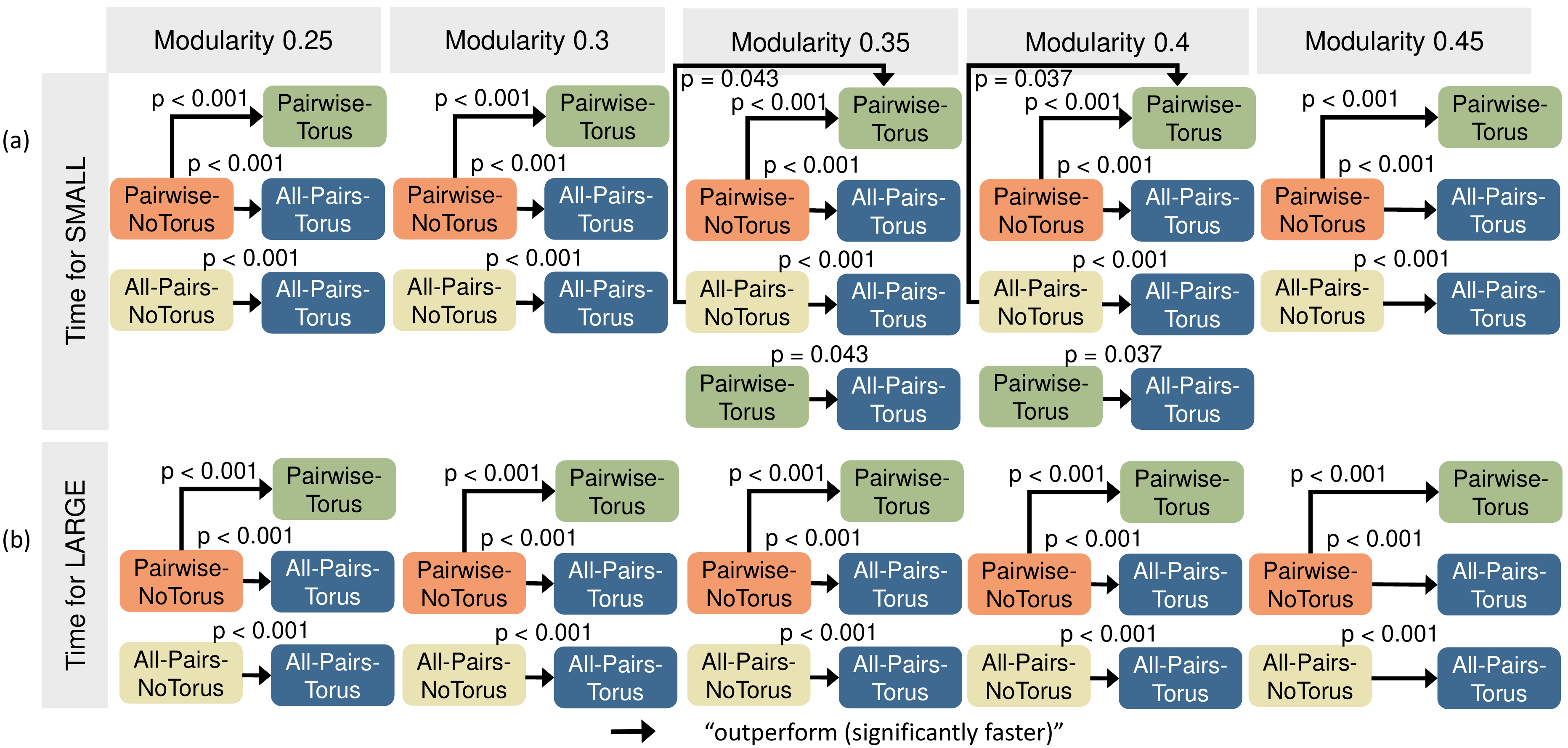}
    \caption{Graphics showing statistically significant results of running time of \dsmall{} graphs: 68-80 nodes, 710-925 links, and \dlarge{} graphs: 126-134 nodes, 2310-2590 links of 200 networks laid out using \toruslayout{} and \nodelinklayout{} for both \tpairwise{} and \twebcola{} algorithms when varying graph modularity between 0.25 and 0.45 for \dsmall{} and \dlarge{} networks.}
    \label{fig:runtimecomparison}
\end{figure}

\begin{figure}
    \centering 
    \includegraphics[width=0.9\textwidth]{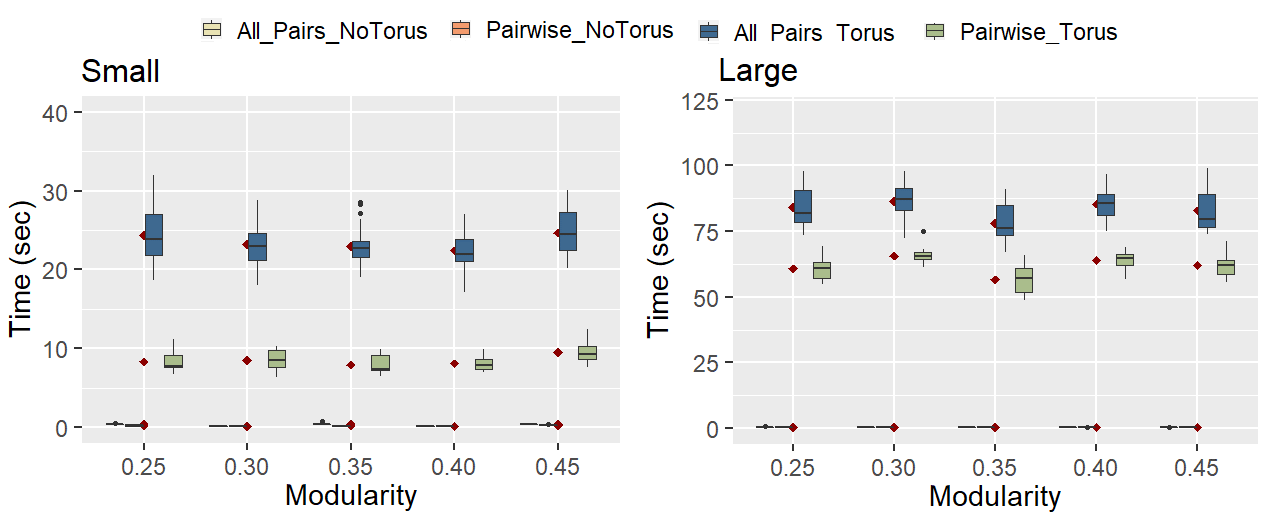}
    \caption{Box plots of running time of 200 networks laid out using \toruslayout{} and \nodelinklayout{} for both \tpairwise{} and \twebcola{} algorithms when varying graph modularity between 0.25 and 0.45 for \dsmall{} and \dlarge{} networks.}
    \label{fig:runtimeboxplotscomparison}
\end{figure}

The graphs in our sample corpus were generated using algorithms designed to simulate real-world community structures in graphs~\cite{brandes2003experiments,fortunato2010community}, using generators from NetworkX~\cite{networkx}.
We generated 200 graphs, grouped by two variables: graph modularity \cite{newman2006modularity} (5 levels from low to high: \textbf{0.25, 0.3, 0.35, 0.4, 0.45}) and graph size (2 levels: \dsmall{}: 68-80 nodes, 710-925 links, 3-8 clusters, and \dlarge{}: 126-134 nodes, 2310-2590 links, 3-8 clusters). This gives us 10 classes. For all classes, the graph density is calculated by the ratio of the number links of the graph to the maximum number of possible links ($\frac{2\times|E|}{|V|\times(|V|-1)}$). This density is fixed at a range of 0.3$\pm$0.01. 

We use a standard \textit{Random Partition Network} model by Fortunato~\cite{fortunato2010community} and \textit{Gaussian Random Partition} model~\cite{brandes2003experiments} to generate our graph corpus. This gives us graphs with clustering information based on the desired range of modularity, density and size. We exclude graphs whose minimum modularity of an individual cluster is $\le$ 0.23, which we found was the minimum to provide visible community structure \cite{fortunato2010community}.

\begin{figure}
    \centering 
    \includegraphics[width=\textwidth]{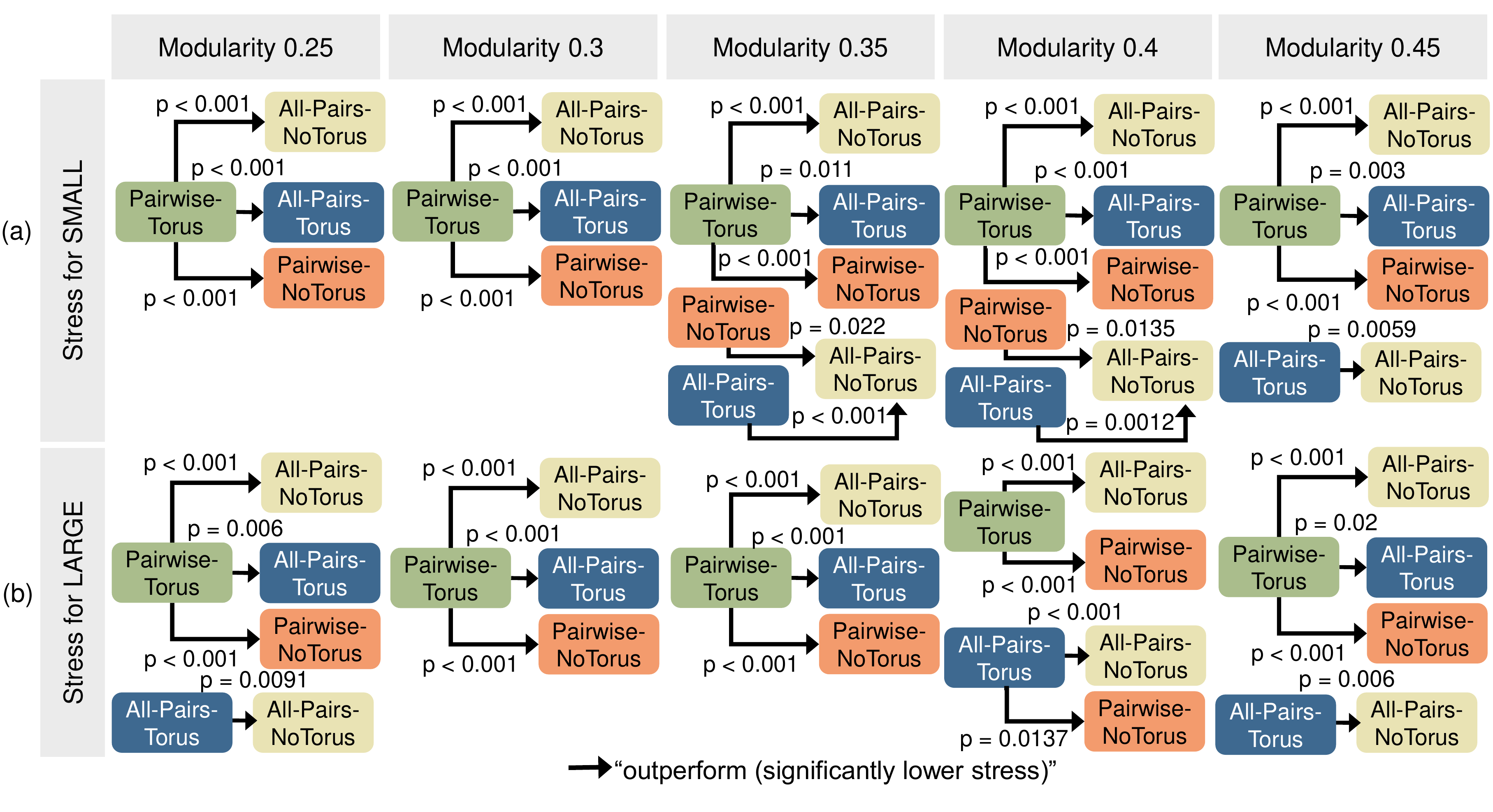}
    \caption{Graphics showing statistically significant results of stress of \dsmall{} graphs: 68-80 nodes, 710-925 links, and \dlarge{} graphs: 126-134 nodes, 2310-2590 links of 200 networks laid out using \toruslayout{} and \nodelinklayout{} for both \tpairwise{} and \twebcola{} algorithms when varying graph modularity between 0.25 and 0.45 for \dsmall{} and \dlarge{} networks.}
    \label{fig:stresscomparison}
\end{figure}

\begin{figure}
    \centering 
    \includegraphics[width=\textwidth]{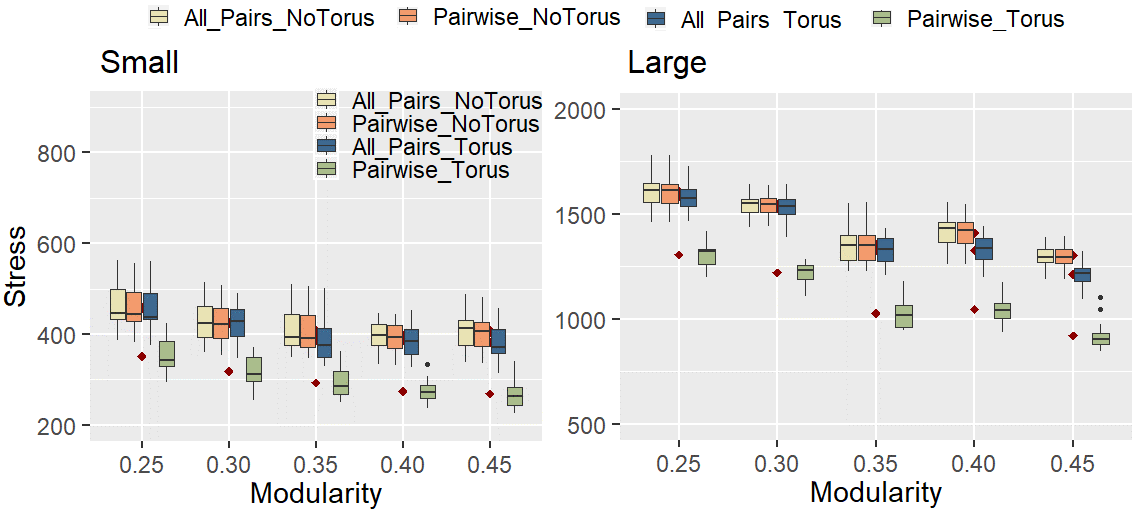}
    \caption{Box plots of stress of 200 networks laid out using \toruslayout{} and \nodelinklayout{} for both \tpairwise{} and \twebcola{} algorithms when varying graph modularity between 0.25 and 0.45 for \dsmall{} and \dlarge{} networks.}
    \label{fig:stresscomparisonboxplots}
\end{figure}

\begin{figure}
    \centering 
    \includegraphics[width=\textwidth]{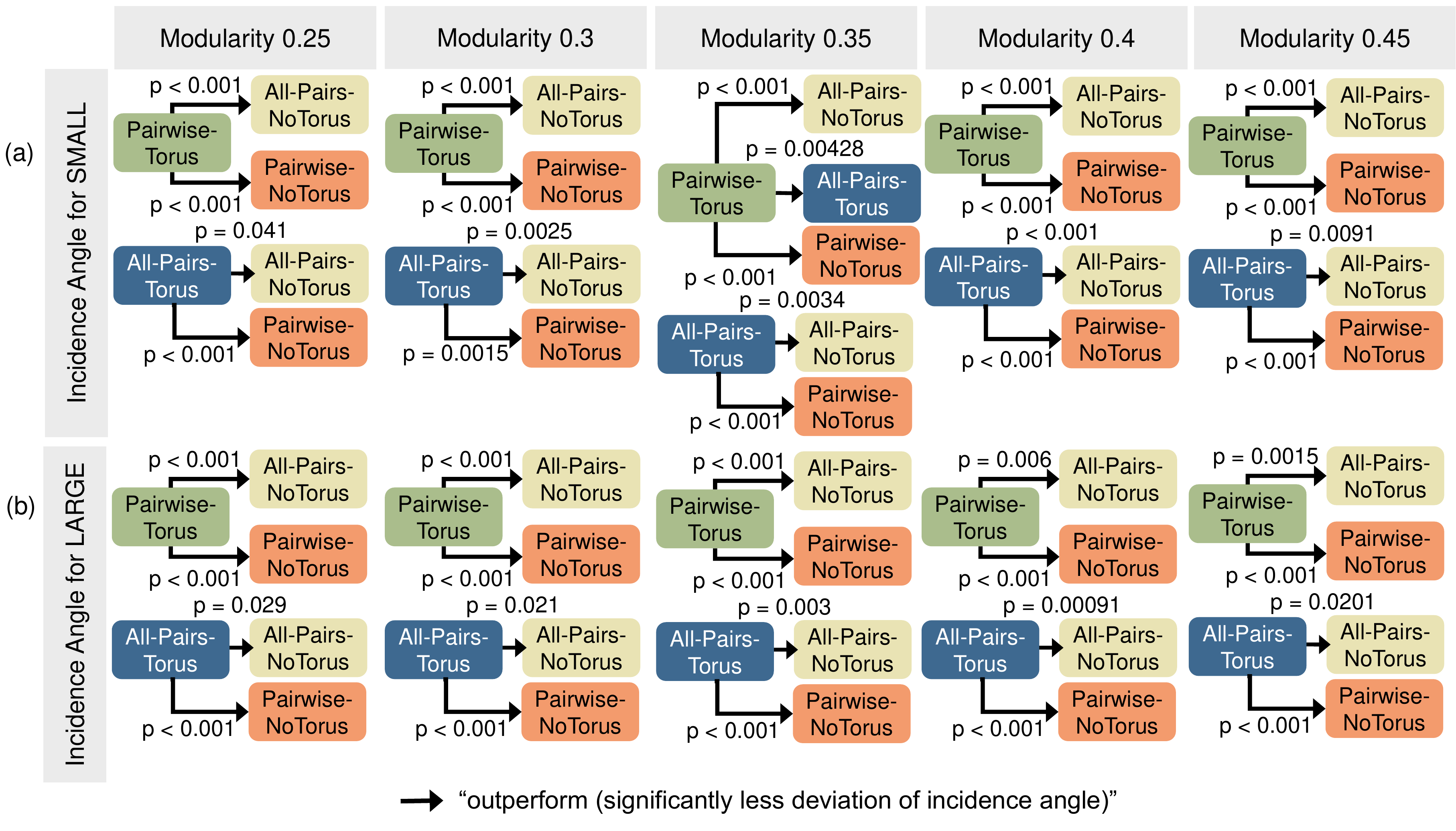}
    \caption{Graphics showing statistically significant results of incidence angle of \dsmall{} graphs: 68-80 nodes, 710-925 links, and \dlarge{} graphs: 126-134 nodes, 2310-2590 links of 200 networks laid out using \toruslayout{} and \nodelinklayout{} for both \tpairwise{} and \twebcola{} algorithms when varying graph modularity between 0.25 and 0.45 for \dsmall{} and \dlarge{} networks.}
    \label{fig:anglecomparison}
\end{figure}
\begin{figure}
    \centering 
    \includegraphics[width=\textwidth]{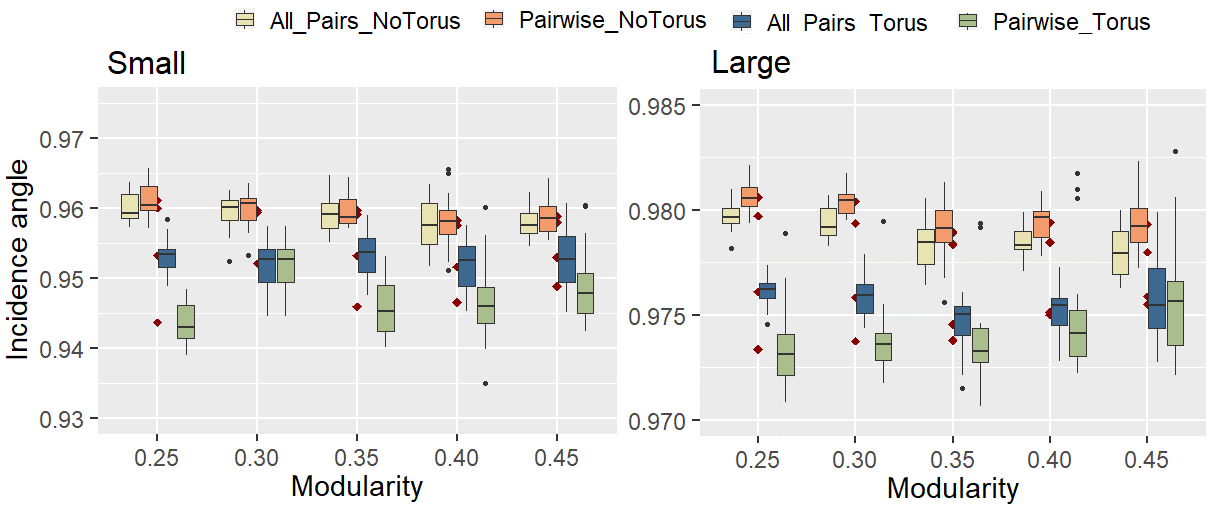}
    \caption{Box plots of incidence angle of 200 networks laid out using \toruslayout{} and \nodelinklayout{} for both \tpairwise{} and \twebcola{} algorithms when varying graph modularity between 0.25 and 0.45 for \dsmall{} and \dlarge{} networks.}
    \label{fig:anglecomparisonboxplots}
\end{figure}

\begin{figure}
    \centering 
    \includegraphics[width=\textwidth]{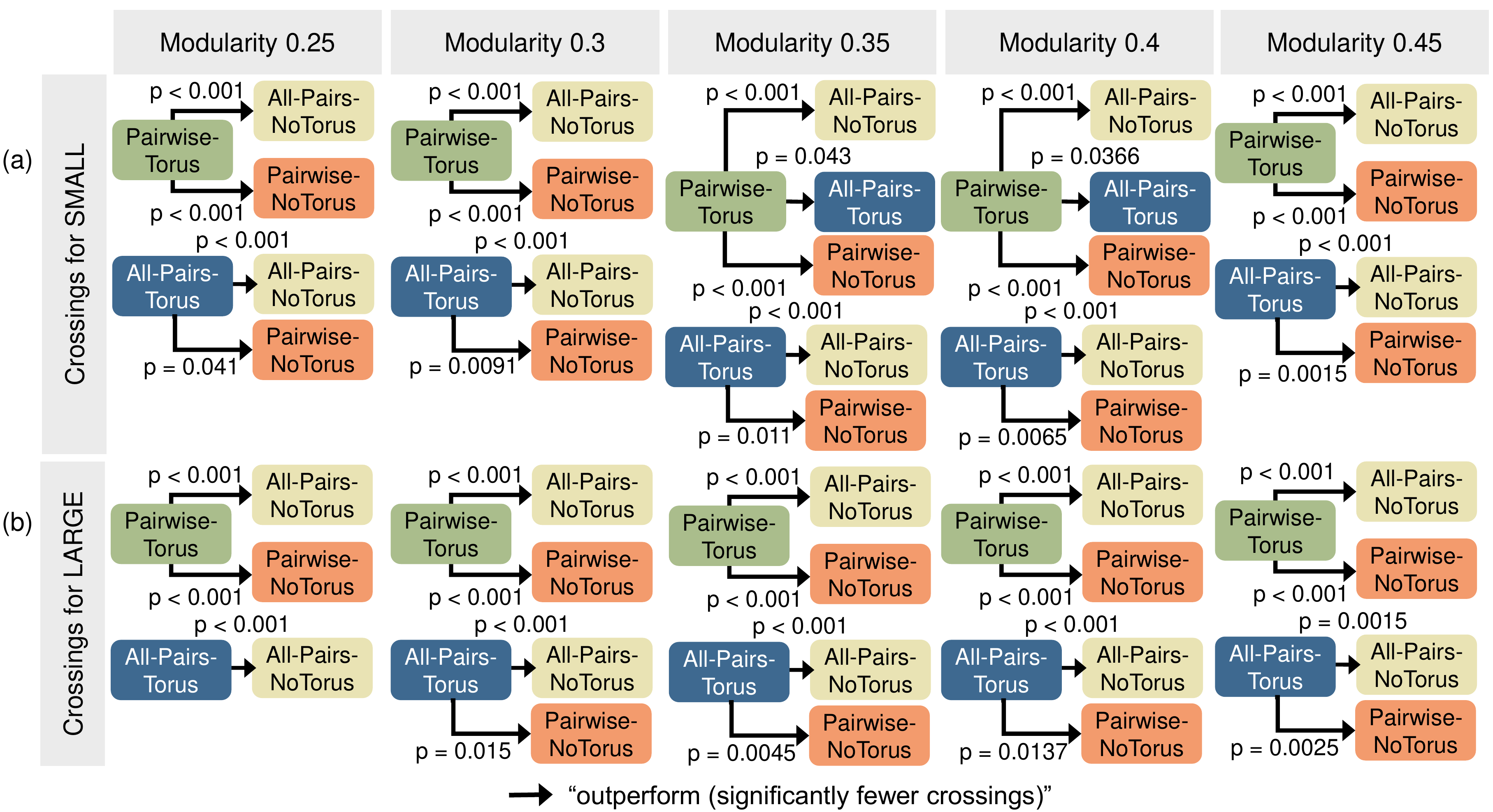}
    \caption{Graphics showing statistically significant results of link crossings of \dsmall{} graphs: 68-80 nodes, 710-925 links, and \dlarge{} graphs: 126-134 nodes, 2310-2590 links of 200 networks laid out using \toruslayout{} and \nodelinklayout{} for both \tpairwise{} and \twebcola{} algorithms when varying graph modularity between 0.25 and 0.45 for \dsmall{} and \dlarge{} networks.}
    \label{fig:crossingcomparison}
\end{figure}
\begin{figure}
    \centering 
    \includegraphics[width=\textwidth]{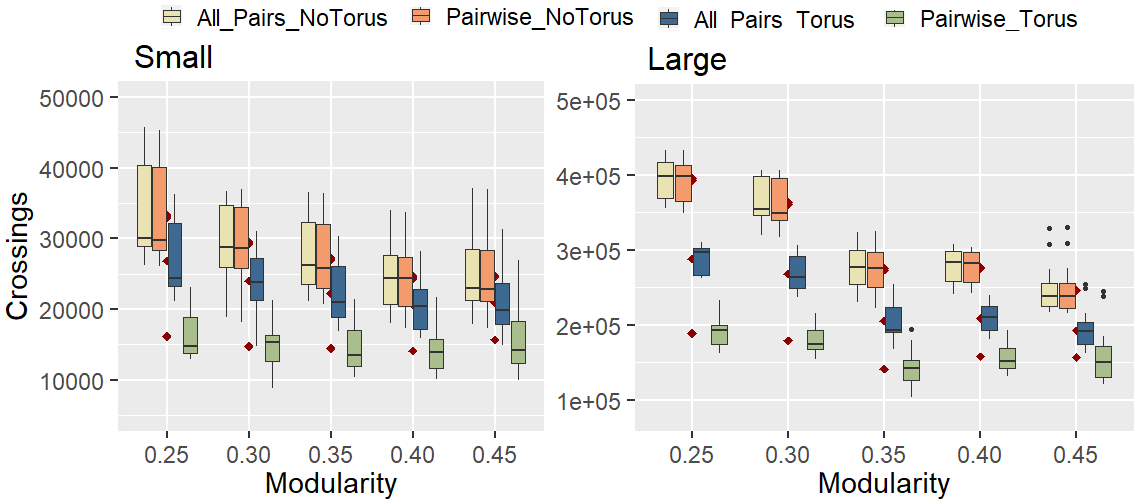}
    \caption{Box plots of link crossings of 200 networks laid out using \toruslayout{} and \nodelinklayout{} for both \tpairwise{} and \twebcola{} algorithms when varying graph modularity between 0.25 and 0.45 for \dsmall{} and \dlarge{} networks.}
    \label{fig:crossingcomparisonboxplots}
\end{figure}
\begin{figure}
    \centering 
    \includegraphics[width=0.9\textwidth]{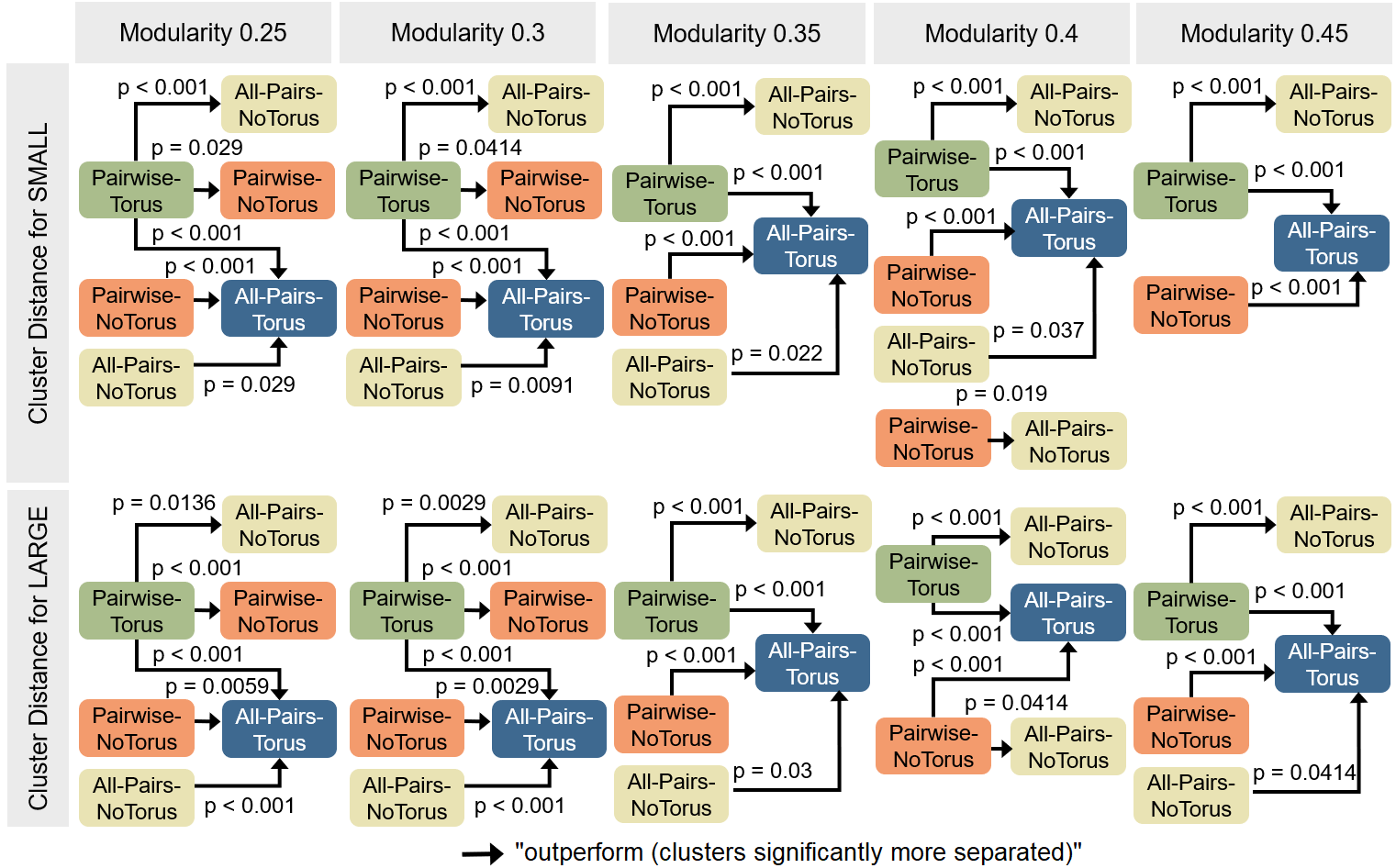}
    \caption{Statistical results of cluster distance of 200 networks laid out using \toruslayout{} and \nodelinklayout{} for both \tpairwise{} and \twebcola{} algorithms when varying graph modularity between 0.25 and 0.45 for \dsmall{} and \dlarge{} networks.}
    \label{fig:clusterdistancecomparison}
\end{figure}
\begin{figure}
    \centering 
    \includegraphics[width=0.9\textwidth]{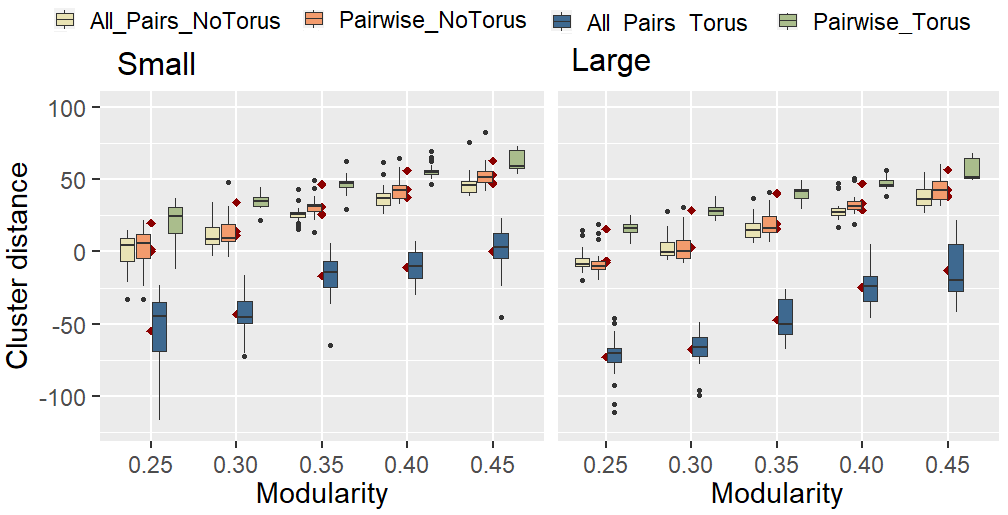}
    \caption{Box plots of cluster distance of 200 networks laid out using \toruslayout{} and \nodelinklayout{} for both \tpairwise{} and \twebcola{} algorithms when varying graph modularity between 0.25 and 0.45 for \dsmall{} and \dlarge{} networks.}
    \label{fig:boxplots_clusterdistancecomparison}
\end{figure}
\subsection{Runtime comparison}
\label{sec:torus2:runtimecomparison}

We compare runtime performance of the \tpairwise{} algorithm with \twebcola{}. We implemented the \tpairwise{} algorithm in JavaScript and D3 \cite{d3js:2020}, and implemented \twebcola{} as per \autoref{sec:torus1:stressminimisation} in \emph{WebCola}\cite{webcola}. 
We ran both algorithms with networks from our graph corpus as described in~\autoref{sec:torus2:graphcorpus}. For each graph, we generated network layouts with 20 random initial node positions controlled by seed within a 1$\times$1 square at the centre for each graph. The configuration of unit link length, stopping criteria, and maximum number of iterations are the same for each technique and described in~\autoref{sec:torus2:algorithm}. We record the time of each technique using Google Chrome browser (version 80), running on an Intel i7-7800X (3.5GHz) CPU and 32GB of RAM.  
Overall, \autoref{fig:torus2:graphicaesthetics} indicates that \tpairwise{}-\toruslayout is significantly faster (p=0.037) to converge than \twebcola{}-\toruslayout{} by 62$\%$ (differences between means) for \dsmalleasy{}. The significant results are also found for \dsmall{} graphs with modularity at 0.35 and 0.4 as shown in~\autoref{fig:runtimecomparison}, which shows graphics of statistically significant results of running time using our graph corpus (\autoref{sec:torus2:graphcorpus}). The corresponding results in box plots are shown in~\autoref{fig:runtimeboxplotscomparison} . 

\autoref{fig:lowmodularityexample}(a) shows mean stress over time of an example graph for \dlargehard{} with 20 runs. An example graph for \dlargeeasy{} is illustrated in~\autoref{fig:highmodularityexample}. The result shows that \tpairwise{} avoids the algorithm getting stuck in local minima of the stress function as opposed to \twebcola{}. Furthermore,  \tpairwise{}-\toruslayout{} reaches lower stress levels than \twebcola{}-\toruslayout{} at convergence. While the convergence time is affected by stopping conditions, in our evaluation, we used the same convergence threshold and maximum number of iterations for all methods. However, step size attenuation also affects convergence time. It is computed differently in \twebcola{} torus, which is based on gradient contribution of all pairs nodes as in Dwyer et al.~\cite{dwyer2008topology} and \tpairwise{}, which we chose 80 iterations as the threshold $\tau$ (\autoref{eqn:annealingschedule}) between exponential and convergence schedule based on experimental results. Experimentally, we found a larger $\tau$ led to a longer time to converge, as seen in~\autoref{fig:torus2:parametersettings_on_running_time}(a), but it did not give much improvement in terms of cluster separation, as shown in \autoref{fig:torus2:parametersettings_on_cluster_distance}. \autoref{fig:lowmodularityexample}(b-e) shows an example for a \dlargehard{} graph. It shows that \tpairwise{}-\toruslayout{} generates higher-quality layouts than \tpairwise{}-\nodelinklayout{}, \twebcola{}-\nodelinklayout{} and \twebcola{}-\toruslayout{} methods at convergence. A similar result can be observed for the example from the \dlargeeasy{} graph, shown in \autoref{fig:highmodularityexample}(b-e).

\subsection{Quality comparison using established graph aesthetics metrics }
\label{sec:torus2:qualitycomparison}
Following~\autoref{sec:torus1:aesthetics:subsec1}, we first used a set of layout aesthetic quality metrics to compare \nodelinklayout{} and \toruslayout{} using \tpairwise{} and \twebcola{}. However, we are more interested in whether the additional spreading afforded by torus-based layout is better at showing clusters and so revealing the network structure. This was not considered in the previous chapter.

The metrics--- \textit{stress}, \textit{minimum incidence angle}, and \textit{number of line crossings}---have been shown important for user performance in readability tasks in the past by Purchase~\cite{purchase2002metrics} and Ware~\cite{ware2002cognitive}. 

\noindent\textbf{Stress}---measures how well the layout captures the structure in the underlying network~\cite{devkota2019stress}. Graphs with lower stress have been found to be more preferred by users \cite{dwyer2009comparison}. We calculate stress as per~\autoref{eqn:stress}. Graphics of statistically significant results are summarised in~\autoref{fig:stresscomparison} with corresponding box plots in~\autoref{fig:stresscomparisonboxplots}. As shown in \autoref{fig:lowmodularityexample} and \autoref{fig:torus2:graphicaesthetics}, \tpairwise{}-\toruslayout{} achieves significantly lower stress ($p < 0.001$) than \tpairwise{}-\nodelinklayout{}, \twebcola{}-\nodelinklayout{} for \dsmalleasy{}, \dlargeeasy{}, \dsmallhard{}, and \dlargehard{}. \tpairwise{}-\toruslayout{} achieves significantly lower stress ($p < 0.001$) than \twebcola{}-\toruslayout{} for \dsmalleasy{}, \dsmallhard{}, and \dlargehard{}. For unwrapped layout, we found \tpairwise{}-\nodelinklayout{} achieves significantly lower stress ($p = 0.0135$) than \twebcola{}-\nodelinklayout{} for \dsmalleasy{}.

\noindent\textbf{Incidence Angle}---Maximising the minimum angle of incidence between links entering a node gives better readability of network connectivity. Following the definition in~\autoref{sec:torus1:aesthetics:subsec1}, our metric for incidence angle measures deviation of minimum-incidence angle for each node from the ideal maximum for that node's degree.  Smaller deviation from this ideal is better. Graphics of statistically significant results are shown in~\autoref{fig:anglecomparison}, with corresponding box plots in~\autoref{fig:anglecomparisonboxplots}.
\autoref{fig:torus2:graphicaesthetics} shows that \tpairwise{}-\toruslayout{} achieves significantly smaller deviation (p$<$0.001) than \tpairwise{}-\nodelinklayout{} for \dsmalleasy{}, \dlargeeasy{}, \dsmallhard{}, and \dlargehard{}. \tpairwise{}-\toruslayout{} achieves significantly smaller deviation than \twebcola{}-\nodelinklayout{} for \dsmalleasy{} (p$<$0.001), \dlargeeasy{} (p=0.006), \dsmallhard{} (p$<$0.001) , and \dlargehard{} (p$<$0.001).
\begin{equation}
    \frac{1}{|V|}\sum_{v \in V} \frac{|\theta_v - \mathit{min}\theta_{v}|}{\theta_v}
\end{equation}

\noindent\textbf{Link Crossings}---The negative effect of line crossings on readability of graphs is well studied by Purchase~\cite{purchase2002metrics} and Huang et al.~\cite{huang2009measuring}. \autoref{fig:crossingcomparison} and \autoref{fig:torus2:graphicaesthetics} show that \tpairwise{}-\toruslayout{} achieves significantly ($p < 0.001$) fewer crossings than \tpairwise{}-\nodelinklayout{} and \twebcola{}-\nodelinklayout{} for \dsmalleasy{}, \dlargeeasy{}, \dsmallhard{}, and \dlargehard{}. Furthermore, \tpairwise{}-\toruslayout{} significantly ($p = 0.0366$) achieves fewer crossings than \twebcola{}-\toruslayout{} for \dsmalleasy{}. 
As shown in the box plots in~\autoref{fig:crossingcomparisonboxplots}, the difference of crossings is much more obvious for our \dhard{} graphs than the \deasy{} graphs.

In accord with the findings for small networks in~\autoref{sec:torus1}, our results show that torus-based layout have clear benefits over traditional node link diagrams for larger networks at least for these metrics.  

\subsection{New cluster readability metrics: cluster distance}
\label{sec:torus2:clusterdistance}
We use a new metric for cluster readability: \textit{cluster distance} to measure how well a layout algorithm is able to separate clusters.  For a given layout, for all pairs of clusters whose convex hulls are not overlapping we compute:

\noindent\textbf{Minimum separation between convex hulls}---measures space between non-overlapping clusters. A larger value indicates more distance between the boundaries of clusters in a given layout. For torus, we first identify a convex polygon in a 3 $\times$ 3 torus coordinate for each cluster. We then use the Gilbert–Johnson–Keerthi (GJK) algorithm~\cite{ong1997gilbert} to determine the minimum distance between convex polygons.

Then, for all pairs of clusters that are overlapping, we compute:

\noindent\textbf{Minimum penetration depth between convex hulls}---To measure the minimum translation vector required to separate the convex hulls of the cluster pair, as a measure of cluster overlap. A larger value indicates more severely overlapping clusters. We use the Expanding Polytope Algorithm (EPA) which is based on Minkowski sum to compute the penetration depth by Van~\cite{van2001proximity}.

Then, for each pair of clusters we define \textit{cluster distance} as the negative minimum penetration depth if they are overlapping, or the minimum separation between convex hulls if they are not overlapping.  Across all pairs we take the average minimum cluster distance as a metric of cluster separateness across the whole graph.



Graphics of statistical results and box plots of cluster distance are summarised in~\autoref{fig:clusterdistancecomparison} and~\autoref{fig:boxplots_clusterdistancecomparison}. Overall, we found:
\begin{itemize}[noitemsep,leftmargin=*]
    \item \tpairwise{}-\toruslayout{} significantly outperformed \twebcola{}- \nodelinklayout{}, \twebcola{}-\toruslayout{} in cluster distance for both low and high modularity at 0.25, 0.3, 0.35, and 0.4 for both \dsmall{} and \dlarge{}.
    \item \tpairwise{}-\toruslayout{} outperformed \tpairwise{}-\nodelinklayout{} in cluster distance for modularity at 0.25, 0.3 for \dsmall{} and \dlarge{}.
    \item We also found \tpairwise{}-\nodelinklayout{} significantly outperformed \twebcola{}-\nodelinklayout{} for cluster distance under modularity at 0.4 for \dsmall{} and \dlarge{}.
    \item As \tpairwise{} significantly outperformed \twebcola{}, we used \tpairwise{} to perform user evaluation for \nodelinklayout{} and \toruslayout{} in the next section.
\end{itemize}

\begin{figure}
    \centering
    \includegraphics[width=0.9\textwidth]{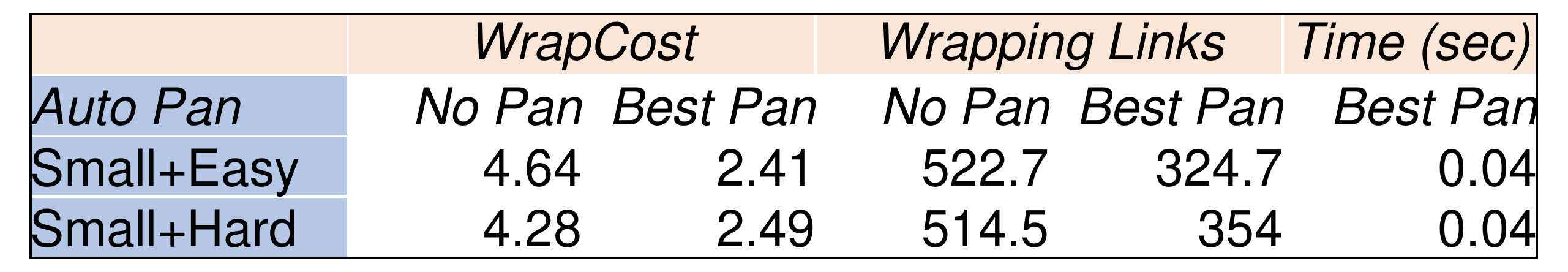}
    \caption{Automatic panning results: mean wrapCost, number of wrapping links across the boundary and running time of 20 random runs for \dsmall{} graphs at high (0.4) and low (0.3) modularity (with layout pre-computed by Algorithm~\ref{algorithm:pairwise}); graph metrics as shown in \autoref{fig:torus2:graphicaesthetics}.}
    \label{fig:torus2:autopanresults}
\end{figure}

\subsection{Automatic panning results}
We conducted a small empirical analysis of the Auto Pan Algorithm \ref{alg:autopan} in terms of number of wrapped links. Results are summarised in Table \autoref{fig:torus2:autopanresults}.
Automatic panning significantly improves the number of wrapped links, while the \emph{wrapcost} penalty prefers wrapping long links over short links, which tends to keep clusters unwrapped.
The difference is visible in \autoref{fig:torus2:autopan}.

\section{User Study 4: Cluster Visualisation Readability}
\label{sec:torus2:userstudy}

The previous section demonstrated the improvement provided by our new algorithm \tpairwise{} with automatic panning, in terms of graph aesthetics and cluster readability metrics. In this section, we investigate if our torus drawings with automatic panning are more effective for people to use than standard unwrapped representations for high-level network topology analysis tasks. 

As mentioned in~\autoref{sec:intro}, visual cluster analysis is an important application of network visualisation. 
The hypothesis that we test in our user study is whether the additional spreading afforded by toroidal layout in terms of the new cluster separation metric  (\autoref{sec:torus2:clusterdistance}) also leads to better human perception of clusters.
We do not use the \twebcola{}-Torus algorithm (\autoref{sec:torus1:stressminimisation}) in the user study because the results of the empirical experiments (\autoref{fig:torus2:graphicaesthetics}, \autoref{fig:clusterdistancecomparison}) overwhelmingly demonstrate that the new \tpairwise{} algorithm is superior.

The particular task that we focus on in this paper is the inspection of community structure, i.e., cluster identification. This study is different from our previous studies in~\autoref{sec:torus1} with small graphs ($\le 15$ nodes, $\le$ 36 links). 
Our study graphs (\autoref{sec:torus2:graphstructure}) are 4.7 to 8.6 times larger in the number of nodes, 21 to 68 times larger in the number of links than that considered in the previous chapter (\autoref{sec:torus1:graph_corpus}). 
In summary, following~\autoref{sec:torus1}, we are interested in answering the following research question that link to \textbf{RG4} (\autoref{sec:intro:RGs:torus}): 

\noindent\textbf{[RQ4.4]}: Do toroidal layouts with interactive wrapping provide more benefit to perception than a standard unwrapped representation for \textit{cluster identification}?



\subsection{Techniques \& Setup}
The techniques in our study are \nodelinklayout{} and \toruslayout{} generated using our new \tpairwise{} algorithm as detailed in~\autoref{sec:torus2:algorithm}. In the study, each trial starts with automatic panning for \toruslayout{}. We did not show any cluster colours in this study to not hint towards any graph structure. Interactive panning using the mouse is enabled only in the \toruslayout{} condition. A user can drag the visualisation such that when the view is panned off one side of the display, it reappears on the opposite side.

\subsection{Tasks \& Dependent Variables}
For our study, we selected two representative network visualisation tasks, inspired by existing task taxonomies by Lee et al.~\cite{lee2006tasktaxonomy} and Saket et al.~\cite{saket2014group}, and that involve the understanding of clusters. For each task, we record task-completion time (\mtime), task-error (\merror), and subjective user confidence in using for the task in terms of rank (1 or 2) (\mconfidence). We record pan distance as the overall distance the participant moves the mouse for each trial. We record user ranking for learnability for cluster related tasks in general (\mlearnability) and overall preference (\moverall).

\begin{figure}
\centering
    \subfigure[\tbelongtocluster\ \nodelinklayout{}]{
        \includegraphics[width=0.4\textwidth]{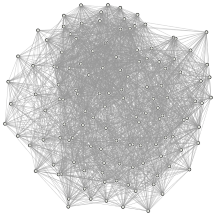}
    }
    \subfigure[\tbelongtocluster\ \toruslayout{}]{
        \includegraphics[width=0.4\textwidth]{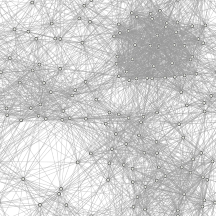}
    }
    \subfigure[\tbelongtocluster\ \nodelinklayout{}]{
        \includegraphics[width=0.4\textwidth]{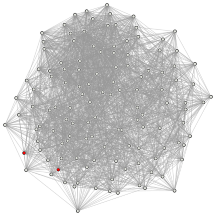}
    }
    \subfigure[\tbelongtocluster\ \toruslayout{}]{
        \includegraphics[width=0.4\textwidth]{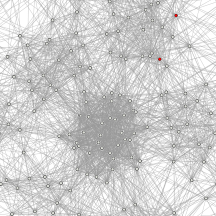}
    }
    \caption{Example of \tclusteridentification\ tasks (a,b) and \tbelongtocluster\ tasks (c,d) for \dlargehard{}: Modularity 0.3, 130 nodes, 2575 links, 5 clusters (a,b); Modularity 0.3, 126 nodes, 2496 links, 5 clusters (c,d)}
    \label{fig:nodeclustertaskgraphs}
\end{figure}

Our tasks are as follows:
\begin{itemize}[leftmargin=*]
    \item \textbf{\tclusteridentification: Identify the number of clusters (\autoref{fig:nodeclustertaskgraphs}(a, b))}
    Participants are required to count the number of clusters they can find in the image. Participants are provided radio buttons to answer 1 to 10. We calculate \merror{} as the absolute difference between the correct answer and the user's response, divided by the correct answer. 
    \item \textbf{\tbelongtocluster: Do the two red nodes belong to the same cluster (\autoref{fig:nodeclustertaskgraphs}(c, d))?}
    We record participants' responses through multiple-choice questions with the values \emph{yes}, \emph{no}, and \emph{not sure} as options. \merror{} is binary with not-sure counting as error. 
\end{itemize}

    

\subsection{Graph Structure}
\label{sec:torus2:graphstructure}

To evaluate effectiveness of toroidal drawings for cluster identification tasks, we used graphs from our graph corpus in~\autoref{sec:torus2:graphcorpus}, with two levels of difficulty (\deasy: modularity=0.4, \dhard: modularity=0.3) and two levels of graph size (\dsmall{} and \dlarge{}), whose graph metrics are summarised in \autoref{fig:torus2:graphicaesthetics}. 
We name these four groups: \dsmalleasy{}, \dlargeeasy{}, \dsmallhard{}, and \dlargehard{}. In each group, we randomly selected 5 graphs for trials for each task. The number of clusters for \tclusteridentification{} ranged between 4 and 7. For \tbelongtocluster{}, the number of clusters ranged between 5 and 7, as we found in pilot studies that fewer than five clusters is too easy for \tbelongtocluster{}. We ran the \tpairwise{} layout algorithm 20 times for a chosen graph for both \nodelinklayout{} and \toruslayout{}. We then selected one layout at random and used it for both \nodelinklayout{} and \toruslayout{} for each study trial. 

\subsection{Hypotheses}
Our hypotheses were pre-registered with the Open Science Foundation: \url{https://osf.io/7vbr4}.

\noindent\textbf{For \tclusteridentification:}
\begin{itemize}[noitemsep,leftmargin=*]
\item \textbf{C4.1}: \toruslayout{} has better task effectiveness (in terms of time and error) than \nodelinklayout{} independent of difficulty levels.

\item\textbf{C4.2}: For \deasy{} graphs, \nodelinklayout{} has better task effectiveness (in terms of time and error) than \toruslayout{} (requires certain panning and mental wrapping).

\item\textbf{C4.3}: For \dhard{} graphs, \toruslayout{} has better task effectiveness (in terms of time and error) than \nodelinklayout{}.

\item\textbf{C4.4}: Participants will report more confidence in \toruslayout{} than \nodelinklayout{}.
\end{itemize}

\noindent\textbf{For \tbelongtocluster:}
\begin{itemize}[noitemsep,leftmargin=*]
\item \textbf{N4.1}: \toruslayout{} has better task effectiveness (in terms of time and error) than \nodelinklayout{} independent of difficulty levels.

\item \textbf{N4.2}: \toruslayout{} has better task effectiveness (in terms of time and error) than \nodelinklayout{} for both \deasy{} and \dhard{} tasks.

\item \textbf{N4.3}: Participants will report more confidence in using \toruslayout{} than \nodelinklayout{}.
\end{itemize}

\noindent\textbf{For participant preference}
\begin{itemize}[noitemsep,leftmargin=*]
\item\textbf{P4.6}: Overall, participants prefer \toruslayout{} over \nodelinklayout{}.
\end{itemize}

\subsection{Experimental Design}
We use a within-subject design (explained in~\autoref{sec:related:evaluationmethods}) with 2 techniques (\toruslayout, \nodelinklayout) $\times$ 2 tasks (\tclusteridentification, \tbelongtocluster) $\times$ 2 level of difficulty (\deasy, \dhard) $\times$ 2 sizes (\dsmall, \dlarge) $\times$ 5 recorded repeats. This leaves us with a total of 80 recorded trials per participant. We blocked the study by tasks, i.e., participants would do both techniques with the same task before moving to the second task. For each task block, we counterbalanced the order of the techniques using a full-factorial design. The order of each level of difficulty and size in each technique was the same: \dsmalleasy{}$\rightarrow$\dlargeeasy{}$\rightarrow$ \dsmallhard{}$\rightarrow$\dlargehard{}. The order of trials for each technique within each level was randomised.

\subsection{Participants and Procedures}
We recruited 32 participants from local institutes through university's email list and snowballing. 19 were males, 13 were females. The age of participants was between 20 and 50 (mean = 31.5). 22 people reported seldom or never using network diagrams while 10 people often used network diagrams in their work or study. 

While run entirely online due to COVID-19 health-concerns, the experimenter supervised each participant through remote video conferencing software to give proper instructions, assure the participants' engagement, and help with eventual questions. The experimental software was loaded in a participant's Google Chrome browser. Each participant shared their screen with the experimenter. The experimenter ensured that each participant used a monitor with resolution no less than 1366 $\times $768 pixels. Each study trial used a stimuli with a size of 650 $\times$ 650 pixels. Each trial was correctly loaded in a participant's browser, before the recording started. The experimenter trained each participant to identify a cluster as a graph structure whose links within a cluster are relatively more than the links between clusters. Before each new task and technique, there were 2 training trials. When a participant gave an answer in these training trials, an image with different clusters highlighted in different colours appeared, outlining the cluster. Participants first had to complete all training trials correctly before proceeding to the recorded trials. Each recorded trial had a timeout of 20 seconds to prevent participants from trying to perform precise link counting. For \toruslayout{} technique, short animations demonstrating interactive torus wrapping were shown.

\begin{figure}
    \centering
    \subfigure[\tclusteridentification\ error]{
        \includegraphics[width=0.8\textwidth]{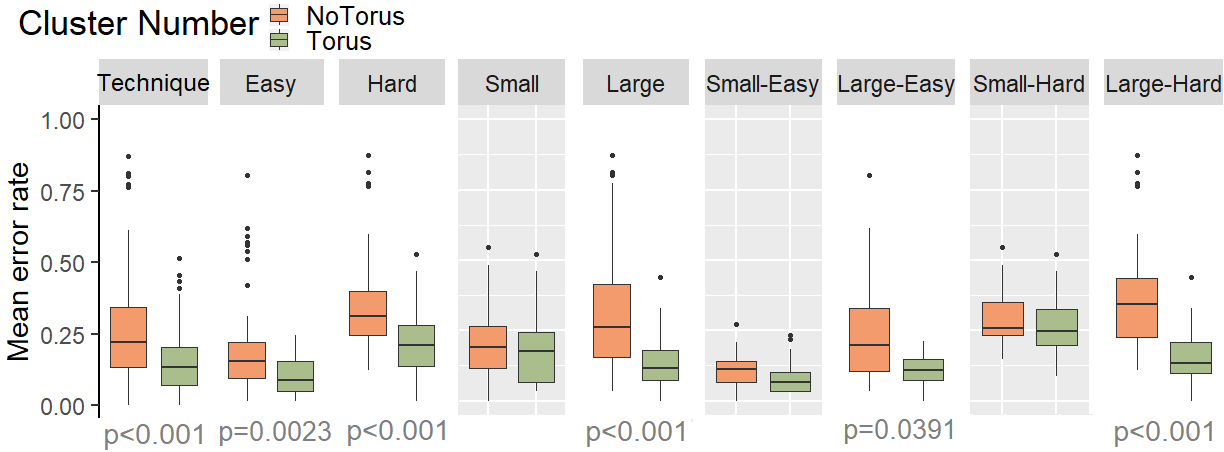}
    }
    \subfigure[\tbelongtocluster\ error]{
        \includegraphics[width=0.8\textwidth]{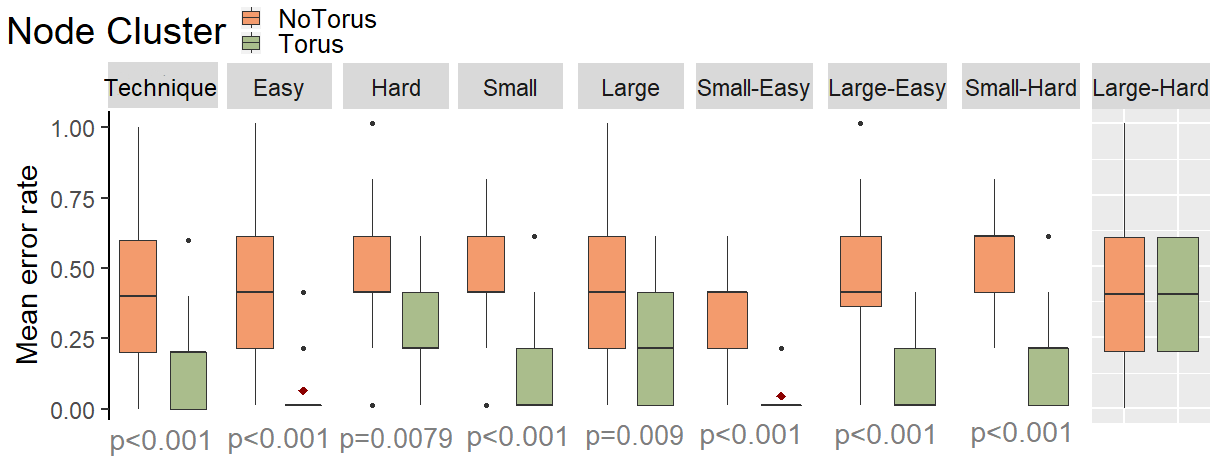}
    }
    \subfigure[\tclusteridentification\ time]{
        \includegraphics[width=0.8\textwidth]{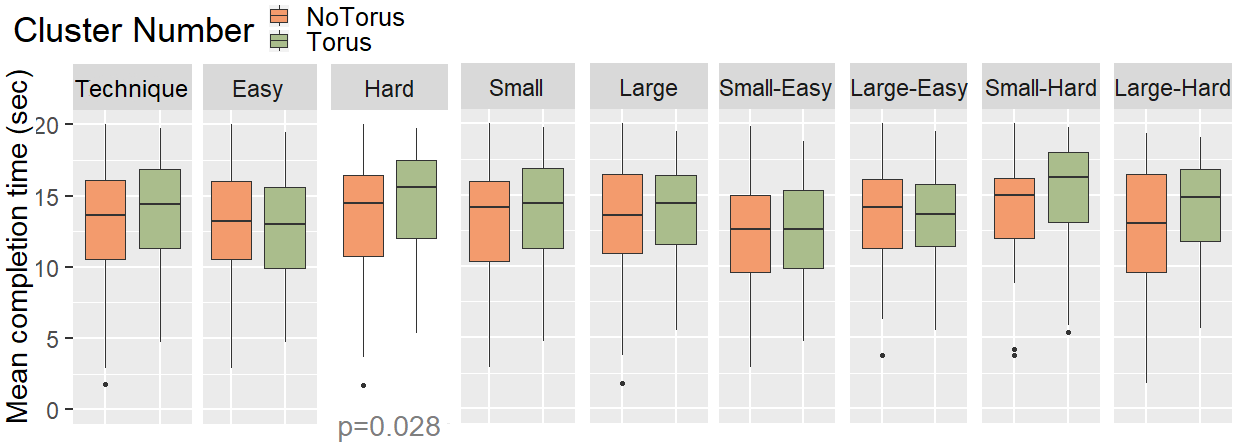}
    }
    \subfigure[\tbelongtocluster\ time]{
        \includegraphics[width=0.8\textwidth]{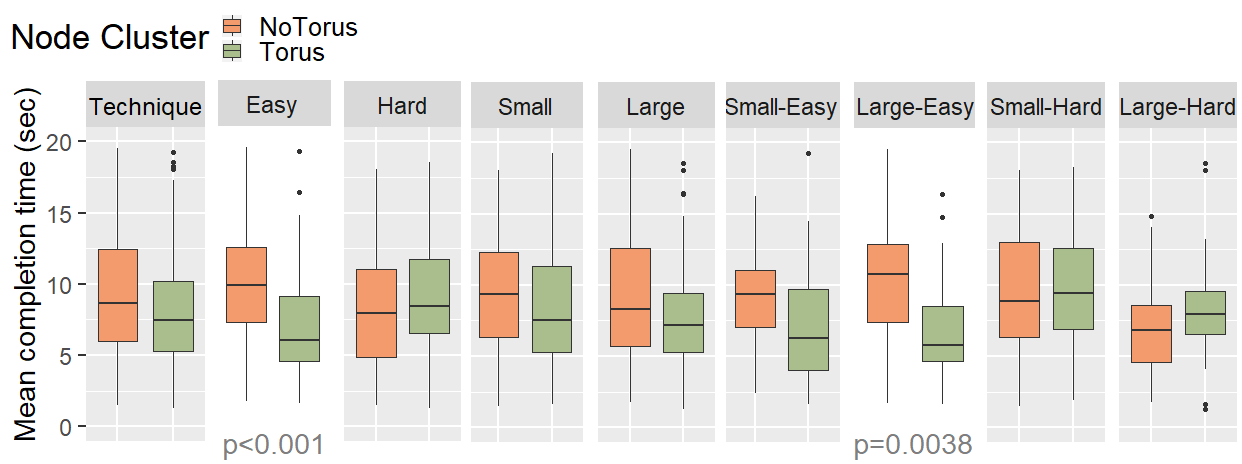}
    }
    \caption{User evaluation results of mean error and time between \nodelinklayout{} and \toruslayout{} split by task,  difficulty and size. Columns of significant results are shown in white background.}
    \label{fig:torus2:Taskscomparison}
\end{figure}


\begin{figure}
    \centering
    \includegraphics[width=0.7\columnwidth]{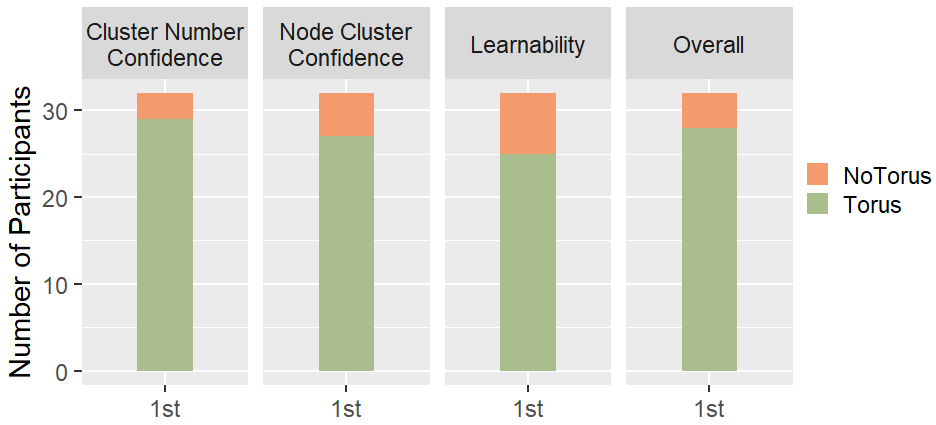}
    \caption{Subjective user rank of \nodelinklayout{} and \toruslayout{} by  the number of participants who ranked the condition as their first preferred choice.} 
    \label{fig:userpreferencecomparison}
\end{figure}


\subsection{Results}

All of the participants completed the training and recorded trials. Therefore we recorded performance for 2,560 trials. Since \merror{} was not normally distributed, we used Friedman's non-parametric test and Tukey's posthoc multiple pairwise comparison with Bonferroni correction~\cite{field2012discovering} to identify significant differences between \nodelinklayout{} and \toruslayout{}. The residuals of \mtime{} were normally distributed, visually checked with Q-Q plots and a histogram, supported by a Shapiro-Wilk test~\cite{shaphiro1965analysis}. The variances of the \mtime{} were equal by Levene's test. We therefore performed a 3-way repeated measures ANOVA to test significant difference in \mtime{} between \nodelinklayout{} and \toruslayout{}. We used paired t-test for posthoc pairwise comparisons with p-value adjustment using Bonferroni correction. Wilcoxon's non-parametric signed rank test was used to analyse paired significance of subjective user rank. Confidence intervals indicate 95\% confidence for mean values for all the pairwise comparisons.



Overall, for \tclusteridentification, we found the following significant results. The corresponding box plots can be found in~\autoref{fig:torus2:Taskscomparison}(a).
\begin{itemize}[leftmargin=*]
    \item For \merror, we found \toruslayout{} ($.15, SD = .1$) significantly outperforming \nodelinklayout{} ($.26, SD = .17$) for technique ($p < .001$), as shown in \autoref{fig:torus2:Taskscomparison}(a).; \toruslayout{} ($.09, SD = .05$) resulted in significantly less errors than \nodelinklayout{} ($.18, SD = .16$) for \deasy{} ($p = .0023$); \toruslayout{} ($.2, SD = .11$) significantly outperformed \nodelinklayout{} ($.33, SD = .16$) for \dhard{} ($p < .001$);  \toruslayout{} ($.12, SD = .08$) has significantly less errors than \nodelinklayout{} ($.31, SD = .21$) for \dlarge{} ($p < .001$); \toruslayout{} ($.1, SD = .05$) significantly outperformed \nodelinklayout{} ($.25, SD = .2$) for \dlargeeasy{} ($p = .0391$); and \toruslayout{} ($.15, SD = .09$) significantly outperformed \nodelinklayout{} ($.37, SD = 2$) for \dlargehard{} ($p < .001$).
    
    \item For \mtime, we found \nodelinklayout{} ($13.7s, SD = 4.65$) was significantly faster than \toruslayout{} ($15.1s, SD = 3.88$) for only \dhard{} ($p = .028$), as shown in~\autoref{fig:torus2:Taskscomparison}(c).
    
    \item Participants report significantly more confidence in using \toruslayout{} than \nodelinklayout{} (p$ < $0.001), as seen in~\autoref{fig:userpreferencecomparison}.
\end{itemize}

    
    

For \tbelongtocluster, we found the following significant results. The corresponding box plots can be found in~\autoref{fig:torus2:Taskscomparison}(b). 
\begin{itemize}[leftmargin=*]
    \item For \merror, \toruslayout{} ($.16, SD = .2$) was found significantly outperforming \nodelinklayout{} ($.43, SD = .21$) for technique ($p < .001$), as seen in~\autoref{fig:torus2:Taskscomparison}(b).; \toruslayout{} ($.05, SD = .09$) resulted in significantly less errors than \nodelinklayout{} ($.39, SD = .22$) for \deasy{} ($p < .001$); \toruslayout{} ($.28, SD = .21$) significantly outperformed \nodelinklayout{} ($.47, SD = .2$) for \dhard{} ($p = .0079$); \toruslayout{} ($.09, SD = 14$) had significantly less errors than \nodelinklayout{} ($.44, SD = .18$) for \dsmall{} ($p < .001$); \toruslayout{} ($.23, SD = .22$) significantly outperformed \nodelinklayout{} ($.42, SD = .24$) for \dlarge{} ($p = .009$); \toruslayout{} ($.03, SD = .07$) was found significantly outperforming \nodelinklayout{} ($.33, SD = .16$) for \dsmalleasy{} ($p < .001$); \toruslayout{} ($.06, SD = .1$) significantly outperformed \nodelinklayout{} ($.45, SD = .25$) for \dlargeeasy{} ($p < .001$); and \toruslayout{} ($.16, SD = .16$) significantly outperformed \nodelinklayout{} ($.55, SD = .14$) for \dsmallhard{} ($p < .001$), as shown in~\autoref{fig:torus2:Taskscomparison}(b).
    
    \item \toruslayout{} ($7.22s, SD = 3.92$) was significantly faster than \nodelinklayout{} ($9.84s, SD = 3.78$) in time for \deasy{} ($p < .001$); and \toruslayout{} ($7.1s, SD = 3.74$) was significantly faster than \nodelinklayout{} ($10.5s, SD = 4.09$) for \dlargeeasy{} ($p = .0038$), as shown in  \autoref{fig:torus2:Taskscomparison}(d).
    
    \item Participants report significantly more confidence in using \toruslayout{} than \nodelinklayout{} ($p < .001$), as shown in~\autoref{fig:userpreferencecomparison}.
\end{itemize}

Overall, \toruslayout{} improved task effectiveness in terms of error rate by 42.3\% on average, compared with \nodelinklayout{} for \tclusteridentification{} task. For \tbelongtocluster{} task, \toruslayout{} improved the task effectiveness in terms of error rate by 62.7\% and time by 32.3\% on average (i.e., difference between means), as opposed to \nodelinklayout{}. Participants reported that \toruslayout{} is significantly easier to learn (\mlearnability) ($p < .001$) than \nodelinklayout{} for our cluster related tasks. \toruslayout{} was significantly preferred (\moverall) ($p < .001$) over \nodelinklayout{} in  overall rank, as shown in \autoref{fig:userpreferencecomparison}. Based on these results, for \textbf{[RQ4.4]} we rejected hypothesis C4.2 and accepted C4.4, N4.3, P4.6. We accepted C4.1, C4.3, N4.1 for error, and N4.2 for error and time for \deasy{}. 



\subsection{Qualitative User Feedback}
\begin{figure}
    \centering
    \subfigure{
    \includegraphics[width=0.45\textwidth]{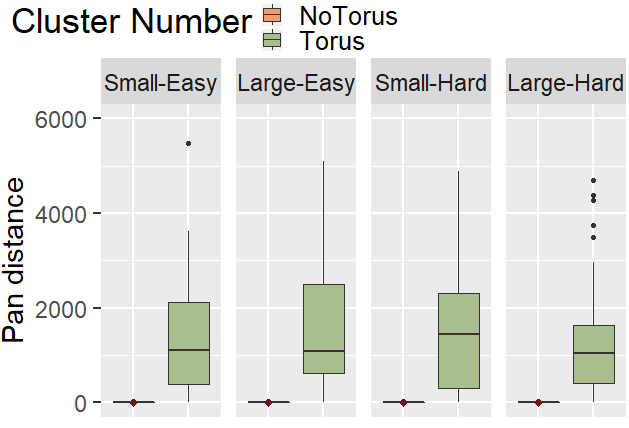}
    }
    \subfigure{
    \includegraphics[width=0.45\textwidth]{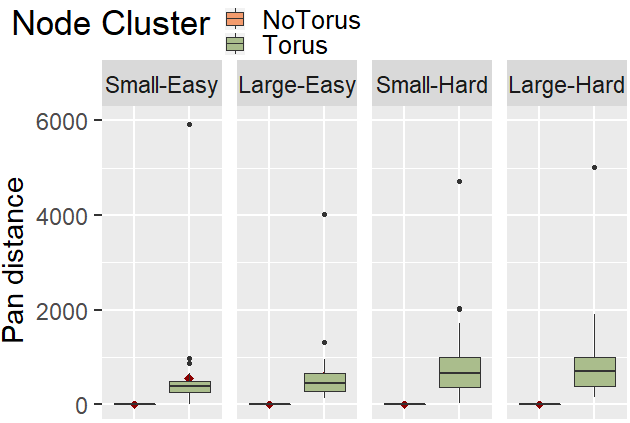}
    }
\caption{Box plots of panning distance results of \nodelinklayout{} and \toruslayout{}.}
\label{fig:torus2:pandistanceresults}
\end{figure}

For understanding clusters, the majority of participants mentioned that the \toruslayout{} with panning provided for less overlap between nodes and was better at spreading nodes and clusters more spatially, thus reducing visual clutter and increasing readability. While our automatic panning optimises the view a user sees first, interaction further helps to find the best view possible. Some participants reported that \tclusteridentification{} required more panning to identify the graph structure. For \tbelongtocluster, on the other side, participants reported that they rarely used interactive panning, except for verifying their answer. This is supported by our measurements that \tbelongtocluster{}, required less user interaction (majority less than 1000 pixels per trial), as shown in~\autoref{fig:torus2:pandistanceresults}. There are 4 participants who reported they favoured \nodelinklayout{} as it gives a full overview of the networks and did not require to identify the same piece of cluster wrapped around top-bottom or left-right. 



\section{Discussion}
\label{sec:torus2:discussion}


Our study results indicate that \nodelinklayout{} is sometimes faster than \toruslayout\ when counting the number of clusters for HARD (Fig.~\ref{fig:torus2:Taskscomparison}(c)), but the error rate in the \nodelinklayout{} condition was generally much higher (Fig.~\ref{fig:torus2:Taskscomparison}(a)).  We believe participants were sometimes faster with \nodelinklayout{} because they were simply guessing the answer. Our new study is intended to be complementary to the studies described in~\autoref{sec:torus1} by evaluating an important task not considered in that study.
Since our new layout algorithm produces high-quality torus layout with fewer crossings, larger incidence angles and  less stress, we would expect that the results from the previous chapter for low-level connectivity understanding and path following tasks would also be reproducible for larger networks using our new algorithm. However, we leave such an evaluation to future work.

\section{Conclusion and Future Work}
\label{sec:torus2:conclusion}


In this chapter, we have explored new network layout algorithms that take advantage of the toroidal topology such that when wrapping specific links around the boundaries, it allows further relaxation of node positions to reduce visual clutter, compared with our torus layout algorithm in~\autoref{sec:torus1} and standard non-wrapped layouts. We addressed~\textbf{RG4} (\autoref{sec:intro:RGs:tools}).


We presented a new algorithm based on \tpairwise{} gradient descent for torus-based layout of networks that is more robust in terms of graph aesthetics and running time than \twebcola{} torus layout method described in~\autoref{sec:torus1}. Furthermore, we improved the resulting layouts by using a novel algorithm that minimises the number of link wrappings across the boundary by centring elements of interest in the display. 

Using this algorithm, we have investigated whether interactive toroidal wrapped layouts provide advantages over traditional flat network layouts for higher-level tasks of understanding network structures. Both an analysis of graph aesthetics metrics and a user study have clearly shown that they do: participants were able to more accurately determine the number of clusters in a graph and more quickly and accurately determine whether two nodes are in the same cluster with the toroidal wrapped layouts. Furthermore, participants preferred the toroidal wrapped layouts to the standard non-wrapped flat node-link. This is the first  demonstration that torus-based layouts provide real-world benefits over traditional node-link layouts. 

Our research also reveals several directions for future research, including exploration of the proposed torus layout approach for understanding complex multidimensional data that needs cleaner representations such as MDS (described in \autoref{sec:discussion:futurework:mds}), investigating if the proposed layout methods could also provide benefits for understanding specific real-world complex and dense network structures, such as biological and social network applications (detailed in \autoref{sec:discussion:networkvis}) and evaluate real-world usefulness with domain experts. Furthermore, improving the network layout algorithm's runtime performance is a major direction (\autoref{sec:discussion:networkvis}).

%
\chapter{Spherical Wrapping for Geographical Data Visualisations}
\label{sec:spheremaps}

\cleanchapterquote{The fact I couldn’t move the pictures was frustrating
[for the static map projections] and I think I didn’t get many of the guesses right.}{Anonymised user study participant, P4}{Direction estimation tasks}

We have investigated the utility of interactive cylindrical and toroidal wrapping of visualisations which have shown benefits for understanding cyclic time series data (\autoref{sec:cylinder}) and understanding the structure of network diagrams (\autoref{sec:torus2}). Crucially, these visualisations are interactive and the viewer can pan the projection so that it ``wraps'' around the plane. Inspired by these results, we now investigate the utility of projections of the surface of a different geometric object, the sphere, onto a 2D plane. This chapter addresses \textbf{RG5}. As discussed in~\autoref{sec:related:maps}, a plethora of map projection techniques exists with various affordances for wrapping (panning). However, existing studies exclusively focus on readability of static maps. 
In this chapter, we investigate the effect of 2D interactive spherical wrapping on different map projections for geographic comprehension tasks. 

Beside geographic data, spherical projections have applications for non-geographic data. For example, there can be advantages to laying out networks (a type of abstract data which has no inherent geometry) on the surface of a sphere such that there is no arbitrary edge to the display or privileged centre (\autoref{sec:related:sphere}). We explore the readability of networks laid out spherically and then projected to a flat surface using interactive spherical wrapping and rotation in the next chapter.

\begin{figure}
	\includegraphics[width=\textwidth]{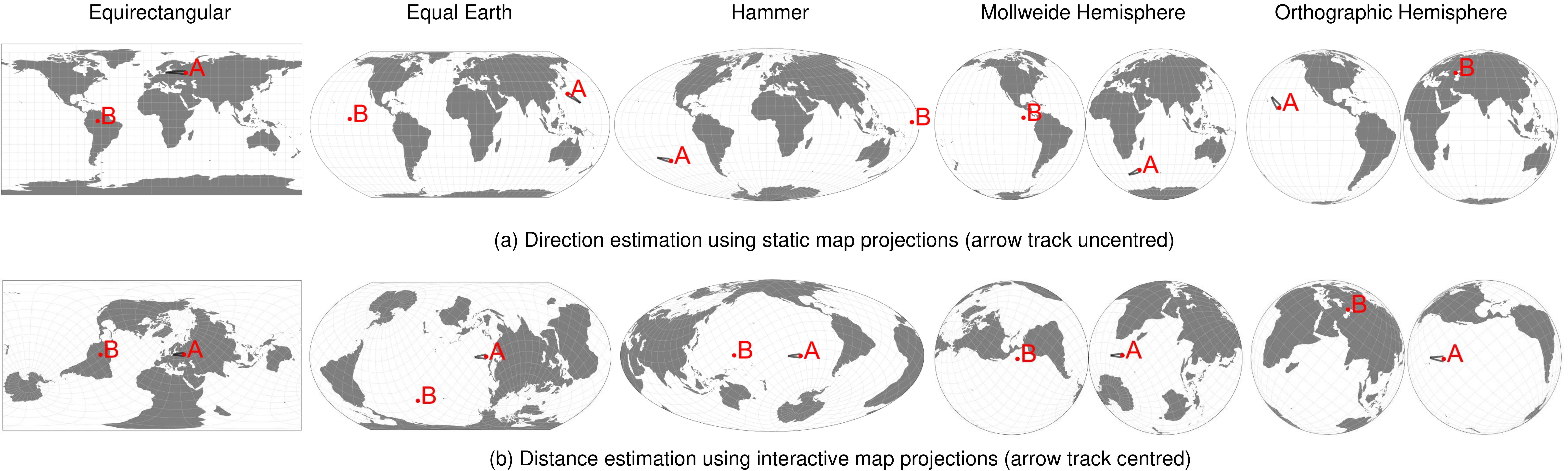}
	\caption{Top row: Direction estimation task of geographic data using the five different spherical projections evaluated in Study 1, with (a) and without (b) interaction.}
  	\label{fig:spheremaps:teaser}
\end{figure}


\section{Map Projection Techniques}
\label{sec:techniques}
In this section, we investigate the effect of interaction on readability of different map projection techniques and identify the projection techniques which best support geographic comprehension tasks, such as distance, area and direction estimation (\autoref{sec:mapstudy}).

We chose five representative map projections for the study.
To explore map readability, we aim to cover a wide range of distortion properties such as preservation of area, distance, shape, direction as discussed in \autoref{sec:related}, and user preference~\cite{avric2015user}. 
We demonstrate the key characteristics of the map projections in \autoref{fig:geo-techniques} and discuss the details as follows:

\begin{figure}
    \centering
	\includegraphics[width=\textwidth]{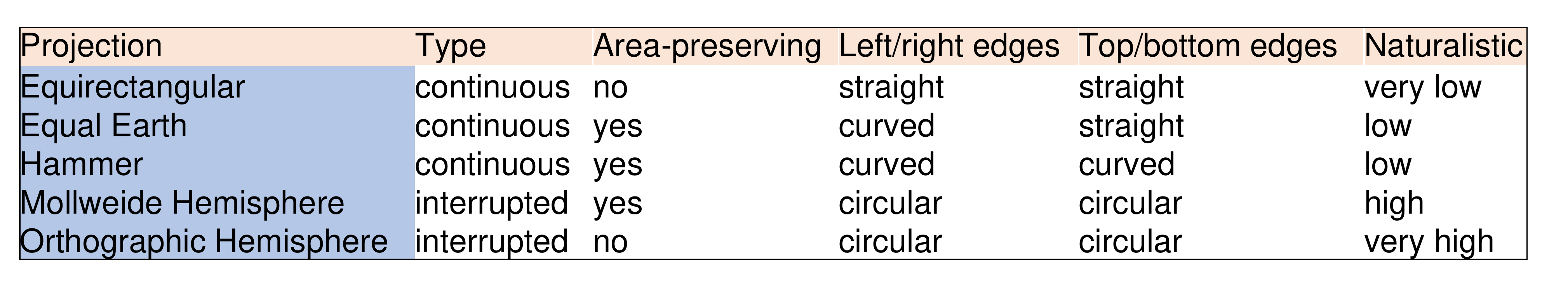}
	\caption{Key characteristics of the five map projections tested in Study 1.}
	\label{fig:geo-techniques}
\end{figure}

\noindent\textbf{\tequirectangular{}} projects the earth onto a space-filling rectangle with the north and south poles extending along the top and bottom edges, respectively. Rectangular projection is the most widely used projection, with common variations including Mercator or Plate Carrée~\cite{jenny2017guide, avric2015user}. \tequirectangular{} projections preserve distances along all meridians and are useful when differences in latitude are measured~\cite{jenny2017guide}. However, it does not preserve the relative size of areas. Furthermore, it has been found confusing for tasks requiring understanding how the edges connect to each other, such as predicting the path of air plane routes crossing the top and bottom edge of the map~\cite{hennerdal2015beyond}. 

\noindent\textbf{\tequalearth{}} is similar to \tequirectangular{} but relaxes the rectangle with rounded corners, diminishing the strength of horizontal distortion near the poles. Furthermore, the similarly shaped Robinson projection has received good subjective ratings from viewers~\cite{avric2015user}. Unlike Robinson projection, \tequalearth{} preserves the relative size of the areas well.

\noindent \textbf{\thammer{}} further diminishes horizontal stretching by projecting the earth onto an ellipse such that the poles are points at the top and bottom. It preserves the relative size of areas and generally has similar properties to \tequalearth~\cite{yang2018maps, jenny2017guide}. The similarly shaped Mollweide projection has been shown less confusing than \tequirectangular{} when judging the continuity of air plane routes as described above~\cite{hennerdal2015beyond}. Furthermore, map readers prefer to see the poles as points rather than lines~\cite{avric2015user}. It has also been found pleasing to many cartographers than other projections due to its elliptical shape~\cite{jenny2017guide}.

\noindent \textbf{\tmollweide{}} projects the earth onto two circles (hemispheres). This again helps diminish horizontal stretching but introduces the cost of new tears (interruption) that a viewer needs to mentally close. 

\noindent \textbf{\torthographic{}} is also hemispheric. It has a naturalistic appearance as it shows the globe (from both sides) seen from an infinite distance. However, compared with \tmollweide{}, it is not area-preserving. 


\subsection{Interaction techniques}
For each projection, we created both \dstatic{} and  \dinteractive{} versions. With \dinteractive{}, a user can freely move regions of interest to the centre of the projection, thus reducing their distortion. Examples of the effect of interaction are shown in the top row in \autoref{fig:spheremaps:teaser}.

We used D3 libraries for creating all the map projection techniques. For \dinteractive{}, we follow Yang et al.~\cite{yang2018maps} and use versor dragging which controls three Euler angles. This allows the geographic start point of the gesture to follow the mouse cursor~\cite{Davies:2013ug}.

We implemented \torthographic{} with two Orthographic map projections with one showing the western and the other showing the eastern hemisphere. They are placed close together as shown in the top row of \autoref{fig:spheremaps:teaser}. When one hemisphere is dragged, the Orthographic projection of the other hemisphere automatically adjusts three-axis rotation angles such that it shows the correct opposite hemisphere. 

Spherical rotation of all projections is demonstrated in the OSF repository\footnote{Interactive examples can be found in \url{https://observablehq.com/@kun-ting}.}
\section{User Study 5: Map Projection Readability}
\label{sec:mapstudy}

The goal of our first study was to understand the effectiveness of (i) different projections for geographic data as well as the (ii) benefit of interactively changing the centre point of these projections. 

\subsection{Techniques}
The techniques are \tequirectangular, \tequalearth, \thammer, \tmollweide, and \torthographic, described in \autoref{sec:techniques}. Each technique is given both static images without the ability to rotate, and with interactive spherical rotation, using the mouse. A user can rotate the visualisation such that when the view is panned off one side of the display, it either reappears on the opposite side (left-right) or is horizontally mirrored (top-top, bottom-bottom), as shown in the top-row of \autoref{fig:spheremaps:teaser}. The area of the rectangular bounding box of each map projection condition was fixed at 700 $\times$ 350 pixels.

\subsection{Tasks \& Datasets}
We selected three representative geographic data visualisation tasks. They were also used in existing map projection studies~\cite{yang2018maps,hennerdal2015beyond,hruby2016journey,carbon2010earth}. 

\begin{figure*}
    \centering
    \includegraphics[width=\textwidth]{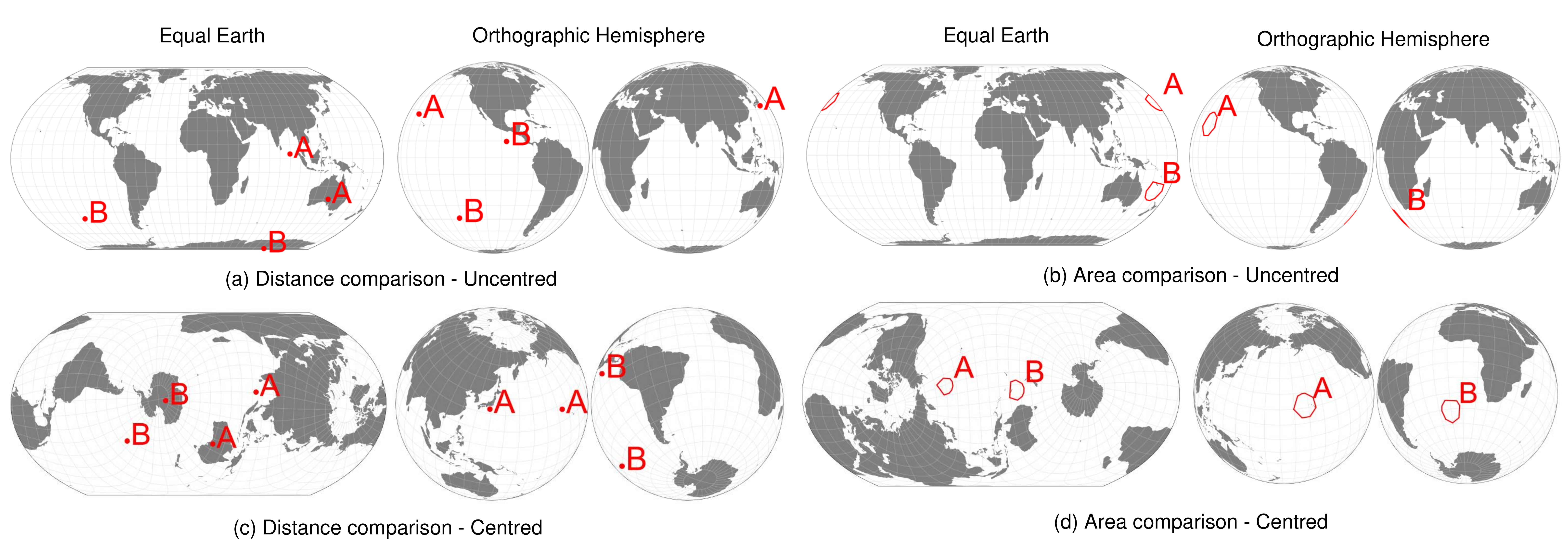}
    \caption[]{Sample trials of Equal Earth and Orthographic Hemisphere for distance and area comparison tasks. \dstatic{} (a,b) may result in uncentred view. \dinteractive{} (c,d) allows a user to drag the map to find the best angle (centred) to answer the task. Direction tasks are shown in \autoref{fig:spheremaps:teaser}}
    \label{fig:Taskscomparison}
\end{figure*}

\noindent\textbf{\tdistancecomparison:} \textit{Which pair of points (pair \textit{A} or pair \textit{B}) represents the greater geographical distance on the surface of a globe?} Participants had to compare the true geographical distance (\textit{as the crow flies}) between two pairs of points on the projection. Participants were provided with radio buttons to answer \textit{A}, \textit{B}, or \textit{not sure}. The results of \textit{not sure} were counted towards the number of incorrectly answered trials. We created data sets for two difficulties through extensive piloting, based on the difference of point distance across point pairs: 10\% difference (easy) and 5\% difference (difficult). The geographic point pairs were randomly chosen, constrained by their individual angular distance (measured in differences in geographical coordinates) between 40\textdegree{} (approx. 4444\textit{km}) and 60\textdegree{} (approx. 6666\textit{km})~\cite{yang2018maps}, and a minimum 60\textdegree{} distance across point pairs. We did not set any upper-bound to not to bias any projection. We created an additional quality control trial with 40\% difference of node distance to test a participant's attention. An example is shown in \autoref{fig:Taskscomparison}(a,c).

\noindent\textbf{\tareacomparison:} \textit{Which polygon (\textit{A} or \textit{B}) covers the greater geographical area on the surface of a globe?} Participants had to compare the size (area) of the polygons. Participants were provided with radio buttons to answer \textit{A}, \textit{B}, or \textit{not sure}. \merror{}, defined in~\autoref{sec:mapstudy:dependentvar}, is binary with not-sure counted as error. Again, we created data sets for two difficulties based on the difference in area they cover: 10\% difference (easy) and  7\% difference (difficult). Eight geographic points of convex polygons were randomly chosen using the same method as Yang et al.~\cite{yang2018maps}, constrained by the individual geographic area between 40 and 60, and a minimum 60\textdegree{} angular distance between centroids of pairwise polygons. There is no upper-bound for the same reason as above. We create an additional quality control trial with 40\% difference in area to test a participant's attention. An example is shown in \autoref{fig:Taskscomparison}(b,d).

\noindent \textbf{\tdirectionestimation:} \textit{Does the trajectory of dot A hit or miss dot B on the surface of a globe?} Participants had to assess whether the trajectory, indicated by an arrow track, from point \textit{A} hits or misses point \textit{B}. Participants were provided with radio buttons to answer \textit{Hit}, \textit{Miss}, or \textit{not sure}. We randomly created pairs of geographic points (A, B, and arrow head) with a minimum angular distance of 60\textdegree{} between A and B. There is no upper-bound. For trials where the trajectory of A misses B, the angular distance between trajectory and B was constrained to 40\textdegree{}. Examples of this task are shown in the first two rows of \autoref{fig:spheremaps:teaser} for each projection technique, where A misses B for equal earth and orthographic hemisphere. For this task, there was only one level of difficulty. We create an additional quality control trial with dot A at the centre and arrow track hitting to a dot B aligned horizontally at the centre to test a participant's attention. 

\subsection{Hypotheses}
Our hypotheses were pre-registered with the Open Science Foundation:~\url{https://osf.io/vctfu}. 

\noindent\textbf{M5.1}: \textit{\dinteractive{} has a lower error rate} than all static map projections across all tasks. Our intuition is that interaction allows regions of interest to be centred and, thus, their distortion reduced.

\noindent \textbf{M5.2}: \textit{\dinteractive{} projections have longer task-completion time than static across all projections and tasks.} Users will spend time interacting to find an optimal centre point for each projection to solve the task.

\noindent \textbf{M5.3}: \textit{For interactivity, users prefer \dinteractive{} to \dstatic{} across all tasks.} Intuition as above.

\noindent \textbf{M5.4}: \textit{\dinteractive{} \torthographic{} has a lower error rate} than other interactive non-hemisphere (\tequirectangular{}, \tequalearth{}, \thammer{}) projections across all tasks.
This is inspired by a virtual globe study by Yang et al.~\cite{yang2018maps}. Our intuition is that an interactive view of the 3D globe will have similar benefits to the VR representation.

\noindent \textbf{M5.5}: \textit{Users prefer \dinteractive{} \torthographic{} over all other \dinteractive{} projections across tasks.} Intuition as above.





\subsection{Experimental Setup}

Design is within-subject per task, where each participant performed \textit{one} task (\tdistancecomparison, \tareacomparison, \tdirectionestimation) on all projections (\tequirectangular, \tequalearth, \thammer, \tmollweide, \torthographic) in both \dstatic{} and \dinteractive{} and in all levels of difficulty (\textit{easy}, \textit{hard}). Each of these 10 conditions for \tareacomparison{} and \tdistancecomparison{} was tested in 12 trials with two difficulty levels (6 easy, 6 hard). For \tdirectionestimation{}, we reduced the number of trials to 8 as pilot participants reported the direction estimation was too difficult for long distance. 
Similar to our prior visualisation crowd-sourced studies described in~\autoref{sec:cylinder:userstudy} and Brehmer et al.~\cite{brehmer2018visualizing}'s studies, we randomly inserted a quality control trial with low difficulty in addition to normal trials to each condition to test participants' attention. The study was blocked by interactivity. Within each block, the order of the map projection techniques  was counterbalanced using William et al.'s Latin-square design~\cite{williams1949experimental}. This technique resulted in 10 possible orderings for the 5 projections while the order of projections in each block was the same. Each recorded trial had a timeout of 20 seconds, inspired from pilot studies.  

\subsection{Participants and Procedures}

We crowd-sourced the study via the \textit{Prolific Academic system}~\cite{palan2018prolific}. Participants on Prolific Academic have been reported to produce data quality comparable to Amazon Mechanical Turk~\cite{peer2017beyond}. Many visualisation studies have used this platform, including our prior studies described in~\autoref{sec:cylinder:userstudy} and Satriadi et al.~\cite{satriadi2021quantitative}'s study. 
We hosted an online study on the Red Hat Enterprise Linux (RHEL7) system. We set a pre-screening criterion on performance that required a minimum approval rate of 95\% and a minimum number of previous successful submissions of 10. We also limited our study to desktop users with larger screens. We paid \pounds5 (i.e.\ \pounds7.5/h), considered to be a good payment according to the Prolific Academic guidelines.

We recorded 120 participants who passed the attention check trials, completed the training and recorded trials. This comprised 4 full counterbalanced blocks of participants (10 $\times$ 4 $\times$ 3 tasks). 57 of our participants were females, 63 were males. The age of participants was between 18 and 55. 7 participants rank themselves as \textit{regularly} using GIS or other tools to analyse geographical data. 105 occasionally read maps, e.g., using Google Map or GPS systems. 8 had very little experience with any sort of map. 

Before starting the study, each participant had to complete a tutorial explaining projection techniques and tasks. 
The tutorial material contained \textit{Tissot's indicatrix}~\cite{snyder1987map}, a set of circular areas placed on both the poles of the equator, indicating the type and magnitude of area, shape, and angular distortion in a given projection. The setting has been inspired by Yang et al.~\cite{yang2018maps}. 
An online demonstration of the study is available: \url{https://observablehq.com/@kun-ting/gansdawrap}

\subsection{Dependent Variables}
\label{sec:mapstudy:dependentvar}
We measured \textbf{task-completion-\emph{time}} (\mtime{}) for each trial in milliseconds, counted between the first rendering of the visualisation and the mouse click of the \textit{answer trial} button, which hid the visualisation and showed an interface for the participants to input their answers. We measured the \textbf{\emph{error rate}} (\merror{}) as the ratio of incorrect over all answers.
We asked participants to \textbf{\emph{rank}} (\mpref{}) each map projection individually within the \textit{static} and the \textit{interactive} block according to their perceived effectiveness. We also ask participants to provide their justifications for the rankings as \textbf{\emph{qualitative feedback}.} After they completed both blocks, we recorded participants' preference of the interactivity between \dstatic{} and \dinteractive{} individually for each projection, and their overall preference between \dstatic{} and \dinteractive{}.

\subsection{Statistical Analysis Methods}
\label{sec:study-1-stats}

\begin{figure}
    \centering
    \includegraphics[width=1.1\textwidth]{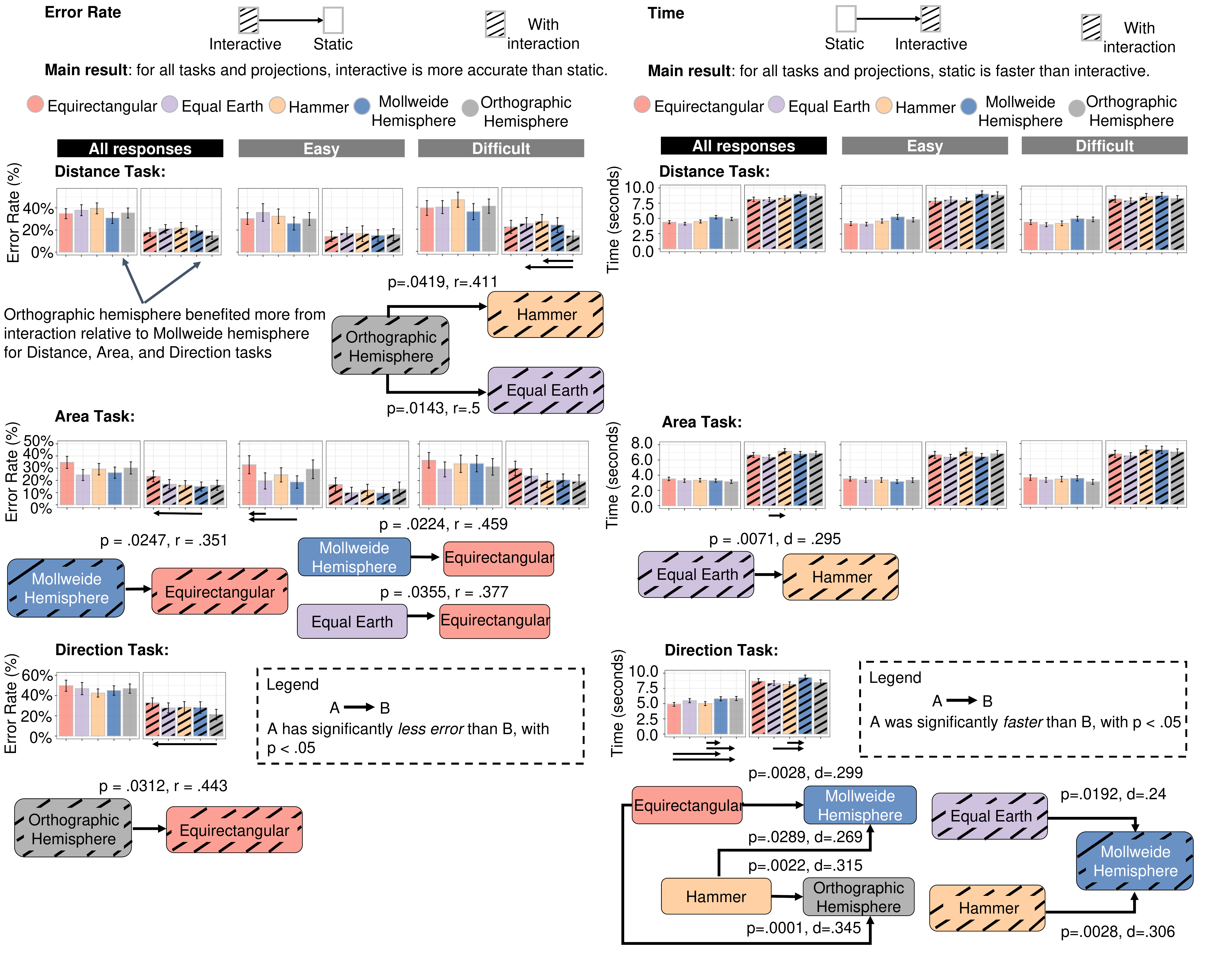}
    \caption{Error rate (left) and Time (right) results. Significant differences between projections are shown as arrows. Significant differences between interactive and non-interactive conditions are \textbf{omitted} to improve readability. Error bars indicate 95\% confidence
intervals. Effect size results for Cohen's $r$ and Cohen's $d$~\cite{cohen2013statistical} are presented for \merror{} and \mtime{}, respectively. Statistically significant results are highlighted in flow diagrams below the bar charts. Bars and boxes with stripe patterns refer to interactive conditions. Overall, equal earth performed well for \mtime{} and \merror{} for some tasks. Orthographic hemisphere benefited more from interaction than the mollweide hemisphere for \merror{}.}
    \label{fig:study-1-results}
\end{figure}

\begin{figure}
    \centering
    \includegraphics[width=\textwidth]{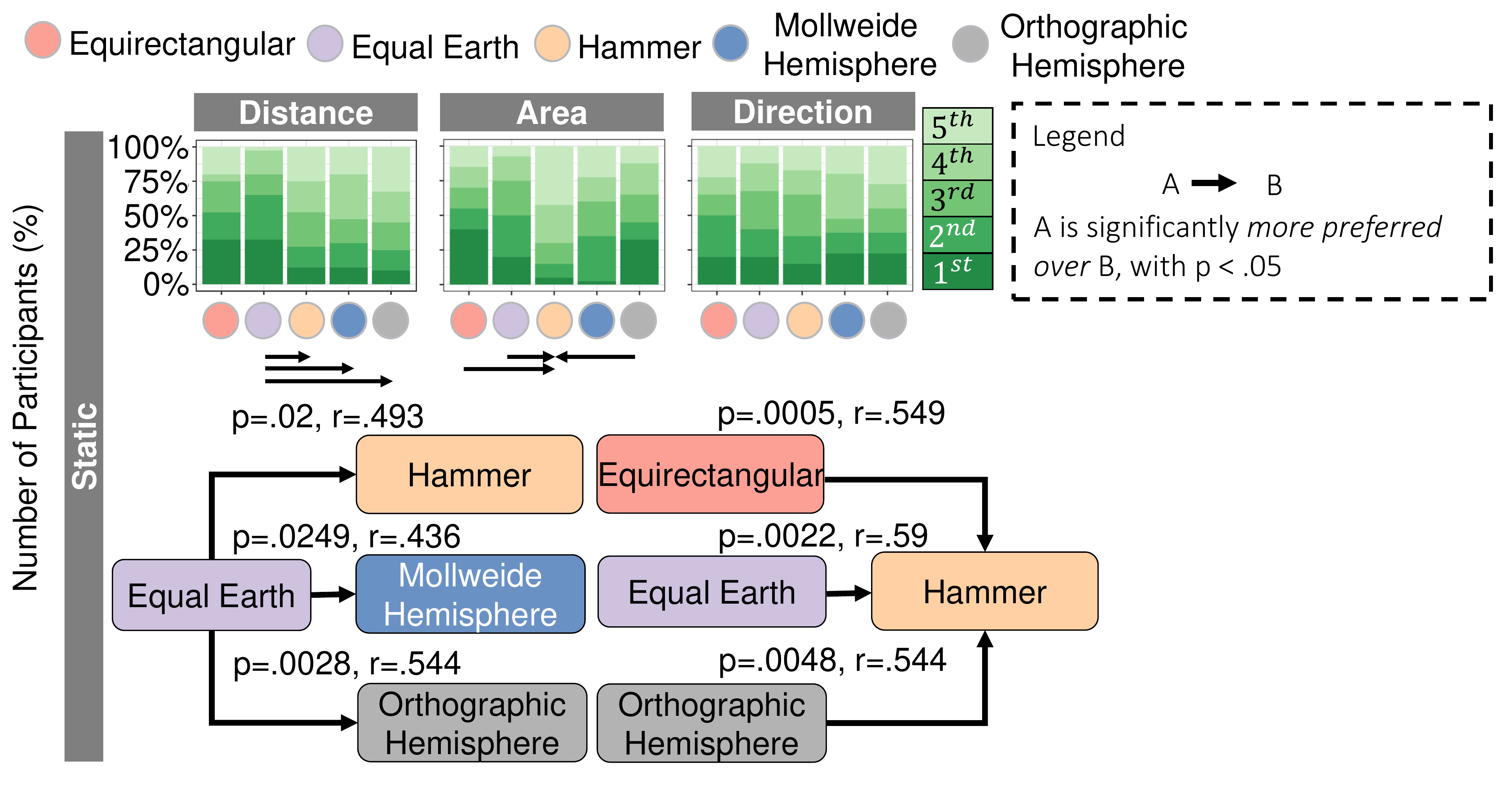}
    \caption{Subjective user rank of map projections within the \emph{static} group for three tested tasks. Lower rank indicates stronger preference. Arrows indicate statistical significance with $p<0.05$. Overall, equal earth was found to be preferred over hemispheric projections for the distance comparison task, while hammer was preferred over by other projections for area comparison task.}
    \label{fig:study-1-ranking-results}
\end{figure}

\begin{figure*}
    \centering
    \includegraphics[width=\textwidth]{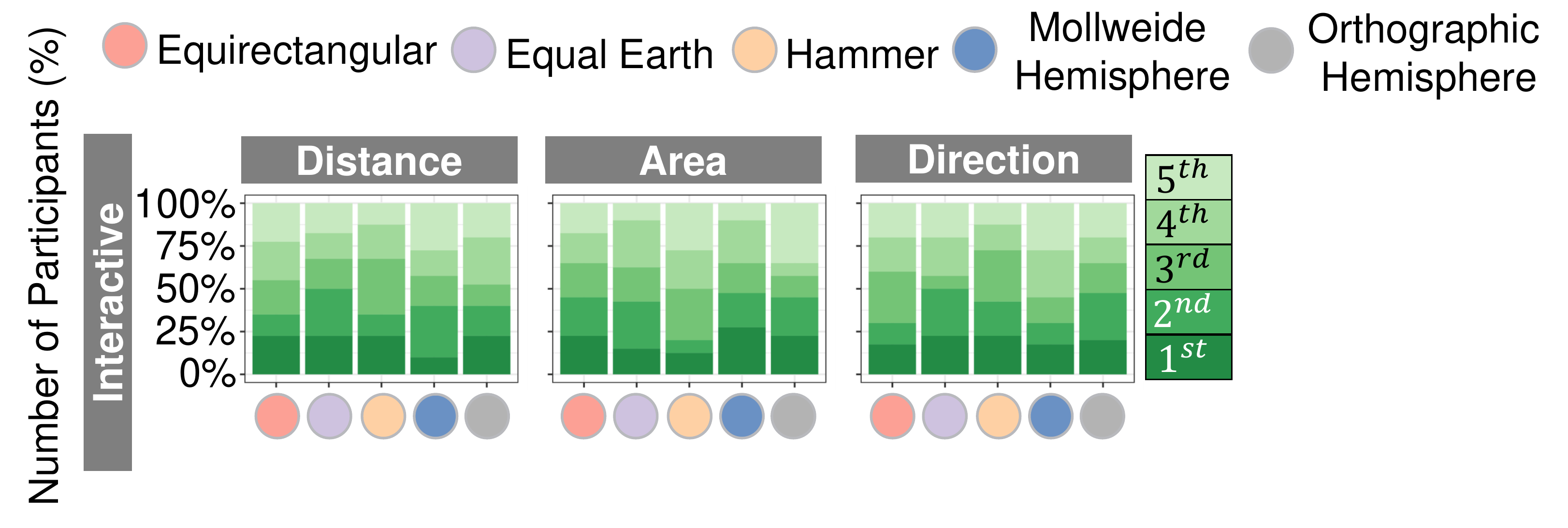}
    \caption{Study 5: Subjective user rank of map projections within the \emph{interactive} group for three tested tasks. Higher rank indicates stronger preference. Interaction makes the preference between \emph{interactive} projections similar (not statistically significant). Effect sizes can be found in \autoref{fig:study-1-ranking-results-effect-size}.}
    \label{fig:study-1-ranking-results-interactive}
\end{figure*}

We used \emph{sqrt}-transformation for \mtime{} to meet the normality assumption. We then used linear mixed modelling to evaluate the effect of independent variables on the dependent variables~\cite{Bates2015}.
We modelled all independent variables (five map projections, two interaction levels and two difficulty levels) and their interactions as fixed effects. 
We evaluated the significance of the inclusion of an independent variable or interaction terms using log-likelihood ratio. 
We then performed Tukey's HSD post-hoc tests for pairwise comparisons using the least square means~\cite{Lenth2016}. 
We used predicted vs. residual and Q---Q plots to graphically evaluate the homoscedasticity and normality of the Pearson residuals respectively.
For \merror{} and \mpref{}, as they did not meet the normality assumption, we used the \emph{Friedman} test to evaluate the effect of the independent variable, as well as a Wilcoxon-Nemenyi-McDonald-Thompson test for pairwise comparisons.
We also used the Wilcoxon signed-rank test for comparing the accuracy of static and interactive map projections. The confidence intervals are 95\% for all the statistical testing. We demonstrate the \emph{error rate} and \emph{time} in \autoref{fig:study-1-results}. We show interactivity and map projection rankings as stacked bar charts in \autoref{fig:study-1-ranking-results}, \autoref{fig:study-1-ranking-results-interactive}, and \autoref{fig:study-1-interactivity-ranking-results}.

Following existing work which reports statistical results with standardised effect sizes~\cite{okoe2018node,TransparentStatsJun2019, yoghourdjian2020scalability} or simple effect sizes with confidence intervals~\cite{besancon:hal-01436206, besanccon2017pressure}, we interpret the standardised effect size for a parametric test using Cohen's d classification, which is 0.2, 0.5, and 0.8 or greater for small, moderate, and large effects, respectively~\cite{cohen2013statistical}. For non-parametric tests, we interpret the standardised effect size for a Wilcoxon's signed-rank test using Cohen's r classification, which is 0.1, 0.3, and 0.5 or greater for small, moderate, and large effects, respectively~\cite{cohen2013statistical, pallant2013spss}.

\subsection{Results}
We report on the most significant findings for \tdistancecomparison{}, \tareacomparison{}, and \tdirectionestimation{} visually in \autoref{fig:study-1-results} and \autoref{fig:study-1-ranking-results}.

In the following, significance values are reported for $p < .05 (*)$, $p < .01 (**)$, and $p < .001 (***)$, respectively, abbreviated by the number of stars in parenthesis. 

\textbf{For \tdistancecomparison{}}, we found \finteraction{} ($***$) and \fmap{} $\times$ \fdifficulty{} ($***$) both had a significant effect on \mtime{}. For \finteraction{}, post hoc analysis shows that
\dstatic{} was faster than \dinteractive{} ($***$).
For \fmap{} $\times$ \fdifficulty{}, post hoc analysis shows that in \deasy{}, \tequirectangular{} and \tequalearth{} were faster than \torthographic{} ($**$) and \tmollweide{} ($***$). \thammer{} was also faster than \tmollweide{} ($***$). In \dhard{}, \tequalearth{} was faster than \torthographic{} ($*$) and \tmollweide{} ($***$).

We found \fmap{} had a significant effect on \emph{error rate} in \dhard{} ($*$). \torthographic{} was more accurate than \tequalearth{} ($*$) and \thammer{} ($*$). We also found \dinteractive{} was more accurate than \dstatic{} in all \fmap{} ($***$). 

In \dstatic{} \fmap{}, participants preferred \tequalearth{} over \thammer{} ($*$), \tmollweide{} ($*$), and \torthographic{} ($**$). 

\textbf{For \tareacomparison{}},  we found \finteraction{} ($***$), \fmap{} $\times$ \fdifficulty{} ($***$) and \fmap{} $\times$ \finteraction{} ($**$)  had a significant effect on \emph{time}. For \tareacomparison{}, \dstatic{} was faster than \dinteractive{} ($***$). For \fmap{} $\times$ \finteraction{} ($**$), \dinteractive{} \tequalearth{} was faster than \dinteractive{} \thammer{} ($**$).

We found \fmap{} had a significant effect on \merror{} in \dinteractive{} ($*$), \deasy{}  \dstatic{} ($*$) and \dhard{} \dinteractive{} ($*$). \tmollweide{} was more accurate than \tequirectangular{} ($*$). 
\dstatic{} \tequalearth{} ($*$) and \dstatic{}\tmollweide{} ($*$) were more accurate than \dstatic{} \tequirectangular{} in \deasy{}.
\dinteractive{} \tmollweide{} and \dinteractive{} \dinteractive{} tended to be more accurate than \dinteractive{} \tequirectangular{}, but not statistically significant (with $p=0.09, 0.07$ respectively).
We also found \dinteractive{} was more accurate than \dstatic{} in all \fmap{} (all $***$). 

In \dstatic{} \fmap{}, participants preferred \tequirectangular{} ($***$), \tequalearth{} ($**$) and \torthographic{} ($**$) than \thammer{}.

\textbf{For \tdirectionestimation{}},  we found \fmap{} ($***$), \finteraction{} ($***$) and \fmap{} $\times$ \finteraction{} ($**$) had a significant effect on \emph{time}. For \tdirectionestimation{}, \dstatic{} was faster than \dinteractive{} ($***$).
In \dstatic{} \fmap{}, \tequirectangular{} and \tequalearth{} were faster than \tmollweide{} and \torthographic{} (all $**$).
In \dinteractive{} \fmap{}, \tequalearth{} was faster than \tmollweide{} ($**$).

We found \fmap{} had a significant effect on \emph{error rate} in \dinteractive{}. \torthographic{} was more accurate than \tequirectangular{} ($*$). We also found \dinteractive{} was more accurate than \dstatic{} in all \fmap{} ($***$). 

We did not find a significant difference in preference for \dinteractive{} groups.

\section{Key Findings and Discussion}
\label{sec:mapstudydiscussion}

\textbf{Interaction improved error rate and is preferred} over static map projections while taking participants longer time to complete (across all tasks).
This main effect was found statistically significant with moderate and large effect sizes in all three tested tasks: \tdistancecomparison{} (\autoref{fig:study-1-effect-size-distance}), \tareacomparison{} (\autoref{fig:study-1-effect-size-area}), and \tdirectionestimation{} (\autoref{fig:study-1-effect-size-direction}). The corresponding error bar graphs are shown in \autoref{fig:study-1-results} and stacked bar chart of user preference of interactivity in \autoref{fig:study-1-interactivity-ranking-results}). For \tdistancecomparison{} and \tdirectionestimation{}, \dinteractive{} is always better (significantly less error and more preferred) than \dstatic{}. For \tareacomparison{}, \dinteractive{} has less error than the corresponding non-interactive projection but not necessarily all other non-interactive projections. \dinteractive{} was significantly preferred (moderate effects) for \thammer{} and \tequalearth{} but not necessarily other projections for \tareacomparison{}. However, \dinteractive{} was significantly preferred in the overall user rank over \dstatic{} (top horizontal bar in~\autoref{fig:study-1-interactivity-ranking-results}) for \tdistancecomparison{} (large effects), \tdirectionestimation{} (large effects) and \tareacomparison{} (moderate effects). \dinteractive{} is always slower (large effects) than \dstatic{} (\autoref{fig:study-1-results}-Time). Therefore, we accept M5.1, M5.2, M5.3. 

These results provide strong evidence that, with interactions, participants were able to find a better projection centre than the default one in a static map. Some participants explicitly mentioned the benefits of having interaction and their preference, e.g., \emph{``when moved [,the interactive conditions makes] it easier to judge when [the areas] were both placed in the middle in the least distorted part of the map.''} (P20, Area-\dinteractive{}), and \emph{``The fact I couldn't move the pictures was frustrating [for the static conditions] and I think I didn't get many of the guesses right.''} (P4, Direction-\dstatic{}).




\textbf{The choice of projections makes less difference and depends on tasks. However, overall, we found that equal earth and orthographi hemisphere performed well, while equirectangular may be a poor choice, organised in the following key findings.}


\textbf{In continuous projections, \tequalearth{} performed well in terms of \merror{} for \tareacomparison{}, \mtime{} for \tareacomparison{} and \tdirectionestimation{}, and \mpref{}-\dstatic{} for both \tdistancecomparison{} and \tareacomparison{}. \tequalearth{} was not significantly worse than any other continuous projections.} We found that for static projections, \tequalearth{} tended to be more accurate (moderate effects) than \tequirectangular{} for Area-Easy. With interaction, \tequalearth{} tended to be faster (small effects) than \tmollweide{} for \tdirectionestimation{}, and \thammer{} for Area-All (\autoref{fig:study-1-results}-Time). Though the time results are statistically significant, the small effect sizes seem to indicate that the choice of projection makes a slight difference~\cite{helske2021can,cockburn2020threats,TransparentStatsJun2019, schafer2019meaningfulness}. For \mpref{}-\dstatic{}, \tequalearth{} was significantly preferred over \thammer{} (moderate effect), \tmollweide{} (moderate effect), and \torthographic{} (large effect) for \tdistancecomparison{} (\autoref{fig:study-1-ranking-results}-Distance). Furthermore, there is a strong evidence with important effects for \mpref{}-\dstatic{} that \tequirectangular{}, \tequalearth{}, and \torthographic{} were significantly preferred over \thammer{} (large effects) for \tareacomparison{} (\autoref{fig:study-1-ranking-results}-Area). With interaction, there are no significant differences in user preference. This result is shown in~\autoref{fig:study-1-ranking-results-interactive}.

User preference of \dstatic{} \tequalearth{} partially confirms Šavrič et al.~\cite{avric2015user}, who found Robinson projection (which is similarl to \tequalearth) was preferred over interrupted projections such as \tmollweide{} and Goode Homolosine. This is also supported by our participants' feedback where continuous maps are preferred over interrupted ones for \dstatic{}. For example, the preference of \dstatic{} \tequalearth{} was supported by our participants' feedback, as they mentioned \emph{``I didn't feel comfortable when the hemispheres were split into two separate circle portions. I preferred [equal earth] when they were a continuous map surface.''} (P18, Distance-\dstatic{}), and \emph{``the map's edges [in equal earth projection] didn't feel as distorted as other maps such as equirectangular.''} (P23, Area-\dstatic{}.) We also identified many positive comments for \tequalearth{}, e.g., \emph{``[equal earth projection] gave me a better sense of global spatial reasoning''} (P34, Distance-\dstatic{}), and \emph{``the equal earth projection was the easiest to visualise as it seemed to have a good balance of not being too distorted without having the difficulty of wrap around visualisations that the hemisphere projections had.''} (P37, Distance-\dstatic{}).

Surprisingly, although \thammer{} is an equal-area projection, participants did not like it for area comparison in static maps (\autoref{fig:study-1-ranking-results}-Area), This result partially differs from existing studies~\cite{avric2015user} where poles represented as points were preferred over poles represented as lines. 
We conjecture this effect is because when the target area is at the edges, the shapes are severely distorted, which makes it difficult to accurately accumulate its represented area, as participants mentioned \emph{``Equirectangular only was distorted from top to bottom, while [hammer was] also distorted on the sides.''} (P31, Area-\dstatic{}). 

Finally, participants disliked \thammer{} for \dstatic{}-Area. For example, participants mentioned \emph{``I think [equirectangular] felt easier to read my first choice vs what I rate number 5 [for hammer].''} (P17, Area-\dstatic{}) and \emph{``the hammer too oval to make sense of.''} (P19, Direction-\dstatic{}).

\textbf{In hemispheric projections, interaction reduced \merror{} of \torthographic{} to a point that it tended to have a lower error rate than some interactive continuous projections for Distance-Hard and Direction, and not significantly slower than any interactive projections across all tasks.} We found \torthographic{} performed well. 
\torthographic{} benefited more from interaction (large effects) for \merror{} than \tmollweide{} (moderate effects) for \tdistancecomparison{}, while \torthographic{} benefited slightly more from interaction with similar effect sizes for \merror{} than \tmollweide{} for \tareacomparison{} and \tdirectionestimation{} (\autoref{fig:study-1-results}-\merror{}). The effect size can be found: \tdistancecomparison{} (\autoref{fig:study-1-effect-size-distance}), \tareacomparison{} (\autoref{fig:study-1-effect-size-area}), and \tdirectionestimation{} (\autoref{fig:study-1-effect-size-direction}). This is also supported by a strong evidence that \dinteractive{} \torthographic{} has a lower error rate (large effects) than \tequalearth{} and a lower error rate (moderate effects) than \thammer{} for Distance-Hard (\autoref{fig:study-1-results}-Error). By contrast, even with interaction, \tmollweide{} was still slower (small effects) than non-hemisphere for \tdirectionestimation{} (\autoref{fig:study-1-results}-Time). \dinteractive{} \torthographic{} did not have a significantly lower error rate for \tareacomparison{} than any interactive continuous projections, nor was there any significant difference in \mpref{}-\dinteractive{}  (\autoref{fig:study-1-ranking-results-interactive}). Therefore, we reject M5.4, M5.5.

It was surprising that \torthographic{} was comparable to other interactive projections for \tareacomparison{} since it is not an equal-area map projection. Meanwhile, the other non-equal-area map projection, i.e., \tequirectangular{} tended to produce more errors (moderate effects). We believe \torthographic{} were perceived as less distorted than the other projections due to the ``natural'' orthographic distortion, which is similar to viewing the sphere at infinite distance (e.g.\ as if through a telescope), echoed by our participants. For example, one quoted response of \tdirectionestimation{}: \emph{``Orthographic Hemisphere seemed the most accurate to me. The round shape was really useful. The Mollweide Hemisphere was tricky, I didn't like the way the two points had to be on the separate circles for it to be accurate sometimes.''} (P4, Direction-\dinteractive{}). 

Meanwhile, it seems many people were good at mentally connecting both hemispheres: \emph{``I found it easier to create comparisons that didn't have a lot of distortion with the option of having the two side-by-side globes.''} (P18, Direction-\dinteractive{}). 

Despite being hemispheric, \tmollweide{} was found to be less intuitive for some tasks by participants. We conjecture that there is a slight distortion near the edge of two circles which may make it confusing when centring the region of interest, as participants mentioned \emph{``Orthographic hemisphere felt the easiest to compare because it was like looking at a globe, but the strange way of moving the mollewide hemisphere made it confusing, that is why that is my last option.''} (P23, Area-\dinteractive{}), \emph{``The mollweide was a bit difficult to drag and place arrow.''} (P23, Direction-\dinteractive{}.). 

However, there were also participants who did not like the hemispheric projections due to the need to inspect two separated spheres and instead they preferred the continuous maps in the non-hemispheric group. For example, \emph{``I found it a lot easier to identify on the single maps rather than the double because sometimes they were spaced too far away''} (P11, Area-\dinteractive{}), \emph{``it was hard to picture how Mollweide and Orthographic Hemispheres connected.''} (P7, Distance-\dstatic{}), and \emph{``the map layout was easier to see because the targets were all on one map rather than two spheres.''} (P36, Distance-\dinteractive{}).



\textbf{Even with interaction, \tequirectangular{} still tended to perform poorly in terms of \merror{} for \tareacomparison{} and \tdirectionestimation{}}. Overall, for static projections, \tequirectangular{} tended to have a higher error rate (moderate effects) than \tequalearth{} and \tmollweide{} (\autoref{fig:study-1-results}-Error).
To our surprise, with the ability to rotate to centre the region of interest, \tequirectangular{} still tended to be outperformed for \merror{} by \tmollweide{} (moderate effect) for Area-All and by \torthographic{} (moderate effect) for \tdirectionestimation{} (\autoref{fig:study-1-results}-Error). This partially confirms Hennerdal et al.'s static map study where \dstatic{} \tequirectangular{} was found confusing when estimating the airplane route that wraps around~\cite{hennerdal2015beyond}. We conjecture that \tequirectangular{} features the most distortion of all tested projections due to the high level of stretching near the poles, supported by participants' feedback. For example, \dinteractive{} \tequirectangular{} was still found to have a greater distortion near the poles than other projections, as participants mentioned \emph{``The sphere is easiest to visualize in my head, I was thinking about just drawing a line on a ball or other spherical shape. The ones with the biggest distortment on the edges confused me most.''} (P13, Direction-\dinteractive{}) and \emph{``The Mollweide Hemisphere and Orthographic Hemisphere- easier to see a side by side comparison of the areas. The Hammer- less areas of distortion compared to the Equal Earth and Equirectangular. Equal Earth had too much distortion as did the Equirectangular which in my opinion distorted the most.''} (P21, Area-\dinteractive{}).

\section{Limitations and Future Work}
\label{sec:study-1-threats}


The statistically significant results with large effect sizes provide strong evidence that adding the spherical rotation interaction to static maps improves the accuracy, is strongly preferred, but at the cost of longer completion time across all tasks. However, despite being statistically significant, the differences between projections within static or interactive groups are of small sizes for \mtime{}, medium-sized for \merror{} and large-sized for \mpref{}-Static-Area (\autoref{fig:study-1-results}, \autoref{fig:study-1-ranking-results}). Although medium-sized differences in \merror{} are likely to be noticeable in practical applications, these results only allow us to say \tequalearth{} and \torthographic{} performed well for some tasks~\cite{helske2021can,cockburn2020threats,TransparentStatsJun2019, schafer2019meaningfulness}.

Although the results show statistically significant differences between the selected hemispheric and non-hemispheric projections across all tasks for \merror{}, the results do not allow us to say that hemispheric projections are always more accurate than non-hemispheric projections. Arguably, the poor performance of non-hemispheric projections for \tareacomparison{} and \tdirectionestimation{} is entirely based on \tequirectangular{} compared with either \tmollweide{} or \torthographic{}. If the results from the \tequirectangular{} were not considered, the hemispheric and non-hemispheric projections seem to have very similar error rates for \tareacomparison{} and \tdirectionestimation{}. Similarly, the results do not allow us to say that non-hemispheric projections are always faster, as this seems only based on \tmollweide{} being significantly slower (small effects) than some non-hemispheric projections for \dstatic{} and \dinteractive{} for the direction tasks.

Surprisingly, unlike the study by Yang et al.~\cite{yang2018maps}, we did not identify superior performance of \torthographic{} compared to other map projections.
We conjectured that rendering \torthographic{} on a 2D flat display produced different effectiveness in perception and interaction compared to Yang et al's 3D globes in VR.
Meanwhile, although not in all tasks, \torthographic{} demonstrated some advantages for \tdistancecomparison{}, \tdirectionestimation{} and benefited more from interaction than \tmollweide{} for \merror{} (~\autoref{sec:mapstudydiscussion}).

\rev{We also note that in our study design, we experimented with two levels of difficulties for \tdistancecomparison{} and \tareacomparison{}, where the sole parameter we vary is the geographic distance between point pairs or pairs of areas. This allows us to evaluate map projection readability with easy as well as more challenging trials, as we believe increasing the geographic distance between two entities will make the task more challenging. However, there are other parameters that could also affect performance that we did not test, such as the geographic distance between individual pair of data points for \tdistancecomparison{}, size of geographic areas for \tareacomparison{}, and the angular difference between geographic points for \tdirectionestimation{}. We leave testing other geographic comprehension tasks and the other parameters to future work.}

While this chapter exclusively focuses on interactive panning, another possible wrapping approach, as presented in~\autoref{sec:designspace:sphere} is tiled-display, which we detail in~\autoref{sec:conclusion:others}.

\section{Conclusion}
\label{sec:spheremaps:conclusion}
The overwhelming finding in the map projection study (\autoref{sec:mapstudydiscussion}) confirms the benefits of interactive spherical wrapping for geographic comprehension tasks, addressing \textbf{RG5} (\autoref{sec:intro:RGs:sphere}).


The finding that interactive spherical wrapping of map projections led to significantly lower error rate and more preferred over standard map projections across all tasks tested suggests that such interaction should be routinely provided in online world maps, for example in education or in reporting of world events, climate patterns, and so on.  While less clear, our results indicated that \tequalearth{} was the best performing continuous projection, while the straight-forward \torthographic{} was an effective hemispheric projection. Our results also indicate that even with interaction, \tequirectangular{} may be a poor choice for comparing areas and estimating directions.

%
\chapter{Comparing Spherical and Toroidal Wrapping for Network Structured Data}
\label{sec:spherevstorus}

\cleanchapterquote{Torus is the easiest to interpret because the clusters do not get distorted if the map is moved around. Equal Earth is actually easier than it looks to interpret because even though the clusters get distorted, if they are moved around, they actually bunch up together once they are put near the middle of the map.}{Anonymised user study participant, P12}{Cluster identification tasks}

Spherical projection has application beyond geographic data.  
There can be advantages to laying out abstract data (data which has no inherent geometry) on the surface of a sphere such that there is no arbitrary edge to the display or privileged centre~\cite{rodighiero2020drawing}.  
Providing interactive panning has previously been found to be of benefit in understanding network layouts based on torus projection (\autoref{sec:torus2}).

We address \textbf{RG5}. Inspired by the promising results of spherical wrapping on map projections, in this chapter we investigate the utility of the most successful interactive spherical projections from~\autoref{sec:spheremaps}, to compare sphere, torus, and non-wrapped flat layout. We investigate network layouts on the surface of a sphere projected onto a 2D plane. Again we allow the user to interactively pan the projection  around the plane. 

\begin{figure}
\centering
	\includegraphics[width=\textwidth,trim=3cm 0cm 3cm 0cm]{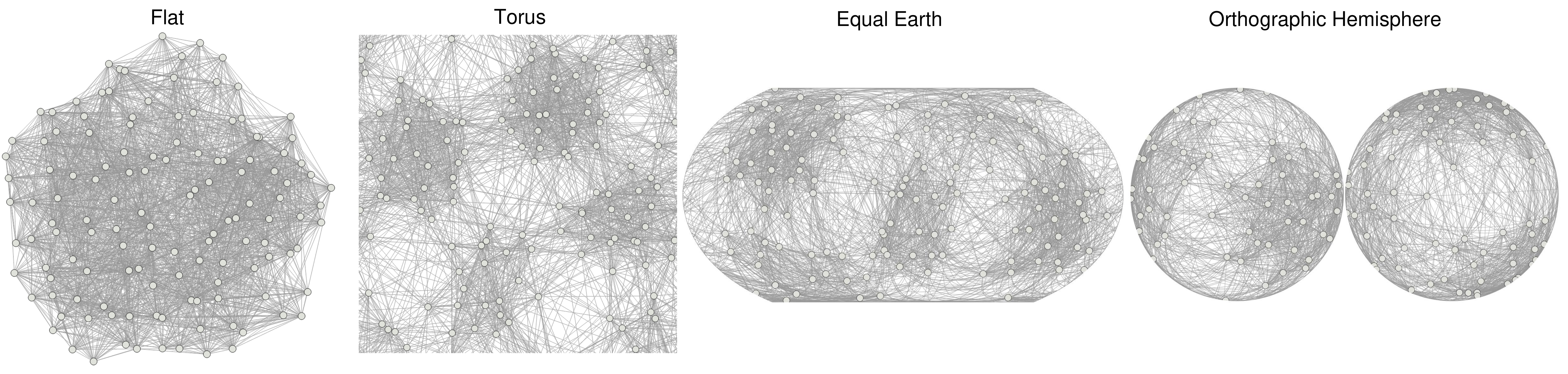}
	\caption{Study 6 compares the two interactive spherical projections found most effective in Study 5, and compares them to toroidal and standard `flat' layout for network data.}
  	\label{fig:sphervstorus:teaser}
\end{figure}


While the relationship between readability of geographic and network data on spherical projections has not been studied directly, there are commonalities in the analysis tasks that might be applicable for each. For example, understanding network clusters may involve comparing the relative size of their boundaries, similar to map area and shape comparison. Network path following tasks require the user to trace links (AKA `edges') between nodes in the network while maps also require understanding how regions connect, and in both cases splits or distortion due to spherical projection may present a challenge.

As we discuss in detail in \autoref{sec:related}, while there has been much work to develop the algorithms for spherical projection of maps and algorithms for layout of networks on a 3D sphere, we find that interactive 2D displays of such spherical layouts of data (whether geographic or network) are not well studied. Also projections of 3D spherical layouts of network data have not been compared to projections of networks arranged on the surface of other 3D geometries, in particular toroidal layouts which have been shown in~\autoref{sec:torus2}.


\section{Spherical Network Layout}
\label{sec:layout}
In Study 5, we found that interaction (panning by spherical rotation) makes spherical geographic projections overwhelmingly more accurate for distance, area and direction tasks.  A question, however, is whether such 2D interactive spherical projections are also useful for abstract data.  As discussed in \autoref{sec:related:sphere}, there have been a number of systems using immersive environments to visualise network data on 3D spherical surfaces or straightforward perspective projections of spheres.  Various advantages have been claimed to the opportunities for embedding a network layout in the surface of a sphere---without boundary---such as centring any node of interest in the layout and layout of non-privileged centre~\cite{brath2012sphere,perry2020drawing, rodighiero2020drawing}

Further, there are obvious disadvantages to projection, since all projections introduce some degree of distortion and discontinuity.  There are therefore three questions, linked to \textbf{RG5} (\autoref{sec:intro:RGs:sphere}: 

\noindent \textbf{RQ5.1:} \textit{Which of the most promising projections from  study 5 are the best for visualising the layout of a node-link diagram}?


\noindent \textbf{RQ5.2:} \textit{Does a spherical projection have advantages in supporting network understanding tasks compared to conventional 2D layouts?}

\noindent \textbf{RQ5.3:} \textit{Does a spherical projection provide perceptual benefits compared
with arrangements on other 3D topologies, such as a torus?}


Before we can answer these questions, we need techniques to create effective layouts of complex network data on a spherical surface and to orient the projections optimally in 2D.

\subsection{Plane, Spherical and Toroidal Stress Minimisation}
\label{sec:layoutalgorithms}
The tasks we investigate are cluster understanding tasks and path following.
Network clusters are loosely defined as subsets of nodes within the graph that are more highly connected to each other than would be expected from chance connection of edges within a graph of that density.  More formally, a clustered graph has disjoint sets of nodes with positive \textit{modularity}, a metric due to Newman which directly measures the connectivity of given clusters compared to overall connectivity \cite{newman2006modularity}.  To support cluster understanding tasks we need a layout method which provides good separation between these clusters.

To support path following tasks, we need a layout method which spreads the network out relatively uniformly according to connectivity. This will help minimise crossings between edges.

We follow previous chapters (\autoref{sec:torus1} and \autoref{sec:torus2}) which adopted a stress minimising approach.  Stress-minimisation is a commonly used variant of a general-purpose force-directed layout and does a reasonable job of satisfying both of these readability criteria~\cite{huang2009measuring,purchase2002metrics}. The \textit{stress} metric ($\sigma$) for a given layout of a graph with $n$ nodes in a 2D plane is defined (following Gansner et al.\ \cite{gansner2004graph}) as:

$$
    \mathit{\sigma}_\mathrm{plane} = \sum_{i=1}^{n-1} \sum_{j=i+1}^n w_{ij} (\delta_{ij} - d_{ij})^2~~~~,~~~~~~~~~d_{ij} = |x_i - x_j|
$$

\noindent where: $\delta_{ij}$ is the ideal separation between the 2D positions ($x_i$ and $x_j$) of a pair of nodes $(i,j)$ taken as the all-pairs shortest path length between them; $d_{ij}$ is the actual distance between nodes $i$ and $j$ (in the plane this is Euclidean distance); and $w_{ij}$ is a weighting which is applied to trade-off between the importance of short and long ideal distances, we follow the standard choice of $w_{ij} = 1/d_{ij}^2$.

We follow previous recent work by Perry et al.~\cite{perry2020drawing}, in adapting stress-based graph layout to a spherical surface by redefining $d_{ij}$ to arc-length on the sphere surface, or (assuming a unit sphere):

$$
 \mathit{\sigma}_\mathrm{sphere} = \sum_{i=1}^{n-1} \sum_{j=i+1}^n w_{ij} (\delta_{ij} - d_{ij})^2~~~~,~~~~~~~~~d_{ij} = \mathrm{arccos} ( x_i \cdot x_j )
$$

\noindent where $x_i$ and $x_j$ are the 3D vector offsets of nodes $i$ and $j$ respectively from the sphere centroid and $(\cdot)$ is the inner product.  For the layout to be reasonable, the ideal lengths $\delta$ must be chosen such that the largest corresponds to the largest separation possible on the unit sphere surface, which is $\pi$.  Thus, we set the ideal length of all edges to $\pi/\mathit{graph diameter}$.

The other layout against which we compare is a projection of a 3D torus, which, as discussed in \autoref{sec:torus2}, has been shown to provide better separation between clusters than a flat (conventional) 2D layout.  We use the same layout method which is also based on stress in the 2D plane but which, for each node pair, requires selecting the stress term from the set $A$ of 9 possible torus adjacencies for that pair which contributes the least to the overall stress (as shown in~\autoref{fig:threebythreetile}):

$$
 \mathit{\sigma}_\mathrm{torus} = \sum_{i=1}^{n-1} \sum_{j=i+1}^n w_{ij} \mathrm{arg~min}_{\alpha\in A} (\delta_{ij} - d_{ij\alpha})^2 , d_{ij\alpha} = |x_i - x_{j\alpha}|
$$

While Perry et al.\ follow the multi-dimensional scaling literature in using a \textit{majorization} method to minimise $\sigma_\mathrm{sphere}$, we follow \autoref{sec:torus2} in using the stochastic pairwise gradient descent approach developed by Zheng et al.\ \cite{zheng2018graph} which can be adapted straightforwardly and effectively to obtain layout for all of $\sigma_\mathrm{plane},\sigma_\mathrm{sphere},\sigma_\mathrm{torus}$.

Note that the layouts which result from minimising $\sigma_\mathrm{plane}$ and $\sigma_\mathrm{torus}$ are already in 2D.  For the spherical layout obtained by minimising $\sigma_\mathrm{sphere}$, we use either \torthographic{} or \tequalearth{} projections to generate the stimuli for our study. The detailed pseudocode of our algorithms are available in \url{https://github.com/Kun-Ting/gansdawrap}.

\subsection{Auto-pan Algorithms}
\label{sec:autopan}

\begin{figure*}[t]
    \centering
     \subfigure[\torthographic{} - No pan]{
    \includegraphics[width=0.4\textwidth]{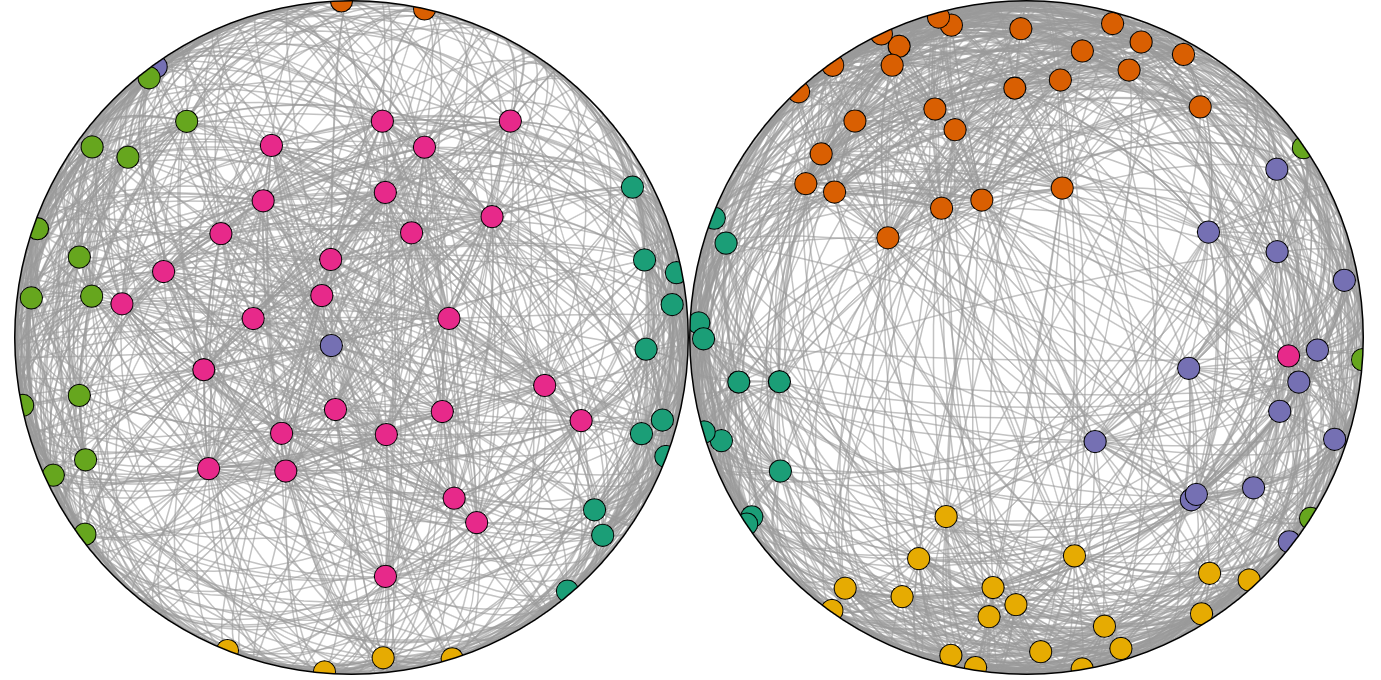}
    }
    \subfigure[\torthographic{} - Best pan]{
    \includegraphics[width=0.4\textwidth]{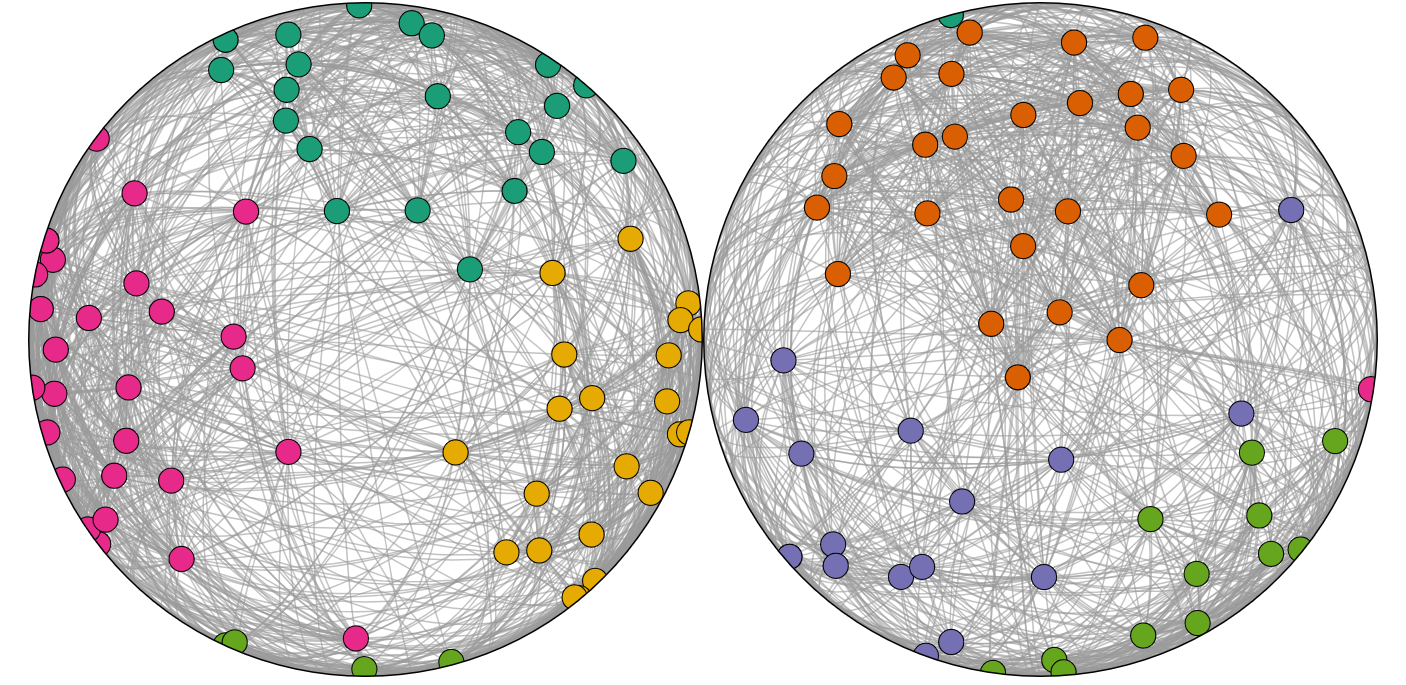}
    }
    \subfigure[\tequalearth{} - No pan (left), edge pixel mask (inset)]{
    \includegraphics[width=0.3\textwidth]{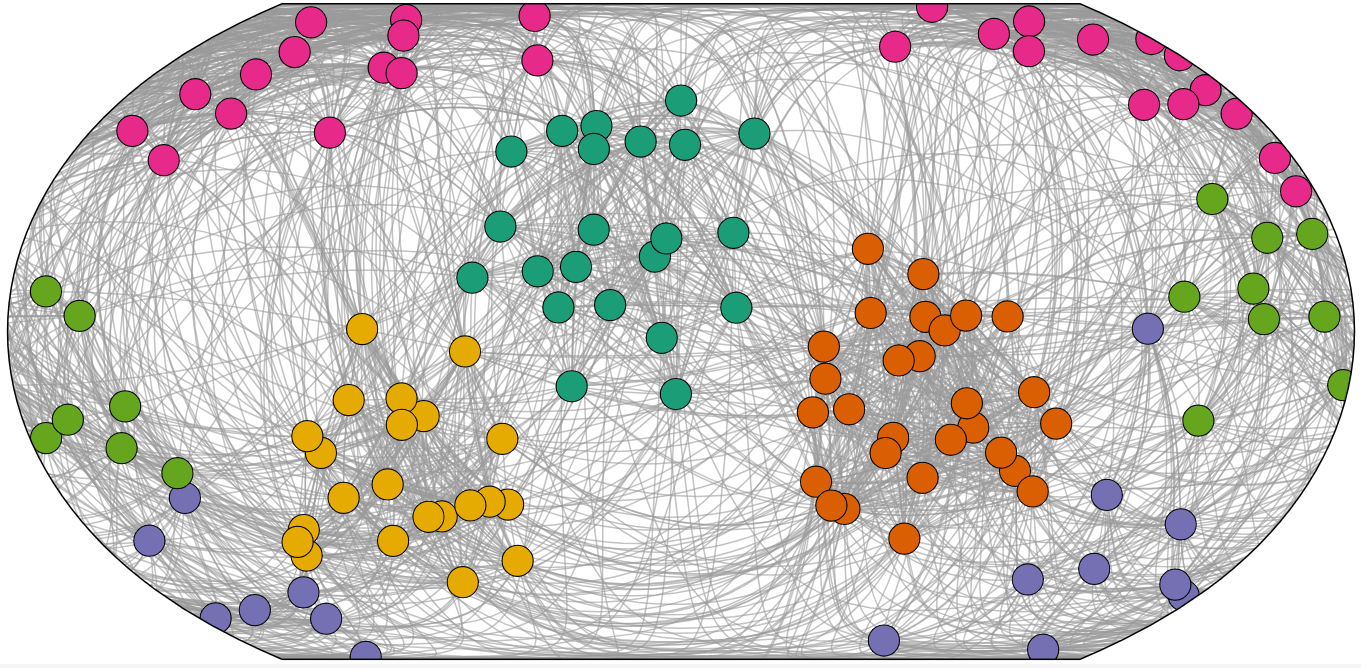}
    \includegraphics[width=0.15\textwidth,trim=1cm 0cm -1cm 0cm ]{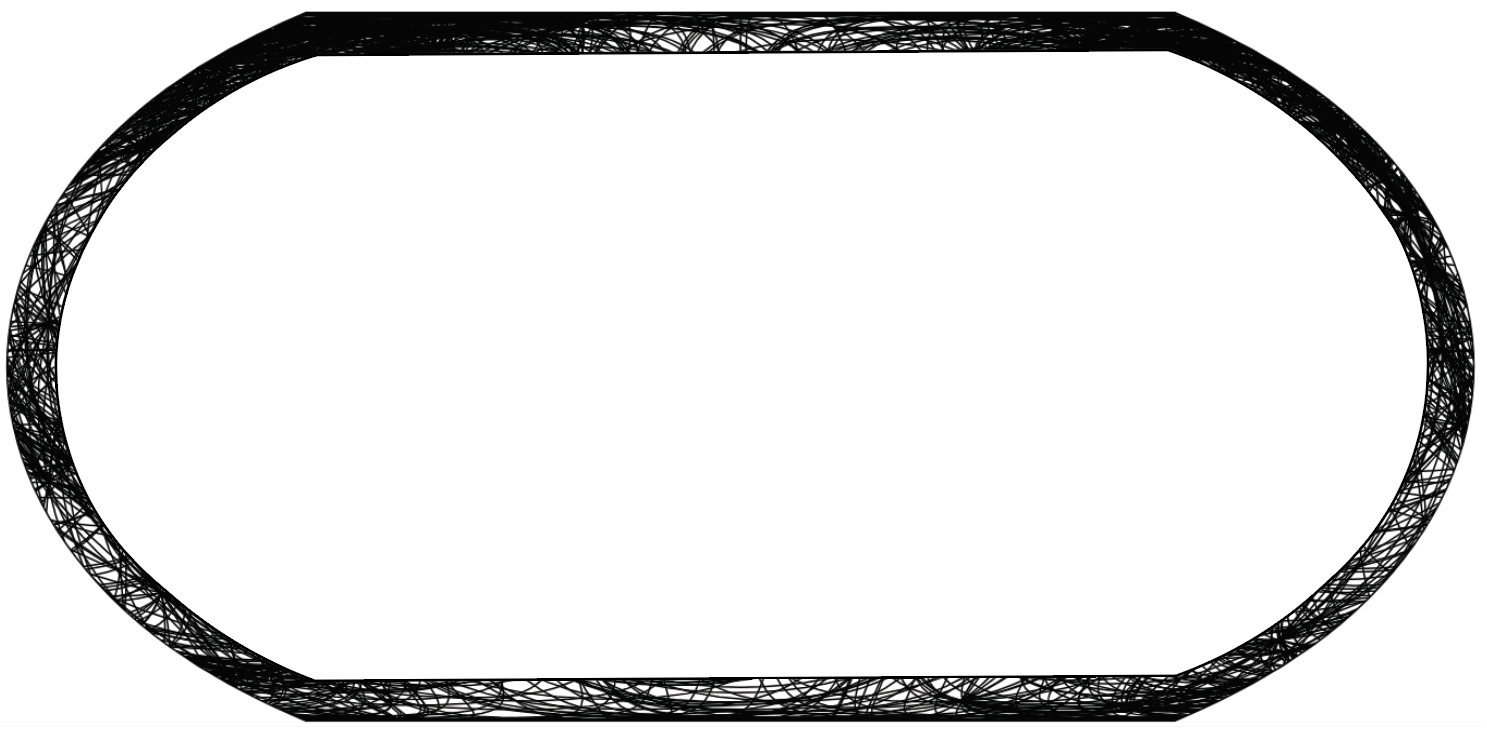}
    }
    \subfigure[\tequalearth{} - Best pan (left), edge pixel mask (inset)]{
    \includegraphics[width=0.3\textwidth]{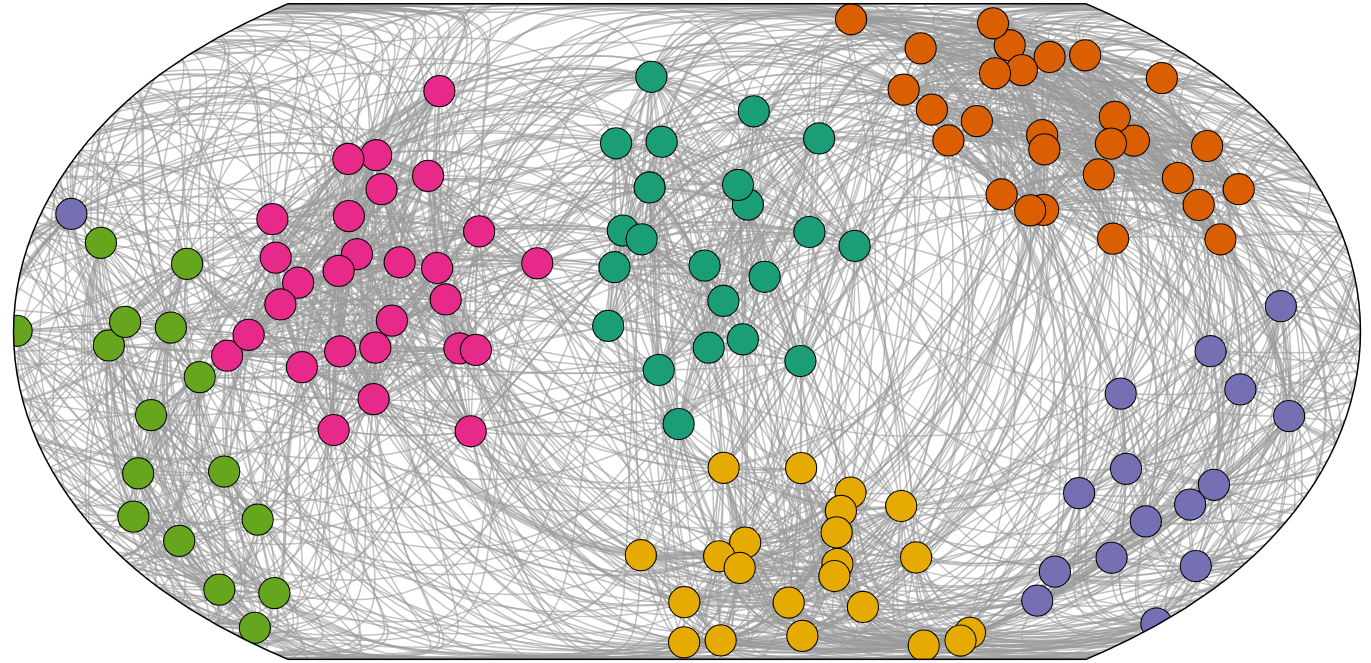}
    \includegraphics[width=0.15\textwidth,trim=1cm 0cm -1cm 0cm]{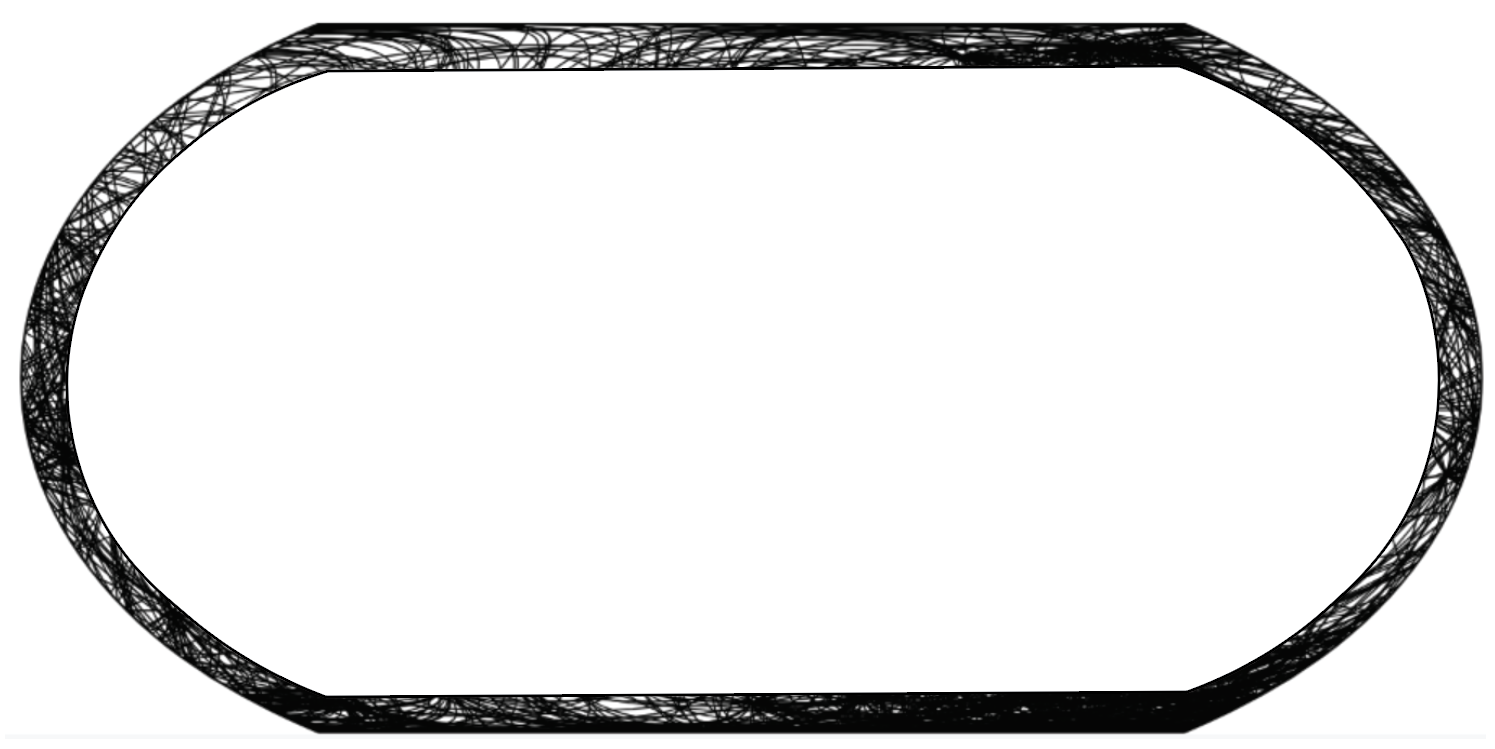}
    }
    \caption[]{
    Before and after demonstration of our auto-pan algorithms of a graph with six clusters differentiated by colour (note: study graphs did not have colour).
    Without auto-pan, clusters can be split across the hemispheres in \torthographic{} (a) or at the boundaries in \tequalearth{} (c).  Auto-pan reduces the number of wrapped edges and thereby brings the clusters together (b) and (d).
    }
    \label{fig:autopan}
    \vspace{-1em}
\end{figure*}

For toroidal network layout, \autoref{sec:torus2:autopanalgorithm} introduced an algorithm to automatically pan the toroidal layout horizontally and vertically to minimise the number of edges which wrap around at the boundaries.  Spherical projections can also suffer when too many edges are split across the boundaries.  Furthermore, edges are more distorted in spherical projections when they are near the edges.  Therefore, for fair comparison with toroidal layouts, it was necessary to find a method to auto-rotate the sphere to reduce the numbers of such edges.  However, while the toroidal auto-panning algorithm can be done with horizontal and vertical scans (linear time in the number of edges), the spherical layout does not permit such a trivial search algorithm.  We therefore develop heuristics to perform auto-rotate for the spherical projections.  For both, we choose a simple stochastic method of randomly selecting a large number (e.g., 1000 iterations) of three-axis spherical rotation angle triples $(\lambda,\phi,\gamma)$ and choosing the triple for which edges crossing (or near) boundaries is minimised.

For \torthographic{} projection this crossing number is trivial to count precisely.  Simply, for all pairs of nodes if the nodes are not on the same face, then they must cross a boundary.  A suboptimal \torthographic{} projection rotation, and the result of autorotation to minimise this crossing count is shown in \autoref{fig:autopan}(a) and \autoref{fig:autopan}(b), respectively.

For \tequalearth{} projection the edges are curved and so determining those that cross the boundary for a given geo-rotation is more complex.  Further, in this projection edges near the periphery are significantly more distorted than those near the centre, so even visually determining if an edge that comes close to the boundary continues on the same side of the projection or wraps around to the other side is not easy.  Thus, instead of counting boundary crossings, we penalise all edges which come close to the boundaries.  To compute the penalty we analyse the periphery of a monochrome bitmap of the projected edge paths.  The penalty is then simply the number of black pixels.  Example masked bitmaps are shown for sub-optimal and more-optimal rotations of an \tequalearth{} projection in \autoref{fig:autopan}(c) and \autoref{fig:autopan}(d), respectively.

\subsection{Automatic Panning Results}
We conducted a small empirical analysis of the Auto-Pan algorithm to assess the numbers of links wrapped for \torthographic{} and the number of pixels (higher values indicate less wrappings) at the boundary of \tequalearth{} projections compared to 10 random rotations. Across 10 study graphs (\dsmalleasy{}, \dsmallhard{}) used in our cluster understanding tasks, we found for \torthographic{}, the mean crossing count for random rotations was 262.16. With automatic panning, this number was improved by $25.6\%$ and was reduced to 208.7.  For \tequalearth{}, our automatic panning increased the number of pixels at the boundary region by $12.1\%$ (Section A.2: Table 2).  We found these auto-pan algorithms resulted in a noticeable improvement in keeping clusters from being separated, as evidenced in \autoref{fig:autopan}.


\section{User Study 6: Spherical Network Projection Readability}
\label{sec:networkstudy}

In Study 5, we found \tequalearth{}, had advantages in terms of error rate, time, and subjective user feedback. \torthographic{}, despite being an interrupted projection, benefited more from interaction than \tmollweide{} and performed well in terms of error rate for distance and direction tasks (\autoref{sec:mapstudydiscussion}).
Based on these findings, we chose interactive \tequalearth{} and \torthographic{}, to understand their performance in visualising networks. 
We compare these spherical projections to standard flat graph layout (\tnodelink{}) and a projection of a \ttorus{} geometry (\autoref{sec:torus2}).  
Our study was run in a similar way to Study 5 and included a new set of 96 participants through the Prolific platform.


\subsection{Techniques \& Setup}
The techniques in our study are \tnodelink{}, \ttorus{}, \tequalearth{}, and \torthographic. Layouts are computed as described in \autoref{sec:layout}.
All the techniques support interactive panning except for \tnodelink{} which does not wrap around.
The area of the rectangular bounding box of each technique condition is the same. For \tnodelink{} and \ttorus{}, the resolution is fixed at $650 \times 650$ pixels, and for \tequalearth{} and \torthographic{}, the resolution is fixed at $900 \times 317$ pixels.
For \clusteridentification{}, we did not color clusters to not reveal any graph structure.

\subsection{Tasks}
\begin{figure*}
    \centering
    \includegraphics[width=\textwidth]{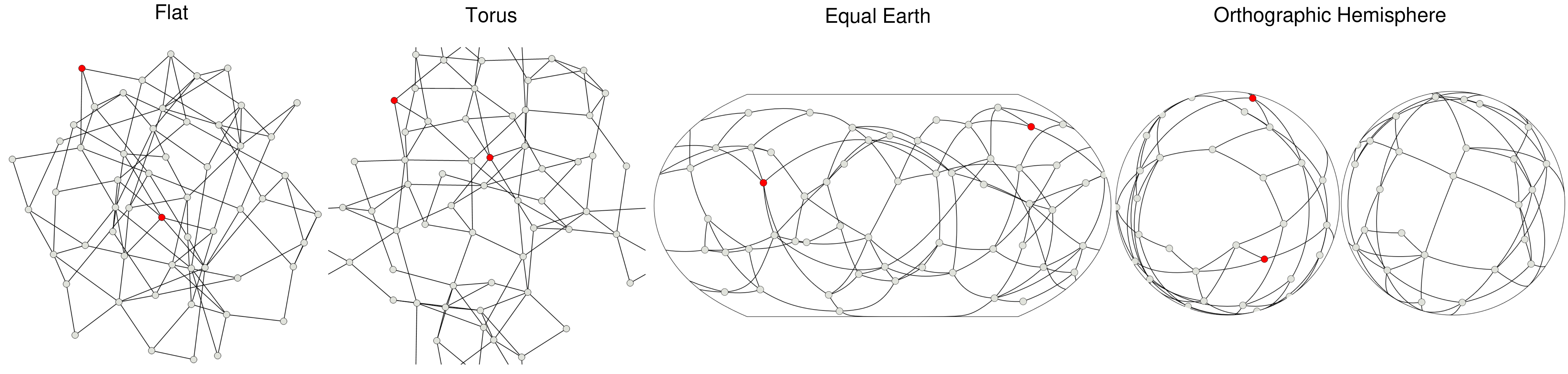}
    \caption[]{Example of one study graph laid out using 4 techniques described in \autoref{sec:layoutalgorithms} for \tshortestpathnumber{}. The shortest path length is 3 in this example. Participants were provided with interactive panning for spherical and toroidal layouts to explore the network.}
    \label{fig:graphtasks}
\end{figure*}
Our study is intended to be complementary to the studies described in~\autoref{sec:torus:study2} and~\autoref{sec:torus2:userstudy} by evaluating a new geometry sphere with important tasks considered in those studies. Inspired by those studies, we selected two representative network visualisation tasks comparing 2D layout of 3D surface topology.

\noindent \textbf{\clusteridentification:} \textit{Please count the number of clusters in this graph.} Participants answered through radio buttons: choices ranged from 1 to 10. We created a quality control trial for each condition with two clusters with clearly marked boundaries to assess participants' attention.  Participants who did not answer these trials correctly were excluded. An example of 5 clusters from \dlargeeasy{} graphs (\autoref{sec:networkdatasets}) using 4 different layouts is shown in \autoref{fig:sphervstorus:teaser}-Bottom.

\noindent \textbf{\shortestpathnumber:} \textit{What is the shortest path length between the red nodes?}: Participants were required to count the smallest number of links between two red nodes. Radio buttons allowed them to answer between 1 to 6. Again, we created a quality control trial for each layout condition, with shortest path length two and links on the path highlighted in red. An example of \deasy{} graphs  (\autoref{sec:networkdatasets}) using 4 different layouts is shown in \autoref{fig:graphtasks}.

\subsection{Data Sets}
\label{sec:networkdatasets}
We prepared a separate graph corpus for each task, full details and stimuli are presented in our supplementary material. For \clusteridentification, we use graphs from~\autoref{sec:torus2:graphcorpus}, generated using algorithms designed to simulate real-world community structures in graphs~\cite{brandes2003experiments,fortunato2010community}. 
Graphs are grouped by two variables: difficulty in terms of graph modularity~\cite{newman2006modularity} (\deasy: modularity=0.4, \dhard: modularity=0.3) and size (2 levels: \dsmall{}: 68-80 nodes, 710-925 links, and \dlarge{}: 126-134 nodes, 2310-2590 links). The number of clusters is between 4 and 7. 
%
For \shortestpathnumber{} the clustered graphs were too dense, so we generated sparser graphs using scale-free models~\cite{barabasi1999emergence,watts1998collective}.
We chose graphs with two levels of density (\deasy{}: 0.075, and \dhard{}: 0.11) with 50 to 57 nodes.
The shortest path length varied between 1 and 4.

\subsection{Hypotheses}

Hypotheses were pre-registered with the Open Science Foundation:~\url{https://osf.io/equhp}.

\noindent \textbf{G5.1}: \textit{\tequalearth{} and \ttorus{} have better task effectiveness for \clusteridentification{} (in terms of time and error) than \torthographic} (RQ5.1, RQ5.3) \textit{or \tnodelink{}} (RQ5.2). While \torthographic{} performed well in Study 5, our pilot studies for \clusteridentification{} revealed that cuts and distortion of clusters at the borders made them hard to count. For (RQ5.2), the inspiration was based on our prior cluster readability studies in~\autoref{sec:torus2:userstudy}.

\noindent\textbf{G5.2}: \textit{Participants will prefer \tequalearth{} and \ttorus{} to \torthographic{}} (RQ5.1, RQ5.3) \textit{or \tnodelink{}} (RQ5.2) \textit{for \clusteridentification.} Our early pilots indicated this preference --- perhaps for the same reasons as above and inspiration from our prior studies~\autoref{sec:torus2:userstudy}.
    
\noindent\textbf{G5.3}: \textit{\ttorus{} has better task effectiveness (in terms of time and error)} for \shortestpathnumber{} than \tnodelink{} \tequalearth{} or \torthographic{} (RQ5.3). This assumption is based on pilot studies and prior studies indicating curved links might hamper path tracing tasks~\cite{du2017isphere, xu2012user}.
    
\noindent\textbf{G5.4}: \textit{Participants will prefer \ttorus{} to \tnodelink{}, \tequalearth{} or \torthographic{}} (RQ5.3) \textit{for \shortestpathnumber{}.} Again, it was assumed due to distortion of links.

\subsection{Experimental Design}

We use a within-subjects design for each task with 4 techniques. Each participant was randomly assigned one of the tasks by the experimental software. For \clusteridentification, we used 2 levels of difficulty (Easy, Hard) $\times$ 2 sizes (Small, Large) $\times$ 5 repetitions. We randomly inserted one additional quality control trial to each layout condition. Each recorded trial had a timeout of 20 seconds to prevent participants from trying to perform precise link counting. This leaves us with a total of 80 recorded trials per participant. We counterbalanced the order of the techniques using a full-factorial design. The order of each level of difficulty and size in each technique was the same: \dsmalleasy{}$\rightarrow$\dlargeeasy{}$\rightarrow$ \dsmallhard{}$\rightarrow$\dlargehard{}. The order of trials for each technique within each level was randomised. 

For \shortestpathnumber, we used 2 levels of difficulty (Easy, Hard) $\times$ 8 repetitions. There were 2 repetitions of each shortest path length per level. One additional quality control trial was added for each layout condition. This leaves us with a total of 64 trials per participant. Each recorded trial had a timeout of 30 seconds, informed by pilot studies. The order of each level of difficulty in each technique was the same: \deasy{}$\rightarrow$ \dhard{}. The order of trials for each technique within each level was randomised. 

\subsection{Participants and Procedures}

Setup and inclusion criteria for participants were the same as for Study 5. 
We recorded 96 participants (37 female, 59 male, age range [18,50]) who passed the attention check trials, completed the training and recorded trials. This comprised 2 fully counterbalanced blocks of participants (24 $\times$ 2 $\times$ 2 tasks). They ranked their familiarity with network diagrams as: 9 often seeing network diagrams; 62 occasionally; and 25 never. 

For \clusteridentification{}, each participant was presented with a tutorial explaining the concept of network clusters. For \ttorus \tequalearth, and \torthographic, animated videos were given to demonstrate the interactive panning or rotation. Each participant then completed training trials,  similar to the recorded trials. 
For \shortestpathnumber, each participant was presented with a tutorial explaining the concept. For \ttorus, \tequalearth, \torthographic, animated videos were given to demonstrate the interactive panning or rotation. 
Training trials were similar to the recorded trials. 

Specific instructions given for each task is available in the supplementary material\footnote{An online demonstration of the study is available: \url{https://observablehq.com/@kun-ting/gansdawrap}}.

\subsection{Dependent Variable and Statistical Analysis Methods}

We recorded task-completion time (\mtime), task-error (\merror), and subjective \textit{confidence} as \mpref{} (the smaller the more confident). We calculated \merror{} as the normalised absolute difference between the correct answer and response. We used the same statistical analysis methods and standardised effect sizes from the first study (\autoref{sec:study-1-stats}).

\begin{figure*}
    \centering
    \includegraphics[width=\linewidth]{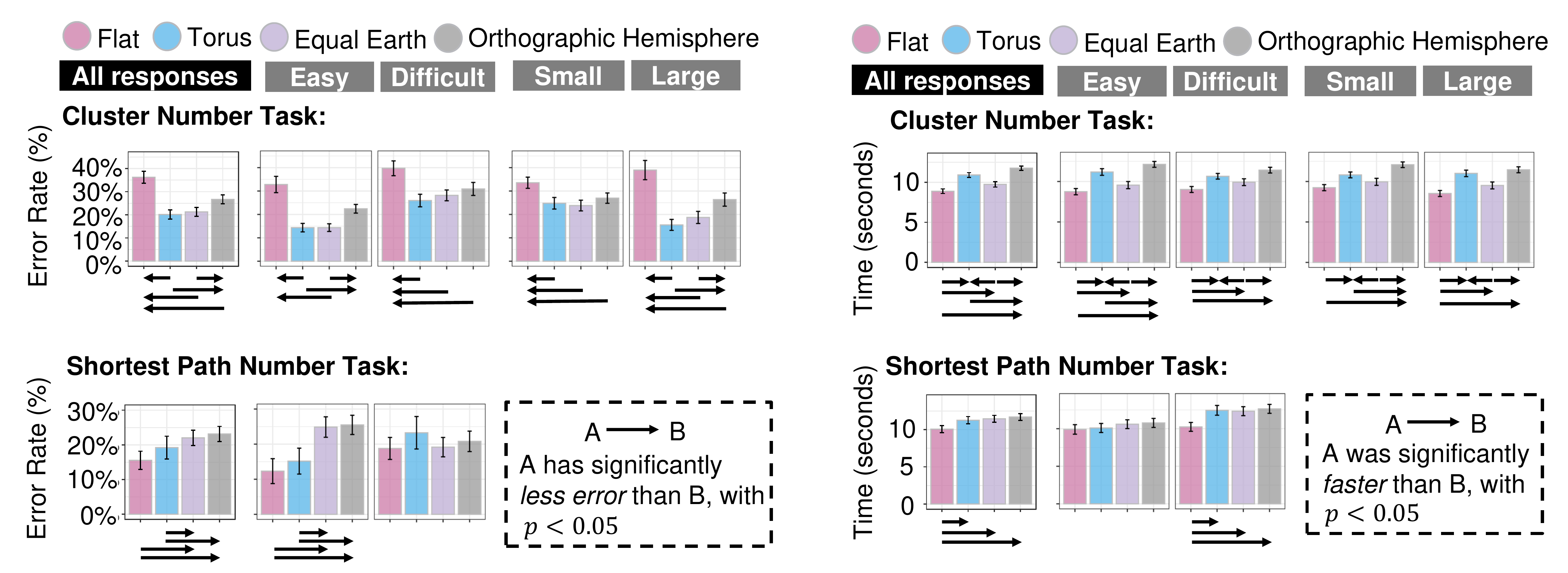}
    \caption{Error rate (left) and Time (right) results of Study 6. Error bars are 95\% confidence intervals.} Significant differences between projections are shown as arrows. Detailed statistical results are available in~\autoref{fig:study-2-results-graphics}. Effect sizes are available in~\autoref{fig:study-2-results-effect-size}.
    \label{fig:study-2-results}
\end{figure*}


\begin{figure}
    \centering
    \includegraphics[width=\linewidth]{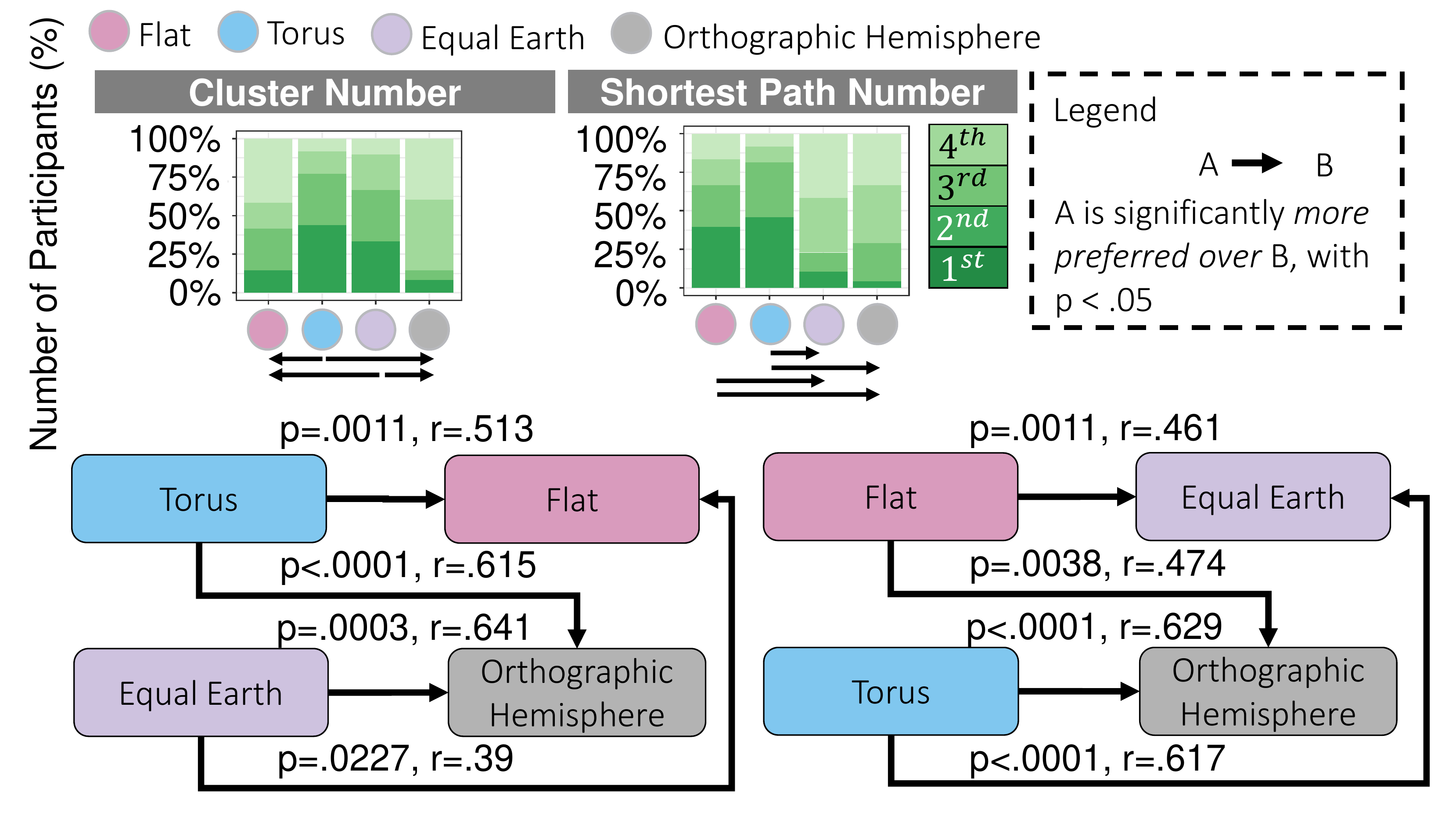}
    \caption{User confidence ranking of each condition for two tested tasks. Arrows indicate statistical significance with $p<0.05$. Effect size results of Cohen's r~\cite{cohen2013statistical} are presented along with the arrows.}
    \label{fig:study-2-ranking-results}
     \vspace*{-6pt}
\end{figure}

\subsection{Results}
\label{sec:networkstudyresults}

    

In the following, significance values are reported for $p < .05 (*)$, $p < .01 (**)$, and $p < .001 (***)$, respectively, abbreviated by the number of stars in parenthesis. 

For \tclusteridentification{}, we found Layout had a significant effect on error rate in All Responses ($***$), Easy ($***$), Hard ($***$), Small ($***$), and Large ($***$). \ttorus{} ($***$), \tequalearth{} ($***$) and \torthographic{} ($*$) were more accurate than \tnodelink{}. \ttorus{} ($*$) and \tequalearth{} ($**$) were more accurate than \torthographic{}. We also found \ttorus{} and \tequalearth{} were more accurate than \tnodelink{} for Easy ($***$) and \torthographic{} for Easy ($***$).  We found \ttorus{}, \tequalearth{}, and \torthographic{} were more accurate than \tnodelink{} for Hard ($***$), Small ($***$). We also found \ttorus{} ($***$), \tequalearth{} ($***$) and \torthographic{} ($*$) were more accurate than \tnodelink{} for Large. \ttorus{} ($*$) and \tequalearth{} ($**$) were more accurate than \torthographic{} for Large.

We found Layout ($***$), Layout $\times$ Difficulty ($***$) and Layout $\times$ Size ($*$) had a significant effect on time. In Layout, \tnodelink{} is faster than \ttorus{}, \tequalearth{} and \torthographic{} (Layout $***$). \ttorus{} was found faster than \torthographic{} (Layout $***$). \tequalearth{} was faster than \ttorus{} and \torthographic{} (Layout $***$). In Layout $\times$ Difficulty, \tnodelink{} is faster than \ttorus{} (Easy/Hard $***$), \tequalearth{} (Easy $*$, Hard $***$), and \torthographic{} (Easy/Hard $***$). \tequalearth{} is faster than Torus (Easy $***$, Hard $*$) and \torthographic{} (Easy/Hard $***$). We also found \ttorus{} was faster than \torthographic{} (Easy $*$). In Layout $\times$ Size, \tnodelink{} was faster than \ttorus{} (Small/Large $***$), \tequalearth{} (Large $***$), and \torthographic{} (Small/Large $***$). \tequalearth{} was found faster than \ttorus{} (Small $**$, Large $***$), and \torthographic{} (Small/Large $***$). \ttorus{} was also found faster than \torthographic{} (Small $***$).

For user ranking, participants preferred \ttorus{} to \tnodelink{} ($**$) and \torthographic{} ($***$). Participants preferred \tequalearth{} to \tnodelink{} ($*$) and \torthographic{} ($***$).

For \tshortestpath{}, we found Layout had a significant effect on error rate in all ($***$). We found for All Responses, \tnodelink{} was more accurate than \tequalearth{} ($**$) and \torthographic{} ($***$). \ttorus{} was more accurate than \tequalearth{} ($**$) and \torthographic{} ($**$). In Easy, we found \tnodelink{} was more accurate than \tequalearth{} ($***$) for Easy, and \torthographic{} ($***$) for Easy. \ttorus{} was also more accurate than \tequalearth{} ($***$) for Easy and \torthographic{} ($**$) for Easy. We did not find a significant difference in Hard.

We found Layout had a significant effect on time ($***$). \tnodelink{} was faster than \ttorus{}, \tequalearth{} and \torthographic{} ($***$). In Hard, we also found \tnodelink{} was faster than \ttorus{}, \tequalearth{} and \torthographic{} ($***$). We did not find a significant difference in Easy.
For user ranking, participants preferred \tnodelink{} to \tequalearth{} ($**$) and \torthographic{} ($**$). \ttorus{} was also found preferred to \tequalearth{} ($***$) and \torthographic{} ($***$).

\section{Key Findings and Discussion}
\label{sec:networkstudykeyfinding}
We report on the most significant findings for \clusteridentification{}, \shortestpathnumber{} visually in \autoref{fig:study-2-results} and \autoref{fig:study-2-ranking-results}. 


\textbf{For \clusteridentification{}, \tequalearth{} and \ttorus{} performed equally well and \tequalearth{} was slightly faster than \ttorus{}, while they both significantly outperformed \torthographic{} in terms of \merror{}, \mtime{}, and \mpref{}}.
The poor performance of \torthographic{} compared with \tequalearth{} and \ttorus{} was found statistically significant with moderate and large effect sizes for \merror{} (All, Easy, Large), \mtime{} (Easy, Large), and large effects for \mpref{} (\autoref{fig:study-2-results}-top and \autoref{fig:study-2-ranking-results}).
The statistics graphics of significant statistical results of \merror{} and \mtime{} are shown in~\autoref{fig:study-2-results-graphics}. Effect sizes of \merror{} and \mtime{} results are shown in~\autoref{fig:study-2-results-effect-size}. Effect sizes of \mpref{} results are shown in~\autoref{fig:study-2-results-rank-effect-size}.  
These results confirm \textbf{G5.1, G5.2} for (RQ5.1, RQ5.3).  
Surprisingly, while results from Study 5 (\autoref{sec:mapstudydiscussion}) showed that interactive \torthographic{} tended to be more accurate than \tequalearth{} for distance comparisons and not worse than any other projections for area comparisons, it turned out to be significantly worse than \tequalearth{} and \ttorus{} for reading \clusteridentification. 

While it has the advantage of being a straightforward mapping from the sphere, \torthographic{} is an interrupted and non-area-preserving projection.  Therefore, we conjecture that this discontinuity caused participants to struggle to make out cluster boundaries as compared with continuous representations in \ttorus{} and \tequalearth{}. Furthermore, there is no link distortion in \ttorus{}. This was supported by participants' feedback, e.g., \emph{``The `Equal Earth' method felt much easier to distinguish every [individual] cluster.''} (P20), and \emph{``Orthographic Hemisphere uses 2 maps so it is much more difficult to interpret than the others [...] making it harder to isolate clusters.''} (P12). 
Furthermore, \tequalearth{} and \ttorus{} were found to be easier to make out cluster boundaries, while the separated maps in \torthographic{} made it confusing, as participants mentioned \emph{``It was easier seeing the boundaries of the set number of clusters in [torus] representation.''} (P11), and \emph{``I found understanding the movement of the orthographic diagram quite challenging.''} (P4). 

Some participants explicitly mentioned that interaction improves the readability of clusters in \ttorus{} and \tequalearth{}, while \torthographic{} is confusing although it looks naturalistic, e.g., \emph{``Torus is the easiest to interpret because the clusters do not get distorted if the map is moved around. Equal Earth is actually easier than it looks to interpret because even though the clusters get distorted, if they are moved around, they actually bunch up together once they are put near the middle of the map. Orthographic Hemisphere uses 2 maps so it is much more difficult to interpret than the others due to this making it harder to isolate clusters.''} (P12), \emph{``Second place is Torus because you can very easily navigate. Third place is equal earth because you can see almost all of the clusters at the same time, but it's a bit confusing because of distortion. Last place is Ortographic Hemisphere because it is very confusing''} (P16), and \emph{``Orthographic was the most convenient one due to familiar shape of earth.''} (P6). 

We also note that although being statistically significant, the small effect sizes indicate that \tequalearth{} is slightly faster than \ttorus{}~\cite{helske2021can,cockburn2020threats}.

\textbf{\tequalearth{}, \torthographic{} and \ttorus{} significantly resulted in less error than \tnodelink{}. \tequalearth{} and \ttorus{} were significantly preferred over \tnodelink{} but \ttorus{} and \torthographic{} took longer time than \tnodelink{}.}
This was found statistically significant with moderate and large effect sizes (\autoref{fig:study-2-results}-top, \autoref{fig:study-2-ranking-results}, Section A.2: Figure 15, Figure 16, and Figure 17), leading us to reject \textbf{G5.1} for (RQ5.2) and confirm \textbf{G5.2} for (RQ5.2). These results provide strong evidence that with automatic panning and interaction, participants were able to better identify the high-level network structures using spherical projections. 
They also confirm the results presented in~\autoref{sec:torus2:userstudy} where toroidal layouts with automatic panning significantly outperformed \tnodelink{} in terms of error for cluster understanding tasks. Some participants explicitly mentioned that good separation of clusters in continuous surfaces such as \tequalearth{} and \ttorus{} helped understanding. 

For example, some participants explicitly mentioned that good separation of clusters in continuous surfaces such as \ttorus{} and \tequalearth{} helped understanding, e.g., \emph{``Distance between the clusters as well as perspective played a huge role in my choices. The more further apart the clusters were the easier it was for me to count them.''} (P18), \emph{``I think "equal earth" was the easiest and most understandable."Node-link" was something like a challenge for me,it was very difficult to understand it.''} (P28), \emph{``I could see patterns more easily in the torus representation. It was easier seeing the boundaries of the set number of clusters in that representation.''} (P11), and \emph{``Torus seems to be the easiest, because it was just flat surface with possibility of dragging.''} (P14), while  \tnodelink{} tended to appear tangled, e.g., \emph{``With the [\tnodelink{}], it was sometimes apparent where the clusters were, but I feel like I chose 1 cluster as an answer too frequently with the tightly packed examples.''} (P11).

Participants also mentioned the automatic panning helped them ``see'' without the need to interact e.g., \emph{``[with \tequalearth,] even though if we move the picture [it] is very hard to understand the clusters. [It] turned out to be easiest to find the clusters (by not moving the picture).''}  (P37). 

\textbf{For \shortestpathnumber{}, \tnodelink{} and \ttorus{} performed equally well}, both significantly outperforming \tequalearth{} and \torthographic{} in terms of \merror{} and \mpref{}, while \tnodelink{} is faster than all representations. We found \tnodelink{} and \ttorus{} have a lower error rate than \tequalearth{} and \torthographic{} for all responses (moderate effects) and Easy (large effects) (\autoref{fig:study-2-results}-bottom). 
The statistics graphics of significant statistical results of \merror{} and \mtime{} are shown in~\autoref{fig:study-2-results-graphics}. Effect sizes of \merror{} and \mtime{} results are shown in~\autoref{fig:study-2-results-effect-size}. 
There is a strong evidence that \ttorus{} is more preferred (large effects) over \tequalearth{} and \torthographic{}, while \tnodelink{} is more preferred (moderate effects) over \tequalearth{} and \torthographic{} (\autoref{fig:study-2-ranking-results}). Effect sizes of \mpref{} results are shown in~\autoref{fig:study-2-results-rank-effect-size}.  
\tnodelink{} was the fastest for All Responses (small and moderate effects) and Hard (moderate and large effects). We therefore accept \textbf{G5.3, G5.4} for (RQ5.3) but reject \textbf{G5.3, G5.4} for \tnodelink{}.

We conjecture that although \ttorus{} involves broken links across the boundaries, it appears similar to \tnodelink{} using straight links
while the distortion of paths in \tequalearth{} might hamper path tracing tasks, as participants mentioned \emph{``Flat surface is easier to read, it helps sometimes when you can also move it. Earth-like is just hard to use, especially when it's an equal globe.''} (P9), and \emph{``the warping in [\tequalearth{}] made my eyes hurt a little therefore its in 4th place, and the torus being the most straight forward gets 1st with [\tnodelink{}] in second as they're very similar.''} (P13). On the other hand, \torthographic{} has less link distortion but the interruption between two hemispheres and the strong perspective distortion near the boundary of the maps may make participants confused, supported by participants' feedback.

Although \ttorus{} involves broken links across the boundaries, it appears similar to \tnodelink{} using straight links while the distortion of paths in \tequalearth{} might hamper path tracing tasks: \emph{``I am more familiar with the [\tnodelink{}] and torus shaped representations.''} (P6), and \emph{``The flatter that it seemed, the easier it was.''} (P20).

On the other hand, \torthographic{} has less link distortion but the interruption between two hemispheres and the strong perspective distortion near the boundary of the maps may make participants confused. For example, \emph{``Orthographic Hemisphere was even trickier because I couldn't get some of the dots on the same sphere sometimes and the time ran out.'' (P12)} and \emph{``Orthographic Hemisphere is the last because I found it the most restrictive, since it is difficult to make the red points appear together since the curves of the links don't bend, making it harder to see all the ways they connect.''} (P21).

Although automatic panning (\autoref{sec:autopan}) provides some benefits for reducing split of clusters across the boundaries for \clusteridentification{}, it seems it has less benefits for \shortestpathnumber{}, as participants mentioned they still need to position the image to see the full path to identify the shortest path connecting the nodes within the time limit for \tshortestpath{}. For example, \emph{``Equal earth and orthographic hemisphere would make things more puzzling for me. When I had to wrap the map in order to find the fastest path for the nodes, it would usually take me more time to estimate and find the lines between the two red nodes. I would say that when curves are being shaped it is more complex to find the fastest path. It also made me anxious because I had to answer nearly at the last moment. Torus was almost as the simple node-link diagram. However, I found it easier to find the links between the nodes in a diagram that I had little or no interaction.''} (P35) and \emph{``I was more confident using the Node-Link. It was simpler to use visually. With the other three, there were times when it was difficult to position the map in a way that I could identify the shortest path for connecting the nodes.''} (P45).


Overall, these findings suggest that \ttorus{} presents a general solution being not only less error prone than \tnodelink{} or \torthographic{} for \clusteridentification{}, but also comparable to \tnodelink{} for \shortestpathnumber{}.

\section{Limitations and Future Work}
\label{sec:study-2-threats}


A limitation of our study is that we only tested with one layout algorithm. While many different algorithms exist for \tnodelink{} and it is possible that other layout algorithms could be adapted to spherical and torus embeddings, doing so is not necessarily trivial. 
Also, it should be noted that there are algorithms which can optimise layout for a known set of clusters (i.e.\ where the cluster labelling is known in advance) ~\cite{meulemans2013kelpfusion, collins2009bubble,gansner2009gmap,hu2010visualizing}. However, for the tasks tested in this paper we do not pre-identify the clusters but rather leave cluster identification as the user task. 

\section{Conclusion}
\label{sec:spherevstorus:conclusion}
The study results presented in this chapter confirm the benefits of topologically closed surfaces, such as the surface of a torus or sphere, when using node-link diagrams to investigate network structure.  All of \tequalearth{}, \torthographic{}, and \ttorus{} outperformed \tnodelink{} for cluster understanding tasks.  While the spherical projections impeded path following tasks, it seems \ttorus{} may be a good general solution, being as accurate as \tnodelink{} for path following.
Although they have been explored in research, 2D projections of toroidal and spherical network layouts are rarely seen in practice.  This may be because until recently effective layout and projection methods for such geometries were not easily available.  We intend to make all our algorithms and extensions of existing open-source tools available for easy consumption in web applications via GitHub and \textit{npm} packages.

A limitation of our work is that there are many more possible spherical map projections than those evaluated here, although we tried to select the most representative techniques.  Further work may extend such evaluation to other projection types.  Another interesting topic for future investigation would be investigating spherical and torus projections of other types of abstract data representation, such as multi-dimensional scaling techniques of high-dimensional data.  Another family of interactive techniques for exploring graphs involve applying spatial distortion around regions of interest to achieve a kind of ``structure aware zooming'', e.g. \cite{wang2018structure}.  It would be interesting to compare the efficacy of these techniques against interactive sphere and torus projections.       
%
\chapter{Discussion, Future Work and Conclusion}
\label{sec:conclusion}

\cleanchapterquote{We shouldn’t abbreviate the truth but rather get a new method of presentation.}{Edward Tufte}{(American statistician and emeritus professor of political science, statistics, and computer science at Yale University.)}

\begin{figure}
    \centering
	\includegraphics[width=0.9\textwidth]{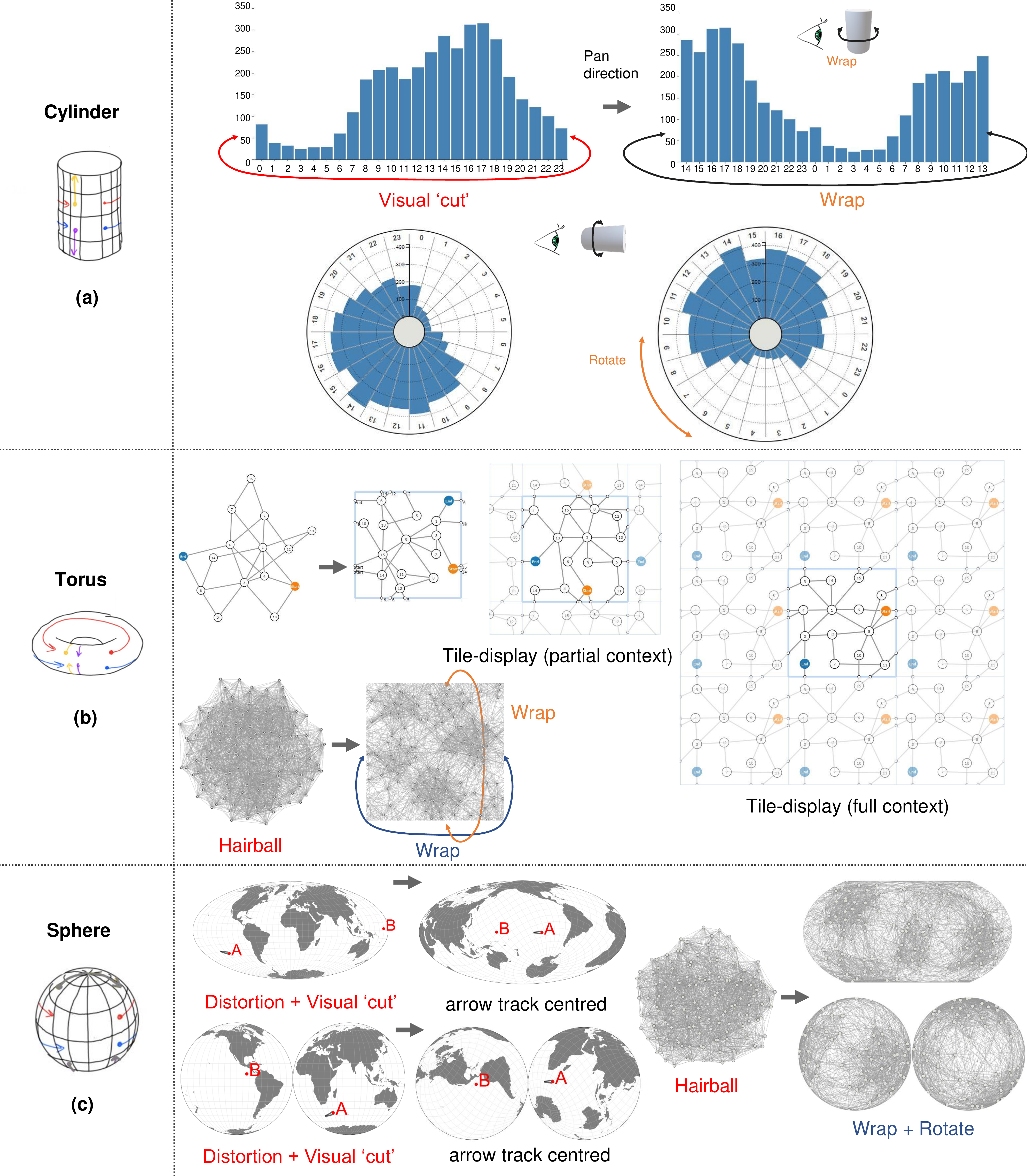}
	\caption{Summary of a common range of non-flat data types that could benefit from being laid out on a 2D wraparound topology.}
	\label{fig:discussion:itsawrap}
\end{figure}
In this final chapter we first summarise the primary research contributions described in the previous chapters (\autoref{sec:conclusion:contributions}).  We then reflect on the utility of wrapped visualisation with regard to some practical examples of network visualisations (\autoref{sec:discussion:networkvis}) and consider other possibilities for future work (\autoref{sec:conclusion:futurework}). 

\section{Contributions}
\label{sec:conclusion:contributions}
In this thesis, we argued that interactive wrapped visualisations provide perceptual benefits compared to standard visualisation techniques for presentation of a range of common data types that are ``not flat'', in the sense that they can not be easily presented on a 2D screen or surface without loss of information (as described in~\autoref{sec:intro}).  Examples of such data types that we have considered include cyclical data, geographic maps and networks of relational data. \autoref{fig:discussion:itsawrap} illustrates example visualisations of these data types that result in extremely promising task performance (in terms of accuracy) or graph layout aesthetics when we employ the concept of wrapped visualisation. 

To investigate this new class of interactive wrapped visualisation and its effectiveness, we presented the first  wrapped visualisation design space that unifies the notion of embedding data on 3D cylinder, sphere and torus topologies in order to create wrapped 2D projections. Our design space  encompasses sophisticated panning interactions with the 2D projections and suggests novel applications for multidimensional scaling plots and matrices (\autoref{sec:designspace}).  We have also created interactive web tools and network layout algorithms that realise the ideas from our design space to provide practical visualisations.
The tools and layout algorithms are open-sourced  to assist future research and development.

Through a series of six user studies with more than 350 participants investigating cyclical time series, networks, and geographic maps, our evaluation results show significant benefits for a number of tasks and visualisation types when compared to their standard unwrapped counterparts in task performance (specifically accuracy, but sometimes at the cost of completion time) and also in terms of user preference. 

Specifically, we first (a) explored visual representation of real-world cyclical time series data presented in bar charts and concentric radial (or ``polar'') charts that can be mapped to a cylindrical topology and explored with novel cylindrical wrapping (which allows the mark to be panned left or right to connect the visual cut split across the left-and-right boundary at the centre) or rotation interaction (a technique that allows for rotating and centring bars for a more aligned comparison). Our results provide the first empirical evidence that interactive wrapping of cyclical time series in linear bar charts provides promising results in terms of accuracy over standard bar charts without such interaction for trend identification, while interactive wrapping of bar charts outperformed standard bar charts, polar charts, and rotatable polar charts in terms of accuracy for pairwise value comparisons (\autoref{sec:cylinder}).

Next, (b) we applied and evaluated ideas of torus wrapping, originally developed in theoretical graph drawing research~\cite{mohar2001graphs,kocay2016graphs}, about embeddings of graphs on surfaces with topologies that allow for better ``unfolding''. With a graph corpus of 72 random networks (up to 15 nodes, 36 links) with scale-free or small-world properties (defined in~\autoref{sec:related:network:forcedirected}), we showed that torus-based layouts created by our layout algorithm afford better aesthetics \rev{than standard non-wrapped force-directed layouts (\autoref{sec:torus1})} in terms of conventional measures like more equal edge length, lower ``stress'' (defined in~\autoref{sec:related:mds}, less deviation of adjacent incident link (edge) angles from the ideal minimum angle (determined by a complete angle divided by the number of links connected to the node) between edges connecting to the same node, and fewer crossings. Using these layouts, we showed in two controlled user studies that torus layout with either additional context (tile repetition that shows network connectivity at the sides of the diagram, as seen in~\autoref{fig:discussion:itsawrap}(b)) or interactive panning provided significant performance improvement (in terms of error and time) over torus layout without either of these improvements, to the point that it is comparable to standard flat node-link layout, confirming benefits of both wrapping approaches for torus described in our design space (\autoref{sec:designspace:dimensions}--wrapping dimension). 

To evaluate realistically larger networks, we further refined our torus-based layout algorithm that better displays high-level network structures like clusters (subsets of nodes that are more tightly connected within the subset than outside the subsets). Our extensive experimental analysis with a large graph corpus (200 random graphs with community structures, up to 128 nodes and 2590 links, described in~\autoref{sec:torus2}) shows that our new toroidal layout
algorithm is considerably more robust, consistently producing
high-quality layouts in terms of improved aesthetics than \rev{standard non-wrapped force-directed node-link layouts} and our previous torus layouts (\autoref{sec:torus1}) including less stress, fewer link crossings, and greater cluster separation. Through our various design prototypes and studies we also established that wrapped visualisations (e.g., torus or sphere wrapping) benefit from an automatic method for choosing an initial centre (a technique we called ``autopanning''). In~\autoref{sec:torus2} and~\autoref{sec:spherevstorus} we presented new techniques for automatic panning of spherical and torus network visualisations. Our user evaluation showed that interactive torus wrapping with automatic pan improve task results in error by 62.7\% and time by 32.3\% (differences between means) over standard flat layouts for larger networks for cluster understanding tasks, and that it led to reduced panning interaction by users in some tasks for torus (\autoref{sec:torus2}). 


Finally, (c) we investigated another wrapping geometry, sphere, which has been widely considered for maps and networks but mostly on a static 2D plane (i.e.\ without support for interactive panning). We demonstrated how interactive spherical wrapping and rotation of geographic maps provides more accurate area,  distance, and direction estimations (\autoref{sec:spheremaps}). We also showed that interactive spherical wrapping and rotation of networks provide benefits over \rev{standard non-wrapped force-directed node-link layouts} for cluster identification tasks, while the spherical wrapping view impedes path following tasks.  Meanwhile, wrapped torus-based network visualisations work well overall, specifically better than non-wrapped force-directed node-link and interrupted spherical projects for cluster identification and being comparable to non-wrapped flat node-link for path following tasks (\autoref{sec:spherevstorus}).

The investigation results suggest that interactive wrapping of these common data visualisations (charts, maps, networks) around topologically closed surfaces (cylinder, sphere, torus) could be more routinely applied to common types of data visualisations, e.g., cylindrical wrapping for comparing values and inspecting trends or patterns in online charts with temporal periodicity (e.g. daily, weekly or monthly), boundaryless views of online social networks without central or peripheral nodes, visual network cluster identification (e.g., social networks with highly connected community structures) using toroidal or spherical wrapping, and online world maps (e.g., for educational training purposes or international travel guide) with interactive spherical wrapping.

\section{Discussion}
\label{sec:conclusion:discussion}
\rev{It should be noted that the benchmark comparisons and network data studies described in (b) and (c) of the previous section compared only one variant of force-directed 2D layout (\autoref{sec:torus1:stressminimisation}) with various wrapped alternatives based on a similar stress minimisation approach.  This choice followed a careful study design methodology of trying to ensure that the layout approach used in each condition was as similar as possible, with the only difference between the techniques under comparison being whether and how they are topologically wrapped. 
However, there are many different layout algorithms for 2D node-link visualisations such as orthogonal link layout by Batini et al.~\cite{batini1986layout,kieffer2015hola}, layered layout by Sugiyama et al.~\cite{sugiyama1981methods}, network layout with data aggregation by Yoghourdjian et al.~\cite{yoghourdjian2018graph} and it is possible that other layout algorithms could be adapted to spherical and torus embedding. 
There are also algorithms which can optimise layout for a known set of clusters (i.e.\ where the cluster labelling is known in advance) ~\cite{meulemans2013kelpfusion, collins2009bubble,gansner2009gmap,hu2010visualizing}. While these approaches might bring back some accuracy benefits while eliminating the (costly) need for interaction, for the tasks tested in this thesis we do not pre-identify the clusters but rather leave cluster identification as the user task as there are many real-world data analysis scenarios where pre-identified clusters are not available or can not be easily computed.
We leave the exploration of other layout algorithms as future work. 
}

\rev{Another interesting issue is that for spherical wrapping, the use of interactive wrapping and rotation changes the orientation of the maps. In the map projection study in~\autoref{sec:spheremaps}, we did not constrain the wrapping interaction, i.e.\ users could rotate the map in any direction. The resulting projected visualisation might therefore not be as intuitive for users after rotation, as north and south poles are no longer at the north and south, and may not fit with their conception of spatial data (maps) and non-spatial (networks).
But for networks, such distortions may be less harmful as there is often no meaning to the spatial information, for example, there are no logical start and end points of the data, nor even a sensible ``inside'' or ``outside''. Higher distortion might even help make clusters more visible by increasing their separation.}

\rev{We also note that there are additional variations of parameters possible for further study, beyond those considered in our existing controlled experiment.  These could include parameters such as the level of difficulty in the distance comparison tasks for geographic comprehension, as described in individual limitations sections in previous chapters.}

In the following, we reiterate the value of the results and insights from the perspective of different data types (networks, maps, time series). For networks, we also provide an informal reflection on practical applications that leads us to identify limitations of our existing work and to call for more formal studies, particularly with application domain experts, in the future.

\section{Network Visualisations}
\label{sec:discussion:networkvis}
With respect to abstract data such as networks, the analysis of our experimental results show that the additional ``spreading out'' that is possible in torus-based layout improves graph layout aesthetics and achieves greater cluster separation than traditional flat node-link layout, and furthermore that the sphere may provide a better use of available space. To further explore the real-world usefulness of wrapping layout, we apply the torus and sphere topology-based layout algorithms described in~\autoref{sec:torus2} and~\autoref{sec:spherevstorus} to other types of real-world networks.  Note that the discussion below is based on our own informal reflections and observations. We recommend that future work explores these applications using more formal user-centred study.



In~\autoref{sec:torus2} and \autoref{sec:spherevstorus}, we show that torus- and sphere-based network layouts provide benefits for larger complex networks for cluster identification tasks while torus works equally well to non-wrapped flat node-link layouts for sparse and smaller networks. To further investigate this question from the perspective of real-world utility, we explore networks that can be too complex when presenting using conventional layout techniques to reveal meaningful patterns. We argue that the additional flexibility afforded in torus or sphere topology may untangle dense structure to benefit readability. 

Preliminarily, we explored a number of real-world social networks.  These particular networks were chosen according to a number of criteria. First, they are all used as examples in seminal social network visualisation papers.  Second, they all have dense small-world~\cite{watts1998collective}, bipartite (defined in~\autoref{sec:discussion:bipartite}) or community structure~\cite{girvan2002community} which is difficult to represent in typical non-wrapped flat network diagrams~\cite{saket2014node,wang2015ambiguityvis}, as described in~\autoref{sec:related:networks}. 

\subsection{Football network}
\label{sec:discussion:football}
\begin{figure}
    \centering
    \subfigure[Football-\nodelinklayout{}]{
    \includegraphics[width=0.49\textwidth]{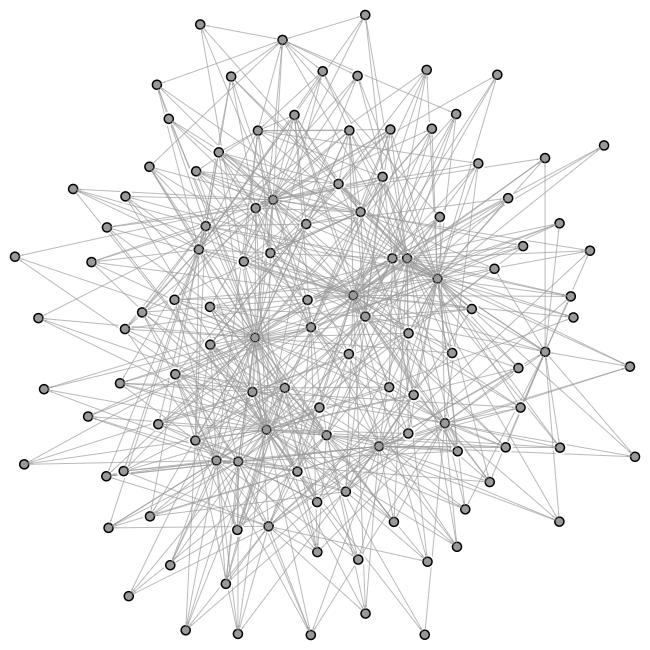}
    }
    \subfigure[Football-\toruslayout{}]{
    \includegraphics[width=0.49\textwidth]{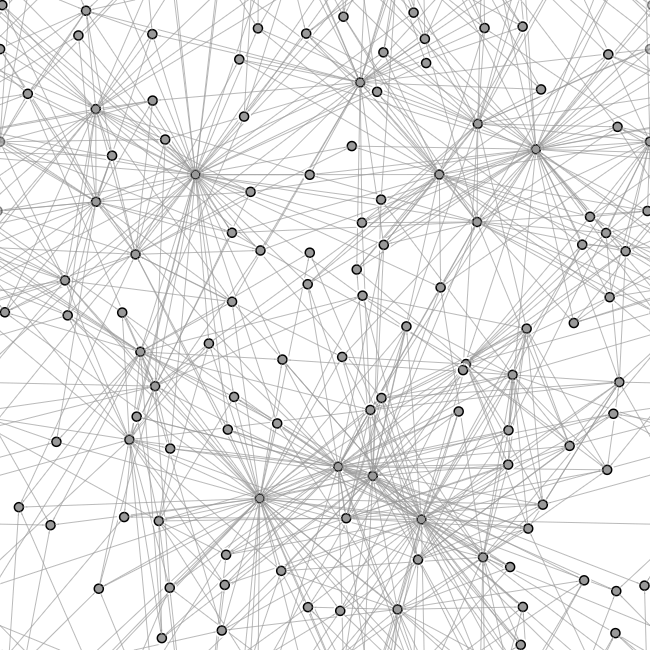}
    }
    \subfigure[Football-\tequalearth{}]{
    \includegraphics[width=0.7\textwidth]{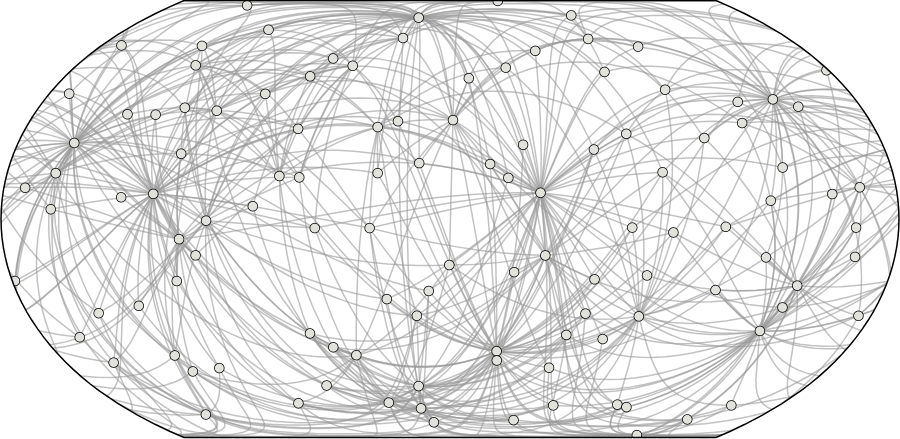}
    }
  \caption{Non-wrapped node-link, toroidal, and spherical representations of an American College football network (115 nodes, 669 links)~\cite{girvan2002community}, laid out using the algorithms discussed in~\autoref{sec:torus2} and~\autoref{sec:spherevstorus}.  The toroidal and spherical layout in (b) and (c) provide better use of the available space to provide greater separation between hub nodes.}
  \label{fig:discussion:football}
\end{figure}

This football network represents American football games between Division IA colleges during the regular season in fall in 2000 by Girvan and Newman~\cite{girvan2002community}. This dataset contains 115 nodes of teams (college names), 669 links of regular season games between the two teams they connect, and 9-14 clusters, i.e., subsets of nodes where nodes are highly connected within the subset than outside the subsets (also known as communities defined by Newman~\cite{newman2006modularity}). However, such networks usually have scale-free or small-world properties, and thus when representing the entire network, many links cross the diagram leading to a densely connected layout such that it can not be untangled sufficiently to understand connectivity between teams, as seen in~\autoref{fig:discussion:football}(a). 

This network has been previously evaluated in social network visualisation literature. For example, Wang et al.~\cite{wang2015ambiguityvis} proposed methods to highlight visual ambiguities as a result of highly connected network structure that leads to visual cluster overlap. They encoded cluster information in the layouts based on pre-computed community detection to better show cluster boundaries. 
Compared with Wang et al.'s approach, our torus and sphere-based layouts do not take any advantage of pre-identified cluster information.  



Using our layout approaches discussed in~\autoref{sec:torus2} and~\autoref{sec:spherevstorus}, \autoref{fig:discussion:football}(b-c) shows that toroidal and spherical representations of the football network have a better use of available space to provide greater separation between hub nodes. This example also shows that the toroidal or spherical layouts may result in potential advantages for better distribution of node positions across the plane that lead to reduced visual clutter, and thus may support hubs identification tasks. 

\subsection{Recipe network}
\label{sec:discussion:recipe}
\begin{figure}
    \centering
    \subfigure[Recipe-\nodelinklayout{}]{
    \includegraphics[width=0.49\textwidth]{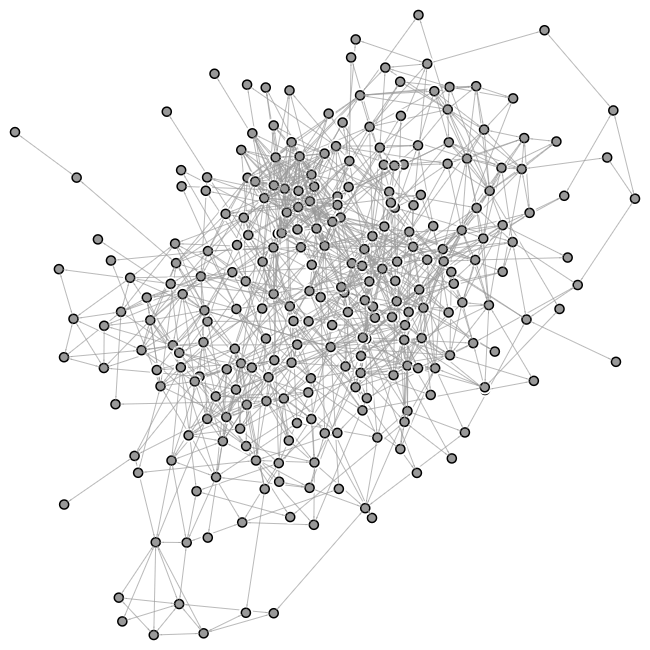}
    }
    \subfigure[Recipe-\toruslayout{}]{
    \includegraphics[width=0.49\textwidth]{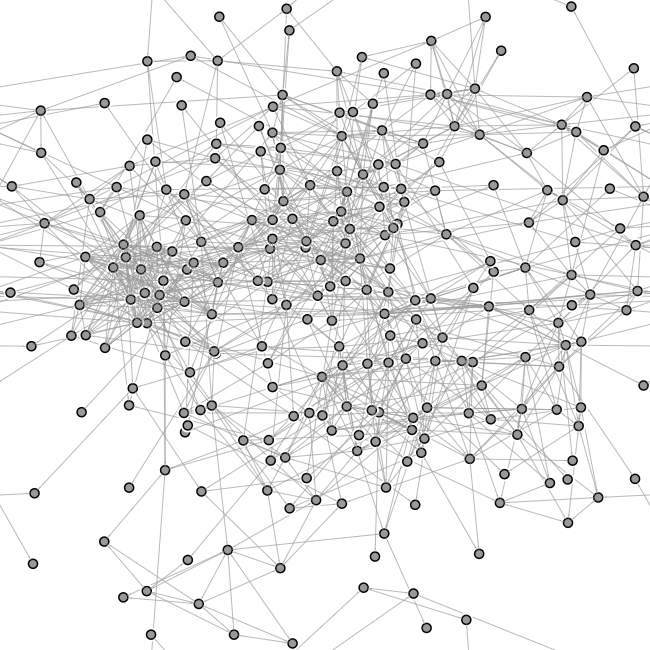}
    }
    \subfigure[Recipe-\tequalearth{}]{
    \includegraphics[width=0.7\textwidth]{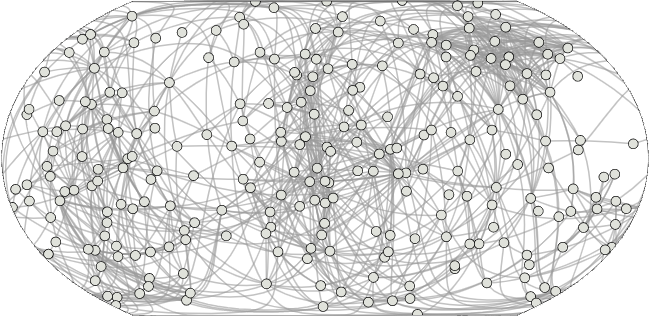}
    }
  \caption{Non-wrapped flat node link, torus, and sphere representation of recipe network, laid out using the algorithms discussed in~\autoref{sec:torus2} and~\autoref{sec:spherevstorus}. The dataset was originally created by Ahn et al.~\cite{ahn2011flavor} and derived by Okoe et al.~\cite{okoe2018node}. It has 258 nodes, 1090 links, and 8 clusters. The toroidal and spherical layout in (b) and (c) make it easier than the node-link representations in (a) to identify that there are three main clusters in the network, which supports cluster understanding tasks.}
  \label{fig:discussion:recipe}
\end{figure}
This recipe network represents cooking ingredients (nodes) connected by weighted links (edges) based on how frequent they are used together in recipes by Ahn et al.~\cite{ahn2011flavor}. This dataset was derived by Okoe et al.~\cite{okoe2018node} who removed disconnected components and removed low-weight edges, resulting in 258 nodes, 1090 edges, and 8 clusters. Variants of this network have been evaluated in user-centred studies previously in network visualisation literature for its usability by a wide range of participants by Saket et al.~\cite{saket2014node} and Okoe et al.~\cite{okoe2018node}. For example, Okoe et al.\ evaluated readability of node-link visualisations against adjacency matrix, a common alternative to node-link visualisation, finding that node-link performed better than matrix for low-level graph exploration tasks such as path following tasks in terms of accuracy and speed. However, similar to Saket et al.~\cite{saket2014node} and other existing work~\cite{meulemans2013kelpfusion, collins2009bubble,gansner2009gmap,hu2010visualizing} which can optimise layout for a known set of clusters (i.e.\ where the cluster labelling is known in advance), Okoe et al.'s network layout also relied on pre-identified cluster information. However, for the layouts and tasks tested in this thesis we do not pre-identify clusters but rather base on a simple generic approach like multidimensional scaling (described in~\autoref{sec:discussion:futurework:mds}) or force-directed (described in~\autoref{sec:intro}) in general, which implicitly group nodes by finding suitable node positions that minimise the difference between pairwise node distance and their graph theoretic distance (e.g., computed by an all-pairs shortest path method).



In our example, we show the layouts where we treat all the links (edges) equally weighted.~\autoref{fig:discussion:recipe}(a) shows that for the recipe network in typical force-directed node-link layout, there are many links crossing the diagram leading to a dense, cluttered ``hairball'', making it hard to determine the major clusters or structures in the underlying network's data. The toroidal and spherical layout in~\autoref{fig:discussion:recipe}(b) and~\autoref{fig:discussion:recipe}(c) show that the layout appears more spread out with the clusters more distinct, making it easier than the traditional flat node-link representations to identify that there are three main clusters in the network. This example shows that the toroidal or spherical layouts result in potential advantages for cluster identification and supports cluster understanding tasks.

\subsection{News co-mention network}
\label{sec:discussion:bipartite}
\begin{figure}
    \centering
    \subfigure[News-co-mention-\nodelinklayout{}]{
    \includegraphics[width=0.49\textwidth]{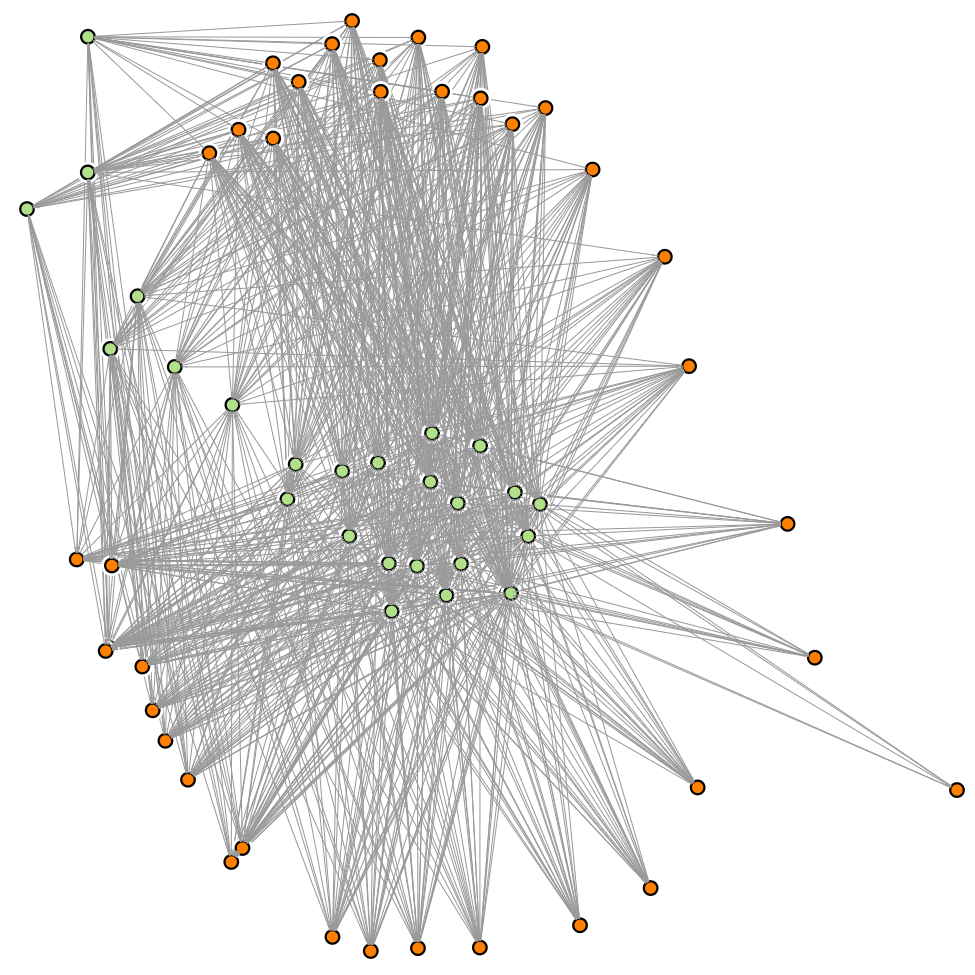}
    }
    \subfigure[News-co-mention-\toruslayout{}]{
    \includegraphics[width=0.49\textwidth]{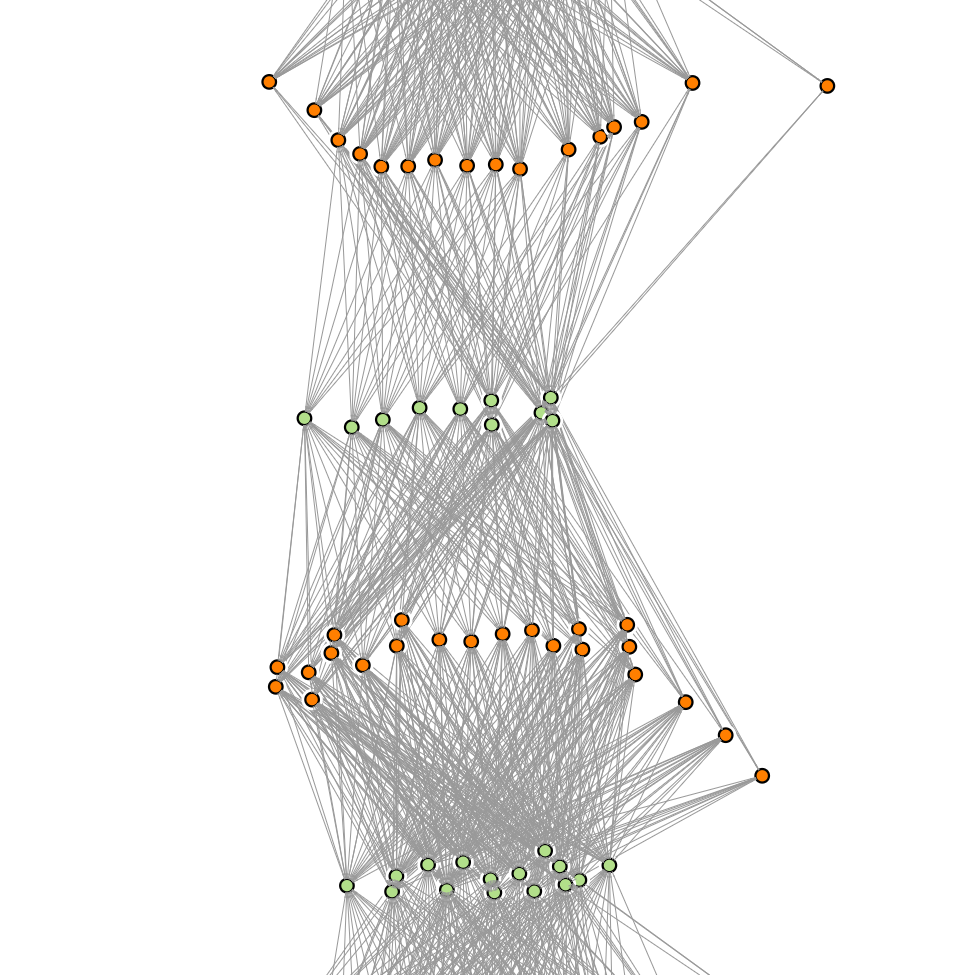}
    }
    \subfigure[News-co-mention-\tequalearth{}]{
    \includegraphics[width=0.7\textwidth]{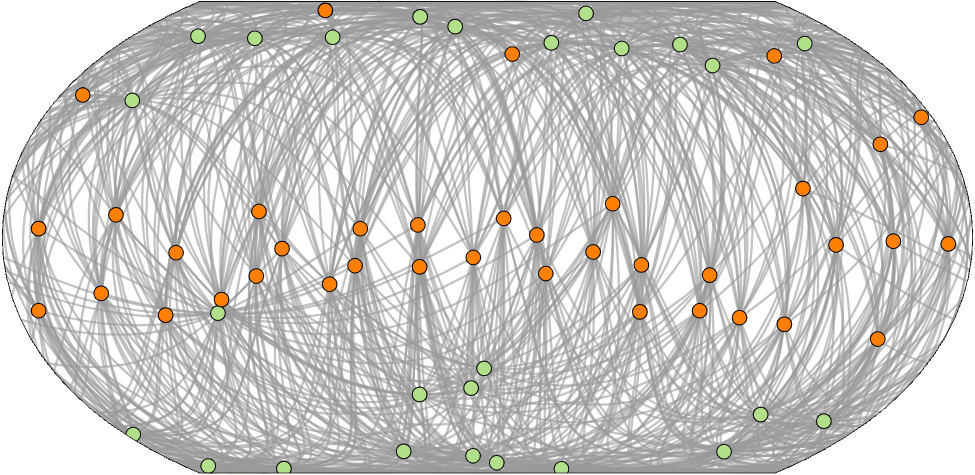}
    }
  \caption{(a-b) shows node-link and toroidal representations of a news co-mention bipartite network (65 nodes, 899 links). The dataset is credited to Faulkner et al.~\cite{faulkner2020measuring}. There are two types of nodes: group and topic. The link indicates the topic is mentioned in the group. colours indicate group (orange) or topic (light green). The toroidal layout in (b) reveals a visual pattern of four bands of alternating group and topic nodes, where groups of nodes alternate their positions and wrap around in one vertical direction. However, this is somehow confusing when seeing four groups rather than two groups in a bipartite network, but it is potentially easier than the flat layout to identify the pattern that reveals its underlying data structure.
  \label{fig:discussion:newscomention}
  }
\end{figure}
Compared with previous two examples showing typical community structures in social networks where tasks such as hubs or cluster identification may be beneficial for network layouts that wrap around topologies, the next example we consider has a different structure, i.e., it is \textit{bipartite}. A bipartite network models relationships between two distinct sets of things (nodes), i.e., the intersection of the two sets is the empty set, where relationships (links) can only occur between the sets (i.e. from a node in one set to a node in the other set), not within the same set (i.e. from a node in one set to another node in the same set). This type of network is ubiquitous in a variety of domains, e.g. to represent  relationships between two different types of things, such as the relationships between women and events in Davis Southern Women social network~\cite{davis2009deep}, between users and their preferred movies for recommendation systems~\cite{jannach2010recommender}, or between groups and topics such as the news co-mention network studied by Faulkner et al.~\cite{faulkner2020measuring}.

\autoref{fig:discussion:newscomention} shows the news co-mention network studied by Faulkner et al.~\cite{faulkner2020measuring}. It is a bipartite network of two sets: groups of people and topics with which they are associated in the news network. There are a total of 39 groups and 26 topics leading to 65 nodes. There are 899 associated connections (links) between groups and topics or issues. A link indicates the topic is mentioned in the group. Colours indicate group (orange) or topic (light green). The flat layout in (a) is based on a standard force-directed layout and does not do a good job of separating group nodes from topic nodes.  That is, we must carefully inspect the links to see that there are no links between topics and no links between groups. However, the toroidal layout in \autoref{fig:discussion:newscomention}(b) is much clearer as it separates nodes into four rows, where from the top row to the bottom it represents topics, groups, and then repeating topics, and groups, in an alternate fashion. Links at the bottom row wrap around to the top row, showing the connectivity between groups (bottom row) and topics (top row). We have found that such bipartite networks are typically arranged by the toroidal layout algorithm such that they only wrap on one side (i.e. cylindrical wrapping).  Within this cylindrical arrangement, we also find that dense-bipartite networks such as this one are arranged into several bands, with the nodes from the two partitions divided into two bands each. A viewer can use touch or mouse drag to interactively wrap the network vertically to observe the connectivity between any two adjacent rows. By contrast, the spherical layout in~\autoref{fig:discussion:newscomention}(c) does not reveal any obvious structure. However, since such spherical projections are really cylindrical (wrapping in one dimension rather than two dimensions, as we saw in our design space - \autoref{sec:designspace}), we conjecture that a similar banded layout should be possible on the sphere with a modified force model. 


It seems such torus layouts of a bipartite network may reveal a visual pattern of alternating sets of groups and topics that may be potentially better to identify bipartite relationships than the flat layout. However, a potential disadvantage is that representing bipartite relationships in four rows may cause confusion to understand that there are only two sets. We would expect that a trained user would be able to easily recognise such alternating layers as a bipartite structure (and would suggest a future study to evaluate this hypothesis).



\subsection{Product space network}
\label{sec:discussion:productspace}
\begin{figure}
    \centering
    \subfigure[ProductSpace-\nodelinklayout{}]{
    \includegraphics[width=0.49\textwidth]{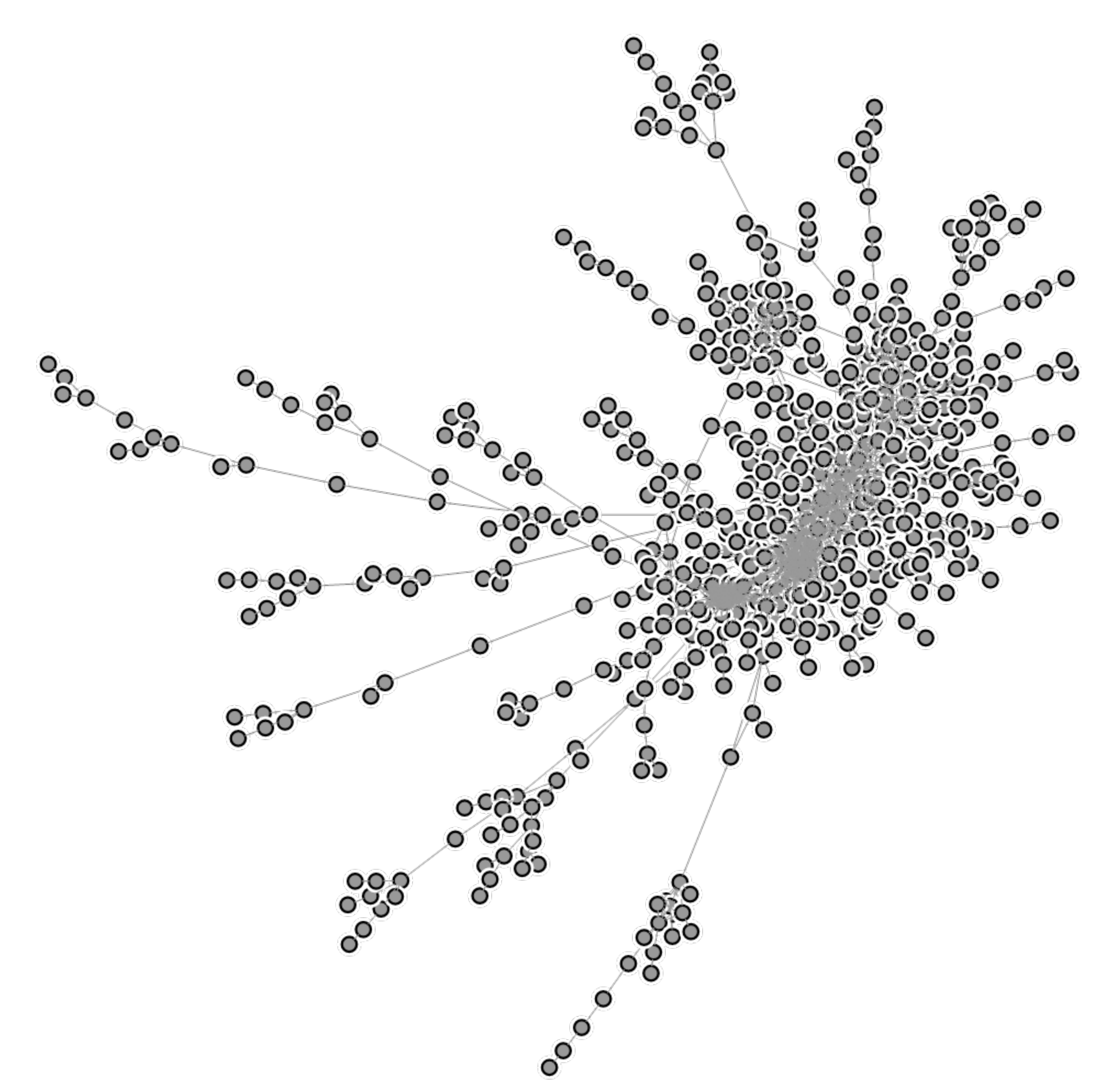}
    }
    \subfigure[ProductSpace-\toruslayout{}]{
    \includegraphics[width=0.49\textwidth]{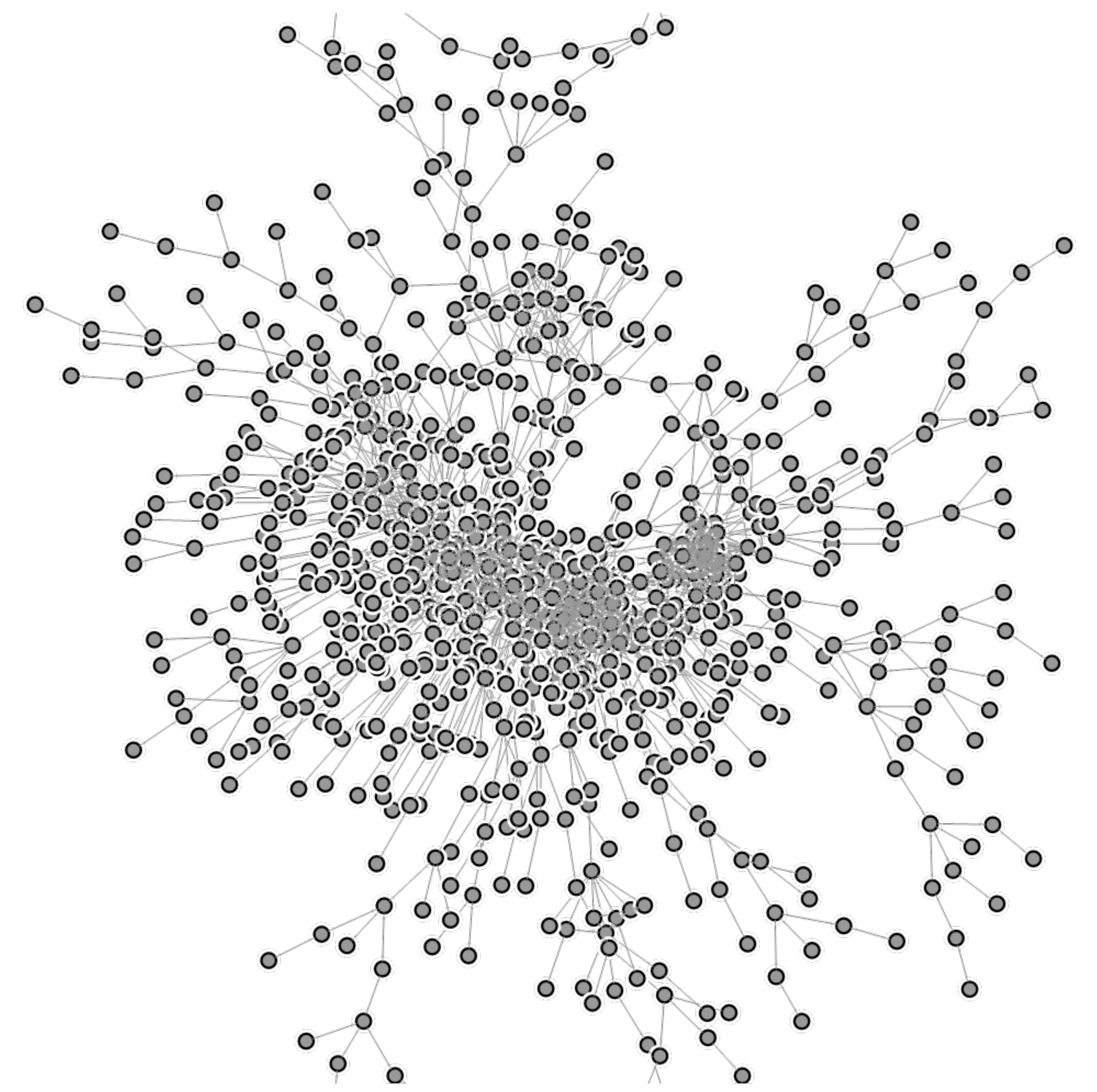}
    }
    \subfigure[ProductSpace-\tequalearth{}]{
    \includegraphics[width=0.7\textwidth]{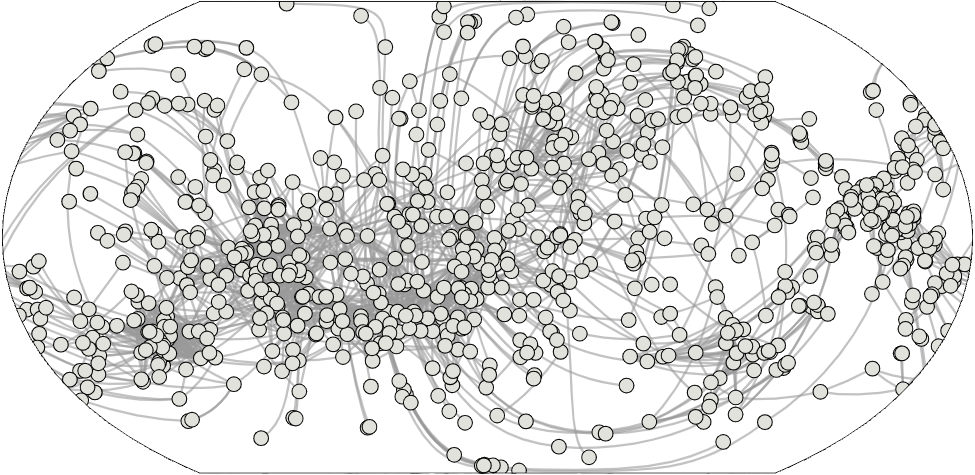}
    }
  \caption{Product space network laid out using standard flat, and our torus and equal earth visualisations. The data sets are from the original product space network created by Harvard Growth Lab~\cite{DVN/FCDZBN_2019}.}
  \label{fig:discussion:atlasraw}
\end{figure}
In previous subsections we have seen examples where 2D network layouts that wrap around continuous 3D torus or sphere surfaces may provide benefits over their non-wrapped flat representations. This subsection gives a counterexample that shows wrapped layouts may not be suitable for complex networks whose subsets contain many sparse tree-like structures. We explore the product space network derived from international trade data over decades, created by the Harvard’s Growth Lab using Harmonised System (HS) format~\cite{DVN/FCDZBN_2019}. This network represents 866 products (nodes) connected by 2532 weighted links (edges) based on the similarities of know-how required to produce the two products where they connect. Tools that visualise such networks have been presented by Simoes et al.~\cite{simoes2011economic} and its network visualisations widely used by various communities~\cite{obeng2020export,cicerone2020promoting} and social dynamics research~\cite{li2019china}.

\autoref{fig:discussion:atlasraw}(a) shows that in the typical non-wrapped node-link representation it is straightforward to observe that nodes further away from the dense structure have a very clear tree-like structure. By contrast, it is almost impossible to discern any high-level network structure from the dense structure due to tightly connected nodes within the structure. The torus (\autoref{fig:discussion:atlasraw}(b)) and spherical (\autoref{fig:discussion:atlasraw}(c)) layouts further untangles the dense structure. However, not only are nodes from the dense structure more ``spread out'' than non-wrapped layout, but the tree structures are also pushed further towards the edge of the display that ``wrap around''. As shown in~\autoref{fig:discussion:atlasraw}(b) it sometimes becomes ambiguous whether the tree structures extending from the top boundary are wrapped to the bottom, and the tree structures extending from the bottom wrapped to the top actually connect or are distinct, due to the visual overlap when these tree branches are spread out and wrapped. Similarly, \autoref{fig:discussion:atlasraw}(c) shows that the tree branches wrapped across the left-and-right boundary may lead to visual ambiguities in their connectivity without introducing artefacts to explicitly highlighting each distinct tree structure.

Since visualising such a network that is mostly fairly sparse, but has a dense core on toroidal or spherical topology does not really seem to take advantage of the torus topology's ability to reduce dense clutter (as we saw in~\autoref{sec:torus2}), we propose that another way to use toroidal layout for such a graph is to analyse just the dense part in isolation - perhaps as a detailed view that complements a more conventional flat layout of the whole graph.  To test this, we extracted the largest bi-connected subset (i.e. the largest subgraph where removing any one node leaves the subgraph connected) of the original product space network. This leaves us a smaller network with 385 nodes and 2031 links. 

\begin{figure}
    \centering
    \subfigure[ProductSpace-\nodelinklayout{}]{
    \includegraphics[width=0.49\textwidth]{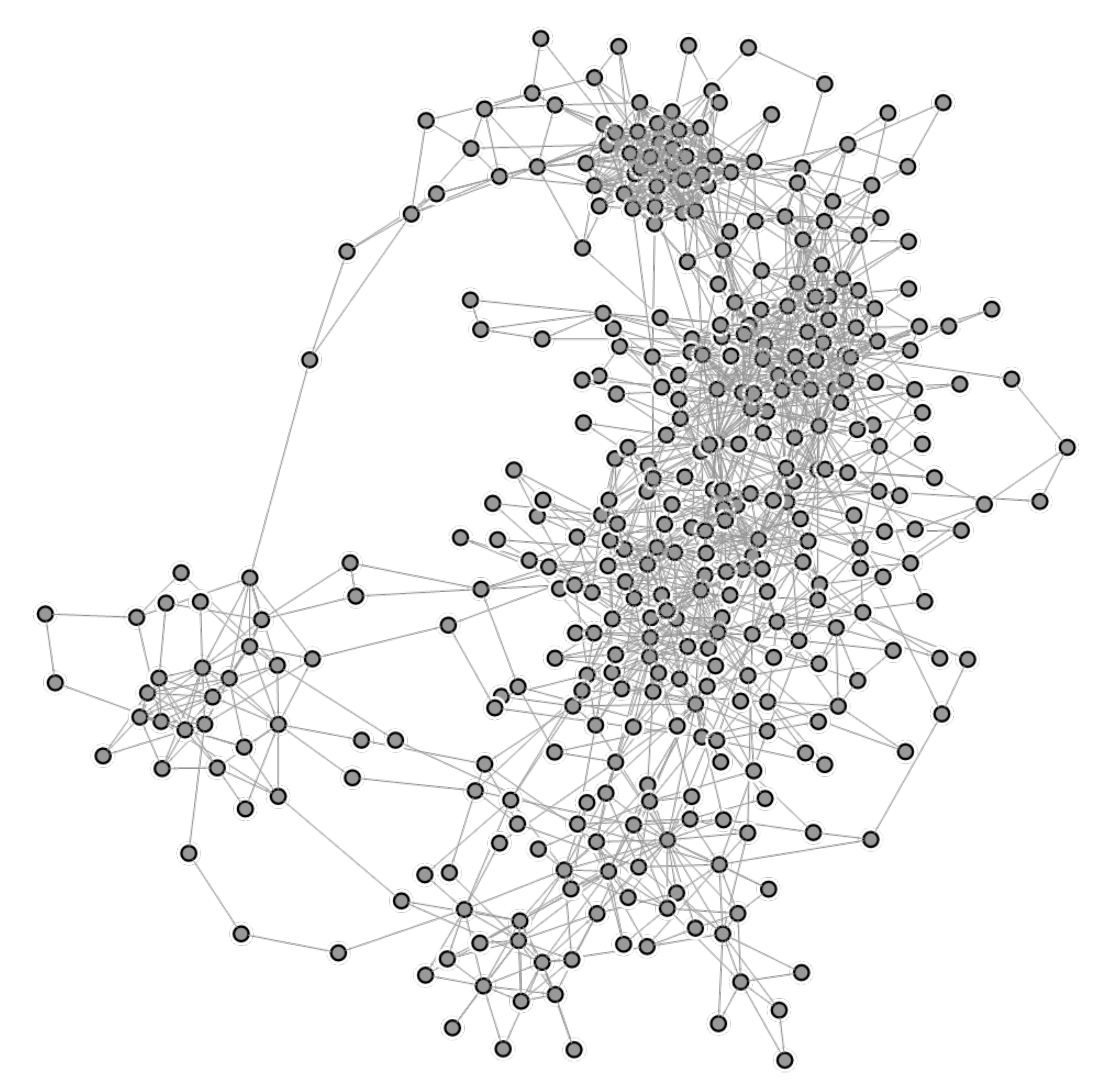}
    }
    \subfigure[ProductSpace-\toruslayout{}]{
    \includegraphics[width=0.49\textwidth]{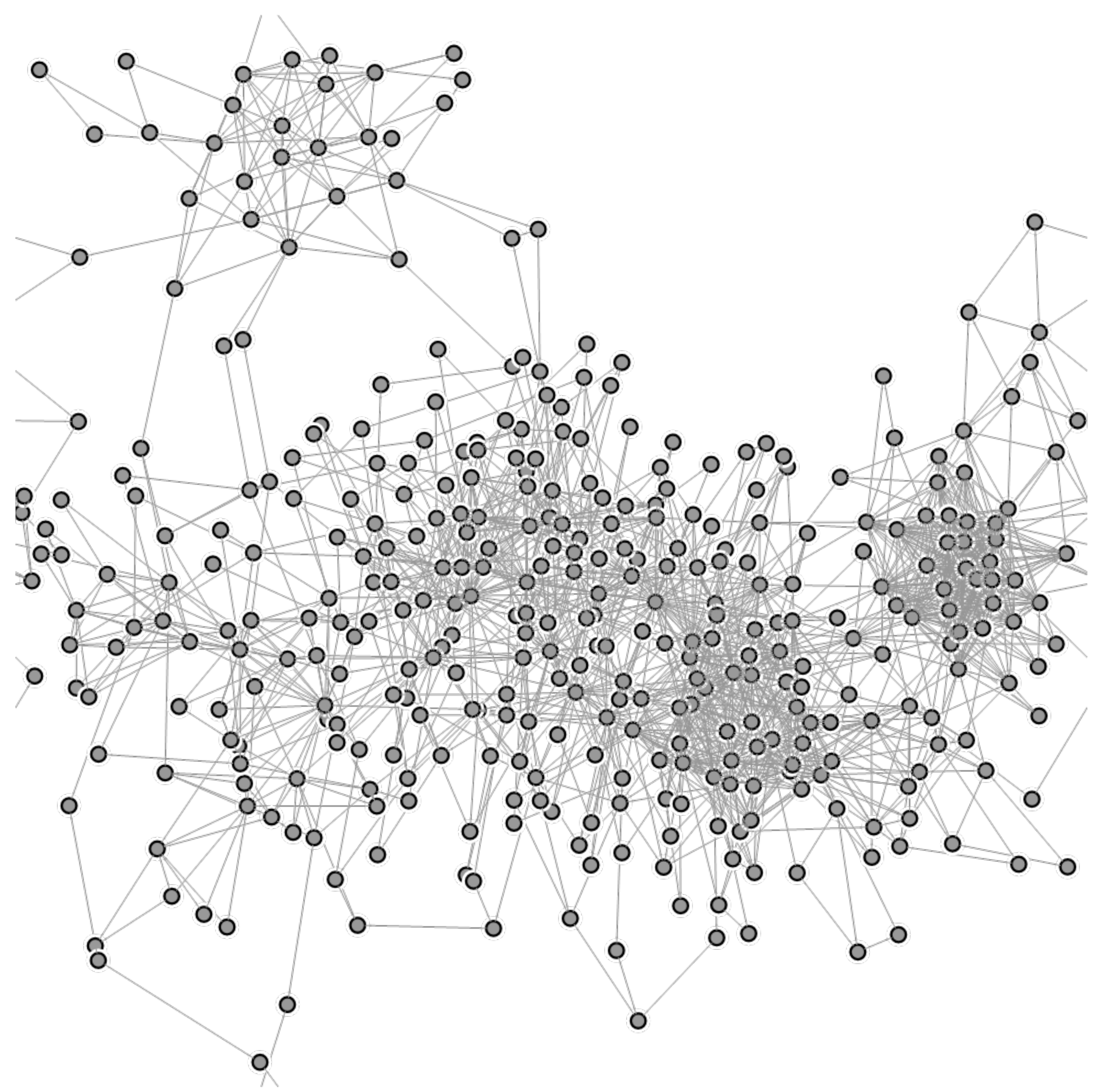}
    }
    \subfigure[ProductSpace-\tequalearth{}]{
    \includegraphics[width=0.7\textwidth]{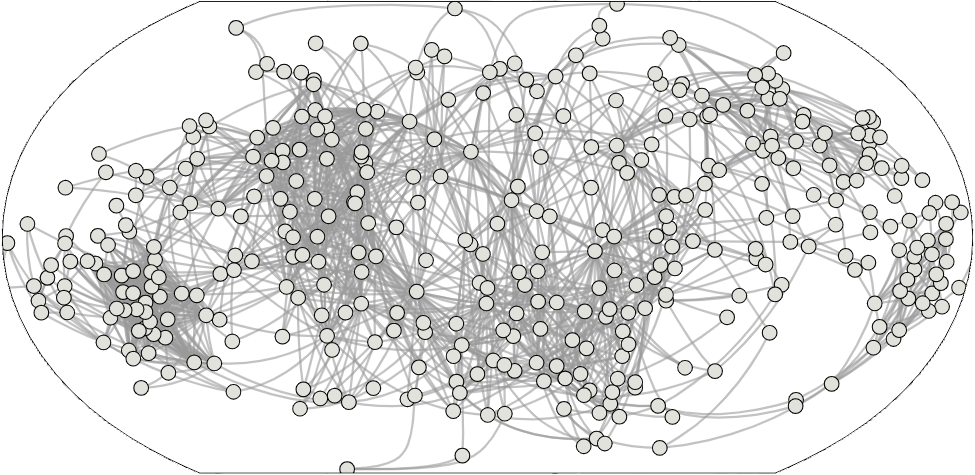}
    }
  \caption{Product space data network of derived datasets from Harvard Growth Lab}
  \label{fig:discussion:atlasderived}
\end{figure}

\autoref{fig:discussion:atlasderived} shows that compared with non-wrapped flat layout in (a), sphere layout in (c) appears more spread out with clusters more distinct. However, there is a limited benefit in the torus layout (b) as opposed to the non-wrapped (a), as the dense structure remains dense. Adapting our layout algorithm from~\autoref{sec:torus2} to considering more advanced layout methods such as weighted link wrapping such that a dense structure could be more aggressively wrapped, while leaving the sparse tree structure unwrapped with a low weight of link wrapping may do a better job of revealing the dense structure while preserving the tree structure.  Schemes for automating, or making interactive, the selection of a dense subgraph within a larger graph for detailed inspection in a toroidal or spherical ``lens" is an interesting direction for future research.

In summary, in above examples, we have seen networks which have dense small-world (\autoref{sec:discussion:football},~\autoref{sec:discussion:productspace}), bipartite (\autoref{sec:discussion:bipartite}) or community structure (\autoref{sec:discussion:recipe}) which is difficult to represent in typical non-wrapped flat network diagrams. While wrapped toroidal or spherical layouts of these examples may suggest interesting applications, the exploration also shows some limitations of toroidal or spherical layouts for product space or bipartite networks. We would like to further improve the algorithms and investigate if torus or spherical representations could be practically usable by domain experts. While this section describes an informal reflection of practical applications of interactive wrapping, we consider a more formal user-centred study with domain experts to further investigate the effectiveness and usability. 

Meanwhile, developing effective network layout algorithms is key to real-world practical network visualisations and assists future research in wider communities. While our refined layout algorithm proposed in~\autoref{sec:torus2} converges significantly more quickly than our first toroidal layout approach in~\autoref{sec:torus1}, the need to compare 9 different alternatives for link wrappings at each iteration means that it is still considerably slower than the corresponding layout algorithms for traditional node-link diagram layout. Improving its speed is a major direction for future research. 

Another future direction is to explore whether the extra flexibility in wrapping networks in 2D plane of topologically closing surfaces also provide benefits for more complex multivariate network visualisation that involve representation of both node-link and their attributes~\cite{nobre2019state} such as socio-semantic of internet community networks encode social connections (e.g., friendship, co-citation, co-occurrence, mentorship) among social actors and cultural associations (e.g., similarities between opinions, cultural schemes, key concepts of publications)~\cite{roth2020socio,roth2021quoting,roth2021socio} and protein-protein interaction networks that are highly connected with dense structure.


\section{Other possibilities for future work}
\label{sec:conclusion:futurework}
In this section, we discuss other possibilities for future work, including wrapping visualisations for other data types discussed in our design space that have not been evaluated, such as multidimensional scaling plots, self-organising maps, and cyclical time series with multiple levels of periodic dimensions.  Apart from the wrapping approach using interactive panning, possibilities of static wrapped display using tile repetition for geographic maps are also presented.

\subsection{Interactive wrapping for multidimensional scaling}
\label{sec:discussion:futurework:mds}
\begin{figure}
    \centering
    \includegraphics[width=1\columnwidth]{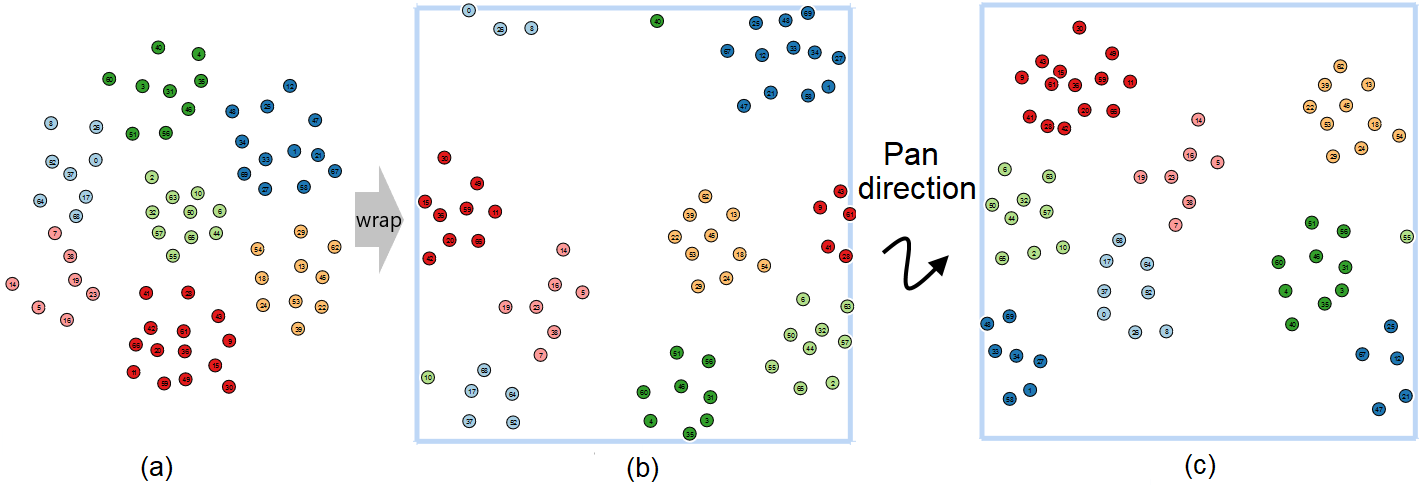}
    \caption{Examples of torus panning on an MDS using the Hepta data set~\cite{thrun2020clustering} with already known clustering information:
    (a) traditional (unwrapped) MDS plot on a 2D-plane; (b, c) two configurations wrapped MDS plot laid-out on a 2D torus, using the algorithm presented in~\autoref{sec:torus2}. The dataset was adapted from a Benchmark data set from Fundamental Cluster Problem Suite (FCPS).}
    \label{fig:discussion:mds}
\end{figure}
As introduced in~\autoref{sec:related:mds}, many real-world data structures contain objects that have multiple attributes (e.g., a patient has temperature, weight, height, blood pressure, etc). 
In such multidimensional data analysis, one important goal for dimension reduction methods such as multidimensional scaling (MDS) (introduced in~\autoref{sec:related:mds}) is to seek to optimise the pairwise low-dimensional distance (e.g. Euclidean distance) between data points to as well as possible to represent the pairwise similarity or dissimilarity in the original data; and ultimately, to reveal the underlying structure for convenient perception of clusters, correlation between data points, ratings, and similarities between statistics data. 

This process, known as stress minimisation~\cite{martins2012multidimensional,bian2020implicit}, solves the same type of problem we consider for optimising node placement to simulate their graph theoretic distance in this thesis. However, like force-directed node placement problem, stress minimisation is also a non-trivial problem, and thus many heuristics algorithms exist, such as standard gradient descent, stochastic gradient descent by Bottou~\cite{bottou2010large}, stress majorisation by Gansner et al.~\cite{gansner2004graph} and Leeuw and Mair~\cite{de2009multidimensional}.

For complex data (such as larger and denser datasets), MDS plots need a cleaner presentation to discern density distribution or identifying high-level structures such as clusters~\cite{van2008visualizing,lu2019doubly}. Based on the similarity between MDS and network layout algorithms explained in more detail in~\autoref{sec:related:mds}, and the promising results of improved layout quality (in terms of aesthetics) when wrapping a network on a torus topology, and improved task performance (in terms of accuracy) when introduced with interactive wrapping for cluster identification tasks for sphere or torus topology, we argue that it is also likely to lead to interesting applications for MDS.

\autoref{fig:discussion:mds} shows MDS plots of a common Hepta data set which has been evaluated by Thrun~\cite{thrun2020clustering}. It has already known clustering information. Using the network layout algorithms proposed in~\autoref{sec:torus2}, it shows that MDS can potentially wrap around a 2D projected torus where the arrangement of data points have possible additional spreading and make better use of available space, while exploring with interactive panning. Future work should evaluate the effect of torus wrapping on perception, e.g. using interactive wrapping or tile display.

\subsection{Interactive wrapping for self-organising maps}
\label{sec:discussion:futurework:soms}
\begin{figure}
    \centering
	\subfigure[TorusSOM]{
    \includegraphics[width=0.3\textwidth]{gfx/related/torus_prior_work2.png}
    }
	\subfigure[SphereSOM]{
    \includegraphics[width=0.5\textwidth]{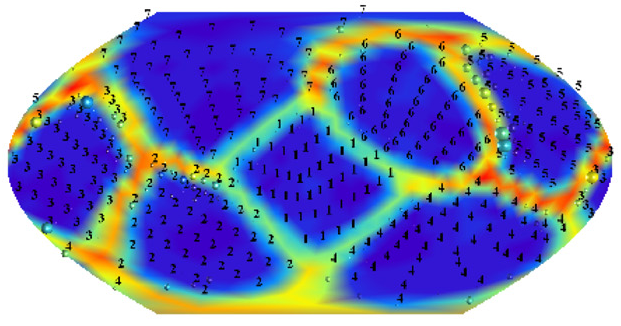}
    }
	\caption{SOM rendered on (a) a torus layout with tile repetition connecting regions of data wrapped across the boundaries. Image is credited to~\cite{ultsch2003maps}; (b) a 2D spherical projection where regions split across the left-and-right boundaries. The image at (a) and (b) are credited to Ultsch et al.~\cite{ultsch2003maps} and Wu et al.~\cite{wu2006spherical}, respectively.}
	\label{fig:discussion:som}
\end{figure}
Another example of arrangements of abstract data that seeks to optimise the layouts to better understand their relationships or the structure is self-organising maps (SOMs). SOMs, described in~\autoref{sec:related} are commonly used as an abstract data structure that represents multidimensional data, i.e., a patient has weight, height, blood pressure, etc in lower dimensions\cite{kohonen1982self,squire2005visualization,wu2006spherical}. 
Discerning high-level structure, such as clusters, is an important task in SOMs. In SOM, data points can be grouped into clusters where data points in spatially nearby clusters have more similar values than data points in distal clusters. 



Many methods exist for preserving the topological structure of multidimensional data in lower dimensions~\cite{kohonen1982self} on self-organising maps of documents. In order to alleviate the ``privileged centre effect'' (or border effect studied by Wu et al.~\cite{wu2006spherical}), where clusters at the centre have more neighbours than those at the boundary and thus are updated more often than those at the border, SOMs have been embedded on a sphere or torus. Potential advantages include no concept of border so that all the clusters may be considered when updating neighbouring relationships. \autoref{fig:discussion:som}(a) shows a toroidal tile repetition of SOM where the clusters wrap around in both horizontal and vertical directions. Similarly, \autoref{fig:discussion:som}(b) shows equal earth representation of SOMs where clusters wrap around horizontally. 

However, above existing research focused on visualising SOMs either visualising in 3D or projecting to a static 2D plane. The effect of interactive wrapping of these representations has not previously been presented. Based on our study results, we might hypothesise that interactive wrapping or tile display (\autoref{fig:discussion:som}(a)) reinforces understanding about the neighbourhood relationship of SOMs wrapped across the edge of the display, leaving exploration of such an effect on usability and readability of SOMs to a future work. 


\subsection{Interactive torus wrapping for cyclical time series}
\label{sec:conclusion:cyclicaltimeseries}
\begin{figure}
	\includegraphics[width=\textwidth]{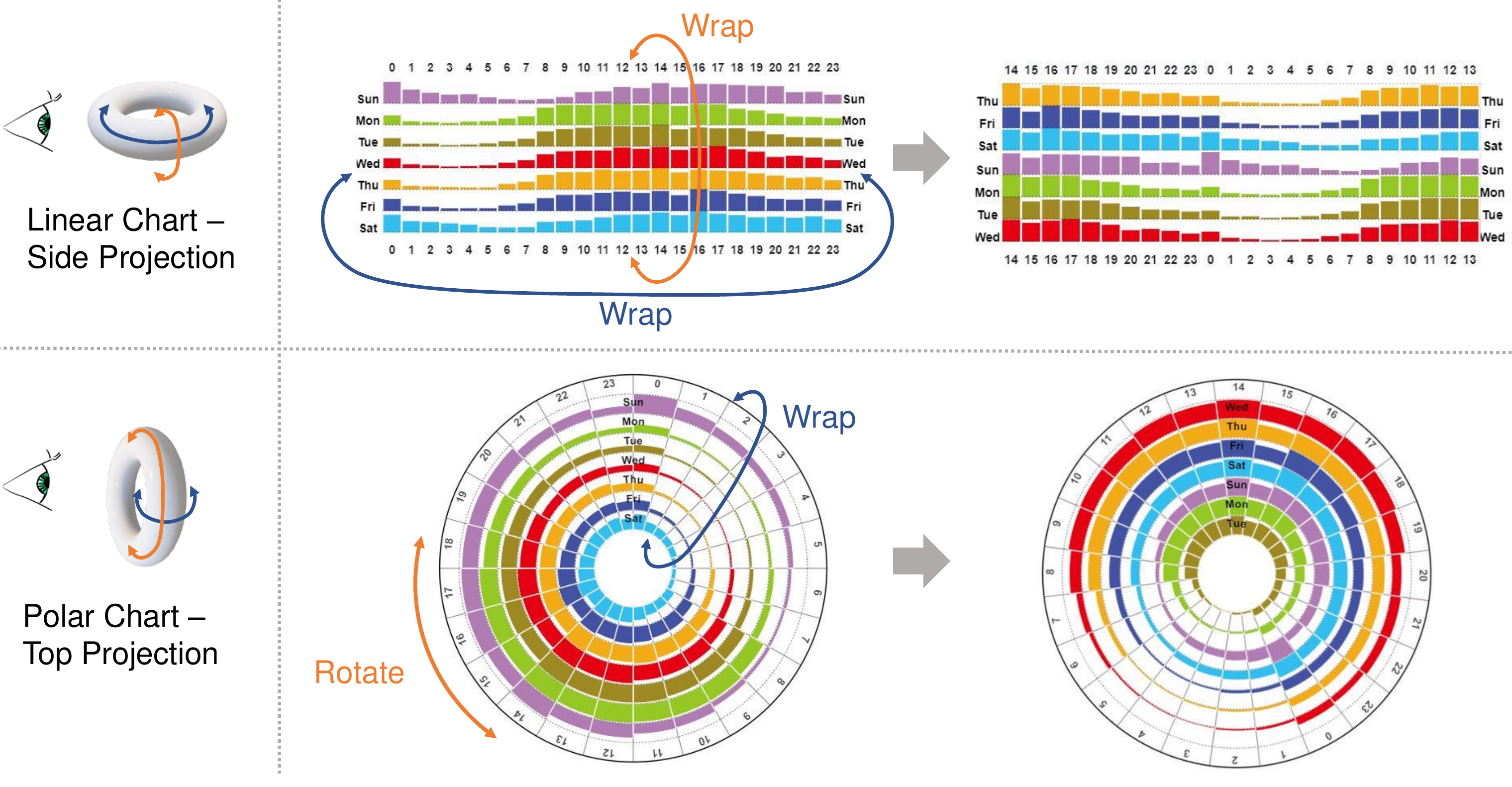}
	\caption{Cyclical time series using torus wrapping and rotation can help to understand the visual cut that are otherwise split across the left-and-right boundary and top-and-bottom boundary. The data source is from Kaggle open dataset~\cite{hourlytrafficaccidentsdataset}.}
	\label{fig:discussion:cyclicaltimeseries}
\end{figure}
\begin{figure}
    \centering
    \includegraphics[width=1\columnwidth]{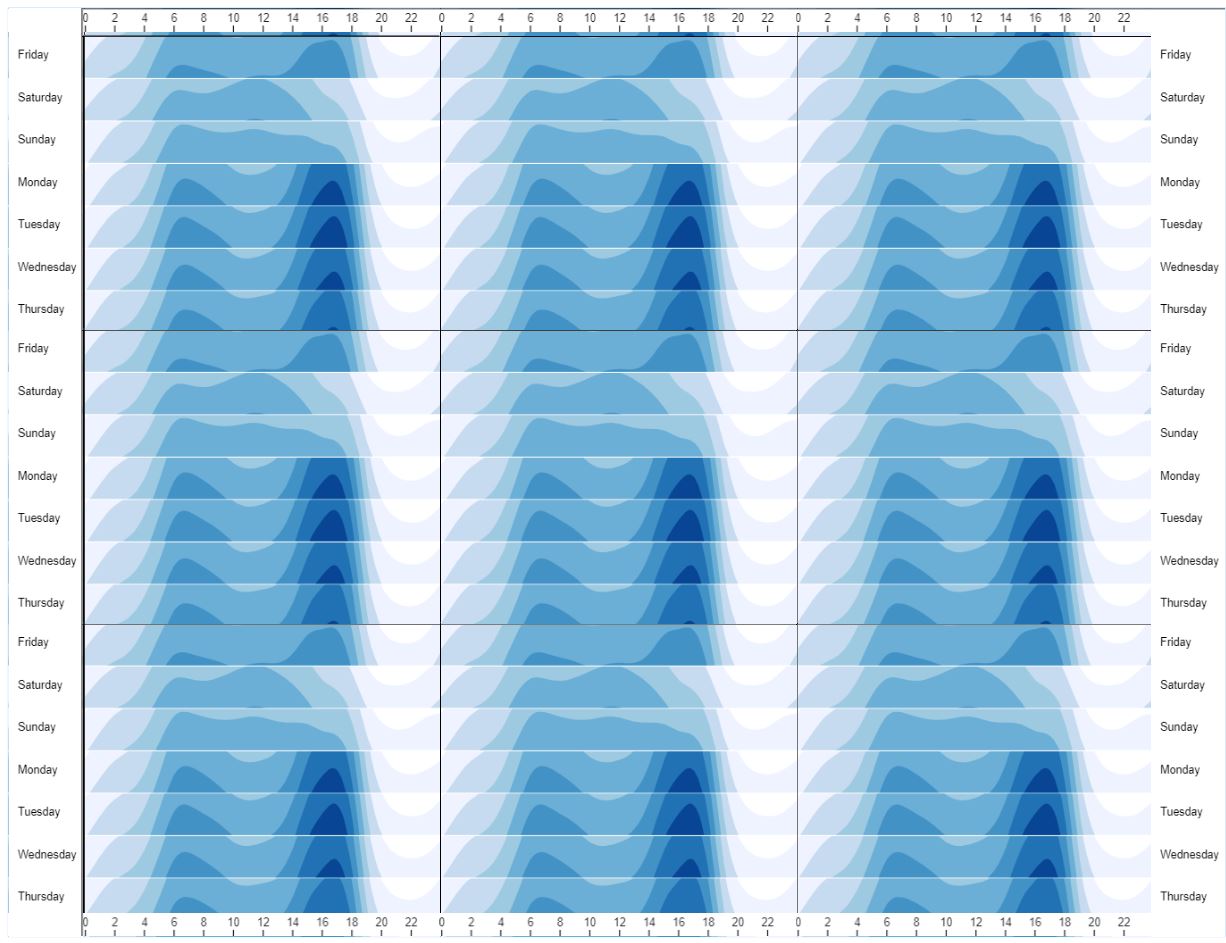}
    \caption{Example of cyclical time series of New South Wales traffic accidents represented on a 2D projected torus topology with $3\times 3$ tile display as described in~\autoref{sec:designspace:torus:torustopology}. Two periodic dimensions such as 24 hours of a day across 7 days a week repeat across the top, bottom, left, and right edges of the original chart (shown in the central tile) without repetition. It allows a viewer to observe continuous ranges of both dimensions across the chart's boundaries with any break in the temporal dimensions.} 
    \label{fig:discussion:torustiles}
\end{figure}
We have shown in~\autoref{sec:cylinder} that one dimensional (cylindrical) wrapping provides considerable benefits for understanding real-world cyclical time series with one periodic dimension (e.g., 24 hours over a day, or 12 months over a year). However, cyclical time series has multiple levels of periodicity (e.g. weekly as well as daily). Effective visual representation of time series data for multiple levels of temporal dimensions to reveal meaningful trends and patterns, and improving their readability remain open questions~\cite{brehmer2016timelines,bach2017descriptive}.



Extending the concept of one-dimensional wrapping, we have introduced two-dimensional toroidal wrapping for cyclical time series data that show multiple levels of periodicity (e.g., daily and weekly, in~\autoref{sec:designspace:torus:torustopology}), including 2D bar chart arrays arranged on a side-projected torus for interactive wrapping, polar chart arrays arranged on a side-projected torus for rotatable panning and wrapping (\autoref{fig:discussion:cyclicaltimeseries}), and furthermore tiled display by replicating linear bar charts of cyclical time series on the sides of the charts both horizontally and vertically (\autoref{fig:discussion:torustiles}) to provide full connectivity wrapped across the boundaries without being hindered (when reading the charts) by an arbitrary visual ``cut'' (defined in~\autoref{sec:intro}).


In 2D linear bar chart arrays, having 0am on the left, 11pm on the right, or Sunday on the top, Saturday on the bottom (\autoref{fig:discussion:cyclicaltimeseries}-upper left) is completely arbitrary. Similarly, in polar chart arrays, having the bars (hours) arranged along layered circumferences of the chart in an order analogous to a natural clock (\autoref{fig:discussion:cyclicaltimeseries}-lower left) across Sunday to Saturday is not the only option. Using interactive wrapping proposed in our design space (\autoref{sec:designspace:torus}), a viewer can move the linear charts horizontally to see a continuous range of bars between, e.g., 11pm and 6am without being split by the cut across the left-and-right boundary, while panning the charts vertically connect the cuts between Friday and Sunday (\autoref{fig:discussion:cyclicaltimeseries}-upper right). Interactively rotating a polar chart (derived from our design space in~\autoref{sec:designspace:torus:torustopology}) along the hour dimension allows for centring the hour period between, e.g., 9am and 8pm for a more aligned comparison during this period, while wrapping the polar charts moves Wednesday from the inside of the polar visualisation to its outside to provide a continuous view between, e.g., Friday and Sunday (\autoref{fig:discussion:cyclicaltimeseries}-lower right). 
 

Based on the promising results of readability of linear bar charts with cylindrical wrapping (\autoref{sec:cylinder:results}), we might hypothesise that, for more complicated case of 2D temporal data exploration in a 2D torus topology, interactive wrapping with two temporal dimensions may still outperform the rotational and wraparound polar chart. In future, we intend to further investigate this hypothesis and also compare with the effect of static tile repetition without such interaction.

\subsection{Other possibilities for future work}
\label{sec:conclusion:others}
\begin{figure}
    \centering
	\includegraphics[width=0.7\textwidth]{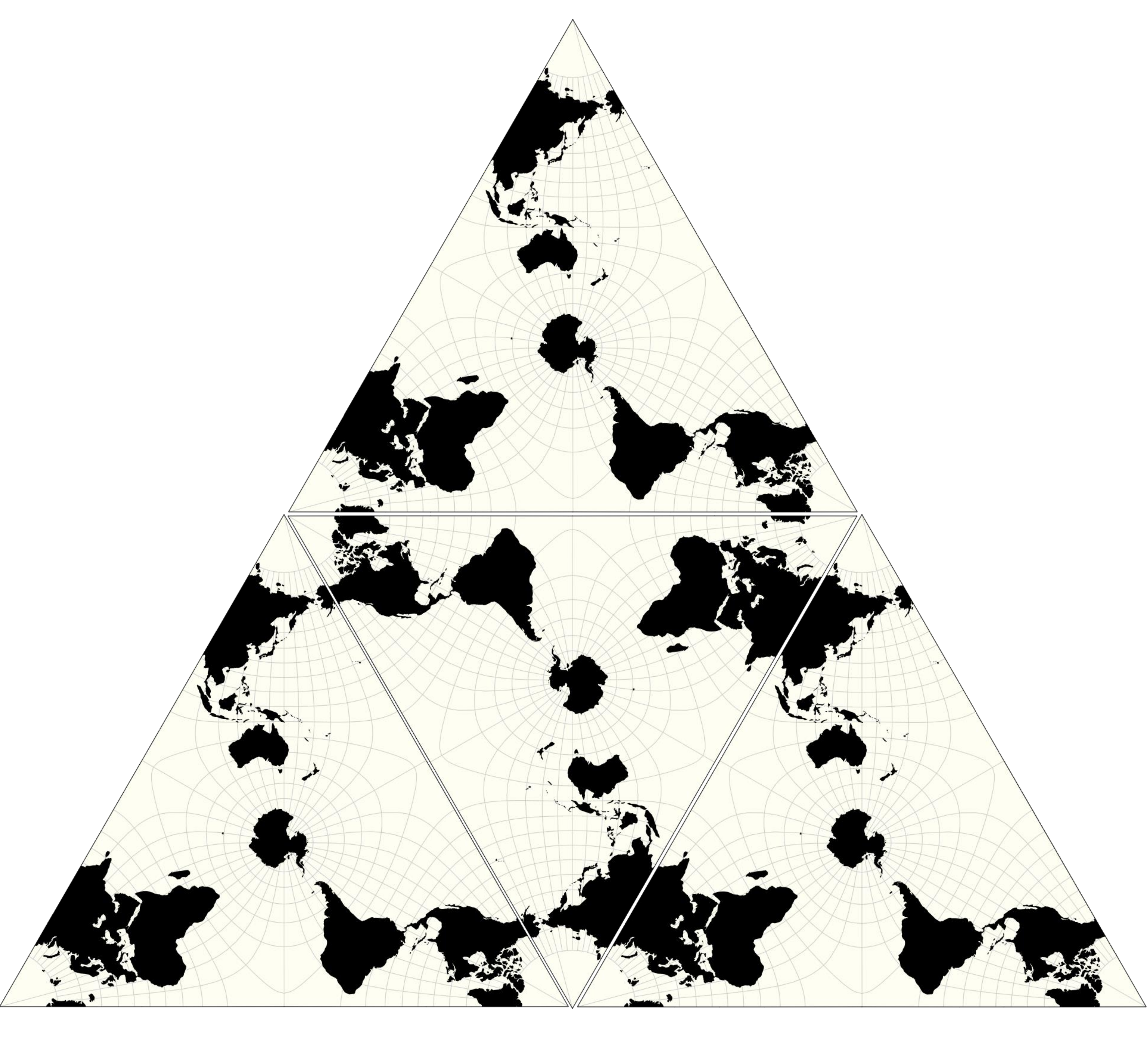}
	\caption{Tile display of maps in Lee\'s tetrahedron projection; image adapted from Lee\'s tetrahedron projection and credited to Lee}
	\label{fig:discussion:tetrahedron}
\end{figure}
\begin{figure}
	\includegraphics[width=0.9\textwidth]{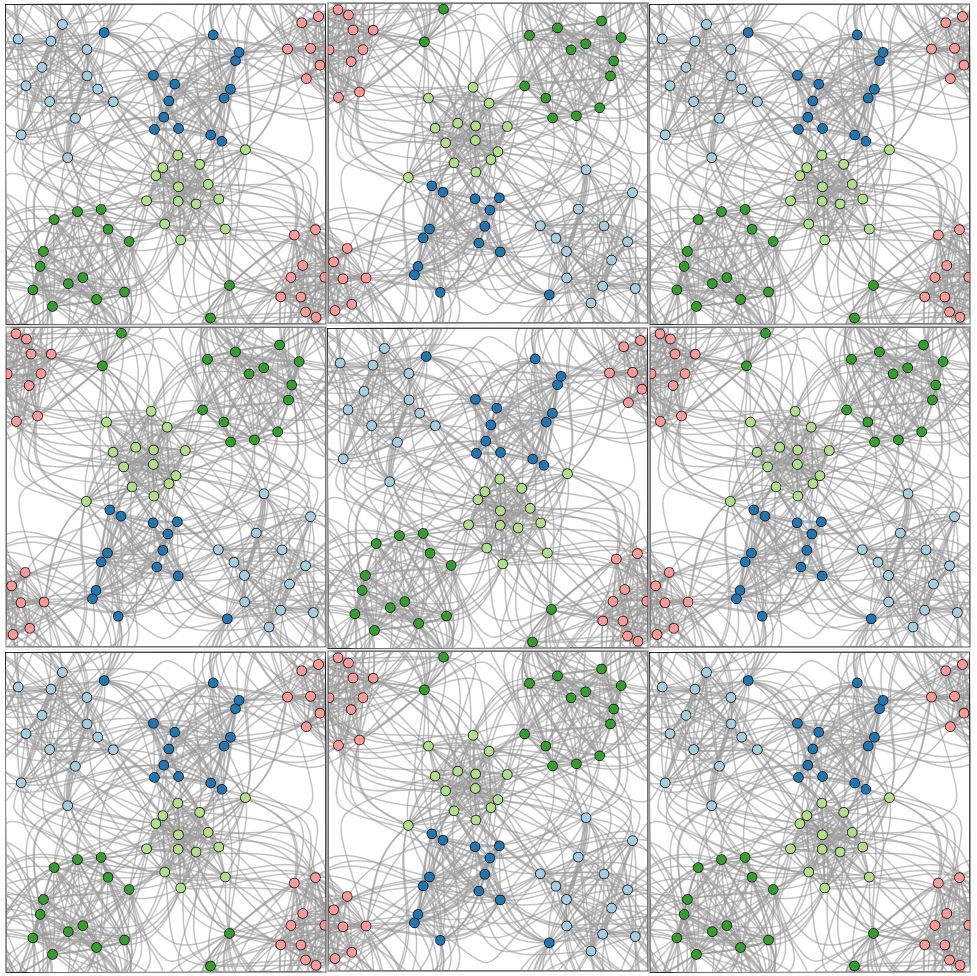}
	\centering
	\caption{Pierce\'s Quincuncial projection of ``Wrapping'' dimension in both spatial directions: horizontally and vertically repeated tiles. The repeated tiles show continuity of data points at the boundaries which preserves the data point movement wrapped across the boundaries.}
	\label{fig:discussion:quincuncialnetworktiles}
\end{figure}

In~\autoref{sec:designspace:dimensions} we presented two approaches of wrapping: tile display and interactive wrapping. Our evaluation results show that adding tile repetition to torus layouts provides significant benefits for understanding link wrapping across the boundaries, while after adding interactive panning, torus layouts without tile display performs equally well with either the tile display or non-wrapped layouts (\autoref{sec:torus1}). However, it has not been shown whether tile displays of other topologies considered in our design space, such as cylinder (\autoref{fig:designspace:mercatortiles}) or sphere (\autoref{fig:discussion:tetrahedron} and \autoref{fig:discussion:quincuncialnetworktiles}), also provide benefits over static/interactive wrapped (without tile repetition) or non-wrapped representations. 

It is also possible to further investigate the aspects of the wrapping method that drive better performance. From a theoretical perspective, it would be interesting to consider what aspects of the wrapping method driving the better performance. Is it getting to see the data from multiple views? Or is it the active aspect of getting to manipulate the data to the desired view? We cannot easily say whether the interactive panning we provided is better than a passive animation of wrapping affording different views. 

While answering these questions is beyond the scope of this thesis. One way to do this would be to repeat the study that was done in~\autoref{sec:cylinder}, \autoref{sec:torus2}, \autoref{sec:spheremaps}, and video record the interactions (active group). Then, ``yoke'' each participant in this group to a new set of participants in a second group, who watches video recordings of the interaction rather than doing the interactions themselves (passive yoked group). If both groups perform similarly well for the interactive wrapped visualisations compared with the static ones, that would suggest that the benefits of this new method is the multiple views, but if the active group does better, that would suggest that interacting with the visualisations is key. We leave this to future work.
%
\section{Wrapping up}
\label{sec:discussion:conclusion}



We began this thesis by reviewing the related research on data visualisation, revisiting existing data that wraps around topologies, network layout algorithms, and empirical evidence (\autoref{sec:related}). Afterwards, we presented a design space exploration which identifies various data types that wrapping visualisations makes sense, and our interactive tools design that create novel applications (\autoref{sec:designspace}). We then discussed each of the wrapping topologies and effectiveness evaluation: cylindrical (\autoref{sec:cylinder}), toroidal (\autoref{sec:torus1} -- smaller networks and~\autoref{sec:torus2} -- larger networks), and spherical topologies (\autoref{sec:spheremaps} -- maps and~\autoref{sec:spherevstorus} -- networks). We gave a discussion of the benefits and limitations of interactive wrapping and provide our view on the future of pannable wrapped visualisations and the potential challenges that provide implications to future research (\autoref{sec:conclusion}). 

Finally, we concluded this thesis with a contribution and a clear take-away message: It's a Wrap! Visualisations can be brought out of the box to better analyse the trends, identify patterns, and explore the relationships within the data; ultimately, we hope it helps make better decisions. 


%
{%
\setstretch{1.1}
\renewcommand{\bibfont}{\normalfont\small}
\setlength{\biblabelsep}{0pt}
\setlength{\bibitemsep}{0.5\baselineskip plus 0.5\baselineskip}
\printbibliography[nottype=online]
\newrefcontext[labelprefix={@}]
\printbibliography[heading=subbibliography,title={Webpages},type=online]
}
\cleardoublepage



\appendix\cleardoublepage
%
\chapter{Appendix}
\label{sec:appendix}


\begin{figure}[!ht]
    \centering
    \includegraphics[width=0.9\textwidth]{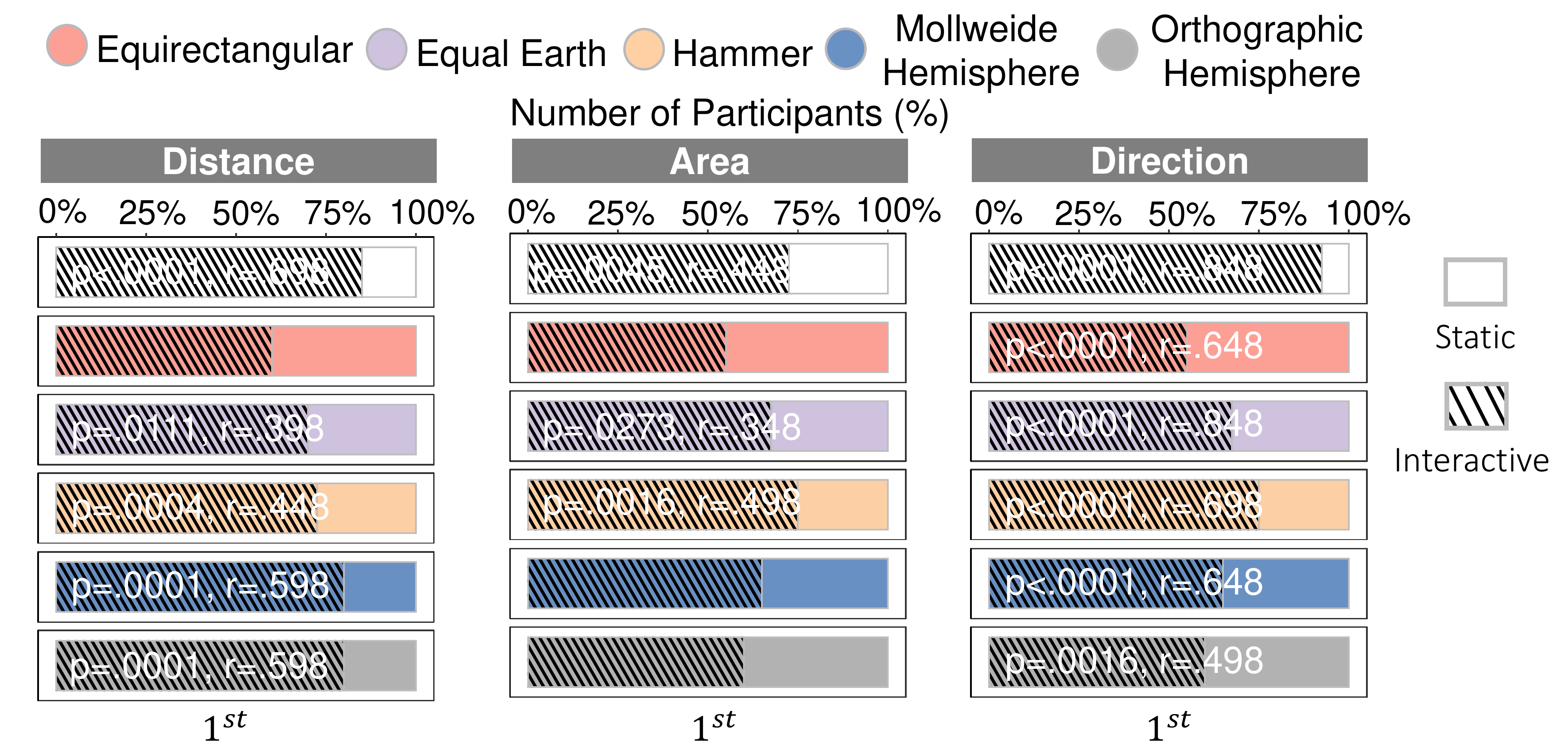}
    \caption{Study 5: User preference rank of interactivity for overall (top) or within each map projection. Statistical significance results with $p<0.05$ and effect sizes of Cohen's r are shown. x-axis shows the number of participants in percentage.}
    \label{fig:study-1-interactivity-ranking-results}
\end{figure}

\begin{figure}
   \centering
    \includegraphics[width=\textwidth]{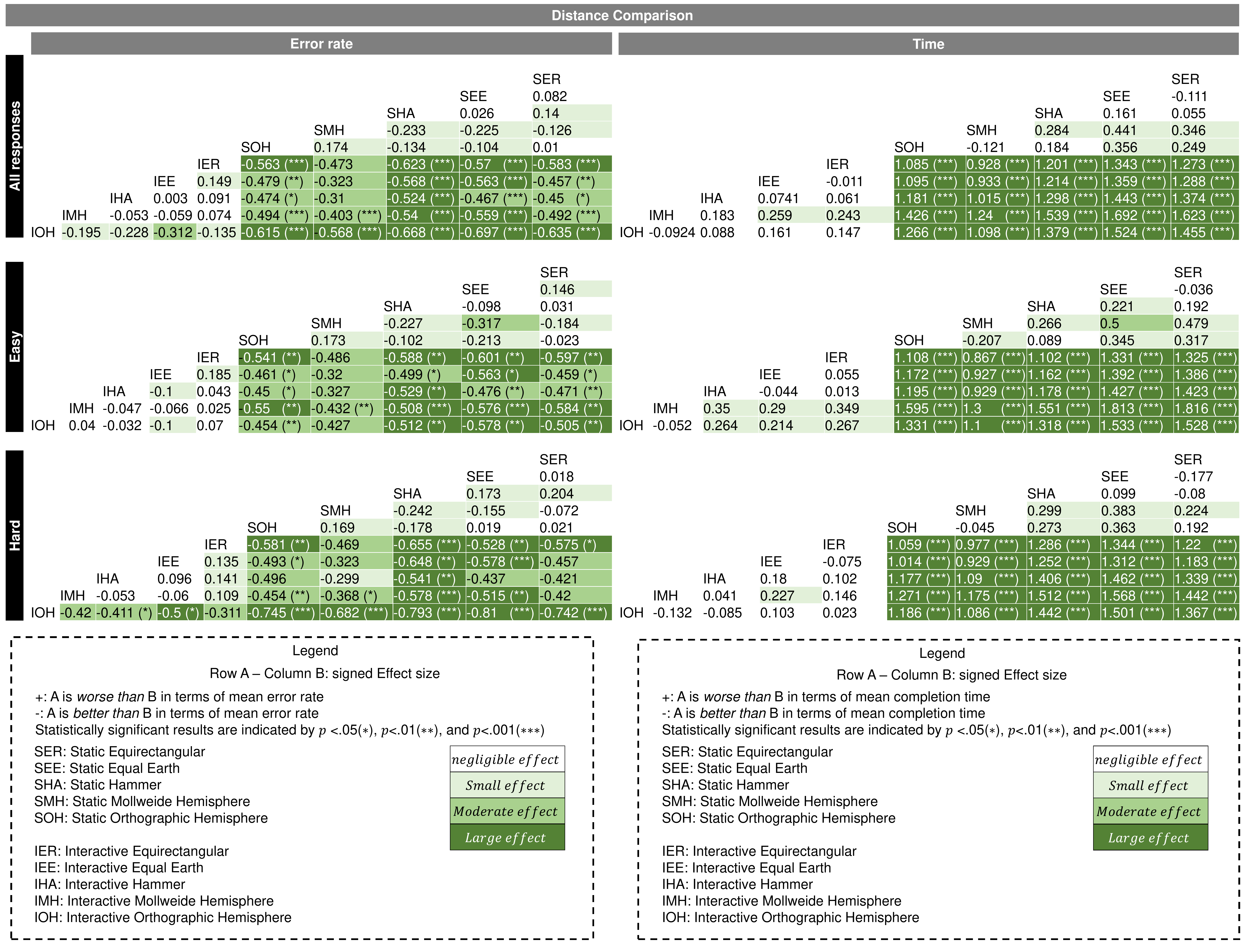}
    \caption{Matrices of pairwise effect size results of \merror{} and \mtime{} of Study 5 for Distance task. Values indicate effect size results of Cohen's r and Cohen's d~\cite{cohen2013statistical} for \merror{} and \mtime{}, respectively.}
    \label{fig:study-1-effect-size-distance}
\end{figure}

\begin{figure}
   \centering
    \includegraphics[width=\textwidth]{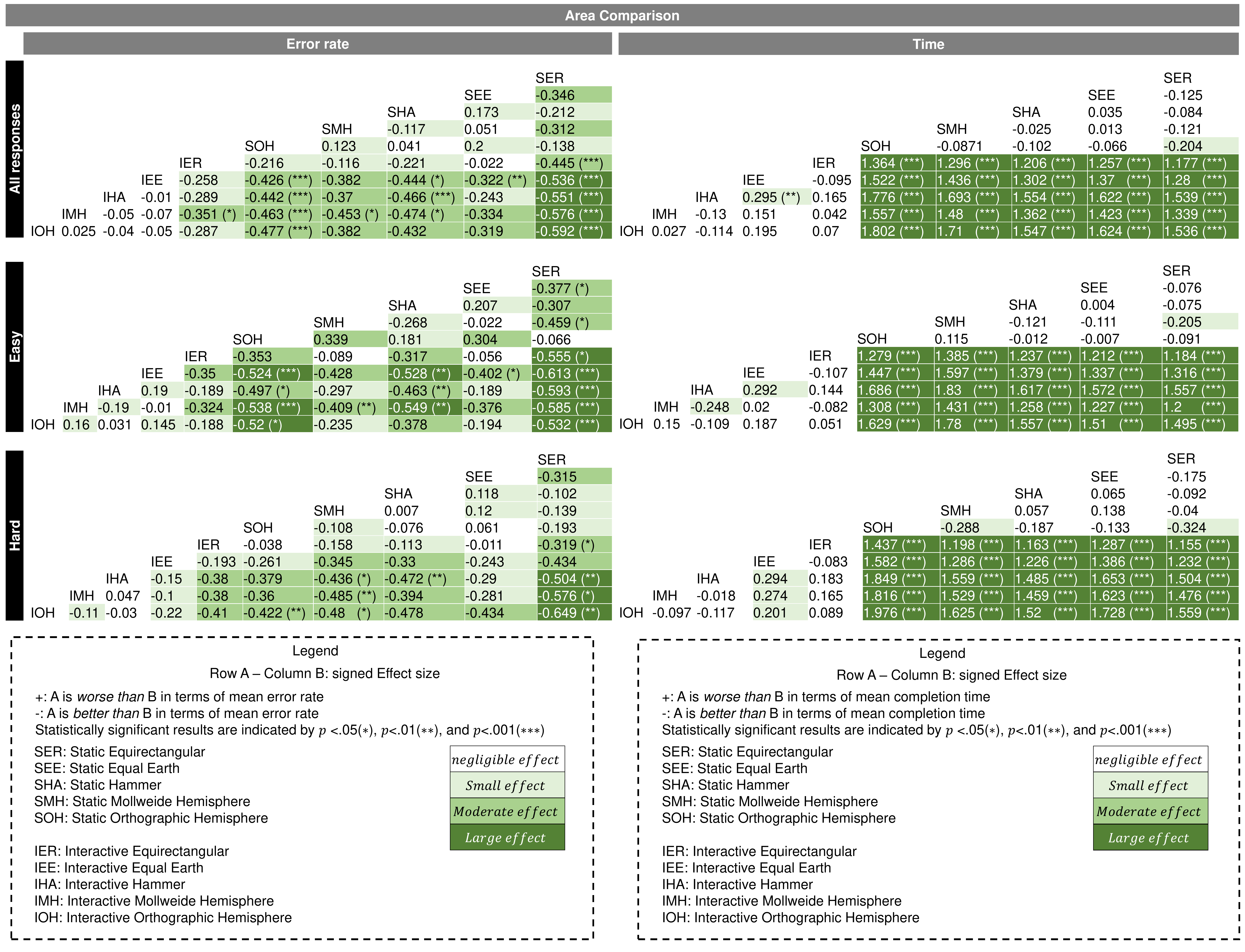}
    \caption{Matrices of pairwise effect size results of \merror{} and \mtime{} of Study 5 for Area task. Values indicate effect size results of Cohen's r and Cohen's d~\cite{cohen2013statistical} for \merror{} and \mtime{}, respectively.}
    \label{fig:study-1-effect-size-area}
\end{figure}

\begin{figure}
   \centering
    \includegraphics[width=\textwidth]{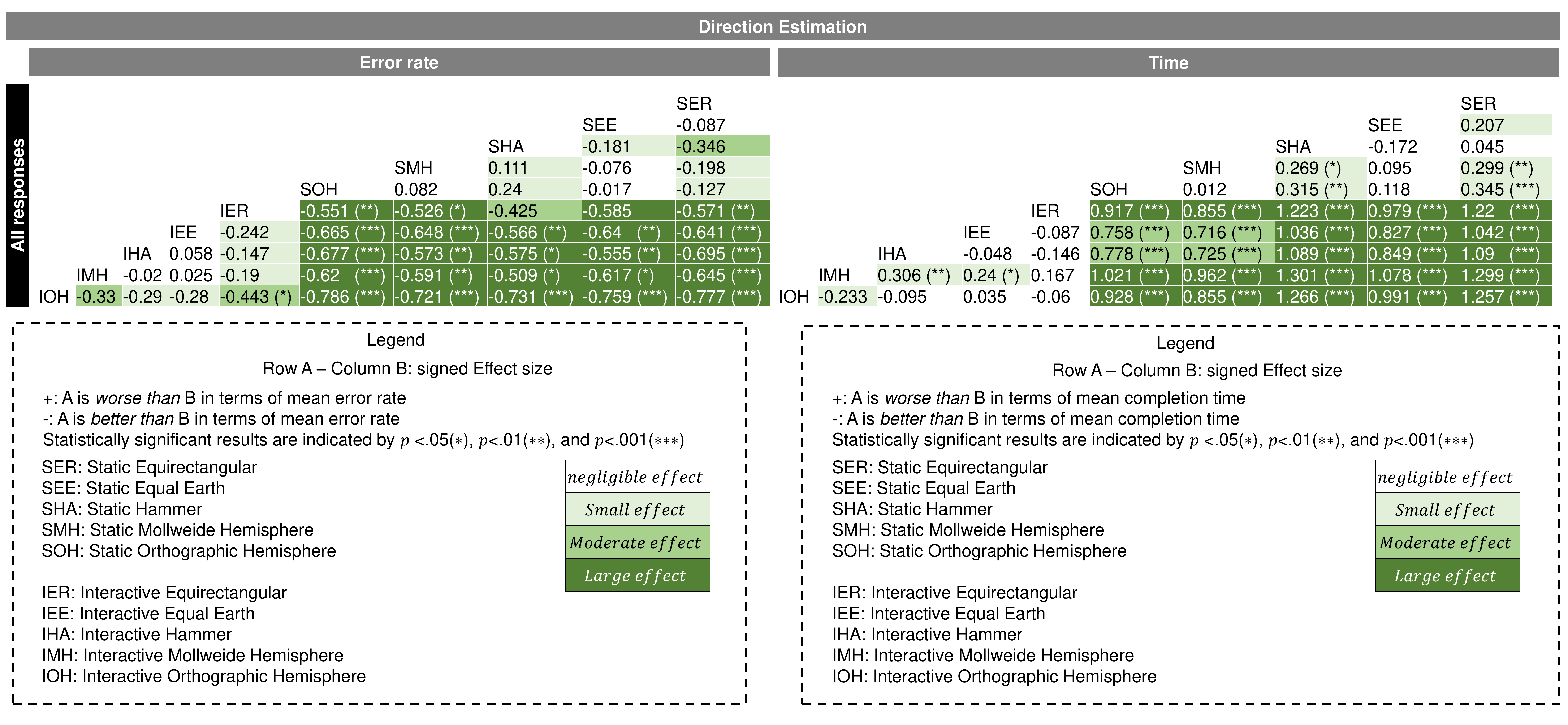}
    \caption{Matrices of pairwise effect size results of \merror{} and \mtime{} of Study 5 for Direction task. Values indicate effect size results of Cohen's r and Cohen's d~\cite{cohen2013statistical} for \merror{} and \mtime{}, respectively.}
    \label{fig:study-1-effect-size-direction}
\end{figure}

\begin{figure}
    \centering
    \includegraphics[width=1\textwidth]{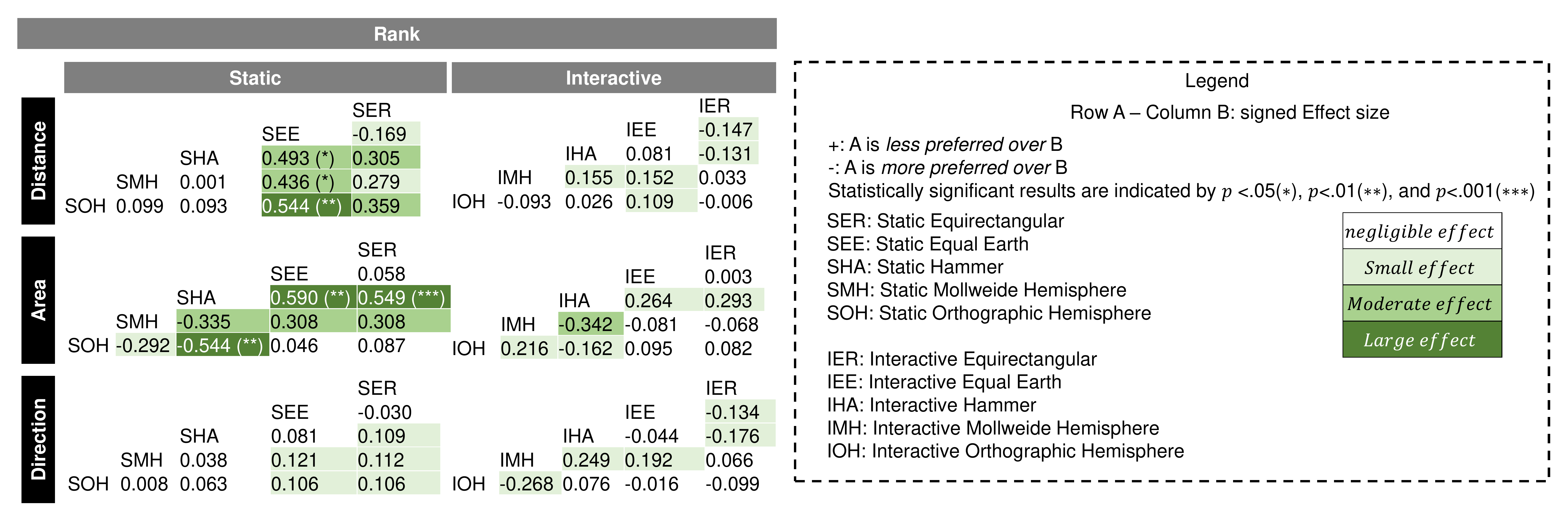}
    \caption{Matrices of pairwise effect size results of subjective user rank of Study 5 map projections within the \emph{static} (left) and \emph{interactive} (right) groups for three tested tasks. Values indicate effect sizes results of Cohen's r.}
    \label{fig:study-1-ranking-results-effect-size}
\end{figure}


\begin{figure}
    \centering
    \includegraphics[width=1\textwidth]{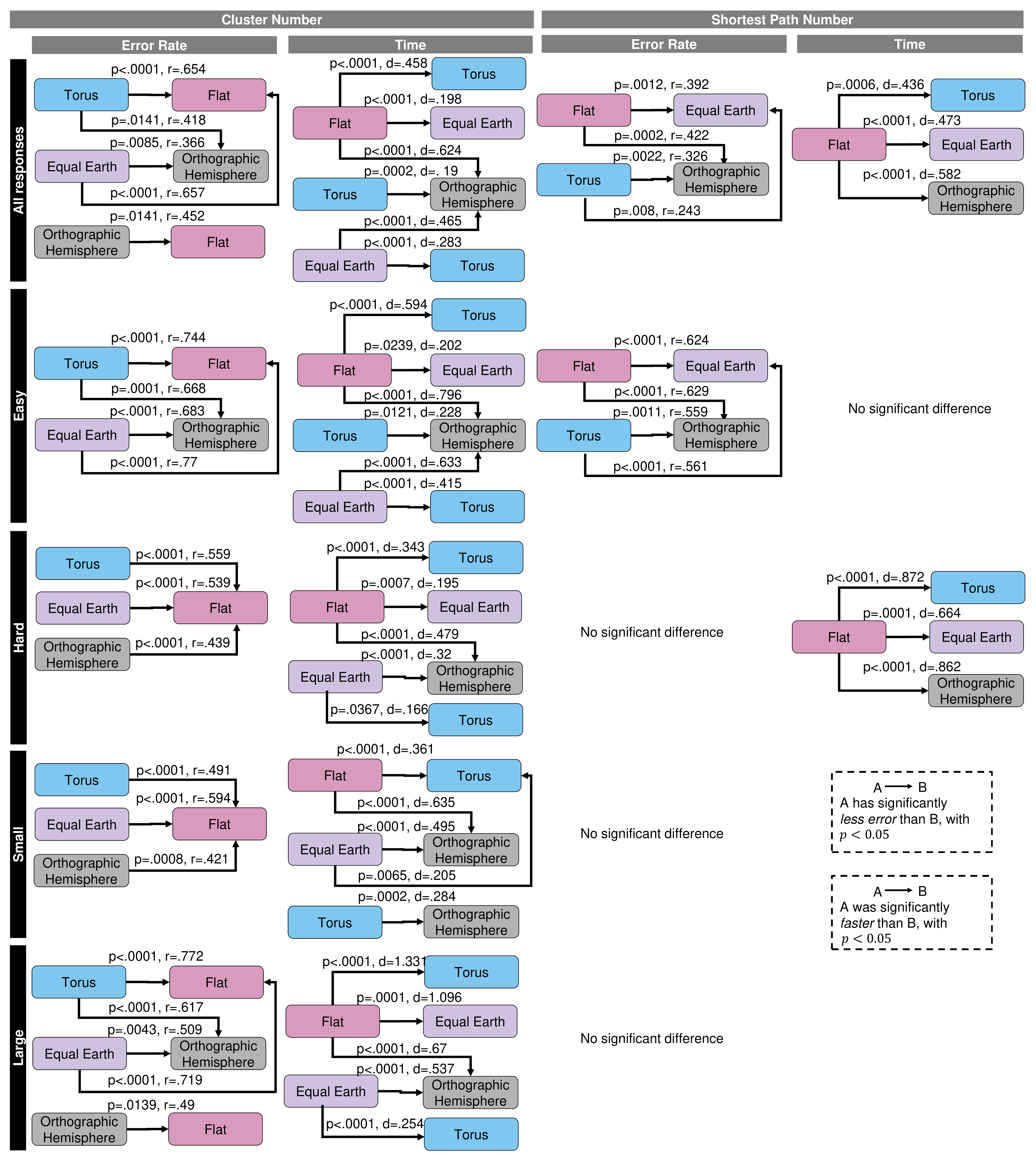}
    \caption{Study 6: Statistically significant results of \merror{} and \mtime{} for Cluster Number task (left) and Shortest Path Number task (right). Effect size results of Cohen’s r and Cohen’s d for \merror{} and \mtime{} are presented along with the arrows.}
    \label{fig:study-2-results-graphics}
\end{figure}

\begin{figure}
    \centering
    \includegraphics[width=\textwidth]{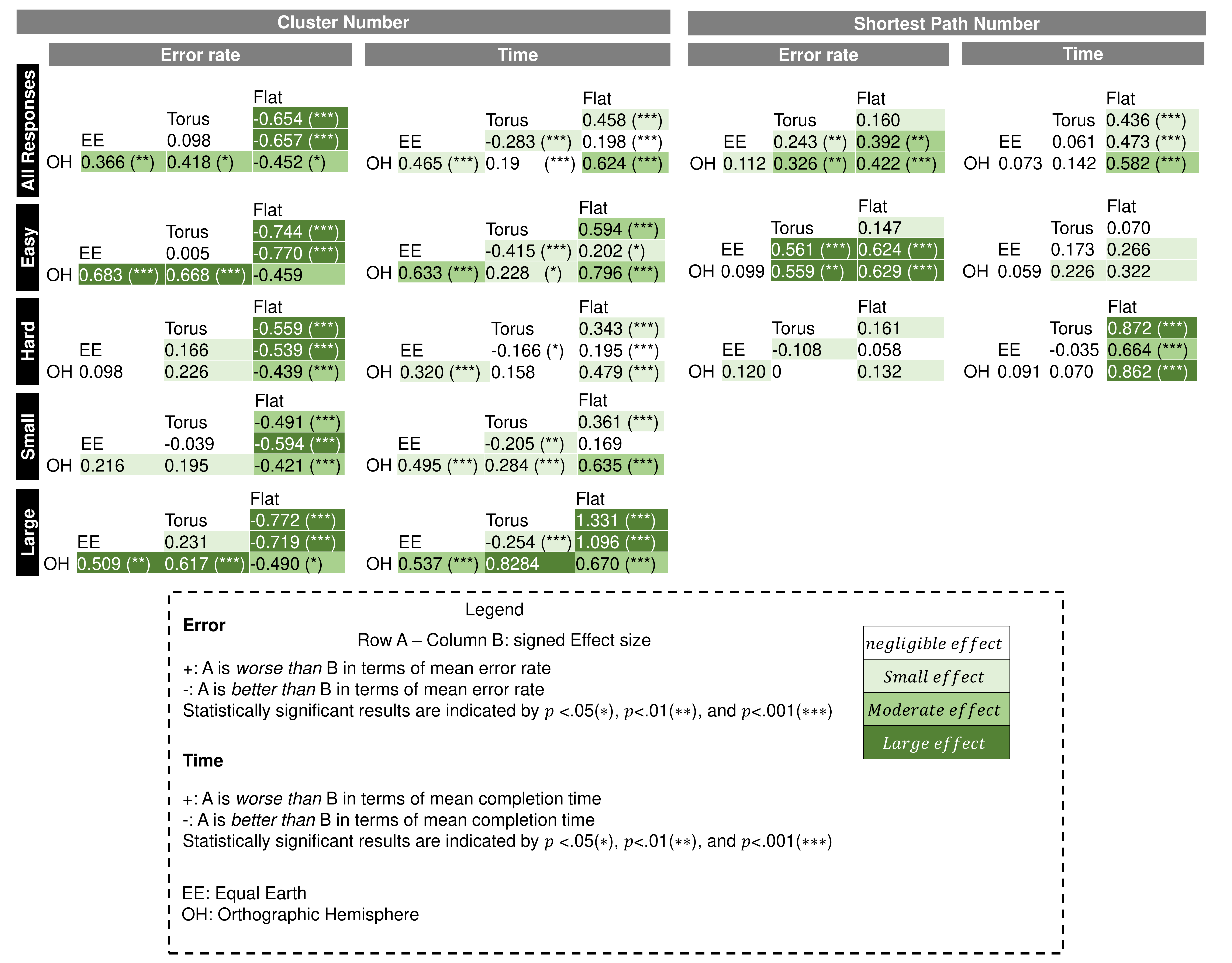}
    \caption{Matrices of pairwise effect size results of \merror{} and \mtime{} of Study 6 for Cluster Number task (left) and Shortest Path Number task (right). Values indicate effect size results of Cohen's $r$ and Cohen's $d$~\cite{cohen2013statistical} for \merror{} and \mtime{}, respectively.}
    \label{fig:study-2-results-effect-size}
\end{figure}

\begin{figure}
    \centering
    \includegraphics[width=\textwidth]{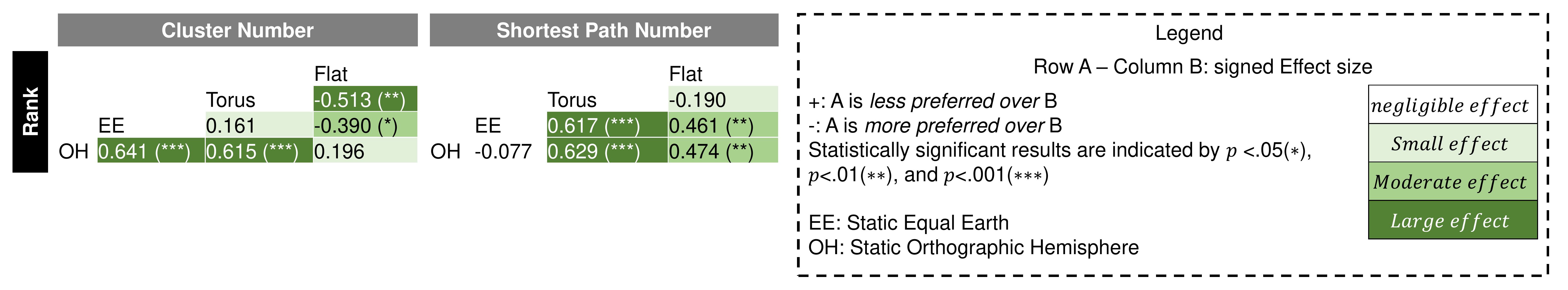}
    \caption{Matrices of pairwise effect size results of \mpref{} of Study 6 for Cluster Number task (left) and Shortest Path Number task (right). Values indicate effect size results of Cohen's r.}
    \label{fig:study-2-results-rank-effect-size}
\end{figure}









\newpage
\mbox{}

\end{document}